\RequirePackage{lineno}
\RequirePackage{fix-cm}

\documentclass[twocolumn,epjc3]{svjour3}  
\pdfoutput=1
\usepackage{atlasphysics}
\usepackage{placeins}
\usepackage{subfigure}
\usepackage{mathrsfs}
\usepackage{amsmath}
\usepackage{slashed}
\usepackage{setspace}
\usepackage{footnote}
\usepackage{multirow}
\usepackage{longtable}
\usepackage{supertabular}
\usepackage{bigstrut}
\usepackage{pdflscape}
\usepackage[abs]{overpic}
\usepackage{cite}
\usepackage{rotating,booktabs}
\usepackage{graphicx}
\usepackage{comment}
\usepackage{epstopdf}
\usepackage{siunitx}

\newcommand*{\numRF}[2]{\num[round-mode=figures,round-precision=#2]{#1}}

\newcommand\twotwo{($2 {\rm j},\,2 {\rm b}$)}
\newcommand\threetwo{($3 {\rm j},\,2 {\rm b}$)}
\newcommand\threethree{($3 {\rm j},\,3 {\rm b}$)}
\newcommand\fourtwodi{\mbox{($\ge4 {\rm j},\,2 {\rm b}$)}}
\newcommand\fourthreedi{\mbox{($\ge4 {\rm j},\,3 {\rm b}$)}}
\newcommand\fourfourdi{\mbox{($\ge4 {\rm j},\ge4 {\rm b}$)}}
\newcommand\fourtwolj{($ 4 {\rm j},\,2 {\rm b}$)}
\newcommand\fourthreelj{($ 4 {\rm j},\,3 {\rm b}$)}
\newcommand\fourfourlj{($ 4 {\rm j},\,4 {\rm b}$)}
\newcommand\fivetwo{($ 5 {\rm j},\,2 {\rm b}$)}
\newcommand\fivethree{($ 5 {\rm j},\,3 {\rm b}$)}
\newcommand\fivefour{($ 5 {\rm j},\ge 4 {\rm b}$)}
\newcommand\sixtwo{(\mbox{$\ge6 {\rm j},\,2 {\rm b}$)}}
\newcommand\sixthree{\mbox{($\ge6 {\rm j},\,3 {\rm b}$)}}
\newcommand\sixfour{\mbox{($\ge6 {\rm j},\ge4 {\rm b}$)}}
\newcommand\htott{$H\to \tau \tau $}
\newcommand\htogaga{$H\to \gamma \gamma$}
\newcommand\htoZZ{$H \to ZZ^{(*)} \to 4 \ell$}
\newcommand\htoWW{$H \to WW^{(*)} \to \ell\nu\ell\nu$}
\newcommand\mt{\ensuremath{m_{t}}} 
\newcommand\tth{\ensuremath{t\bar{t}H}}
\newcommand\ttH{\ensuremath{t\bar{t}H}}
\newcommand\ttbb{\ensuremath{t\bar{t}+b\bar{b}}}

\newcommand\ttcc{\ensuremath{t\bar{t}+c\bar{c}}}
\newcommand\mbb{\ensuremath{m_{\bbbar}}}

\newcommand\htlep{\ensuremath{\HT}}
\newcommand\mll{\ensuremath{m_{\ell\ell}}}

\newcommand\Mjjj{\ensuremath{m_{\rm jjj}}}

\newcommand\hthad{\ensuremath{\HT^{\rm had}}}
\newcommand\whadmass{\ensuremath{m^{\rm min ~ \Delta R}_{\rm uu}}}

\newcommand\ptjetfive{\ensuremath{\pt^{\rm jet5}}}
\newcommand\cent{\ensuremath{{\rm Centrality}}}

\newcommand\mbbmindr{\ensuremath{m_{\rm bb}^{\rm min ~ \Delta R}}}
\newcommand\numjetforty{\ensuremath{N^{\rm jet}_{40}}}
\newcommand\drbbav{\ensuremath{\Delta R^{\rm avg}_{\rm bb}}}
\newcommand\mjjmaxpt{\ensuremath{m_{\rm jj}^{\rm max ~ p_{T}}}}
\newcommand\aplab{\ensuremath{\rm Aplan_{b-jet}}}

\newcommand\mjjmindr{\ensuremath{m_{\rm jj}^{\rm min ~ \Delta R}}}
\newcommand\drlepbbmindr{\ensuremath{\Delta R_{\rm lep-bb}^{\rm min ~ \Delta R}}}
\newcommand\mbjmindr{\ensuremath{m_{\rm bj}^{\rm min ~ \Delta R}}}
\newcommand\mbjmaxpt{\ensuremath{m_{\rm bj}^{\rm max ~ p_{T}}}}
\newcommand\whadpt{\ensuremath{ p_{\rm T,  uu}^{\rm min ~ \Delta R}}}
\newcommand\whaddR{\ensuremath{\Delta R_{\rm uu}^{\rm min ~ \Delta R}}}
\newcommand\mbbmaxpt{\ensuremath{m_{\rm bb}^{\rm max ~ p_{T}}}}
\newcommand\drbbmaxpt{\ensuremath{\Delta R_{\rm bb}^{\rm max ~ p_{T}}}}
\newcommand\mbbmaxM{\ensuremath{m_{\rm bb}^{\rm max ~ m}}}

\newcommand\ptjetthree{\ensuremath{\pt^{\rm jet3}}}

\newcommand\mindijetmass{\ensuremath{m_{\rm jj}^{\rm min~m}}}

\newcommand\mclosest{\ensuremath{m_{\rm jj}^{\rm closest}}}

\newcommand\drbbmaxm{\ensuremath{\Delta R_{\rm bb}^{\rm max~m}}}
\newcommand\mindrlj{\ensuremath{\Delta R_{\rm lj}^{\rm min~\Delta R}}}
\newcommand\maxdrhl{\ensuremath{\Delta R_{\rm hl}^{\rm max~\Delta R}}}
\newcommand\aplaj{\ensuremath{\rm Aplan_{jet}}}
\newcommand\maxdeta{\ensuremath{\Delta \eta_{\rm jj}^{\rm max~\Delta\eta}}}

\newcommand\mindrhl{\ensuremath{\Delta R_{\rm hl}^{\rm min~\Delta R}}}
\newcommand\nhiggsthirty{\ensuremath{\rm{N_{30}^{Higgs}}}}

\newcommand{\ptcth}{\mbox{$p_\mathrm{T}^{\mathrm{cone30}}$}}

\newcommand{\ptcpt}{\mbox{$p_\mathrm{T}^{\mathrm{cone30}} /p_{\mathrm{T}}^e$}}
\newcommand{\htobb}{\mbox{$H\to b\bar{b}$}}

\def\url#1{{\scriptsize\tt#1}}

\newcommand{\SUB}[2]{#1_\textrm{#2}}

\newcommand{\MB}[1]{{\mbox{\mathversion{bold}$#1$}}}

\newcommand{\dif}[1]	 {\textrm{d}#1}

\newcommand{\ShOL}{{\sc SherpaOL}}

\usepackage{preprintcover}

\PreprintCoverPaperTitle{Search for the Standard Model Higgs boson produced in 
association with top quarks and decaying into $\boldsymbol{b\bar{b}}$ in $\boldsymbol{pp}$ collisions
at $\sqrt{\bf s} ~{\bf=}$ $\boldsymbol{8\tev}$ with the ATLAS detector} 

\PreprintIdNumber{CERN-PH-EP-2015-047} 

\PreprintCoverAbstract{A search for the Standard Model Higgs boson produced 
in association with a top-quark pair, $t\bar{t}H$, is presented. The analysis 
uses 20.3~\ifb~of $pp$ collision 
data at $\sqrt{s}=8\tev$, collected with the ATLAS detector at the Large Hadron 
Collider during 2012.  The search is designed for the $H \to b\bar{b}$ decay mode 
and uses events containing one or two electrons or muons. 
In order to improve 
the sensitivity of the search, events are categorised according to their 
jet and $b$-tagged jet multiplicities. A neural network is used to 
discriminate between signal and background events, the latter being dominated 
by $t\bar{t}$+jets production. In the single-lepton 
channel, variables calculated using a matrix element method are included 
as inputs to the neural network to improve discrimination of the irreducible  
\ttbar+\bbbar\ background.  
No significant excess of events above 
the background expectation is found and an observed (expected) limit of 3.4 (2.2) times the Standard Model cross section 
is obtained at 95\% confidence level. 
The ratio of the measured \tth\ signal cross section to the Standard 
Model expectation is found to 
be $\mu = 1.5 \pm 1.1$ assuming a Higgs boson mass of 125$\gev$. }

\PreprintJournalName{Eur.\ Phys.\ J.\ C}

\usepackage[hyperindex]{hyperref} 

\begin{document}

\twocolumn[ \begin{@twocolumnfalse}
		
\title{Search for the Standard Model Higgs boson produced in association with top quarks and decaying into $\boldsymbol{b\bar{b}}$ in $\boldsymbol{pp}$ collisions at $\sqrt{\bf s} ~{\bf=}$ $\boldsymbol{8\tev}$ 
with the ATLAS detector}

\author{The ATLAS Collaboration}

\maketitle
\begin{abstract}
A search for the Standard Model Higgs boson produced in association 
with a top-quark pair, $t\bar{t}H$, is presented. The analysis uses 20.3~\ifb~of $pp$ collision 
data at $\sqrt{s}=8\tev$, collected with the ATLAS detector at the Large Hadron 
Collider during 2012.  The search is designed for the $H \to b\bar{b}$ decay mode 
and uses events containing one or two electrons or muons. 
In order to improve 
the sensitivity of the search, events are categorised according to their 
jet and $b$-tagged jet multiplicities. A neural network is used to 
discriminate between signal and background events, the latter being dominated 
by $t\bar{t}$+jets production. In the single-lepton 
channel, variables calculated using a matrix element method are included 
as inputs to the neural network to improve discrimination of the irreducible  
\ttbar+\bbbar\ background.  
No significant excess of events above 
the background expectation is found and an observed (expected) limit of 3.4 (2.2) times the Standard Model cross section 
is obtained at 95\% confidence level. 
The ratio of the measured \tth\ signal cross section to the Standard 
Model expectation is found to 
be $\mu = 1.5 \pm 1.1$ assuming a Higgs boson mass of 125$\gev$. 
\end{abstract}
\end{@twocolumnfalse} ]

\setcounter{page}{1}
\section{Introduction}
\label{sec:Introduction}

The discovery of a new particle in the search for the
Standard Model (SM)~\cite{Glashow:1961aa,Weinberg:1967aa,Salam:1969aa}
Higgs boson~\cite{Englert:1964aa,Higgs:1964aa,Higgs:1964ab,Guralnik:1964aa}
at the LHC was reported by the ATLAS~\cite{ATLASHiggsDiscovery} and
CMS~\cite{CMSHiggsDiscovery} collaborations in July 2012. There is by
now clear evidence of this particle in the \htogaga, \htoZZ,
\htoWW\ and \htott\ decay channels, at a mass of around 125~\gev, which have 
strengthened the SM Higgs boson 
hypothesis~\cite{ATLASVV,ATLASSpin,ATLAStt,CMSHiggsComb,CMSfermions,CMSSpin} of the observation. 
To determine all properties of the new boson experimentally, it is important to study it in as many production and decay modes as possible.
In particular, its coupling to heavy quarks is a strong focus of current experimental searches.  
The SM Higgs boson production in association with a
top-quark pair ($\ttH$)~\cite{ttH1,ttH2,ttH4,Beenakker:2001rj} with subsequent Higgs decay 
into bottom quarks (\htobb) addresses heavy-quark couplings in both production and decay.
Due to the large measured mass of the top quark, the Yukawa coupling of the top quark ($y_t$) is much stronger 
than that of other quarks. 
The observation of the $\ttH$ production 
mode would allow for a direct measurement of this   
coupling, to which other Higgs production modes are 
only sensitive through loop effects.
Since $y_t$ is expected to be close to unity, it is also argued
to be the quantity that might give insight into
the scale of new physics~\cite{Shaposhnikov2014}.

The \htobb\ final state is the dominant decay mode in the SM for a Higgs boson 
with a mass of 125 GeV. So far, this decay mode has not yet been observed.  
While a search for this decay via the gluon fusion process is precluded by  
the overwhelming multijet background, Higgs boson production 
in association with a 
vector boson ($VH$) ~\cite{ATLASVHbb,CMSVHbb,TEV2012H2bb} or a top-quark 
pair (\ttbar) significantly improves the signal-to-background ratio for this decay. 

\begin{figure*}[ht!]
\begin{center}
\subfigure[]{\includegraphics[width=0.31\textwidth]{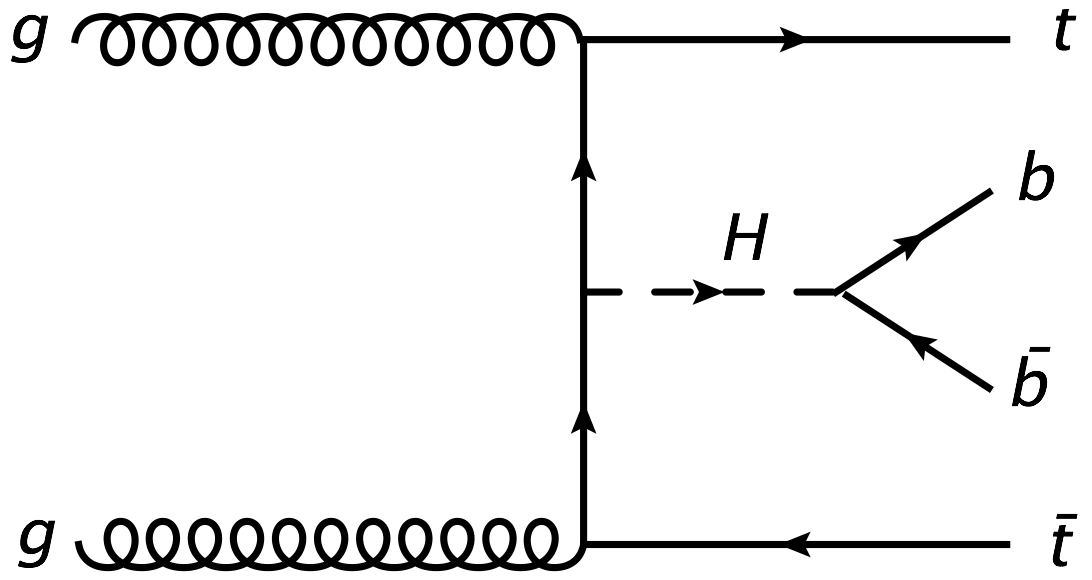}}\label{fig:feyman_a}\hspace{0.5cm}
\subfigure[]{\includegraphics[width=0.31\textwidth]{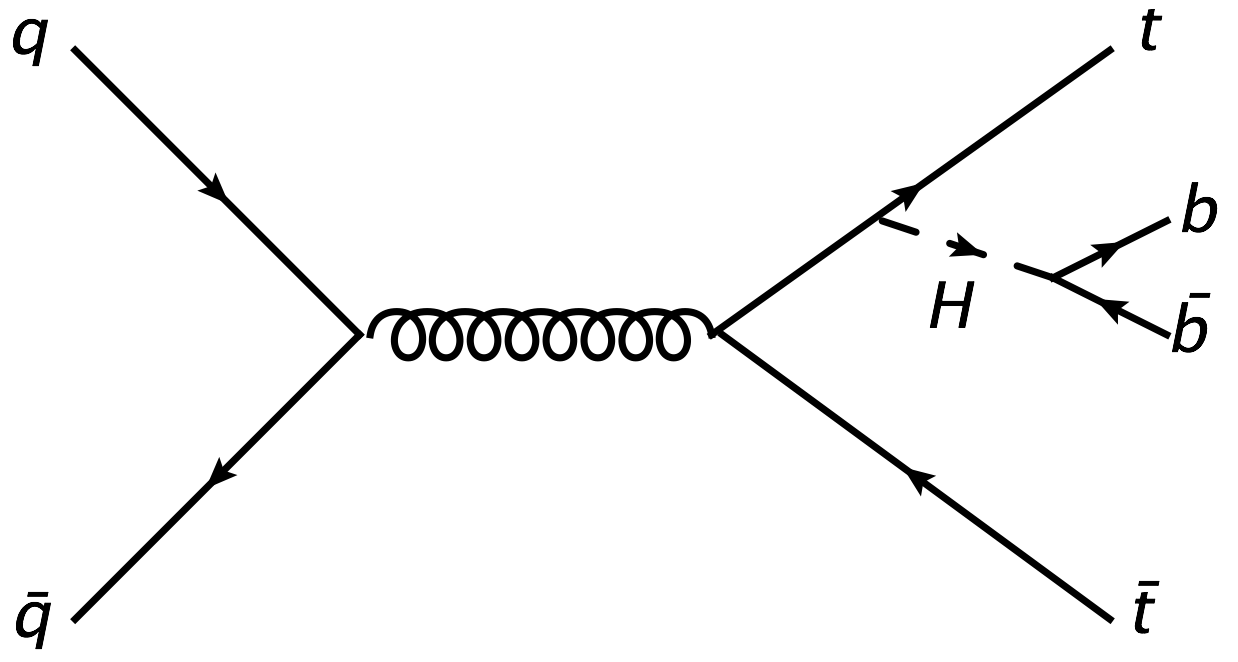}}\label{fig:feyman_b}\hspace{0.5cm}
\subfigure[]{\includegraphics[width=0.31\textwidth]{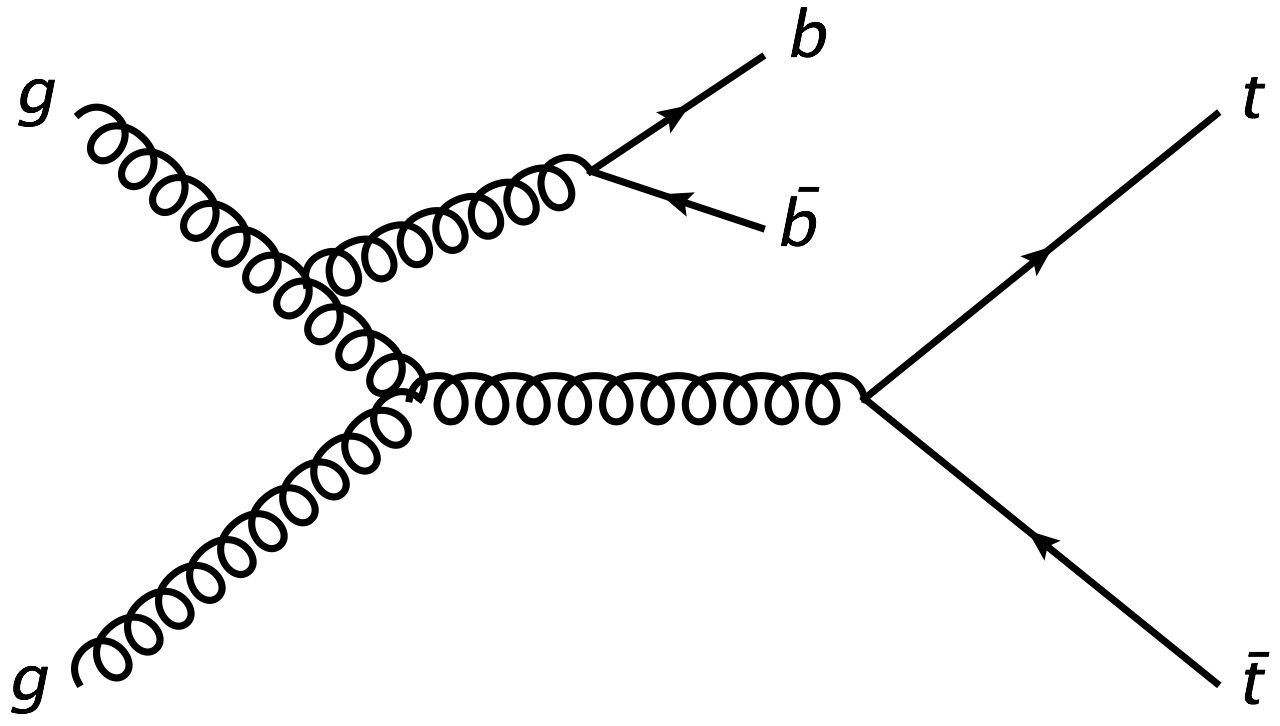}}\label{fig:feyman_c}
\caption{Representative tree-level Feynman diagrams for the production of 
the Higgs boson in association with a top-quark pair (\tth) and the subsequent decay of the Higgs to \bbbar, (a) and (b), and for the main background \ttbar+\bbbar\ (c). }
\label{fig:feyman}
\end{center}
\end{figure*}

This paper describes a search for the SM Higgs boson in the $\ttH$ production
mode and is designed to be primarily sensitive to the $\htobb$
decay, although other Higgs boson decay modes are also treated as signal. 
Figs.~\ref{fig:feyman}(a) and \ref{fig:feyman}(b) show two examples of tree-level diagrams 
for \tth\ production with a subsequent \htobb\ decay. 
A search for the 
associated production of the Higgs boson with a top-quark pair using  
several Higgs decay modes (including \htobb) has recently been published  
by the CMS Collaboration~\cite{CMS8TeVttH} quoting a ratio of the 
measured \tth\ signal cross section to the SM expectation 
for a Higgs boson mass of 125.6$\gev$ of $\mu = 2.8 \pm 1.0$.

The main source of background to this search comes from top-quark pairs  
produced in association with additional jets. The dominant source is \ttbb\ 
production, resulting in the same final-state signature as the signal.  An example is shown in Fig.~\ref{fig:feyman}(c).  
A second contribution arises from $t\bar{t}$ production in association with light-quark ($u$, $d$, $s$) 
or gluon jets, referred to as $t\bar{t}$+light background,    
and from $t\bar{t}$ production in association with $c$-quarks, referred to 
as $t\bar{t}$+$c\bar{c}$. The size of the second contribution depends on the 
misidentification rate of the algorithm used to identify $b$-quark jets. 
 
The search presented in this paper uses 20.3~\ifb\ of data collected with the
ATLAS detector in $pp$ collisions at $\sqrt{s}=8\tev$ during 2012. 
The analysis focuses on final states containing one or two electrons 
or muons from the decay of the \ttbar\ system, referred to as the 
single-lepton and dilepton channels, respectively. Selected events   
are classified into exclusive categories, referred to as ``regions", according to the
number of reconstructed jets and jets identified as $b$-quark jets by the 
$b$-tagging algorithm ($b$-tagged jets or $b$-jets for short). Neural networks (NN) 
are employed in the regions with a significant expected contribution from 
the $t\bar{t}H$ signal to separate it from the background. 
Simpler kinematic variables are used in regions that 
are depleted of the $t\bar{t}H$ signal, and primarily serve
to constrain uncertainties on the background prediction.
A combined fit to signal-rich and signal-depleted regions is 
performed to search for the signal while simultaneously 
obtaining a background prediction.

\section{ATLAS detector}
\label{sec:atlas}

The ATLAS detector~\cite{ATLASTechnicalPaper} consists of 
four main subsystems: an inner tracking system, electromagnetic and hadronic calorimeters, and a muon spectrometer.
The inner detector provides tracking information from
pixel and silicon microstrip detectors in the pseudorapidity\footnote{ATLAS uses a
right-handed coordinate system with its origin at the nominal
interaction point (IP) in the centre of the detector and the $z$-axis
coinciding with the axis of the beam pipe.  The $x$-axis points from
the IP to the centre of the LHC ring, and the $y$-axis points
upward. Cylindrical coordinates ($r$,$\phi$) are used in the
transverse plane, $\phi$ being the azimuthal angle around the beam
pipe. The pseudorapidity is defined in terms of the polar angle
$\theta$ as $\eta = - \ln \tan(\theta/2)$.  Transverse momentum and energy are defined as 
$\pt = \rm{p} \sin \theta$  and $\et = \rm{E} \sin \theta$, respectively.
} range $|\eta|<2.5$ and from a straw-tube 
transition radiation tracker covering $|\eta|<2.0$, all 
immersed in a 2\,T magnetic field
provided by a superconducting solenoid.  
The electromagnetic sampling calorimeter uses lead and liquid-argon (LAr)
and is divided into barrel ($|\eta|<1.475$) and end-cap regions
($1.375<|\eta|<3.2$). 
Hadron calorimetry employs the sampling technique, with either  
scintillator tiles or liquid argon as active media, 
and with steel, copper, or
tungsten as absorber material. The calorimeters cover $|\eta|<4.9$.
The muon spectrometer measures muon tracks within $|\eta|<2.7$ 
using multiple layers of high-precision tracking chambers located in a
toroidal field of approximately 0.5\,T and 1\,T in the central and end-cap 
regions of ATLAS, respectively. The muon
spectrometer is also instrumented with separate trigger chambers
covering $|\eta|<2.4$.

\section{Object reconstruction}
\label{sec:ObjectReconstruction}

The main physics objects considered in this search are electrons, muons,
jets and $b$-jets. Whenever possible, the same object reconstruction is 
used in both the single-lepton and dilepton channels, though some small 
differences exist and are noted below.

Electron candidates~\cite{ElectronPerformance} are reconstructed from
energy deposits (clusters) in the electromagnetic calorimeter that are
matched to a reconstructed track in the inner detector. 
To reduce the background
from non-prompt electrons, i.e. from decays of hadrons (in particular 
heavy flavour) produced in jets, electron candidates are required
to be isolated.  In the single-lepton channel, where such background is 
significant, an $\eta$-dependent isolation cut is made,
based on the sum of transverse energies of cells around the direction of each
candidate, in a cone of size $\Delta R =
\sqrt{(\Delta\phi)^2 + (\Delta\eta)^2} = 0.2$.  This energy sum
excludes cells associated with the electron and is corrected
for leakage from the electron cluster itself. A further 
isolation cut is made on the scalar sum of the track $\pt$ around 
the electron in a cone of size $\Delta R = 0.3$ (referred to as \ptcth).  
The longitudinal impact parameter of the electron track with respect
to the selected event primary vertex defined in Section~\ref{sec:EventSelection}, $z_{0}$, is required to be less
than 2 mm.
To increase efficiency in the dilepton channel, the electron selection 
is optimised by using an improved electron identification method 
based on a likelihood variable~\cite{electronLH} and the electron isolation. 
The ratio of 
\ptcth\ to the \pt\ of the electron is required to be less than 0.12, 
i.e. \ptcpt\ $<$ 0.12. The optimised selection improves the efficiency by
roughly 7\% per electron.   

Muon candidates are reconstructed from track segments in the muon 
spectrometer, and matched with tracks found in the
inner detector~\cite{MuonPerformance}.   
The final muon candidates are refitted using the complete
track information from both detector systems, and are required to satisfy
$|\eta|<2.5$.  Additionally, muons are required to
be separated by $\Delta R > 0.4 $ from any selected jet (see below for 
details on jet reconstruction and selection).
Furthermore, muons must satisfy a $\pt$-dependent track-based isolation
requirement that has good performance under conditions with a high number of  
jets from other $pp$ interactions within the same bunch crossing, 
known as ``pileup'',
or in boosted configurations where the muon is close
to a jet: the track \pt\ scalar sum in a cone of 
variable size $\Delta R < 10\gev/\pt^\mu$ around the muon
must be less than 5\% of the muon $\pt$.
The longitudinal impact parameter of the muon track with respect to 
the primary vertex, $z_{0}$, is required to be less than 2~mm.

Jets are reconstructed from calibrated 
clusters~\cite{ATLASTechnicalPaper,jes} built from energy deposits in the
calorimeters, using the anti-$k_t$
algorithm~\cite{ref:Cacciari2008,ref:Cacciari2006,ref:fastjet} with a
radius parameter $R=0.4$.  Prior to jet finding, a local cluster calibration
scheme~\cite{LCW1,LCW2} is applied to correct the cluster
energies for the effects of dead material, non-compensation and
out-of-cluster leakage. 
The jets are calibrated using energy- and $\eta$-dependent calibration factors,
derived from simulations, to the mean energy of stable particles inside 
the jets. Additional corrections to account for the difference between
simulation and data are applied~\cite{ATLASJetEnergyMeasurement}. 
After energy calibration, jets are required to have
$\pt > 25\gev$ and $|\eta| < 2.5$. 
To reduce the contamination from low-$\pt$ jets due to pileup, the scalar sum of the $\pt$ of tracks 
matched to the jet and originating from the primary vertex
must be at least 50\% of the scalar sum of the $\pt$ of all tracks matched to the jet.
This is referred to as the jet vertex fraction.
This criterion is only applied to jets with $\pt< 50\gev$ and $|\eta|<2.4$.

During jet reconstruction,
no distinction is made between identified electrons and jet candidates.  
Therefore, if any of the jets lie $\Delta R <$ 0.2
from a selected electron, the single closest jet is discarded in order
to avoid double-counting of electrons as jets.  After this, electrons
which are $\Delta R <$ 0.4 from a jet are removed to further suppress 
background from non-isolated electrons.

Jets are identified as originating from the hadronisation of a 
$b$-quark via an algorithm~\cite{ref:ATLAS-CONF-2014-046}
that uses multivariate techniques to combine information from the impact
parameters of displaced tracks with topological properties of
secondary and tertiary decay vertices reconstructed within the jet.
The working point used for this search corresponds to a 70\% efficiency to tag
a $b$-quark jet, with a light-jet mistag rate of 1\%,  
and a charm-jet mistag rate of 20\%, 
as determined for $b$-tagged jets with $\pt >20\gev$ and 
$|\eta|<2.5$ in simulated $t\bar{t}$ events. Tagging efficiencies 
in simulation are corrected to match the results of the calibrations 
performed in data~\cite{ref:ATLAS-CONF-2014-004}. Studies in simulation 
show that these efficiencies do not depend on the number of jets.

\section{Event selection and classification}
\label{sec:EventSelection}
For this search, only events collected using a single-electron or single-muon trigger under stable beam
conditions and for which all detector subsystems were operational are considered. The corresponding
integrated luminosity is 20.3~\ifb. Triggers with different $\pt$ thresholds are combined in a logical OR
in order to maximise the overall efficiency.  The $\pt$ thresholds are 24 
or 60~\gev\ for electrons and 24 or 36~\gev\ for muons.  The triggers with the lower $\pt$
threshold include isolation requirements on the lepton candidate, 
resulting in inefficiency at high $\pt$ that is recovered by the triggers with higher $\pt$ threshold.
The triggers use selection criteria looser than the final 
reconstruction requirements.

Events accepted by the trigger are required to have at least one 
reconstructed vertex with at least five
associated tracks, consistent with the beam collision region in the $x$--$y$ plane.
If more than one such vertex is found, the vertex candidate with the 
largest sum of squared transverse momenta of its associated tracks is taken 
as the hard-scatter primary vertex. 

In the single-lepton channel, events are required to have exactly one
identified electron or muon with $\pt>25$ \gev\  
and at least four jets, at least two of which 
are $b$-tagged. The selected lepton is required to match, with  
$\Delta R < 0.15$, the lepton reconstructed by the trigger.  

In the dilepton channel, events are required to have exactly two
leptons of opposite charge and at least two $b$-jets. 
The leading and subleading lepton must have $\pt>25$ \gev\ 
and $\pt>15$ \gev, respectively. Events in the single-lepton sample 
with additional leptons passing this selection are removed 
from the single-lepton sample to avoid statistical overlap 
between the channels. In the dilepton channel, events are 
categorised into $ee$, $\mu\mu$ and $e\mu$ samples. In the $e\mu$ category, 
the scalar sum of the transverse energy
of leptons and jets, \htlep, is required to be above 130~\gev. 
In the $ee$ and $\mu\mu$ event categories, the invariant mass of the
two leptons, \mll, is required to be larger than 15~\gev\ in events 
with more than two $b$-jets, to suppress
contributions from the decay of hadronic resonances such as the
$J/\psi$ and $\Upsilon$ into a same-flavour lepton pair. In events with 
exactly two $b$-jets, \mll\ is required to be larger than 60~\gev\ 
due to poor agreement between data and prediction at lower \mll. 
A further cut on \mll\ is applied in the $ee$ and $\mu\mu$ categories to reject events close to the $Z$
boson mass: $|\mll - m_Z| > 8$~\gev. 

After all selection requirements, the samples are dominated by $\ttbar$+jets 
background. In both channels, selected events 
are categorised into different regions.
In the following, a given region with $m$ jets of which $n$ are $b$-jets are 
referred to as ``($m {\rm j}, n {\rm b}$)". The regions with a   
signal-to-background ratio $S/B>$ 1\% and $S/\sqrt{B}>0.3$,  
where $S$ and
$B$ denote the expected signal for a SM Higgs boson with $m_H=125\gev$, 
and background, respectively, are referred to as ``signal-rich regions",  
as they provide most of the sensitivity to the signal. 
The remaining regions 
are referred to as ``signal-depleted regions". They are almost purely background-only regions
and are used to constrain systematic uncertainties, thus improving the background prediction 
in the signal-rich regions. The regions are analysed separately  and combined statistically 
to maximise the overall sensitivity. In the most sensitive regions, \sixfour\ in the single-lepton 
channel and \fourfourdi\ in the dilepton channel, $H \rightarrow \bbbar$ decays 
are expected to constitute about 90\% of the signal contribution as shown in 
Fig.~\ref{fig:modes} of Appendix~\ref{sec:SignalPie}. 

In the single-lepton channel, a total of nine independent regions are 
considered: six signal-depleted regions, 
\fourtwolj, \fourthreelj, \fourfourlj, \fivetwo, \fivethree, \sixtwo, and  
three signal-rich regions, \fivefour, \sixthree\ and \sixfour. 
In the dilepton channel, a total of six independent regions are 
considered. The signal-rich regions are \fourthreedi\ and \fourfourdi, while
the signal-depleted regions are \twotwo, \threetwo, \threethree\ and \fourtwodi. 
Figure~\ref{fig:Semi8SoBplots}(a) shows the $S/\sqrt{B}$ and $S/B$ ratios for
the different regions under consideration in the single-lepton channel based on
the simulations described in Sect.~\ref{sec:SimulatedSamples}.  
The expected proportions of different backgrounds in each region are
shown in Fig.~\ref{fig:Semi8SoBplots}(b).  
The same is shown in the dilepton channel in Figs.~\ref{fig:Dil8SoBplots}(a) 
and \ref{fig:Dil8SoBplots}(b).

\begin{figure*}[ht!]
\centering
\subfigure[]{\includegraphics[width=0.445\textwidth]{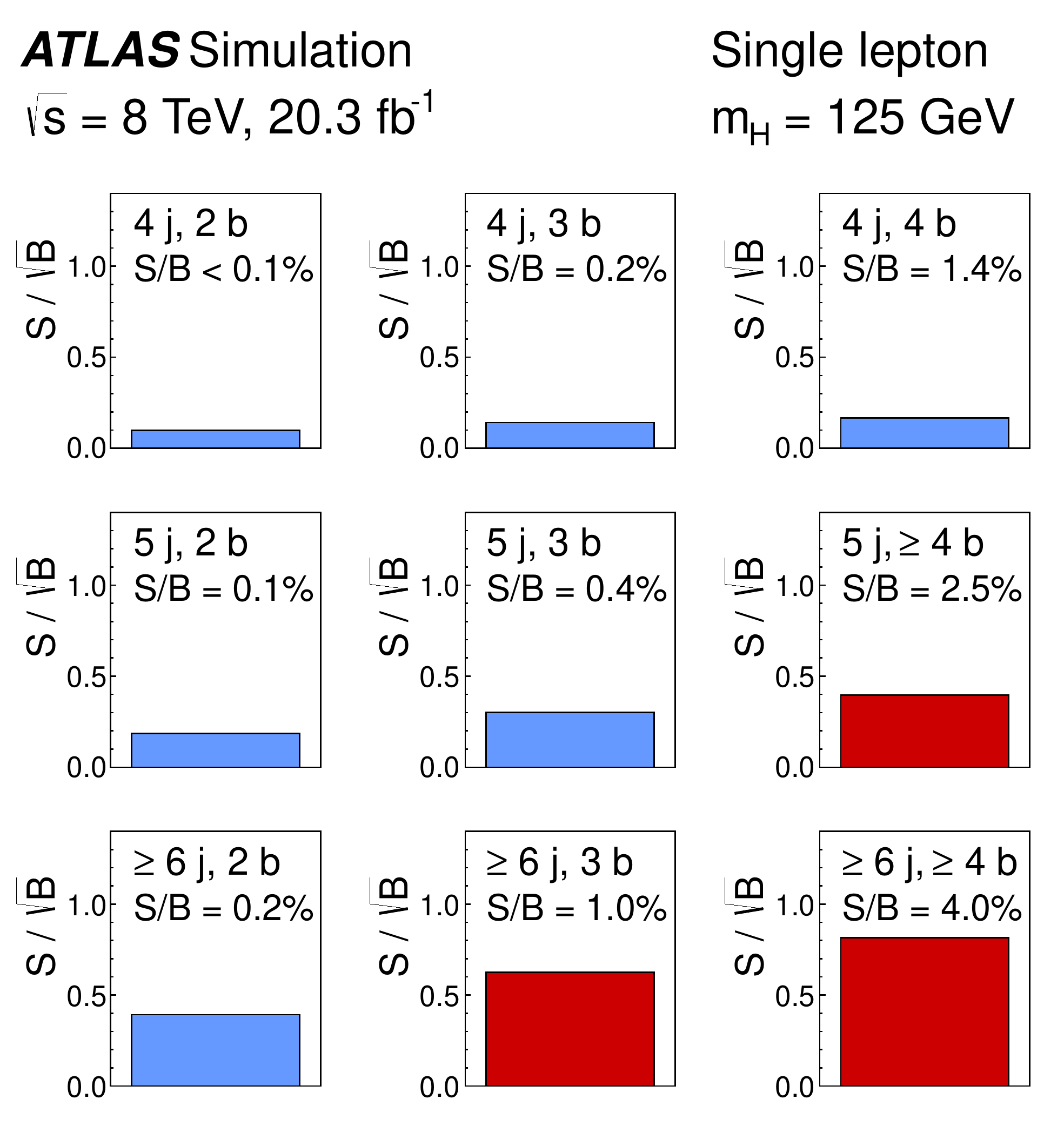}}\label{fig:Semi8SoBplots_a} \hspace{0.25cm}
\subfigure[]{\includegraphics[width=0.515\textwidth]{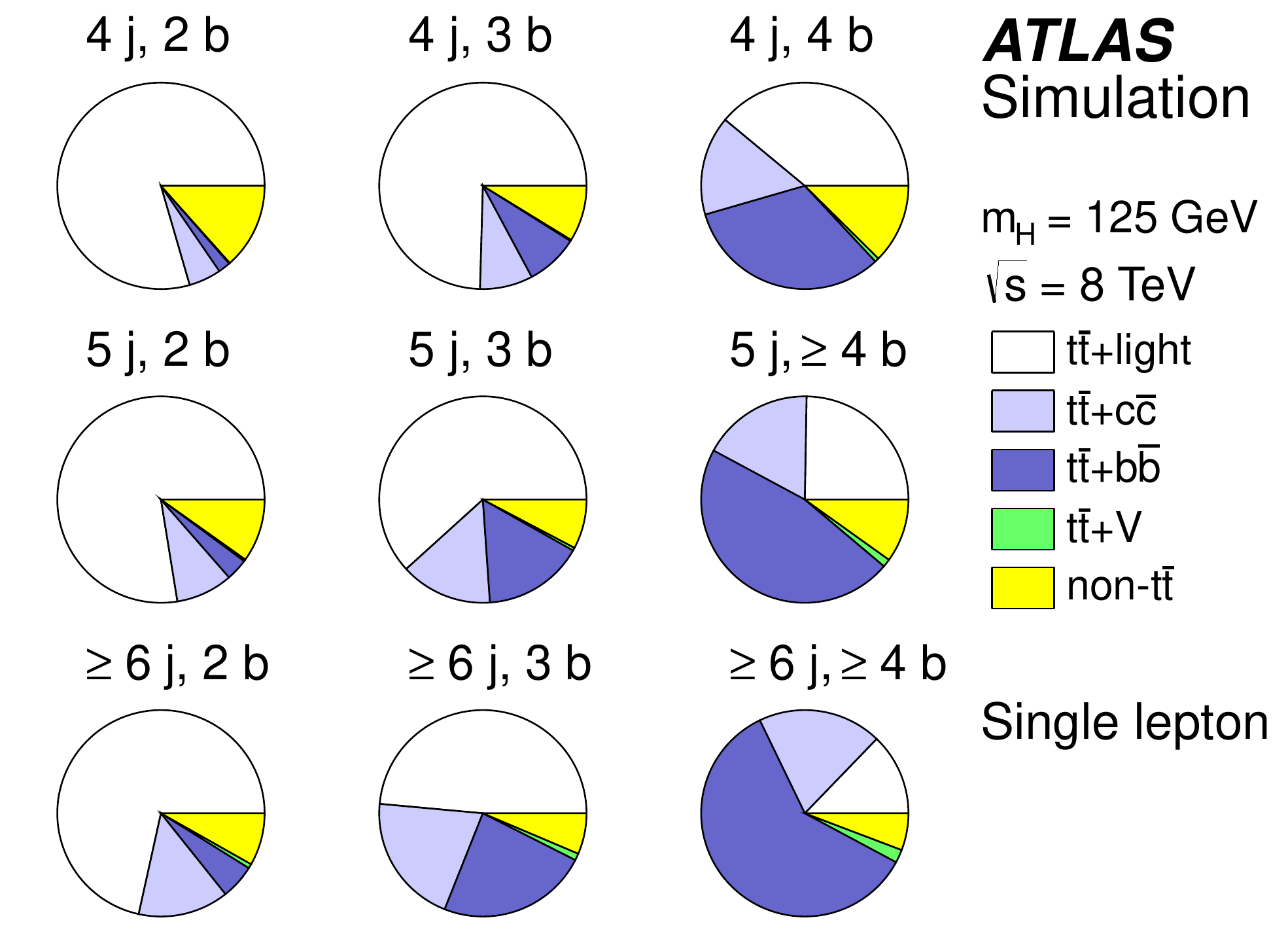}}\label{fig:Semi8SoBplots_b} 
\caption{Single-lepton channel: (a) $S/\sqrt{B}$ ratio for each of the regions 
assuming SM cross sections and
branching fractions, and $\mH=125\gev$. Each row shows the plots for a
specific jet multiplicity (4, 5, $\geq$6), and the columns show the
$b$-jet multiplicity (2, 3, $\geq$4). Signal-rich regions  
are shaded in dark red, while the rest are shown in light blue. 
The $S/B$ ratio for each region is also noted. (b) The fractional 
contributions of the various backgrounds to the total background 
prediction in each considered region. The ordering of the rows and 
columns is the same as in (a).}
\label{fig:Semi8SoBplots}
\end{figure*}

\begin{figure*}[ht!]
\centering
\subfigure[]{\includegraphics[width=0.445\textwidth]{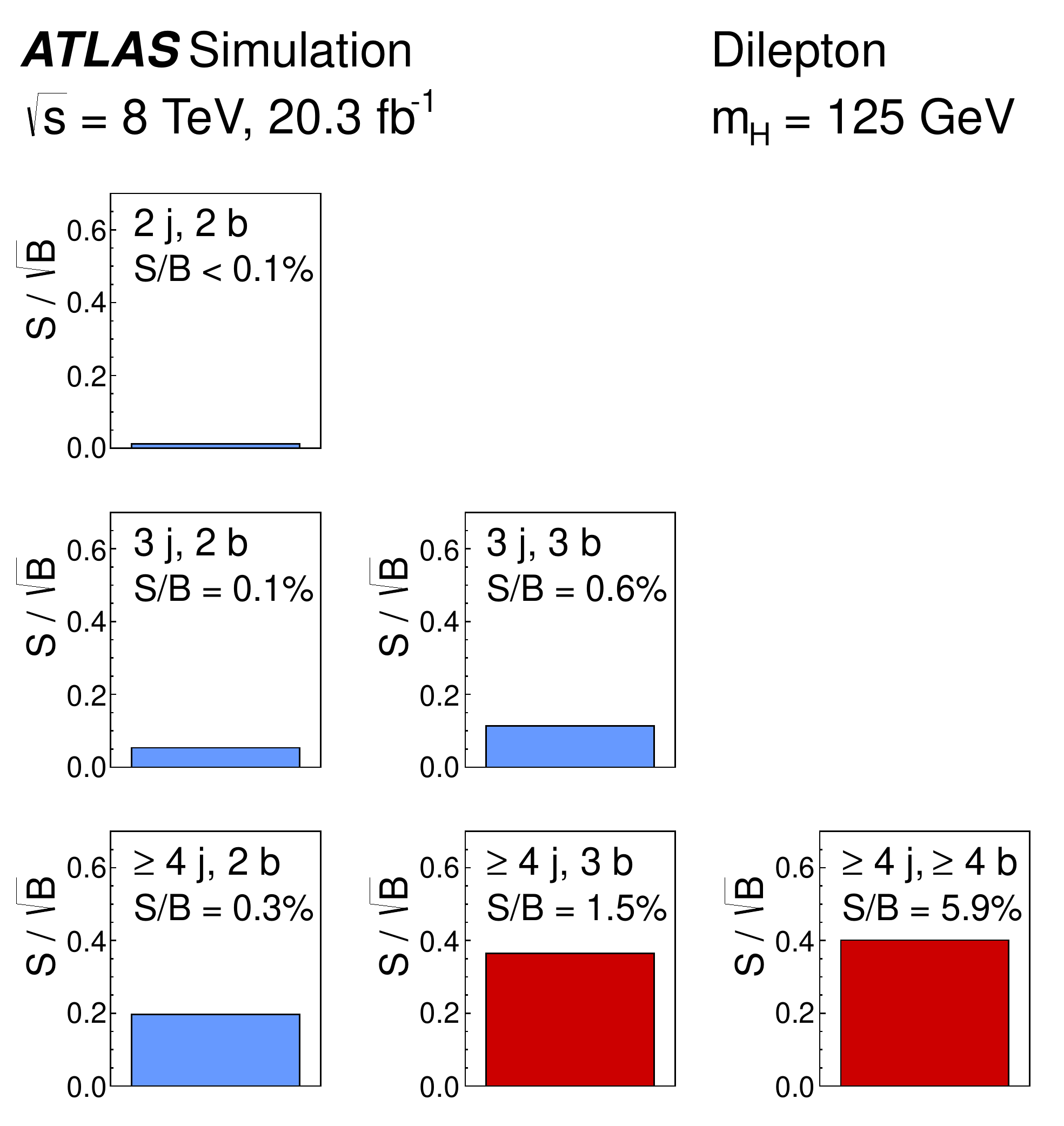}}\label{fig:Dil8SoBplots_a}  \hspace{0.25cm}
\subfigure[]{\includegraphics[width=0.515\textwidth]{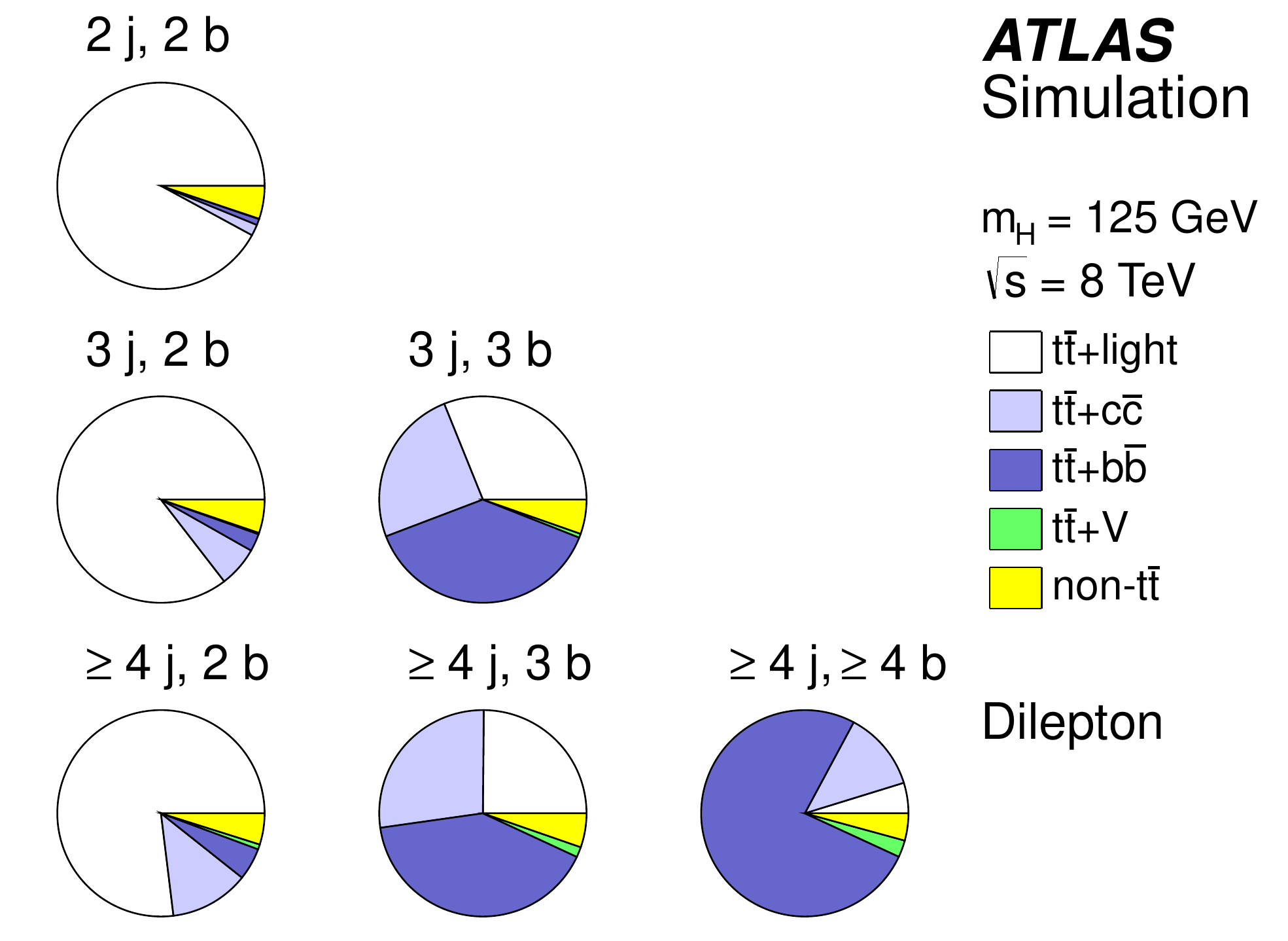}}\label{fig:Dil8SoBplots_b}
\caption{Dilepton channel: (a) The $S/\sqrt{B}$ ratio for each of the regions 
assuming SM cross sections and branching fractions and
$\mH=125\gev$. Each row shows the plots for a specific jet
multiplicity (2, 3, $\geq$4), and the columns show the $b$-jet
multiplicity (2, 3, $\geq$4). Signal-rich regions 
are shaded in dark red, while the rest are shown in light blue. 
The $S/B$ ratio for each region is also noted. (b) The fractional 
contributions of the
various backgrounds to the total background prediction in each 
considered region.  The ordering of the rows and columns is the same as in (a).}
\label{fig:Dil8SoBplots}
\end{figure*}

\section{Background and signal modelling}
\label{sec:SimulatedSamples}

After the event selection described above, the main background in both the single-lepton 
and dilepton channels is $t\bar{t}$+jets production. In the   
single-lepton channel, additional background contributions come from 
single top quark production, followed by the production of a $W$ or $Z$ 
boson in association with jets ($W/Z$+jets),  diboson
($WW$, $WZ$, $ZZ$) production, as well as the associated production of 
a vector boson and a $t\bar{t}$ pair, $t\bar{t}+V$ ($V=W,Z$).
Multijet events also contribute to the selected
sample via the misidentification of a jet or a photon as an
electron or the presence of a non-prompt electron or muon, referred to 
as ``Lepton misID'' background. The corresponding 
yield is estimated via a data-driven method known as 
the ``matrix method''~\cite{ttbar_3pb}.    
In the dilepton channel, backgrounds containing at least two prompt leptons 
other than $t\bar{t}$+jets production arise from $Z$+jets, diboson, 
and $Wt$-channel single top quark production, as well as from 
the $t\bar{t}V$ processes. 
There are also several processes which may contain either non-prompt
leptons that pass the lepton isolation requirements or jets misidentified 
as leptons. These processes include $W$+jets, \ttbar\ production with 
a single prompt lepton in the final state, and single top quark production in $t$- and $s$-channels. 
Their yield is estimated using simulation and cross-checked with a 
data-driven technique based on the selection of a same-sign lepton pair. 
In both channels, the contribution of the misidentified lepton background is
negligible after requiring two $b$-tagged jets.   

In the following, the simulation of each background and of the signal is described in detail.  
For all MC samples, the top quark mass is taken to be $\mt = 172.5$~\gev\ and 
the Higgs boson mass is taken to be $\mH = 125$~\gev.

\subsection{$t\bar{t}$+jets background}
The $t\bar{t}$+jets sample is generated using the {\sc Powheg-Box} 2.0 NLO 
generator~\cite{powheg,powbox1,powbox2} with the {\sc CT10} parton 
distribution function (PDF) set~\cite{ct101}. It is 
interfaced to {\sc Pythia} 6.425~\cite{PythiaManual} with the {\sc CTEQ6L1} PDF set~\cite{cteq6} 
and the Perugia2011C~\cite{Skands:2010ak} underlying-event tune.  
The sample is normalised to the top++2.0~\cite{ref:xs6} theoretical calculation 
performed at next-to-next-to-leading order (NNLO) in QCD  
that includes resummation of next-to-next-to-leading logarithmic (NNLL) soft 
gluon terms~\cite{ref:xs1,ref:xs2,ref:xs3,ref:xs4,ref:xs5}. 

The $\ttbar$+jets sample is generated inclusively, but events are categorised 
depending on the flavour of partons that are matched
to particle jets that do not originate from the decay of the $\ttbar$ system.  
The matching procedure is done using the requirement of $\Delta R < 0.4$.
Particle jets are reconstructed by clustering stable particles excluding 
muons and neutrinos using the anti-$k_t$
algorithm with a radius parameter $R=0.4$, and are required to have $\pt>15\gev$ and
$|\eta|<2.5$.  

Events where at least one such particle jet is matched to a bottom-flavoured hadron are labelled as \ttbar+\bbbar\ events. Similarly, events which are not already categorised as \ttbar+\bbbar, and
where at least one particle jet is matched 
to a charm-flavoured hadron, are labelled as \ttbar+\ccbar\ events. Only hadrons not associated with $b$ and $c$ quarks from top quark and $W$ boson decays are considered.
Events labelled as either \ttbar+\bbbar\  or
\ttbar+\ccbar\  are generically referred to as \ttbar+HF events (HF for ``heavy
flavour'').  The remaining events are labelled as $\ttbar$+light-jet events, 
including those with no additional jets. 

Since {\sc Powheg}+{\sc Pythia} only models \ttbar+\bbbar\ via the parton shower, an 
alternative $\ttbar$+jets sample is generated with the {\sc Madgraph5} 1.5.11 LO 
generator~\cite{madgraph} using the
CT10 PDF set and interfaced to {\sc Pythia} 6.425 for showering and hadronisation.
It includes tree-level diagrams with up to three extra  
partons (including $b$- and $c$-quarks) and uses settings similar to those in  
Ref.~\cite{CMS8TeVttH}. To avoid double-counting of partonic configurations generated by both the
matrix element calculation and the parton-shower evolution, a
parton--jet matching scheme (``MLM matching")~\cite{mlm} is employed.

Fully matched NLO predictions with massive $b$-quarks have become available 
recently~\cite{sherpa_nlo_ttbb} 
within the {\sc Sherpa} with {\sc OpenLoops} 
framework~\cite{Gleisberg:2008ta,open_loops} 
referred to in the following as \ShOL. The \ShOL\ NLO sample is generated 
following the four-flavour scheme using the {\sc Sherpa} 2.0 pre-release and  
the CT10 PDF set.   
The renormalisation scale ($\mu_{\rm R}$) is set to   
$\mu_{\rm R}=\prod_{i=t,\bar{t},b,\bar{b}}E_{\mathrm{T},i}^{1/4} $, 
where $E_{\mathrm{T},i}$ 
is the transverse energy of parton $i$, and the  
factorisation and resummation scales are both set 
to $(E_{\mathrm{T}, t }+E_{\mathrm{T}, \bar{t} })/2$.  

For the purpose of comparisons between 
\ttbar+jets event generators and the propagation of systematic 
uncertainties related to the modelling of \ttbar+HF, as described 
in Sect.~\ref{sec:syst_ttbarmodel}, 
a finer categorisation of different topologies 
in \ttbar+HF is made. 
In particular, the following categories are considered: if 
two particle jets are both matched to an extra $b$-quark or extra $c$-quark 
each, the event is referred to as $\ttbb$ or $\ttcc$; if   
a single particle jet is matched to a single $b$($c$)-quark the event is 
referred to as $t\bar{t}$+$b$ ($t\bar{t}$+$c$); if  
a single particle jet is matched to a \bbbar\ or a \ccbar\ pair, the event  
is referred to as $t\bar{t}$+$B$ or $t\bar{t}$+$C$, respectively.  

Figure~\ref{fig:default_extHFtype} shows the relative contributions 
of the different \ttbar+\bbbar\ event categories to the total \ttbar+\bbbar\ cross 
section at generator level for the {\sc Powheg}+{\sc Pythia}, 
{\sc Madgraph}+{\sc Pythia} and \ShOL\ samples. It demonstrates that 
{\sc Powheg}+{\sc Pythia} is able 
to reproduce reasonably well the \ttbar+HF content of the {\sc Madgraph} 
$\ttbar$+jets sample, which includes a LO \ttbb\ matrix element calculation, 
as well as the NLO \ShOL\ prediction.    

\begin{figure}[ht!]
\begin{center}
\includegraphics[width=0.49\textwidth]{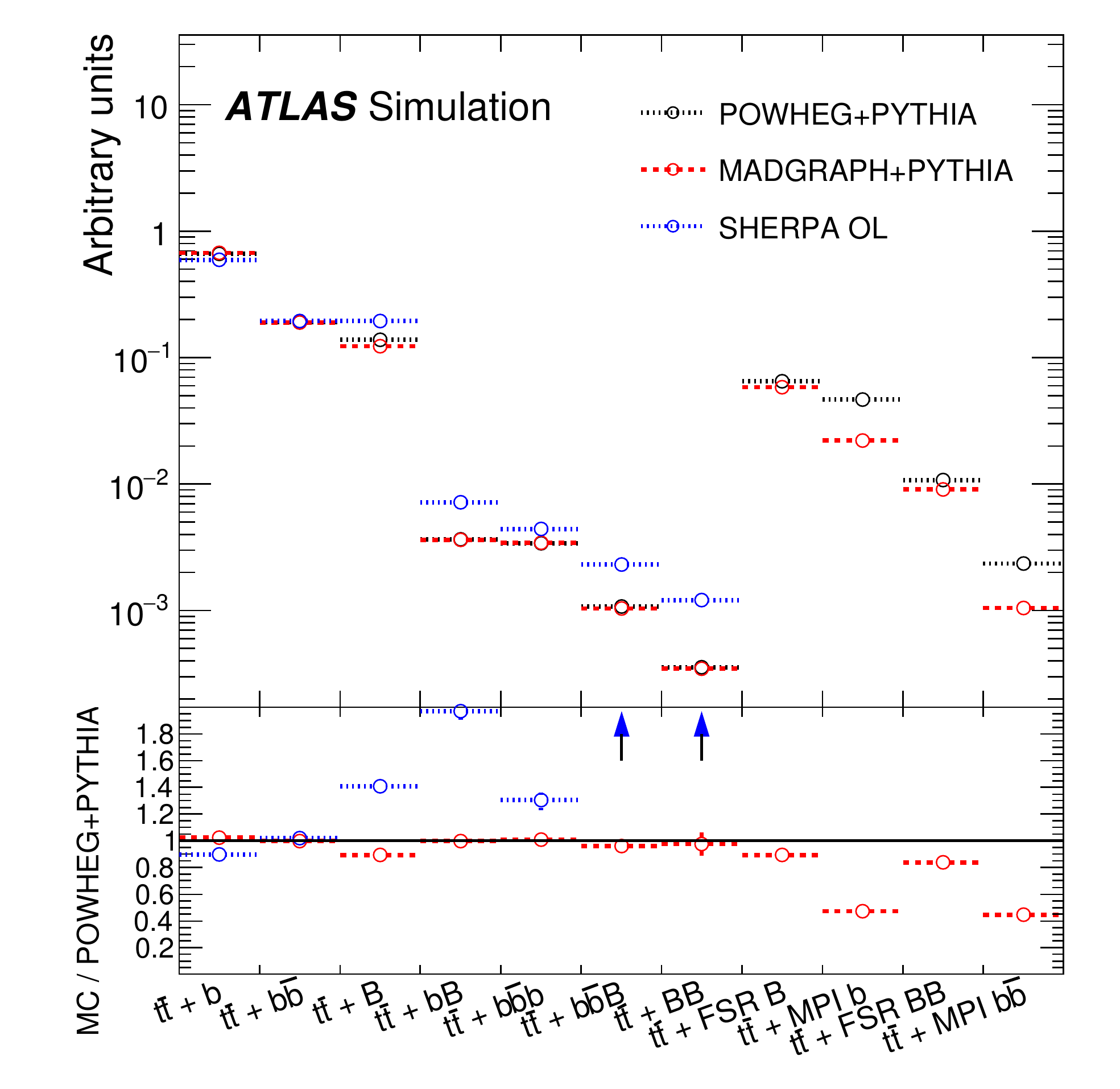}
\caption{Relative contributions of different categories of \ttbar+\bbbar\ 
events in 
{\sc Powheg}+{\sc Pythia}, {\sc Madgraph}+{\sc Pythia} and \ShOL\ samples. 
Labels  
``\ttbar+MPI'' and ``\ttbar+FSR'' refer to events where heavy flavour is 
produced via
multiparton interaction (MPI) or final state radiation (FSR), respectively. 
These contributions are not included in the \ShOL\ calculation. An arrow 
indicates that the point is off-scale. Uncertainties are from the limited 
MC sample sizes.   
}
\label{fig:default_extHFtype}
\end{center}
\end{figure}

The relative distribution across categories is 
such that \ShOL\ predicts a higher contribution of the $t\bar{t}+B$ category, 
as well as every category where the production of a second \bbbar\ pair is required. 
The modelling of the relevant kinematic variables in each category is in 
reasonable agreement between {\sc Powheg}+{\sc Pythia} and \ShOL. 
Some differences are observed in the very low regions of the 
mass and \pt\ of the \bbbar\ pair, and in the \pt\ of 
the top quark and \ttbar\ systems. 

The prediction from \ShOL\ is expected to model the \ttbb\ 
contribution more accurately than both \\ {\sc Powheg}+{\sc Pythia} and {\sc Madgraph}+{\sc Pythia}. 
Thus, in the analysis \ttbar+\bbbar\ events are reweighted from {\sc Powheg}+ \\ {\sc Pythia} to
reproduce the NLO \ttbar+\bbbar\ prediction from \ShOL\ for
relative contributions of different categories as well as their kinematics.  
The reweighting is done at generator level using several kinematic 
variables such as the top quark \pt, \ttbar\ system \pt, $\Delta R$ and \pt\ of the 
dijet system not coming from the top quark decay. In the absence of an NLO calculation
of \ttbar+\ccbar\ production, the {\sc Madgraph}+{\sc Pythia} sample is used to
evaluate systematic uncertainties on the \ttbar+\ccbar\ background. 

Since achieving the best possible modelling of the $\ttbar$+jets background is a key 
aspect of this analysis, a separate reweighting is applied to \ttbar+light and \ttbar+\ccbar\ 
events in {\sc Powheg}+{\sc Pythia} based on the ratio of measured differential 
cross sections at $\sqrt{s}=7\tev$ in 
data and simulation as a function of top quark $\pt$ and $\ttbar$ 
system $\pt$~\cite{topdiff_7TEV}. 
It was verified using the simulation that the ratio derived at 
$\sqrt{s}=7\tev$ is applicable to $\sqrt{s}=8\tev$ simulation.  
It is not applied to the \ttbar+\bbbar\ component   
since that component was corrected to match the best available theory calculation.  
Moreover, the  measured differential cross section is not sensitive to this component. 
The reweighting significantly improves the agreement between simulation and data in the total 
number of jets (primarily due to the \ttbar\ system \pt\ reweighting) and jet \pt\ (primarily due to the top quark \pt\ reweighting). 
This can be seen in Fig.~\ref{fig:ttbarseqrw}, 
where the number of jets and the scalar sum of the jet \pt\ (\hthad) distributions in the exclusive 2-$b$-tag region are 
plotted in the single-lepton channel before and after the reweighting is applied.

\begin{figure*}[ht!]
\begin{center}
\subfigure[]{\includegraphics[width=0.38\textwidth]{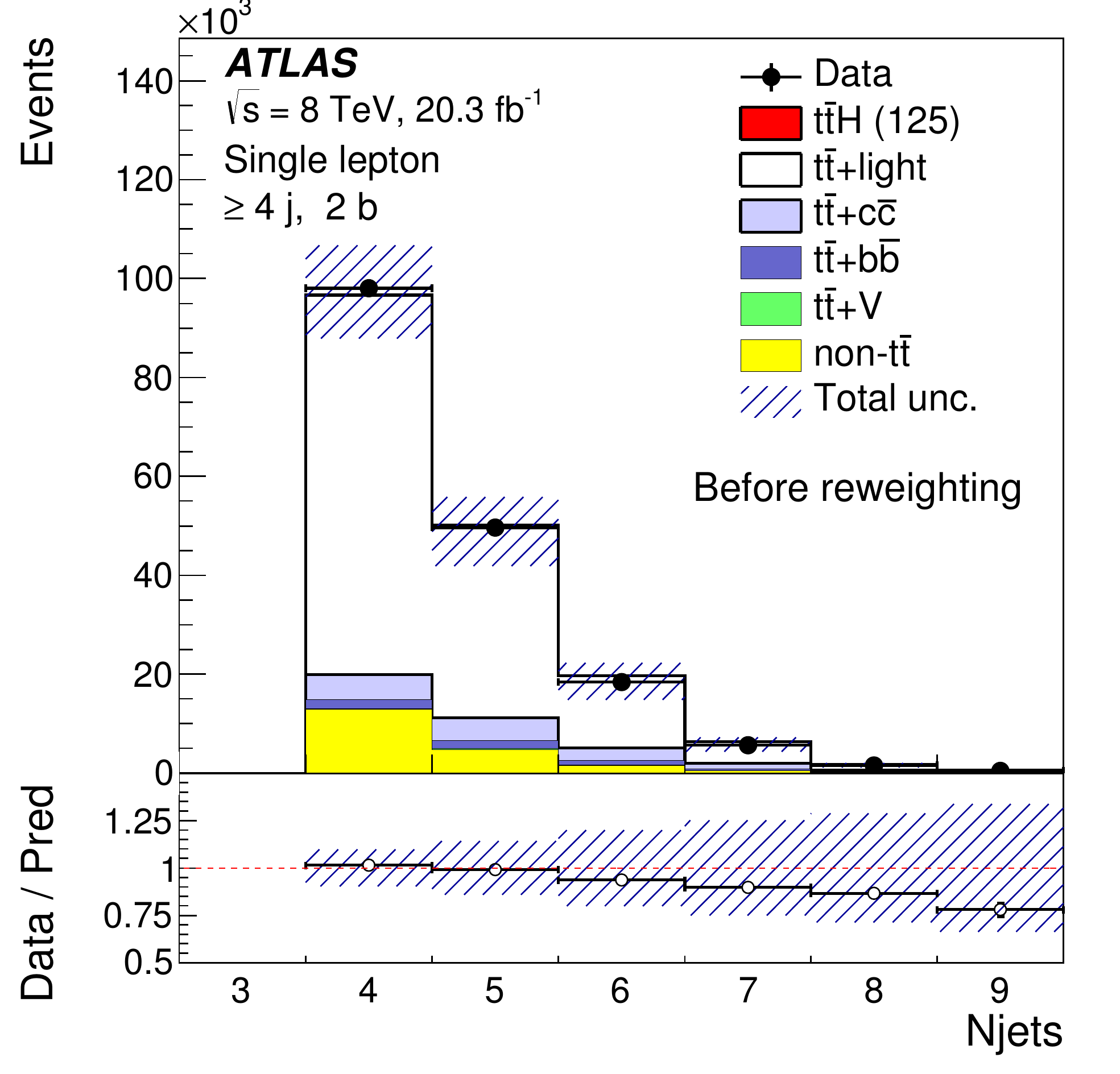}}\label{fig:ttbarseqrw_a}
\subfigure[]{\includegraphics[width=0.38\textwidth]{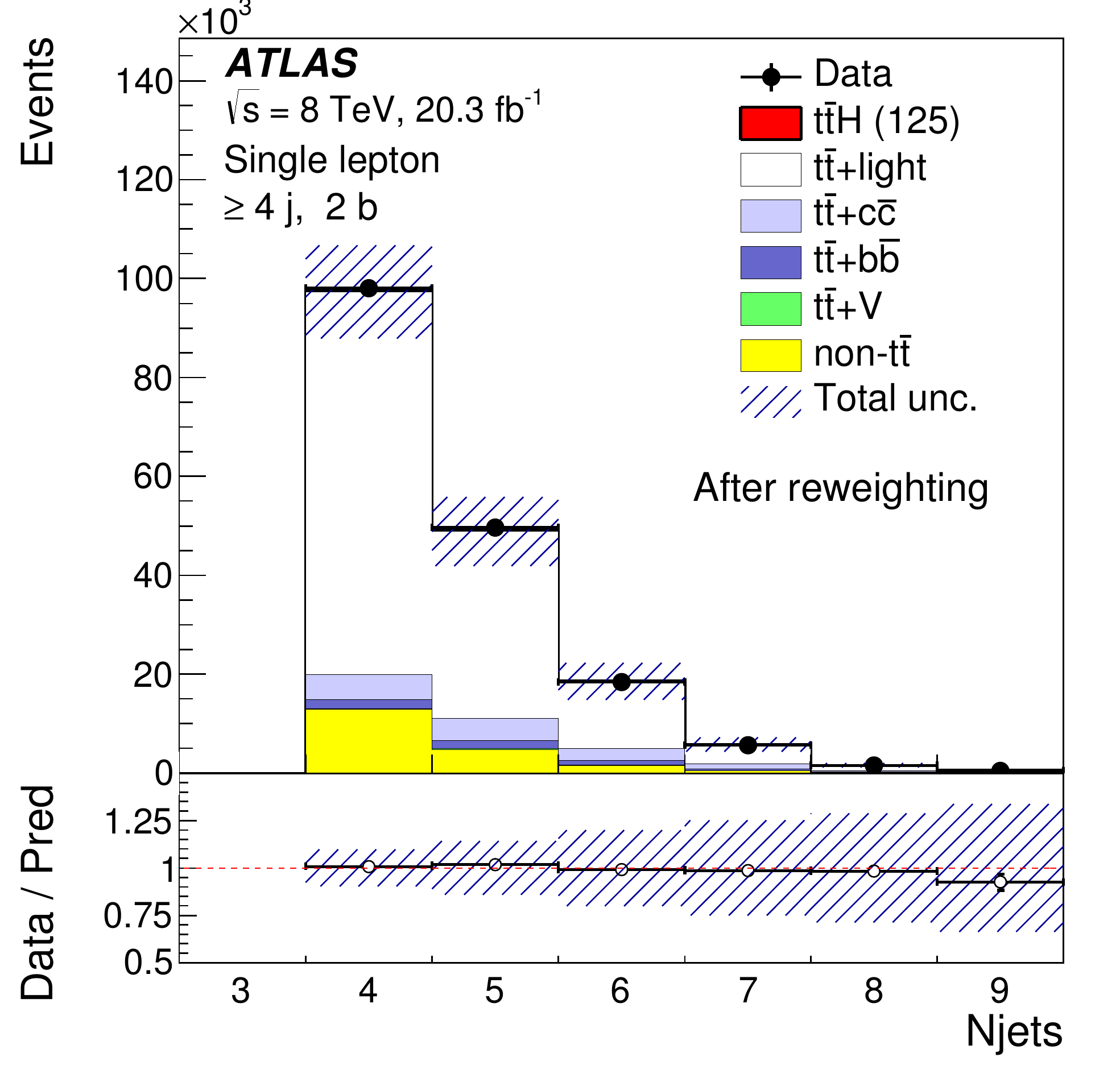}} \label{fig:ttbarseqrw_b} \\
\subfigure[]{\includegraphics[width=0.38\textwidth]{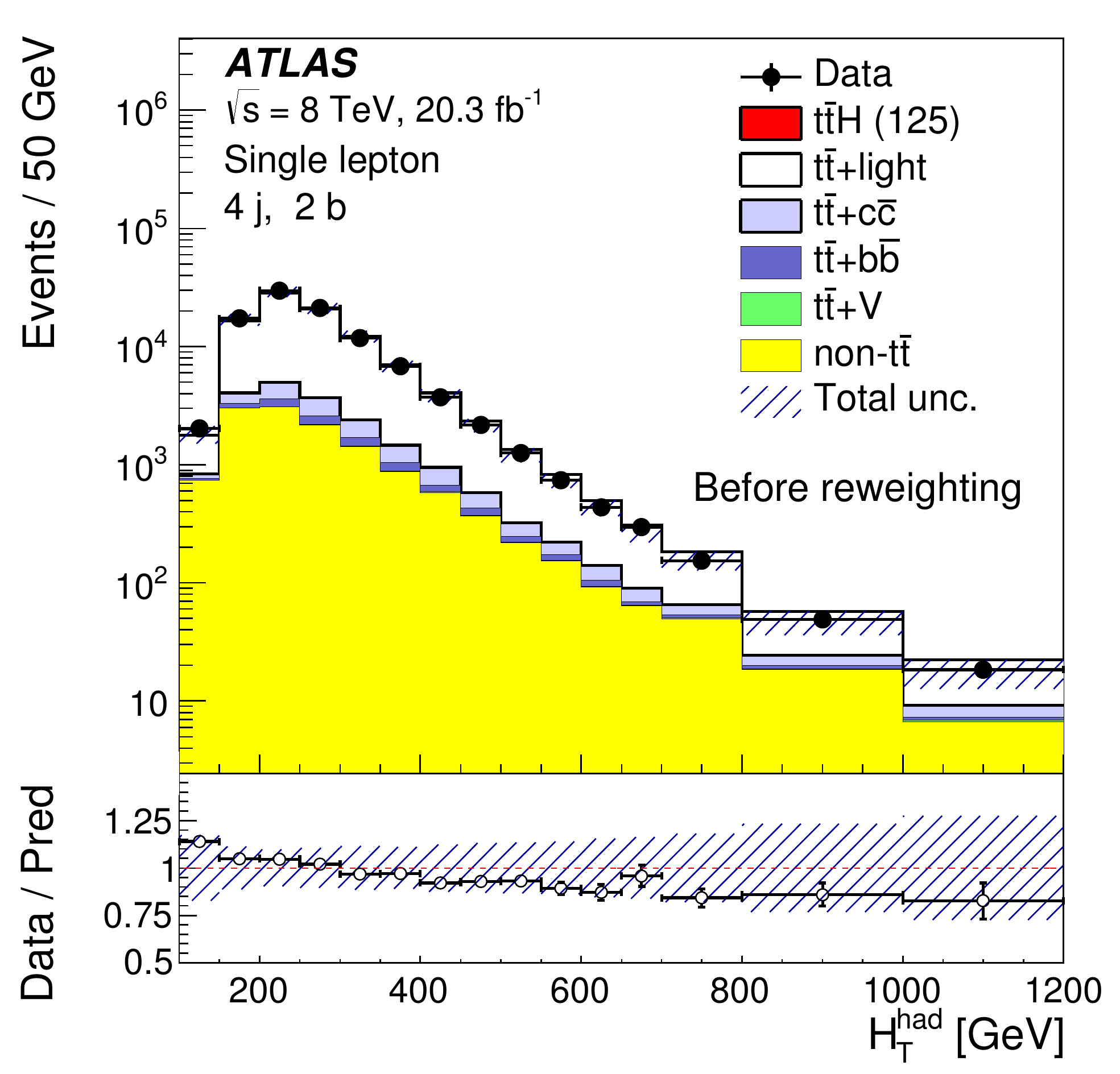}} \label{fig:ttbarseqrw_c}
\subfigure[]{\includegraphics[width=0.38\textwidth]{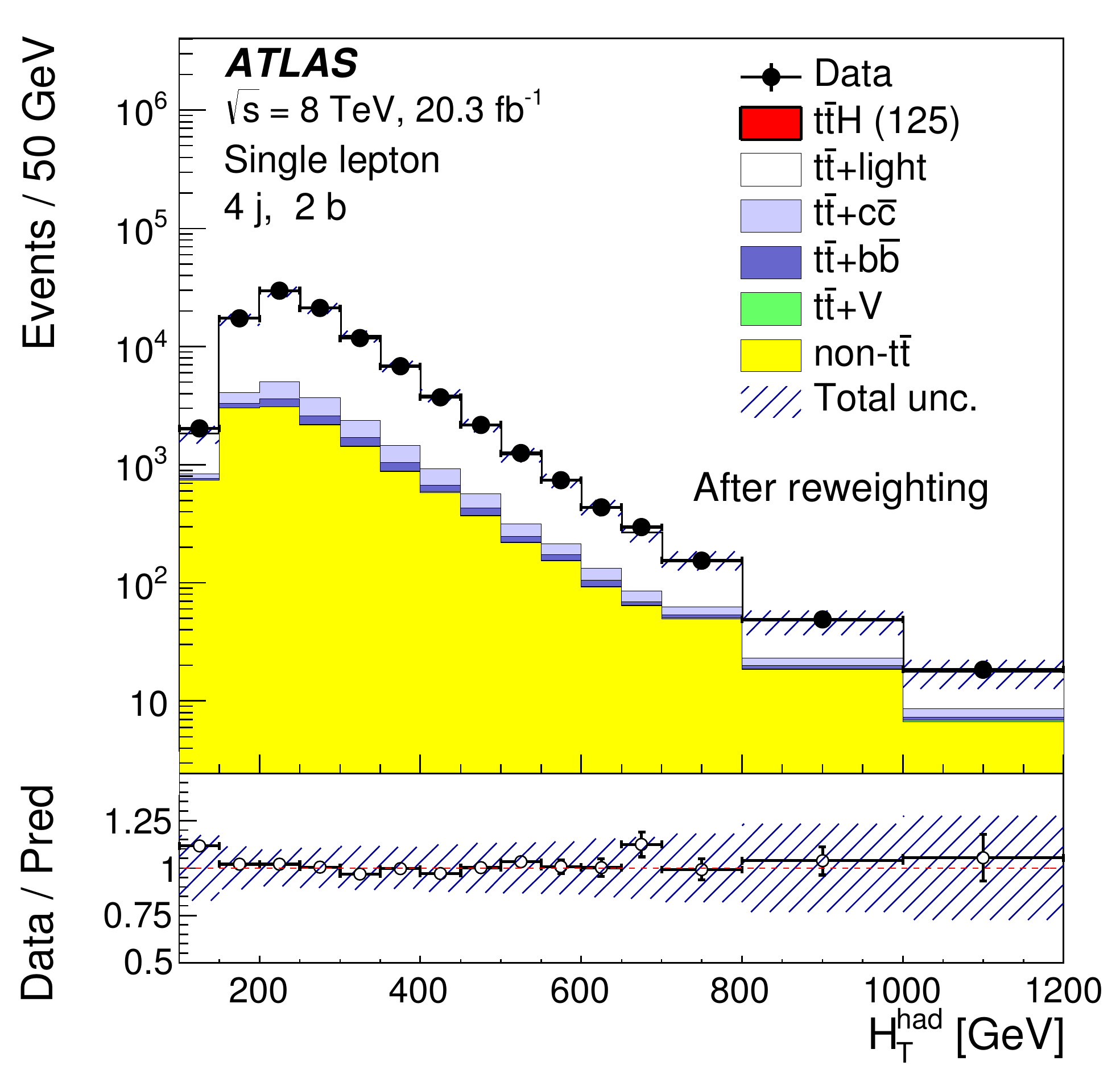}}\label{fig:ttbarseqrw_d}
\caption{The exclusive 2-$b$-tag region of the single-lepton channel before and after the reweighting of the \pt\ of the \ttbar\ system and the \pt\ of the top quark
of the \textsc{Powheg+Pythia} $t\bar{t}$\ sample.   The jet multiplicity distribution (a) before and (b) after the reweighting;  
\hthad\ distributions (c) before and (d) after the reweighting. }
\label{fig:ttbarseqrw}
\end{center}
\end{figure*}

\subsection{Other backgrounds}

The $W/Z$+jets background is estimated from simulation reweighted to account 
for the difference 
in the $W/Z$ \pt\ spectrum between data and simulation~\cite{Zjetsxsec}.  
The heavy-flavour fraction of these simulated backgrounds, i.e. the sum of 
$W/Z$+\bbbar\ and $W/Z$+\ccbar\ processes, is adjusted to
reproduce the relative rates of $Z$ events with no $b$-tags and those with one $b$-tag observed in data.
Samples of $W/Z$+jets events, and diboson production in association with 
jets, are generated using
the {\sc Alpgen 2.14}~\cite{alpgen} leading-order (LO) generator and the 
{\sc CTEQ6L1} PDF set. Parton showers and fragmentation are 
modelled with {\sc Pythia} 6.425 for $W/Z$+jets production and
with {\sc Herwig} 6.520~\cite{herwig} for diboson production. 
The $W$+jets samples are generated with up to five
additional partons, separately for $W$+light-jets,
$Wb\bar{b}$+jets, $Wc\bar{c}$+jets, and $Wc$+jets.
Similarly, the $Z$+jets background is generated with up to five 
additional partons separated in different parton flavours. 
Both are normalised to the respective inclusive NNLO theoretical 
cross section~\cite{vjetsxs}.
The overlap between $WQ\bar{Q}$ ($ZQ\bar{Q}) $($Q=b,c$)
events generated from the matrix element calculation and those
from parton-shower evolution in the $W$+light-jet ($Z$+light-jet) 
samples is removed by an algorithm based on the angular separation
between the extra heavy quarks: if $\Delta R(Q,\bar{Q})>0.4$, the
matrix element prediction is used, otherwise the parton shower
prediction is used. 

The diboson+jets samples are generated with up to three additional partons 
and are normalised to their respecitve NLO theoretical cross sections~\cite{dibosonxs}.

Samples of single top quark backgrounds are generated with 
{\sc Powheg-Box} 2.0 using the {\sc
CT10} PDF set. The samples are interfaced 
to {\sc Pythia} 6.425 with the {\sc CTEQ6L1} set of parton 
distribution functions and Perugia2011C underlying-event tune.  
Overlaps between the \ttbar\ and $Wt$ final states are removed~\cite{mcatnlo_3}.
The single top quark samples are normalised to
the approximate NNLO theoretical cross sections~\cite{stopxs,stopxs_2,stopxs_3}
using the {\sc MSTW2008} NNLO PDF set~\cite{mstw1,mstw2}. 

Samples of $t\bar{t}+V$ are generated with {\sc Madgraph 5} and 
the {\sc CTEQ6L1} PDF set. 
{\sc Pythia} 6.425 with the AUET2B tune~\cite{ATLASUETune2} is used for showering. 
The $t\bar{t}V$ samples are normalised to the NLO cross-section predictions~\cite{ttbarVxs1,ttbarVxs2}.

\subsection{Signal model}
The \ttbar$H$ signal process is modelled
using NLO matrix elements obtained from the HELAC-Oneloop package~\cite{Helac}. 
{\sc Powheg-Box} serves as an interface to
shower Monte Carlo programs. The samples created using this approach are referred 
to as {\sc PowHel} samples~\cite{ttH_NLO}. They are inclusive in Higgs boson decays and are produced using 
the {\sc CT10nlo} PDF set and 
factorisation ($\mu_{\rm F}$) and renormalisation scales set to
$\mu_{\rm F} = \mu_{\rm R} = \mt+\mH/2$.  
The {\sc PowHel} \tth\ sample is showered with      
{\sc Pythia} 8.1~\cite{PythiaManual8} with the {\sc CTEQ6L1} PDF 
and the AU2 underlying-event tune~\cite{ATLASUETune}.  
The $t\bar{t}H$ cross section and Higgs boson decay branching fractions are 
taken from (N)NLO theoretical
calculations~\cite{Dawson:2003zu,Reina:2001sf,Beenakker:2002nc,Beenakker:2001rj,Djouadi:1997yw,Bredenstein:2006rh,Actis:2008ts,Denner:2011mq},
collected in Ref.~\cite{lhcxs}.  In Appendix~\ref{sec:SignalPie}, the relative 
contributions of the Higgs boson decay modes are shown for all regions 
considered in the analysis.

\subsection{Common treatment of MC samples}
All samples using {\sc Herwig} are also interfaced to {\sc Jimmy} 4.31~\cite{jimmy} to simulate the underlying event.  
All simulated samples utilise {\sc Photos 2.15}~\cite{PhotosPaper} to simulate photon radiation and 
{\sc Tauola 1.20}~\cite{TauolaPaper} to simulate $\tau$ decays. 
Events from minimum-bias interactions are simulated with the {\sc Pythia} 8.1 generator with the 
{\sc MSTW2008} LO PDF set  and the AUET2~\cite{ATLASUETune1} tune. They are superimposed on the 
simulated MC events, matching the luminosity profile of the recorded data. 
The contributions from these pileup interactions are simulated both within the same bunch crossing as the 
hard-scattering process and in neighbouring bunch crossings. 

Finally, all simulated MC samples are processed thr- \\ ough a simulation~\cite{atlas_sim}
of the detector geometry and response either using {\sc Geant4}~\cite{geant},
or through a fast simulation of the calorimeter response~\cite{ATLASFastSim}. 
All simulated MC samples are processed through the same reconstruction 
software as the data. Simulated MC events are
corrected so that the object identification efficiencies, energy
scales and energy resolutions match those determined from data control
samples.

Figures~\ref{fig:PRplot_pre}(a) and \ref{fig:PRplot_pre}(b) show a 
comparison of predicted yields to data prior to the fit described in
Sect.~\ref{sec:statmethods} in
all analysis regions in the single-lepton and dilepton channel, respectively.  
The data agree with the SM expectation within the uncertainties of 10-30 \%.
Detailed tables of the event yields prior to the fit and the corresponding 
$S/B$ and $S/\sqrt{B}$ ratios for the single-lepton and dilepton channels
can be found in Appendix~\ref{sec:prefit_tables}. 

\begin{figure*}[!ht]
\centering
\subfigure[]{\includegraphics[width=0.485\textwidth]{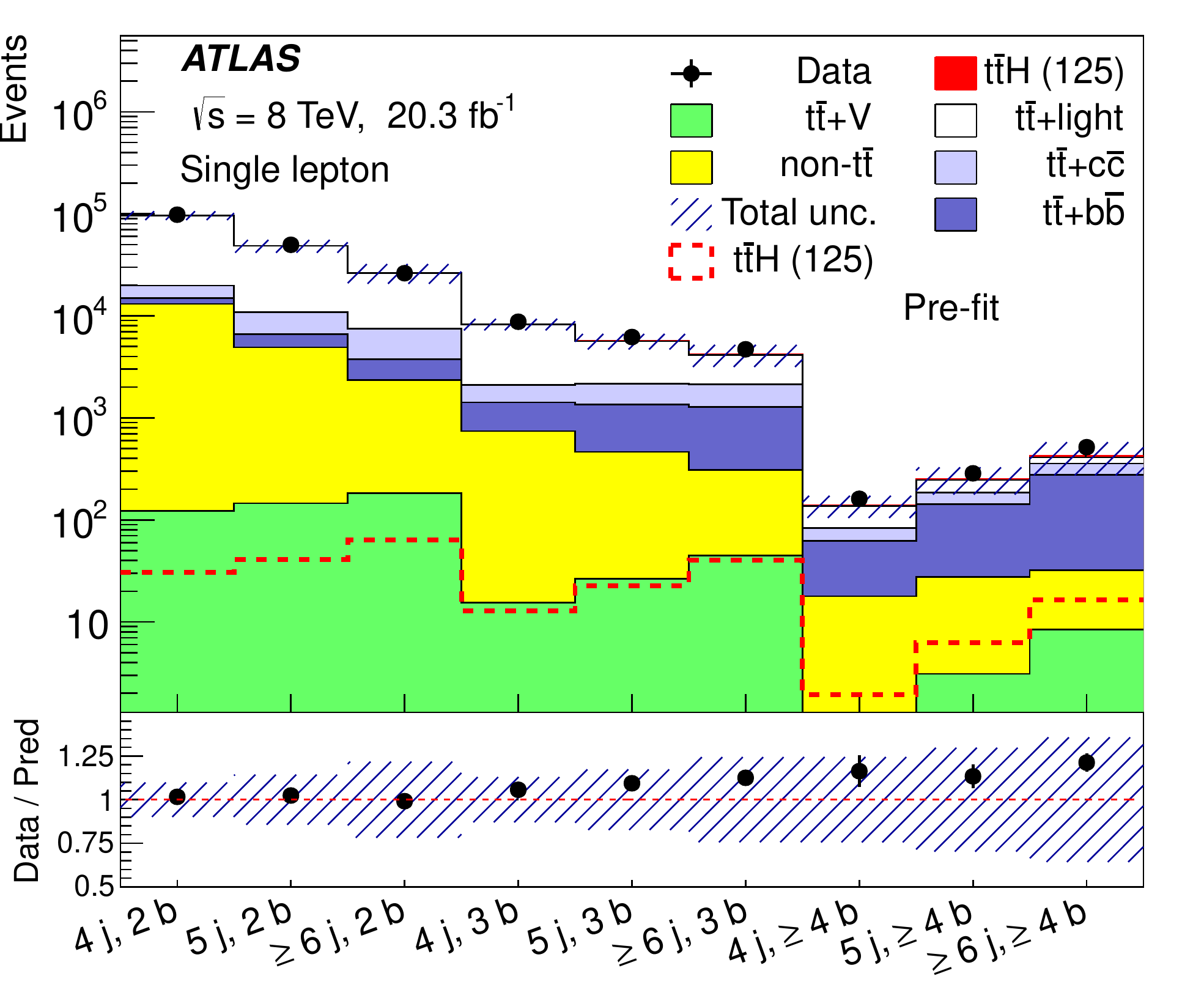}}\label{fig:PRplot_pre_a}  \hspace{0.1cm}
\subfigure[]{\includegraphics[width=0.485\textwidth]{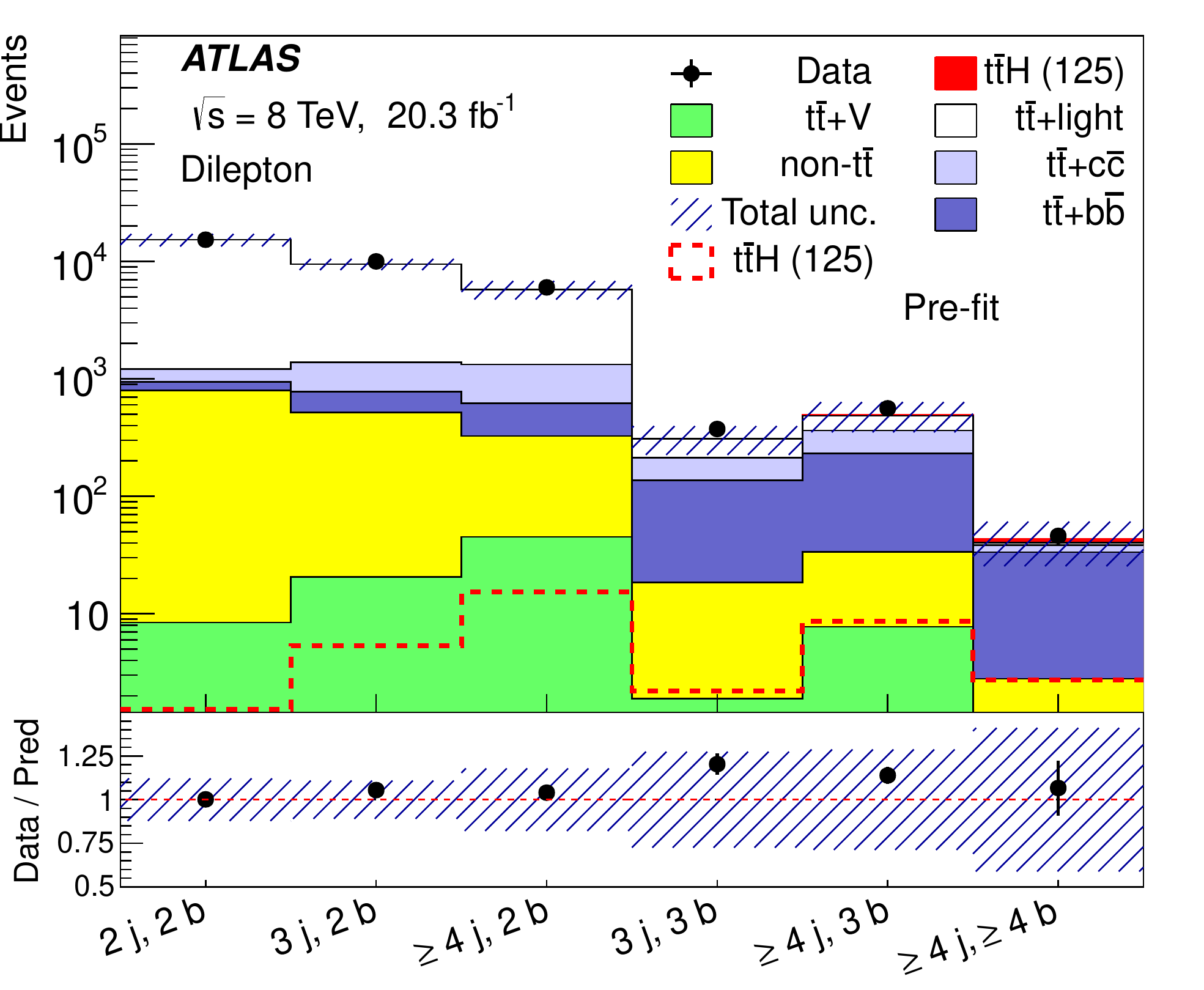}}\label{fig:PRplot_pre_b}
\caption{Comparison of prediction to data in all analysis regions before 
the fit to data in (a)  
the single-lepton channel and (b) the dilepton channel. The signal, normalised to the SM 
prediction, is shown both as a filled red area stacked on the backgrounds and 
separately as a dashed red line. The hashed area corresponds to the total uncertainty on the
yields.}
\label{fig:PRplot_pre}
\end{figure*}

When requiring high jet and $b$-tag multiplicity in the analysis, the
number of available MC events is significantly reduced, leading to large
fluctuations in the resulting distributions for certain samples. 
This can negatively
affect the sensitivity of the analysis through the large statistical uncertainties on
the templates and unreliable systematic uncertainties due to shape fluctuations. 
In order to mitigate this problem, instead of tagging the jets 
by applying the $b$-tagging algorithm, their probabilities to be
$b$-tagged are parameterised as functions of jet flavour, \pt, and \eta.
This allows all events in the sample before $b$-tagging is applied to be used 
in predicting the normalisation and shape after $b$-tagging~\cite{D0btagPRD}. 
The tagging probabilities are derived using an inclusive $t\bar{t}$+jets simulated sample. 
Since the $b$-tagging probability for a $b$-jet coming from top quark decay is slightly 
higher than that of a $b$-jet with the same \pt\ and \eta\ but arising from other sources, they
are derived separately.   
The predictions agree well with the normalisation and shape obtained by 
applying the $b$-tagging 
algorithm directly. The method is applied to all signal and background samples.

\section{Analysis method}
In both the single-lepton and dilepton channels, the analysis uses a neural 
network (NN)   
to discriminate signal from background in each of the regions with significant 
expected \ttH\ signal contribution since the $S/\sqrt{B}$ is very small and the
uncertainty on the background is larger than the signal.  
Those include \fivefour, \sixthree\ and \sixfour\ in the case of the single-lepton channel,
and \fourthreedi\ and \fourfourdi\ in the case of the dilepton channel.  In the dilepton 
channel, an additional NN is used to separate signal from background in 
the \threethree\ channel. Despite a small expected $S/\sqrt{B}$, it nevertheless adds 
sensitivity to the signal due to a relatively high expected $S/B$. In the single-lepton channel, 
a dedicated NN is used in the \fivethree\ region to separate 
\ttbar+light from \ttbar+HF 
backgrounds. The other regions considered in the analysis have lower 
sensitivity, and use $\hthad$ in 
the single-lepton channel, 
and the scalar sum of the jet and lepton $\pt$ (\htlep) in the dilepton 
channel as a discriminant.

The NNs used in the analysis are built using the NeuroBayes~\cite{Neurobayes} package.  
The choice of the variables that enter the NN discriminant is made through 
the ranking procedure implemented in this package based on  
the statistical separation power and the correlation of variables. Several classes of variables were considered: object kinematics, global event variables, 
event shape variables and object pair properties. 
In the regions with $\geq 6$ ($\geq4$) jets, a maximum of seven (five) jets are considered to 
construct the kinematic variables in the single-lepton (dilepton) channel,  first 
using all the $b$-jets, and then incorporating the untagged jets with the 
highest $\pt$. All variables used for the NN training and their pairwise correlations 
are required to be described well in simulation in multiple control regions.    

In the \fivethree\ region in the single-lepton channel, the separation 
between the \ttbar+light and \ttbar+HF events is achieved by 
exploiting the different origin of the third $b$-jet 
in the case of \ttbar+light compared to \ttbar+HF events. 
In both cases, two of the $b$-jets
originate from the \ttbar\ decay. However, in the case of \ttbar+HF events, the 
third $b$-jet
is likely to originate from one of the additional heavy-flavour quarks, whereas in the case
of \ttbar+light events, the third $b$-jet is often matched to a $c$-quark from the
hadronically decaying $W$ boson. Thus, kinematic variables, such as the invariant mass
of the two untagged jets with minimum $\Delta R$, provide discrimination between 
 \ttbar+light and \ttbar+HF events, since the latter presents a distinct peak 
 at the $W$ boson mass which is not present in the former. This and other kinematic 
 variables are used in the dedicated NN used in this region. 
 
In addition to the kinematic variables, two variables calculated using 
the matrix element method (MEM), detailed in Sect.~\ref{sec:ME_MEM},    
are included in the NN training in \sixthree\ and \sixfour\ regions of the 
single-lepton channel. These two variables are
the Neyman--Pearson likelihood ratio ($D1$) (Eq.~(\ref{eq:ME_D1})) and the
logarithm of the summed signal likelihoods (SSLL) (Eq.~(\ref{eq:LL})).
The $D1$ variable provides the best separation between
\tth\ signal and the dominant \ttbar+\bbbar\ background in the \\ \sixfour\ region. 
The SSLL variable further improves the NN performance.  
 
The variables used in the single-lepton and dilepton channels, as well as their ranking in 
each analysis region, are listed in Tables~\ref{tab:varrank} and~\ref{tab:varrankdil}, respectively. 
For the construction of
variables in the \fourfourdi\ region of the dilepton channel, the 
two $b$-jets that are closest in $\Delta R$ to the leptons are considered to originate from the 
top quarks, and the other two $b$-jets are assigned to the Higgs candidate.

\begin{table*}[ht]    
	\centering
	\vspace{0.2cm}
\begin{tabular}{|l|l|c|c|c|c|}
\hline        
\multirow{ 2}{*}{Variable} & \multirow{ 2}{*}{Definition}  & \multicolumn{4}{c|}{NN rank}
\\
\cline{3-6}
& &  $\ge\,6 {\rm j},\ge\,4 {\rm b}$ 
& $\ge\,6 {\rm j},\,3 {\rm b}$ 
& $ 5 {\rm j},\ge\,4 {\rm b}$ 
& $ 5 {\rm j},\,3 {\rm b}$ \\\hline
             $D1$     & Neyman--Pearson MEM discriminant (Eq.~(\ref{eq:ME_D1}))    &   1    &  10  &   -    &  - \\ 
\multirow{2}{*} {\cent}   & Scalar sum of the $\pt$ divided by sum of the $E$ for & \multirow{2}{*} {2}   &  \multirow{2}{*} {2}  &  \multirow{2}{*} {1}  &  \multirow{2}{*} {-} \\ [-0.1cm]
                      & all jets and the lepton   & & & &  \\
\ptjetfive & $\pt$ of the fifth leading jet    &   3    &       7       &       -         &           -     \\ 
 
\multirow{2}{*} {$H1$}  & Second Fox--Wolfram moment computed using &  \multirow{2}{*} {4}  &  \multirow{2}{*} {3}  &  \multirow{2}{*} {2}  &  \multirow{2}{*} {-}      \\ [-0.1cm]
                    & all jets and the lepton      &       &              &                &                 \\ 
 
{\drbbav}  & Average $\Delta R$ for all $b$-tagged jet pairs     &   5    &       6       &       5         &           -       \\ [0.06cm]
 
SSLL  & Logarithm of the summed signal likelihoods (Eq.~(\ref{eq:LL}))       &   6    &       4       &       -         &           -        \\ 
 
\multirow{2}{*} {\mbbmindr} & Mass of the combination of the two $b$-tagged &  \multirow{2}{*} {7}  &  \multirow{2}{*} {12}  &  \multirow{2}{*} {4}  &  \multirow{2}{*} {4}   \\ [-0.1cm]
                      & jets with the smallest $\Delta R$    &    &  &         &          \\ 
 
\multirow{2}{*} {\mbjmaxpt}  & Mass of the combination of a $b$-tagged jet and  &  \multirow{2}{*} {8} &  \multirow{2}{*} {8}  &  \multirow{2}{*} {-}  &   \multirow{2}{*} {-}  \\ [-0.1cm]
                     & any jet with the largest vector sum $\pt$   &     &   &    &          \\ 
 
\multirow{2}{*} {\drbbmaxpt} & $\Delta R$ between the two $b$-tagged jets with the  &  \multirow{2}{*} 9 & \multirow{2}{*} - & \multirow{2}{*} - &  \multirow{2}{*}  -    \\ [-0.1cm]
                      & largest vector sum $\pt$   &  & & &    \\ 
 
\multirow{2}{*} {\drlepbbmindr} & $\Delta R$ between the lepton and the combination & \multirow{2}{*}  {10}  &  \multirow{2}{*}  {11}  &  \multirow{2}{*}  {10}  &  \multirow{2}{*}  {-}  \\ [-0.1cm]
           & of the two $b$-tagged jets with the smallest $\Delta R$   & & & &    \\ 
         
\multirow{2}{*} {\whadmass}  & Mass of the combination of the two untagged jets  &  \multirow{2}{*}  {11}  &  \multirow{2}{*}  9 & \multirow{2}{*} - &  \multirow{2}{*}  2  \\ [-0.1cm]
                      & with the smallest $\Delta R$   &   & & &     \\ 
 
\multirow{2}{*} {\aplab}    & $1.5 \lambda_2$, where $\lambda_2$ is the second eigenvalue of the   & \multirow{2}{*} {12}  & \multirow{2}{*} - & \multirow{2}{*}  8  & \multirow{2}{*}  -  \\ [-0.1cm]
        & momentum tensor~\cite{tensor} built with only $b$-tagged jets    &   & & &    \\ 
 
            {\numjetforty} & Number of jets with $\pt \geq 40\GeV$ &   -    &       1       &       3         &           -        \\ 
 
\multirow{2}{*} {\mbjmindr}   & Mass of the combination of a $b$-tagged jet and &  \multirow{2}{*} -  & \multirow{2}{*} 5  & \multirow{2}{*} -  &  \multirow{2}{*} - \\ [-0.1cm]
                      & any jet with the smallest $\Delta R$  &  &  &  &          \\ 
 
\multirow{2}{*} {\mjjmaxpt}  & Mass of the combination of any two jets with  & \multirow{2}{*} -  &  \multirow{2}{*} - & \multirow{2}{*} 6  &  \multirow{2}{*}  -   \\ [-0.1cm]
                     & the largest vector sum $\pt$   &  &  &  &        \\ 
 
\hthad    & Scalar sum of jet $\pt$    &   -    &       -       &       7         &           -           \\ 
 
\multirow{2}{*} {\mjjmindr}  & Mass of the combination of any two jets with  & \multirow{2}{*} - & \multirow{2}{*} - & \multirow{2}{*} 9 & \multirow{2}{*}  -  \\ [-0.1cm]
         & the smallest $\Delta R$   &   & & &           \\ 
 
\multirow{2}{*} {\mbbmaxpt}  & Mass of the combination of the two $b$-tagged  &  \multirow{2}{*} -  &  \multirow{2}{*} - & \multirow{2}{*} - & \multirow{2}{*} 1   \\ [-0.1cm]
                     & jets with the largest vector sum $\pt$   & & & &            \\ 
 
\multirow{2}{*} {\whadpt}  & Scalar sum of the $\pt$ of the pair of untagged &  \multirow{2}{*} - &  \multirow{2}{*} -  & \multirow{2}{*}  - & \multirow{2}{*} 3  \\ [-0.1cm]
                   & jets with the smallest $\Delta R$    & &  &  &         \\ 
 
\multirow{2}{*} {\mbbmaxM}  & Mass of the combination of the two $b$-tagged &  \multirow{2}{*} - & \multirow{2}{*} -  & \multirow{2}{*}  -  &  \multirow{2}{*} 5  \\ [-0.1cm]
        & jets with the largest invariant mass    &    &  &  &             \\ 
 
\whaddR   & Minimum $\Delta R$ between the two untagged jets    &   -    &       -       &       -         &           6                \\ 
 
\multirow{2}{*} {\Mjjj} & Mass of the jet triplet with the largest vector & \multirow{2}{*}  - & \multirow{2}{*} -  &  \multirow{2}{*}  -  & \multirow{2}{*} 7 \\ [-0.1cm]
                 & sum $\pt$     &   & & &             \\ 
\hline
\end{tabular}
\caption{Single-lepton channel: the definitions and rankings of the variables considered in each of the regions where an NN is used. }
\label{tab:varrank}
\end{table*}

\begin{table*}[ht]
	\centering
\vspace{0.2cm}
\begin{tabular}{|l|l|c|c|c|}
\hline	
\multirow{ 2}{*}{Variable} & \multirow{ 2}{*}{Definition}  & \multicolumn{3}{c|}{NN rank}
\\  
\cline{3-5}
 & & $\ge\,4 {\rm j},\ge\,4 {\rm b}$ 
 & $\ge\,4 {\rm j},\,3 {\rm b}$
 & $3 {\rm j},\,3 {\rm b}$ \\\hline
\maxdeta & Maximum $\Delta \eta$ between any two jets in the event & 	1	&	1	&	1	\\ 
\multirow{2}{*}{\mbbmindr} & Mass of the combination of the two $b$-tagged jets with &\multirow{2}{*}	2 & \multirow{2}{*}8 & \multirow{2}{*} -\\[-0.1cm]
            & the smallest $\Delta R$  &	&  &	\\	
\multirow{2}{*}{\mbb} & Mass of the two $b$-tagged jets from the Higgs candidate &\multirow{2}{*}	3 &\multirow{2}{*}-&\multirow{2}{*}-\\ [-0.1cm]
          & system &	 & &	\\ 
\mindrhl & $\Delta R$ between the Higgs candidate and the closest lepton & 	4	&	5	&	-	\\ 
\multirow{2}{*}{\nhiggsthirty} & Number of Higgs candidates within 30 GeV of the Higgs  &	\multirow{2}{*}5&\multirow{2}{*}2&\multirow{2}{*}5\\ [-0.1cm]
                             & mass of 125 \gev &	& &	\\ 
\multirow{2}{*}{\drbbmaxpt} & $\Delta R$ between the two $b$-tagged jets with the largest &\multirow{2}{*}6&\multirow{2}{*}4	&\multirow{2}{*}8\\ [-0.1cm]
           & vector sum $\pt$ & & &	\\ 
\multirow{2}{*}{\aplaj}    & $1.5 \lambda_2$, where $\lambda_2$ is the second eigenvalue of the   & \multirow{2}{*} {7}  & \multirow{2}{*} 7 & \multirow{2}{*}  -  \\ [-0.1cm]
        & momentum tensor built with all jets    &   & &    \\ 
\mindijetmass & Minimum dijet mass between any two jets   & 8	&	3	&	2	\\ [0.06cm]
\maxdrhl & $\Delta R$ between the Higgs candidate and the furthest lepton &	9	&	-	&	-	\\ 
\multirow{2}{*}{\mclosest} & Dijet mass between any two jets closest to the Higgs & \multirow{2}{*}{10}	&\multirow{2}{*} - &\multirow{2}{*} {10}\\ [-0.1cm]
           & mass of 125 \gev & & &	\\ 
\htlep &  Scalar sum of jet $\pt$ and lepton $\pt$ values &		-	&	6	&	3	\\
\multirow{2}{*}{\drbbmaxm} & $\Delta R$ between the two $b$-tagged jets with the largest &	\multirow{2}{*}-&\multirow{2}{*}9&\multirow{2}{*}-\\ [-0.1cm]
            & invariant mass &	& &	\\ 
\mindrlj & Minimum $\Delta R$ between any lepton and jet   &		-	&	10	&	-	\\ 
\multirow{2}{*}{\cent} & Sum of the $\pt$ divided by sum of the $E$ for all jets and &\multirow{2}{*}- &\multirow{2}{*}-&\multirow{2}{*}7	\\ [-0.1cm]
                 & both leptons & & & 	\\ 
\multirow{2}{*}{\mjjmaxpt} & Mass of the combination of any two jets with the largest &\multirow{2}{*}-&\multirow{2}{*}-&\multirow{2}{*}	9\\ [-0.1cm]
                    & vector sum $\pt$ & & &	 \\ 
\multirow{2}{*} {$H4$} & Fifth Fox--Wolfram moment computed using all jets and 	&\multirow{2}{*}-&\multirow{2}{*}-&\multirow{2}{*}4\\ [-0.1cm]
              & both leptons	& & &	\\ 
\ptjetthree & $\pt$ of the third leading jet  &	-	&	-	&	6	\\ 
\hline
\end{tabular}
\caption{Dilepton channel: the definitions and rankings of the variables considered in each of the regions where an NN is used. }
\label{tab:varrankdil}
\end{table*}

Figures~\ref{fig:Discriminationlj} and~\ref{fig:Discriminationdil} show the 
distribution of the NN discriminant for the \tth\ signal and
background in the single-lepton and dilepton channels, respectively, in the
signal-rich regions. In particular, Fig.~\ref{fig:Discriminationlj}(a) shows
the separation between the \ttbar+HF and \ttbar+light-jet production achieved by a
dedicated NN in the \fivethree\ region in the single-lepton channel.  The distributions in
the highest-ranked input variables from each of the NN regions are shown
in Appendix~\ref{sec:separation}.

\begin{figure*}[ht!]
\centering
\subfigure[]{\includegraphics[width=0.325\textwidth]{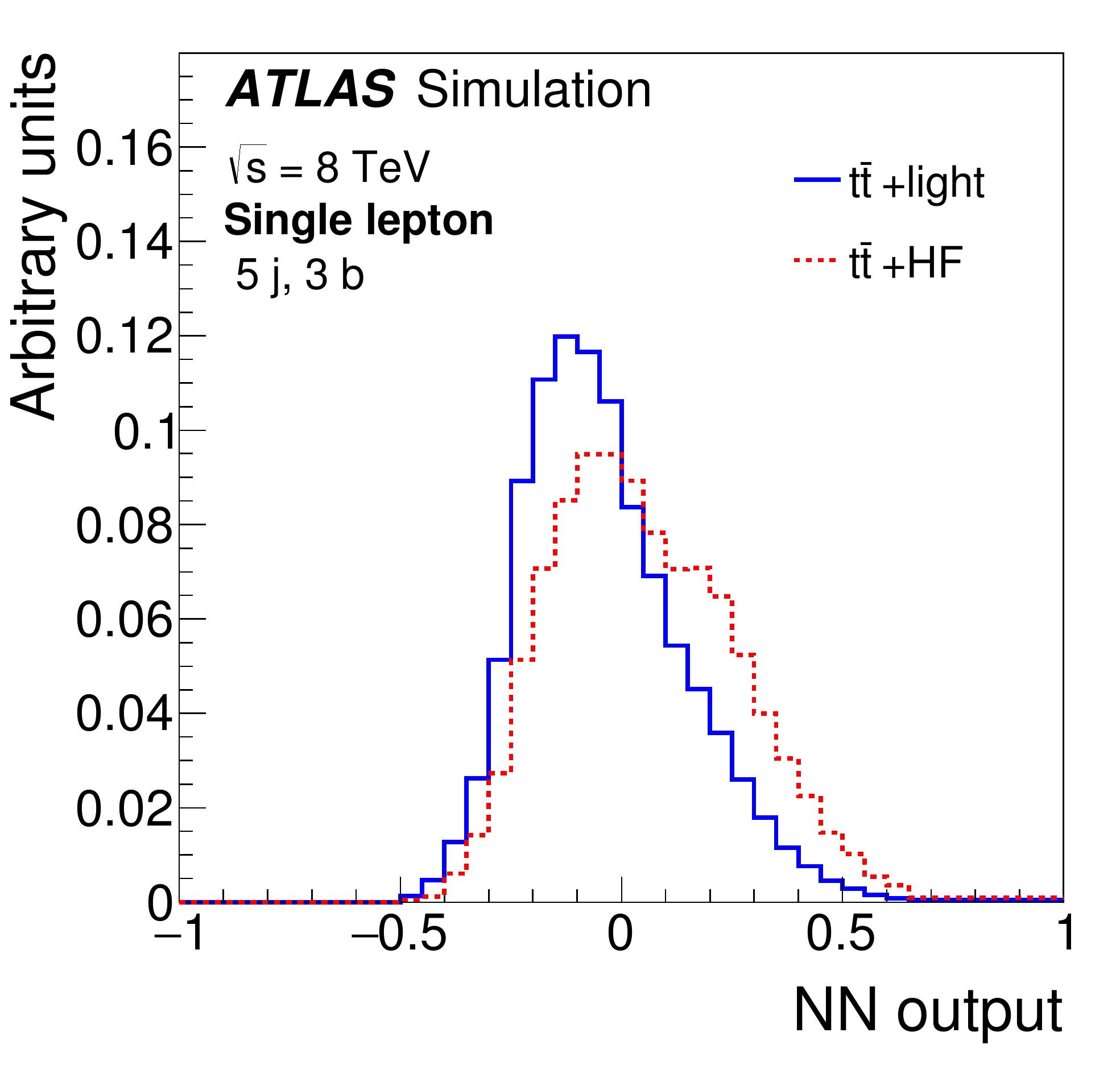}}\label{fig:Discriminationlj_a}
\subfigure[]{\includegraphics[width=0.325\textwidth]{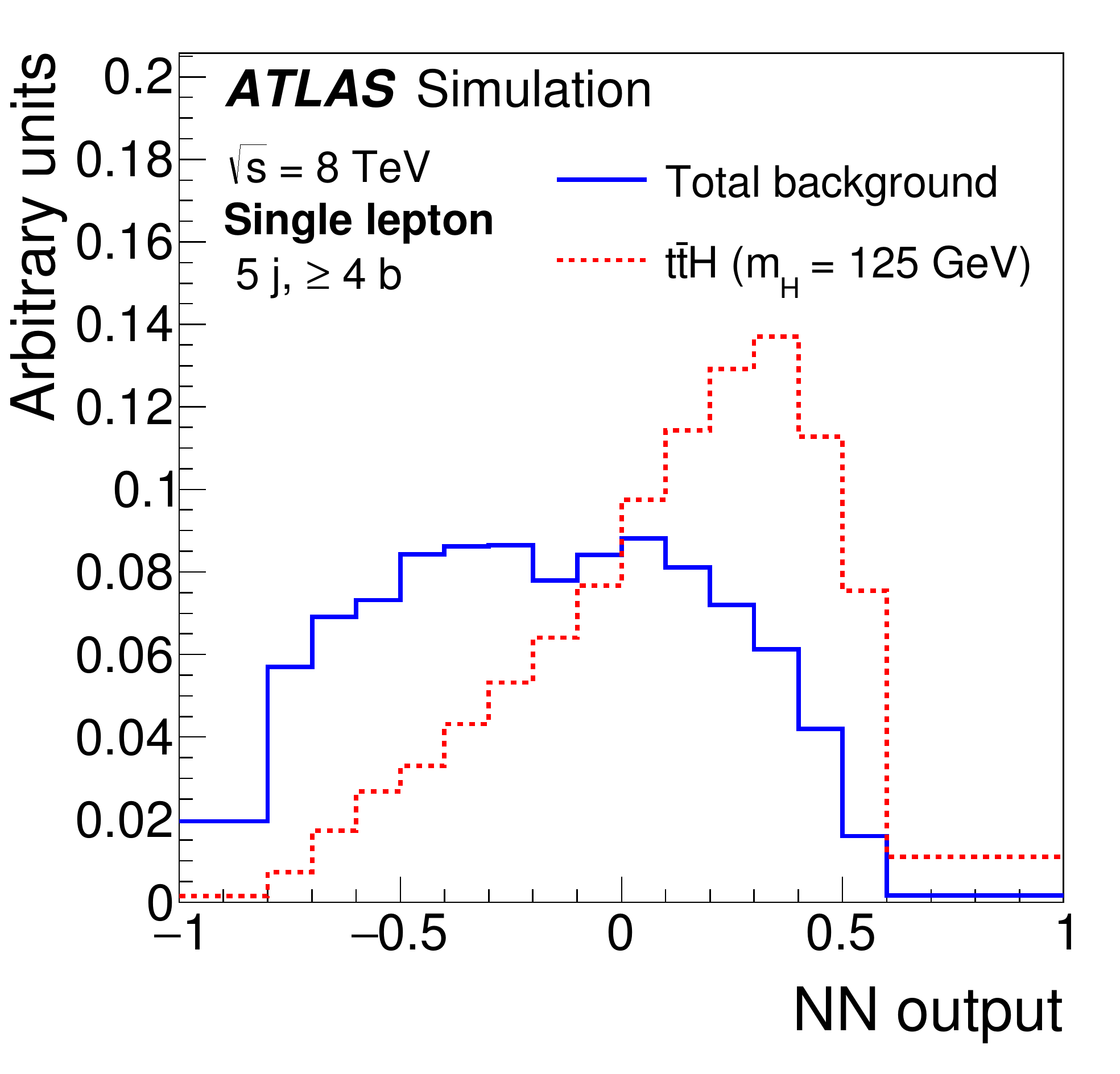}}\label{fig:Discriminationlj_b}\\
\subfigure[]{\includegraphics[width=0.325\textwidth]{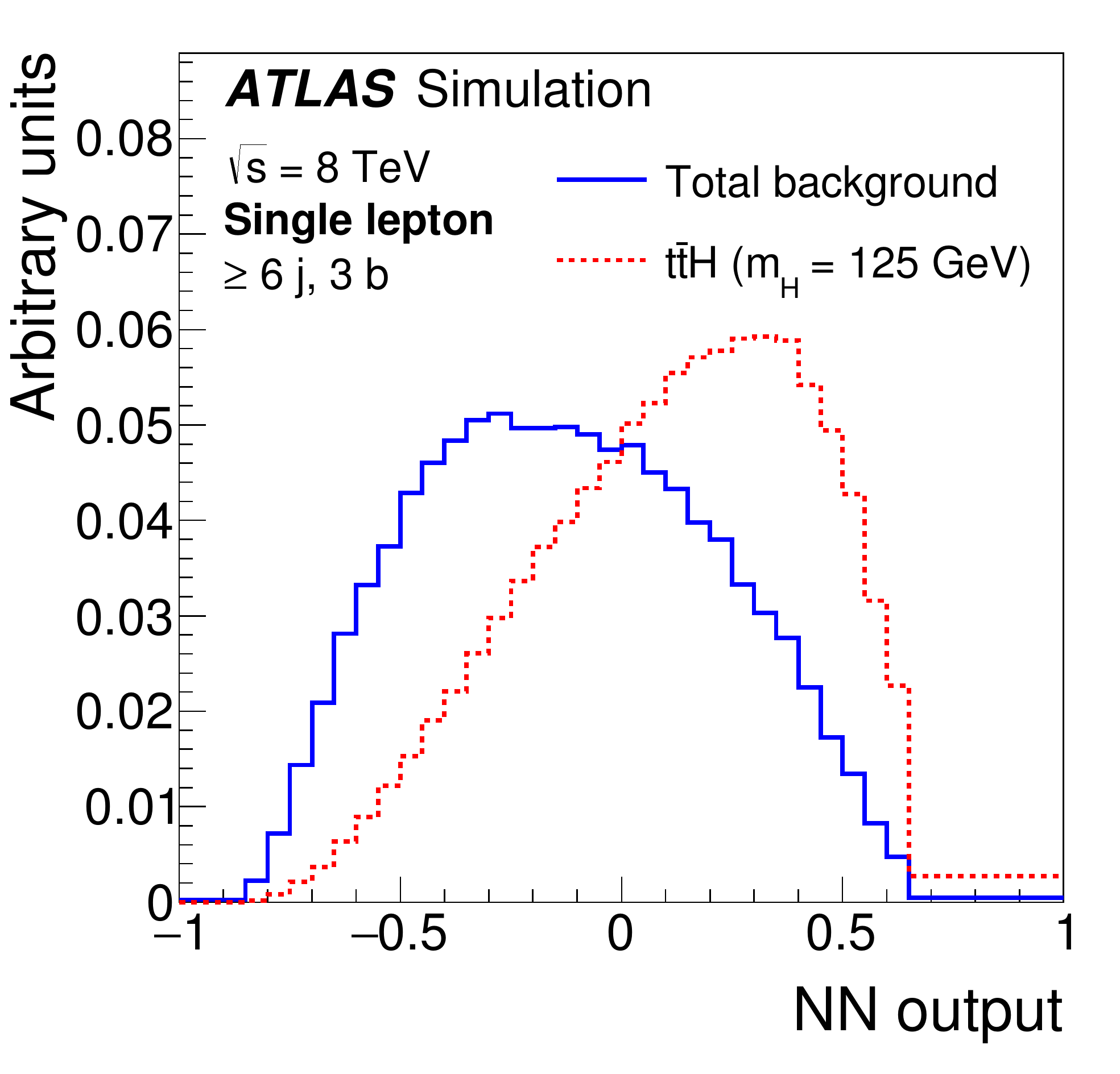}}\label{fig:Discriminationlj_c}
\subfigure[]{\includegraphics[width=0.325\textwidth]{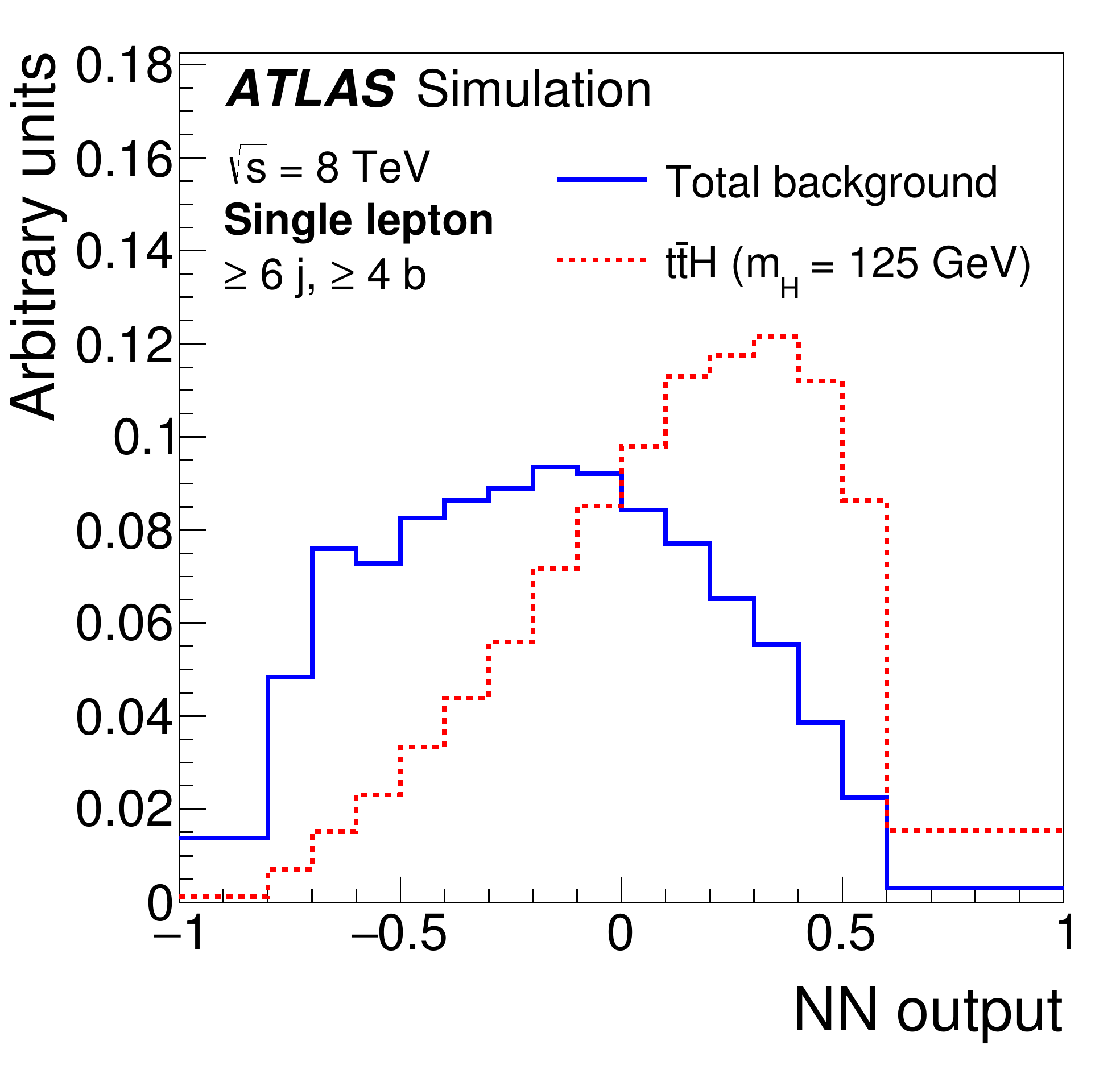}}\label{fig:Discriminationlj_d}
\caption{Single-lepton channel: NN output for the different regions. In the \fivethree\ region (a), 
the \ttbar+HF production is considered as signal and \ttbar+light as background
whereas in the \fivefour\ (b), \sixthree\ (c), and \sixfour\ (d) regions the NN output is for the \tth\ signal 
and total background.  The distributions are normalised to unit area. }
\label{fig:Discriminationlj}
\end{figure*}

\begin{figure*}[ht!]
\centering
\vspace{0.7cm}
\subfigure[]{\includegraphics[width=0.325\textwidth]{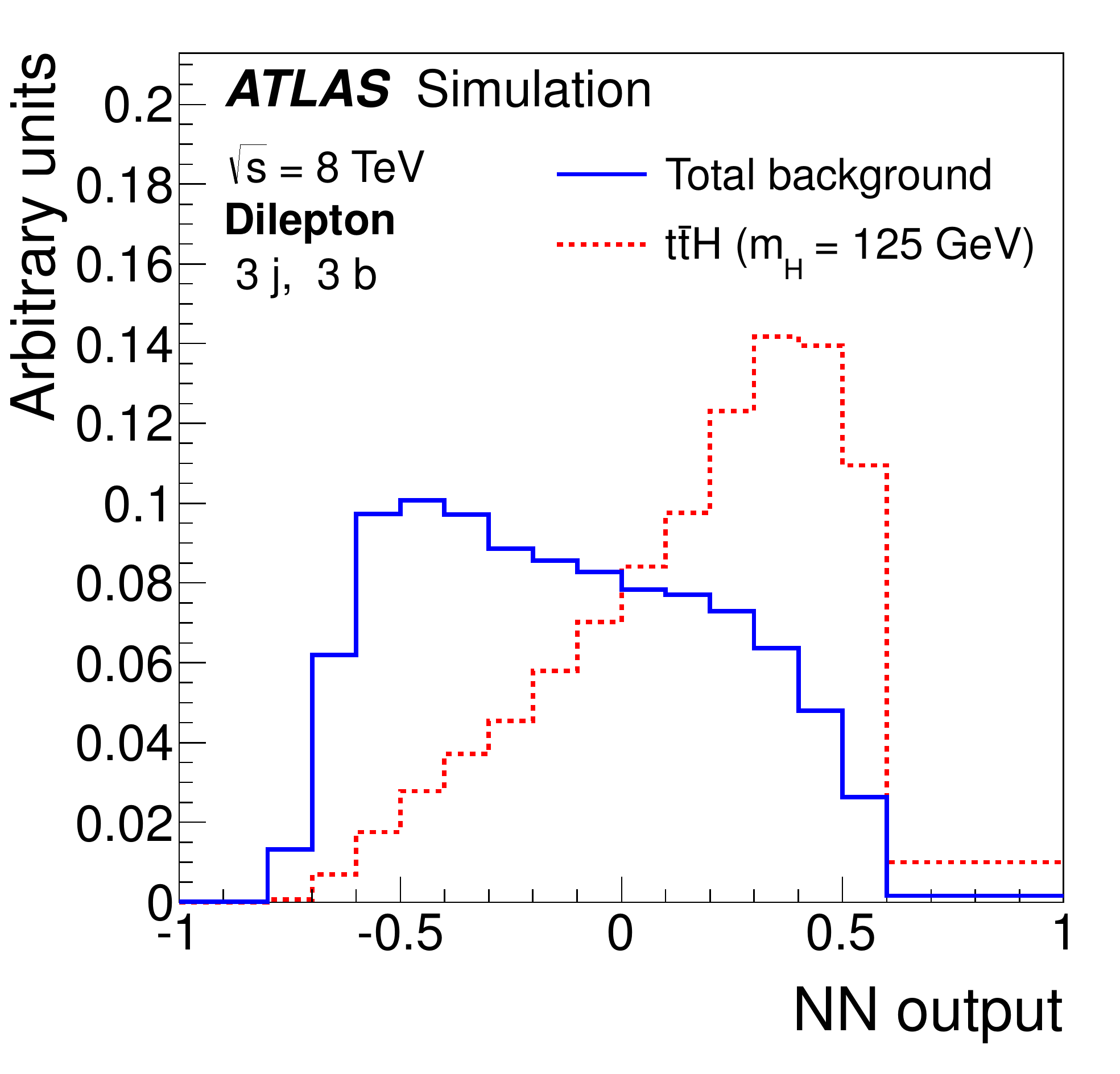}}\label{fig:Discriminationdil_a}
\subfigure[]{\includegraphics[width=0.325\textwidth]{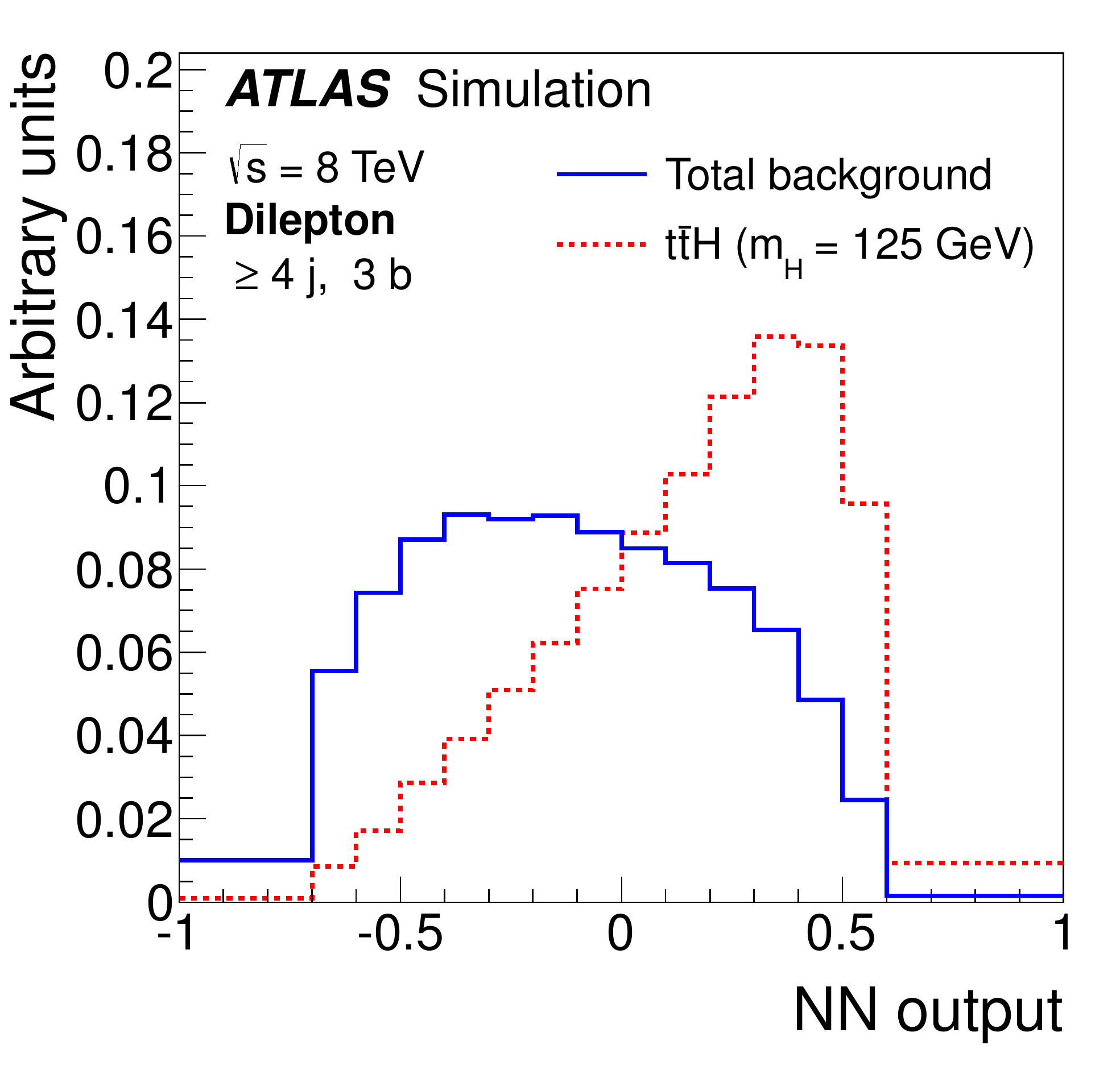}}\label{fig:Discriminationdil_b}
\subfigure[]{\includegraphics[width=0.325\textwidth]{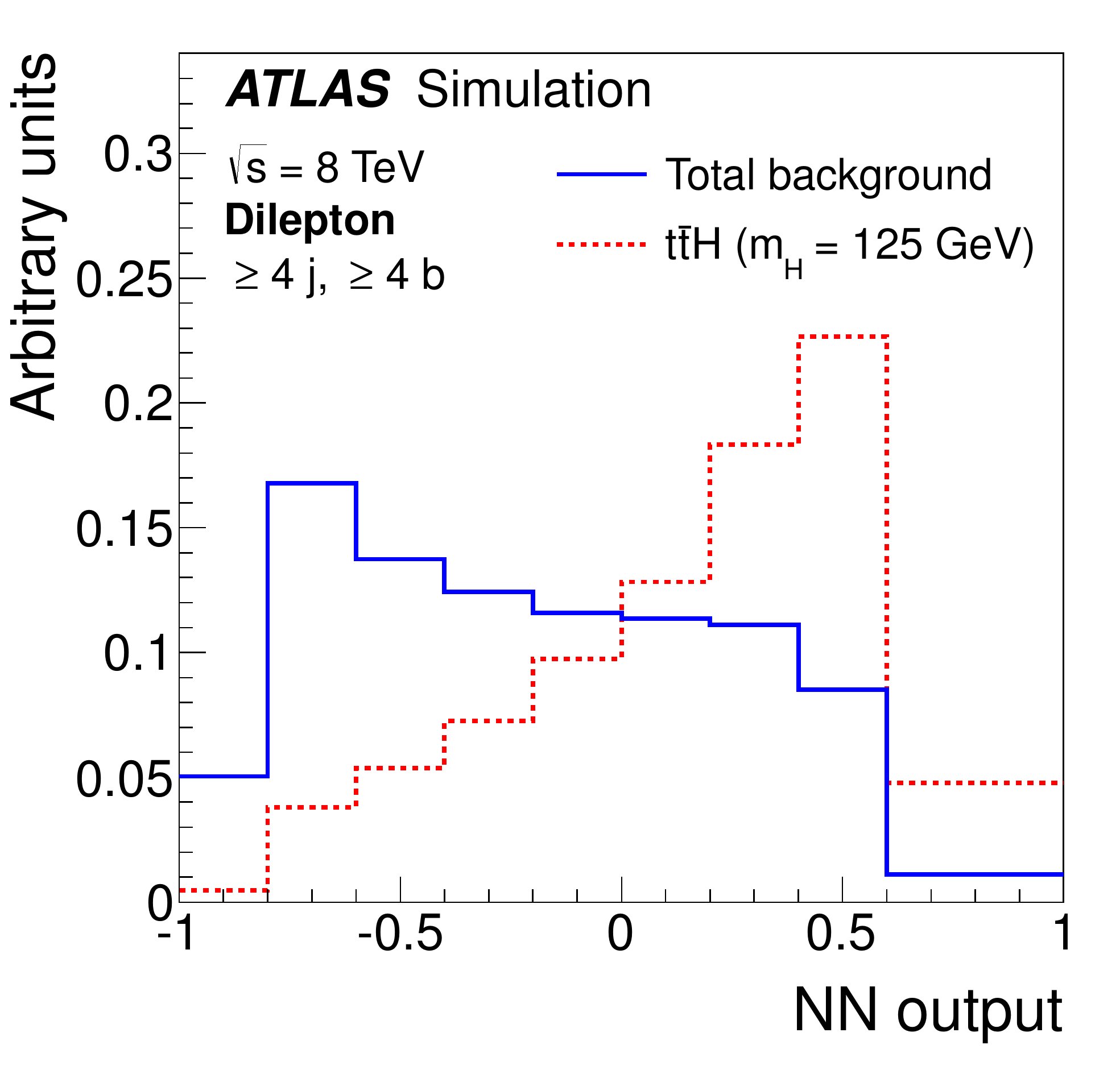}}\label{fig:Discriminationdil_c}
\caption{Dilepton channel: NN output for the \tth\ signal and total background 
in the (a) \threethree, (b) \fourthreedi, and (c) \fourfourdi~regions.  The distributions are normalised to unit area.}
\label{fig:Discriminationdil}
\end{figure*}

For all analysis regions considered in the fit, the \tth\ signal includes all 
Higgs decay modes. They are also included in the NN training. 

The analysis regions have different contributions from various systematic uncertainties, 
allowing the combined fit to constrain them. 
The highly populated \fourtwolj\ and \twotwo\ regions in the single-lepton and dilepton channels, respectively, provide a powerful 
constraint on the overall normalisation of the $t\bar{t}$ background.  
The \fourtwolj, \fivetwo\ and \sixtwo\ regions in the single-lepton channel and
the \twotwo, \threetwo\ and \fourtwodi\ regions in the 
dilepton channel are almost pure in $t\bar{t}$+light-jets background and provide an important constraint on
$t\bar{t}$ modelling uncertainties both in terms of normalisation and shape.
Uncertainties on $c$-tagging are reduced by exploiting the large contribution of $W\to c s$ decays in the
$t\bar{t}$+light-jets background populating the \fourthreelj\ region in the single-lepton channel. 
Finally, the consideration of regions with exactly 3 and $\geq$ 4 $b$-jets in both channels, having different fractions
of \ttbar+\bbbar\ and \ttbar+\ccbar\ backgrounds, provides the ability to constrain
uncertainties on the \ttbar+\bbbar\ and \ttbar+\ccbar\ normalisations.

\section{The matrix element method}
\label{sec:ME_MEM}

The matrix element method~\cite{kondo1} has been used by the D0 and CDF 
collaborations for precision measurements of the top quark 
mass~\cite{d0_topmass,cdf_topmass} and for the observations of single top 
quark production~\cite{d0_stop,cdf_stop}. Recently this technique has 
been used for the \tth\ search by the
CMS experiment~\cite{CMSMEM}. By directly linking 
theoretical calculations and observed quantities, it makes the most complete 
use of the kinematic information of a given event. 

The method calculates the probability density function 
of an observed event to be consistent with physics process $i$ 
described by a set of parameters $\MB{\alpha}$. 
This probability density function $P_i \left ( \MB{x} | \MB{\alpha} \right )$
is defined as

\begin{eqnarray}
\begin{split}
P_i\left(\MB{x}|\MB{\alpha}\right) =  &\frac{(2\pi)^4}{\sigma_i^{\mathrm{exp}}\left(\MB{\alpha}\right)} 
\int \dif{\SUB{p}{A}} \dif{\SUB{p}{B}} \; \MB{f}\left(\SUB{p}{A}\right) \MB{f}\left(\SUB{p}{B}\right) \\ 
& \frac{\left|\mathcal{M}_{i}\left(\MB{y}|\MB{\alpha}\right)\right|^2}{\mathcal{F}} \;
W\left(\MB{y}|\MB{x}\right) \; \dif{\Phi_N}\left(\MB{y}\right)
\end{split}
\end{eqnarray}

\noindent and is obtained by numerical integration
over the entire phase space of the initial- and final-state particles. In this
equation, $\MB{x}$ and $\MB{y}$ represent the four-momentum vectors of all final-state particles at reconstruction and parton level, respectively. 
The flux factor $\mathcal{F}$ and the Lorentz-invariant phase space element 
$\dif{\Phi_N}$ describe the kinematics of the process. The transition matrix 
element $\mathcal{M}_{i}$ is defined by the Feynman diagrams of the hard process. 
The transfer functions $W\left(\MB{y}|\MB{x}\right)$ map the detector quantities 
$\MB{x}$ to the parton level quantities $\MB{y}$. Finally, the cross section 
$\sigma_i^{\mathrm{exp}}$ normalises $P_i$ to unity 
taking acceptance and efficiency into account.

The assignment of reconstructed objects to final-state partons in the hard process 
contains multiple ambiguities. The process probability density is calculated for each allowed assignment permutation 
of the jets to the final-state quarks of the hard process.
A process likelihood function can then be built 
by summing the process probabilities for the $N_{\rm p}$ allowed assignment 
permutation,

\begin{equation}
\mathcal{L}_{i}  \left (\MB{x}|\MB{\alpha}\right )  = \sum_{p=1}^{N_{\rm p}} \strut P_{i}^{\rm p}
\left (\MB{x}|\MB{\alpha} \right ). 
\label{eq:LL}
\end{equation}

The process probability densities are used to distinguish signal from 
background events by calculating the likelihood ratio of the signal and 
background processes contributing with fractions $f_{\textrm{bkg}}$,

\begin{equation}
r_{\textrm{sig}} \left  ( \MB{x} | \MB{\alpha} \right ) = 
\frac{\mathcal{L}_{\textrm{sig}} \left (\MB{x}|\MB{\alpha} \right )}
{\sum\limits_{\textrm{bkg}} f_{\textrm{bkg}} \mathcal{L}_{\textrm{bkg}} 
\left (\MB{x}|\MB{\alpha} \right ) } \label{eq:neyman} . 
\end{equation} 

This ratio, according to the Neyman--Pearson lem- \\ ma~\cite{neymanpearson}, is the 
most powerful 
discriminant between signal and background processes. In the analysis, 
this variable is used as input to the NN 
along with other kinematic variables. 
 
Matrix element calculation methods are generated with {\sc Madgraph 5} in LO. 
The transfer functions are obtained from simulation following a similar procedure as described in 
Ref.~\cite{klfitter}. For the modelling of the parton
distribution functions the {\sc CTEQ6L1} set from the \textsc{LHAPDF} package~\cite{lhapdf} 
is used. 

The integration is performed using \textsc{VEGAS}~\cite{VEGAS}.  Due to the complexity and 
high dimensionality, adaptive MC techniques~\cite{gsl}, 
simplifications and approximations are needed to obtain results 
within a reasonable computing time.
In particular, only the numerically most significant contributing helicity states of a 
process hypothesis for a given event,  
identified at the start of each integration, are evaluated. This does not 
perceptibly decrease the separation power but reduces the calculation time by 
more than an order of magnitude. 
Furthermore, several approximations are made to improve the {\sc VEGAS} convergence rate. 
Firstly, the dimensionality of integration is reduced by assuming that the final-state object 
directions in \eta{} and $\phi$ as well as charged lepton momenta are well measured, and therefore 
the corresponding transfer functions are represented by $\delta$ functions. 
The total momentum conservation and a negligible transverse momentum of the 
initial-state partons allow for further reduction.
Secondly, kinematic transformations are utilised to optimise the integration over the remaining 
phase space by aligning the peaks of the integrand with the integration dimensions. 
The narrow-width approximation is applied to the leptonically decaying $W$ boson. 
This leaves three $b$-quark energies, one light-quark energy, the hadronically decaying 
$W$ boson mass and the invariant mass of the two $b$-quarks
originating from either the Higgs boson for the signal  
or a gluon for the background as the remaining parameters which define the integration phase space.
The total integration volume is restricted based  upon the observed values and the width 
of the transfer functions and of the 
propagator peaks in the matrix elements. Finally, the likelihood 
contributions of all allowed assignment permutations are coarsely integrated, and  
only for the leading twelve assignment permutations is the full integration performed, 
with a required precision decreasing according to their relative contributions.

The signal hypothesis is defined as a SM Higgs boson produced in association
with a top-quark pair as shown in Fig.~\ref{fig:feyman}(a), (b). 
Hence no coupling of the Higgs boson to the $W$ boson is accounted for 
in $| \mathcal{M}_{i}|^{2}$ 
to allow for a consistent treatment when performing the kinematic transformation. 
The Higgs boson is required to decay into a pair of $b$-quarks, while the top-quark pair decays 
into the single-lepton channel. 
For the background hypothesis, only the diagrams of the irreducible 
\ttbb\ background are considered. Since it dominates the most signal-rich analysis regions, 
inclusion of other processes does not improve the separation 
between signal and background. No gluon radiation from the 
final-state quarks is  
allowed, since these are kinematically suppressed and difficult to treat in any kinematic 
transformation aiming for phase-space alignment during the integration process.
In the definition of the signal and background hypothesis the LO diagrams are 
required to have a top-quark pair as an intermediate state resulting in exactly 
four $b$-quarks, two light quarks, one charged lepton (electron or muon) and one neutrino 
in the final state. Assuming lepton universality and invariance under charge 
conjugation,  
diagrams of only one lepton flavour and of only negative charge (electron) are considered.
The probability density function calculation of the signal and background is only performed in the \sixthree\ 
and \sixfour\ regions of the single-lepton channel. Only six reconstructed 
jets are considered in the calculation: 
the four jets with the highest value of the probability to be a $b$-jet returned by the 
$b$-tagging algorithm (i.e. the highest $b$-tagging weight) 
and two of the remaining jets with an 
invariant mass closest to the $W$ boson mass of 80.4 \gev. If a jet is $b$-tagged it cannot be 
assigned to a light quark in the matrix element description. In the case of more than four 
$b$-tagged jets, only the four with the highest $b$-tagging weight are treated as $b$-tagged. 
Assignment permutations between the two light quarks of the hadronically decaying $W$ 
boson and between 
the two $b$-quarks originating from the Higgs boson or gluon result in the same 
likelihood value and are thus not considered.
As a result there are in total 12 and 36 assignment permutations in the \sixfour\ 
and \sixthree\ region, respectively, which need to be evaluated in the coarse integration phase.

Using the \tth\ process as the signal hypothesis and the \ttbb\ process as the 
background hypothesis, a slightly modified version of Eq.~(\ref{eq:neyman}) is used to define 
the likelihood ratio $D1$:

\begin{equation}
 D1=\frac{\mathcal{L}_{t\bar{t}H}}{{\mathcal{L}_{t\bar{t}H}} + 
 \alpha \cdot \mathcal{L}_{t\bar{t}+b\bar{b}}}\
\label{eq:ME_D1} ,
\end{equation}

\noindent where $\alpha=0.23$ is a relative normalisation factor chosen to optimise the
performance of the discriminant given the finite bin sizes of the $D1$ distribution. 
In this definition, 
signal-like and background-like events have $D1$ values close to one and zero, 
respectively.   
The logarithm of the summed signal likelihoods defined by Eq.~(\ref{eq:LL}) 
and the ratio $D1$ are included in the 
NN training in both the \sixthree\ and \sixfour\ regions. 

\section{Systematic uncertainties}
\label{sec:SystematicUncertainties}
				   
Several sources of systematic uncertainty are considered that
can affect the normalisation of signal and background and/or the shape
of their final discriminant distributions.  Individual
sources of systematic uncertainty are considered uncorrelated. 
Correlations of a given systematic effect are maintained across
processes and channels.  Table~\ref{tab:SystSummary} presents a
summary of the sources of systematic uncertainty considered in the analysis, 
indicating whether they are taken to be nor-\\ malisation-only, shape-only, or to
affect both shape and normalisation.  In Appendix~\ref{sec:norm_syst}, the 
normalisation impact of the systematic uncertainties are shown on the \ttbar\ background 
as well as on the \tth\ signal.

In order to reduce the degradation of the sensitivity of the search due to systematic uncertainties,
they are fitted to data in the statistical analysis, exploiting the constraining
power from the background-dominated regions described in
Sect.~\ref{sec:EventSelection}.  Each systematic uncertainty is represented
by an independent parameter, referred to as a ``nuisance parameter'', and
is fitted with a Gaussian prior for the shape differences and a
log-normal distribution for the normalisation. 
They are centred around zero with a width that corresponds to the given uncertainty. 

\begin{table}[ht!]
\centering
\vspace{0.2cm}
\begin{tabular}{lcc}
\hline\hline
Systematic uncertainty & Type  & Comp.\\
\hline
Luminosity                  &  N & 1\\\hline\hline
{\bf Physics Objects}                 &   & \\
Electron                  &  SN & 5 \\
Muon                      &  SN & 6 \\\hline
Jet energy scale            & SN & 22\\
Jet vertex fraction         & SN    & 1\\
Jet energy resolution       & SN & 1\\
Jet reconstruction      & SN & 1\\ \hline
$b$-tagging efficiency      & SN & 6\\
$c$-tagging efficiency      & SN & 4\\
Light-jet tagging efficiency    & SN & 12\\ 
High-\pt\ tagging efficiency  & SN & 1 \\ \hline\hline
{\bf Background Model}                 &   & \\
$t\bar{t}$ cross section    &  N & 1\\
$t\bar{t}$ modelling: \pt\ reweighting   & SN & 9\\
$t\bar{t}$ modelling: parton shower & SN & 3\\
$t\bar{t}$+heavy-flavour: normalisation & N & 2 \\
$t\bar{t}$+$c\bar{c}$: \pt\ reweighting  & SN & 2 \\
$t\bar{t}$+$c\bar{c}$: generator & SN & 4 \\
$t\bar{t}$+$b\bar{b}$: NLO Shape & SN & 8 \\\hline
$W$+jets normalisation      &  N & 3\\
$W$ \pt\ reweighting     &  SN & 1\\
$Z$+jets normalisation      &  N & 3\\
$Z$ \pt\ reweighting     &  SN & 1\\\hline
Lepton misID normalisation  &  N & 3\\
Lepton misID shape          &  S & 3\\\hline
Single top cross section    &  N & 1\\
Single top model            &  SN & 1\\
Diboson+jets normalisation  &  N & 3\\
$t\bar{t}+V$ cross section   &  N & 1\\
$t\bar{t}+V$ model           &  SN & 1\\ \hline\hline
{\bf Signal Model}                 &   & \\
$t\bar{t}H$ scale           & SN & 2 \\
$t\bar{t}H$ generator       & SN & 1 \\
$t\bar{t}H$ hadronisation   & SN & 1 \\
$t\bar{t}H$ PDF   & SN & 1 \\
\hline\hline
\end{tabular}
\caption{\label{tab:SystSummary} List of systematic uncertainties considered. 
An ``N" means that the uncertainty is taken as normalisation-only for all 
processes and channels
affected, whereas an ``S'' denotes systematic uncertainties that are considered shape-only in all processes and
channels.  An ``SN" means that the uncertainty is taken on both shape and normalisation.
Some of the systematic uncertainties are split into several components for a more
accurate treatment. This is the number indicated in the column labelled as ``Comp.". }
\end{table}

\subsection{Luminosity}
\label{sec:syst_lumi}
The uncertainty on the integrated luminosity for the data set used in this
analysis is 2.8\%. It is derived following the same methodology as that detailed 
in Ref.~\cite{lumi13}. This systematic uncertainty is applied to all contributions 
determined from the MC simulation.

\subsection{Uncertainties on physics objects}
\label{sec:syst_objects}

\subsubsection{Leptons}
\label{sec:syst_lepID}
Uncertainties associated with the lepton selection arise from the trigger, 
reconstruction, identification, isolation and lepton momentum scale and resolution.
In total, uncertainties associated with electrons (muons) 
include five (six) components.    

\subsubsection{Jets}
\label{sec:syst_jes}
Uncertainties associated with the jet selection arise from the 
jet energy scale (JES), jet vertex fraction requirement, jet energy resolution and 
jet reconstruction efficiency. Among these, the JES uncertainty has the largest 
impact on the analysis. 
The JES and its uncertainty are derived combining
information from test-beam data, LHC collision data and
simulation~\cite{ATLASJetEnergyMeasurement}.  
The jet energy scale uncertainty is split into 22 uncorrelated sources  
which can have different jet $\pt$ and $\eta$ dependencies. In this analysis,  
the largest jet energy scale uncertainty 
arises from the $\eta$ dependence of the JES calibration  
in the end-cap regions of the calorimeter. It is the second leading 
uncertainty. 

\subsubsection{Heavy- and light-flavour tagging}
\label{sec:syst_btag}
A total of six (four) independent sources of uncertainty affecting 
the $b$($c$)-tagging efficiency are considered~\cite{ref:ATLAS-CONF-2014-004}. 
Each of these uncertainties corresponds to an eigenvector resulting
from diagonalising the matrix containing the information about the total
uncertainty per jet $\pt$ bin and the bin-to-bin correlations. 
An additional uncertainty is assigned due to the extrapolation of the 
$b$-tagging efficiency measurement to the high-\pt\ region.
Twelve uncertainties  
are considered for the light-jet tagging and they depend on jet \pt\ and \eta.      
These systematic uncertainties are taken as uncorrelated between $b$-jets, 
$c$-jets, and light-flavour jets. 

No additional systematic uncertainty is assigned due to the use of 
parameterisations of the $b$-tagging probabilities instead of applying 
the $b$-tagging algorithm directly since the difference between these two
approaches is negligible compared to the other sources.

\subsection{Uncertainties on background modelling}
\label{sec:syst_norm}

\subsubsection{\ttbar+jets modelling}
\label{sec:syst_ttbarmodel}
An uncertainty of +6.5\%/--6\% is assumed for the inclusive $t\bar{t}$
production cross section. It includes uncertainties 
from the top quark mass and choices of the PDF and $\alpha_{\mathrm{S}}$. 
The PDF and $\alpha_{\mathrm{S}}$ uncertainties are calculated using the PDF4LHC 
prescription~\cite{ref:pdf4lhc} with the MSTW2008 68\% CL NNLO, 
CT10 NNLO~\cite{ct102} and NNPDF2.3 5f FFN~\cite{nnpdf} 
PDF sets, and are added in quadrature to the scale uncertainty. 
Other systematic uncertainties affecting the modelling of $t\bar{t}$+jets 
include uncertainties  due to the 
choice of parton shower and hadronisation model, 
as well as several uncertainties related to the reweighting procedure applied 
to improve the {\ttbar} MC model. 
Additional uncertainties are assigned to account for limited knowledge 
of $t\bar{t}$+HF jets production. They are described later in this section.

As discussed in Sect.~\ref{sec:SimulatedSamples}, to improve the 
agreement between data and the \ttbar\ simulation a reweighting 
procedure is applied to \ttbar\ MC events based on the difference in the top quark $\pt$ and $\ttbar$ 
system $\pt$ distributions between data and simulation at $\sqrt{s}=7\tev$~\cite{topdiff_7TEV}.  
The nine largest uncertainties associated with the experimental measurement of 
top quark and \ttbar\ system \pt, representing approximately 95\% of the total experimental 
uncertainty on the measurement, are considered as separate uncertainty sources in the
reweighting applied to the MC prediction. The largest uncertainties on the 
measurement of the differential distributions include radiation modelling in
\ttbar\ events, the choice of generator to simulate \ttbar\ production, 
uncertainties on the components of jet energy scale and resolution, and
flavour tagging. 

Because the measurement is performed for the inclusive \ttbar\ sample and the 
size of the uncertainties applicable to the \ttbar+\ccbar\ component is not known, 
two additional uncorrelated uncertainties are assigned to 
\ttbar+\ccbar\ events, consisting of 
the full difference between applying and not applying the reweightings of 
the \ttbar\ system \pt\ and top quark \pt, respectively.  

An uncertainty due to the choice of parton shower and hadronisation model 
is derived by comparing events produced by {\sc Powheg} interfaced with {\sc Pythia} 
or {\sc Herwig}. Effects on the shapes are compared, symmetrised and applied to the 
shapes predicted by the default model. Given that the change of the parton shower
model leads to two separate effects -- a change in the number of jets and
a change of the heavy-flavour content -- the parton shower uncertainty is represented by three 
parameters, one acting on the \ttbar+light contribution and two others on the \ttbar+\ccbar\  and 
\ttbar+\bbbar\ contributions. These three parameters are treated as 
uncorrelated in the fit.

Detailed comparisons of \ttbar+\bbbar\ production between {\sc Powheg}+{\sc Pythia}
and an NLO prediction of \ttbar+\bbbar\ production based on
\ShOL\ have shown that the cross sections agree within 50\% of each other.
Therefore, a systematic uncertainty of 50\% is applied to the \ttbb\
component of the \ttbar+jets background obtained 
from the {\sc Powheg}+{\sc Pythia} MC simulation. In the absence of an NLO prediction for the 
\ttbar+\ccbar\ background, the same 50\% systematic uncertainty is applied to the \ttbar+\ccbar\
component, and the uncertainties on \ttbb\ and \ttbar+\ccbar\ are treated as uncorrelated.
The large available data sample allows the determination of the 
\ttbar+\bbbar\ and \ttbar+\ccbar\ normalisations with much better precision, 
approximately 15\% and 30\%, respectively (see Appendix~\ref{sec:norm_syst}). 
Thus, the final result does not significantly depend on the exact value of 
the assumed prior uncertainty, as long as it is larger than the precision 
with which the data can constrain it. However, 
even after the reduction, the uncertainties on the \ttbar+\bbbar\ and the 
\ttbar+\ccbar\ background normalisation are still the leading and the third 
leading uncertainty in the analysis, respectively.

Four additional systematic uncertainties in the  
\ttbar+\ccbar\ background estimate are derived from the 
simultaneous variation of factorisation and 
renormalisation scales, matching threshold and $c$-quark mass  
variations in the {\sc Madgraph}+{\sc Pythia} \ttbar\ simulation, and the
difference between the \ttbar+\ccbar\ simulation in 
{\sc Madgraph}+{\sc Pythia} and {\sc Powheg}+{\sc Pythia} since 
{\sc Madgraph}+{\sc Pythia} includes the \ttbar+\ccbar\ process in the matrix 
element calculation while it is absent in {\sc Powheg}+{\sc Pythia}. 

For the \ttbar+\bbbar\ background, three scale uncertainties, 
including changing the functional form of the renormalisation scale to 
$\mu_{\rm R} = (m_t m_{b\bar{b}})^{1/2}$, changing the functional form of the 
factorisation $\mu_{\rm F}$ and resummation $\mu_Q$ scales to 
$\mu_{\rm F} =\mu_{\rm Q} =\prod_{i=t,\bar{t},b,\bar{b}}E_{\rm{T},i}^{1/4}$ and 
varying the renormalisation scale $\mu_{\rm R}$ by a factor of two up and down 
are evaluated. 
Additionally, the shower recoil model 
uncertainty and two uncertainties due to the PDF choice
in the \ShOL\ NLO calculation are quoted. 
The effect of these variations on the contribution of 
different \ttbar+\bbbar\ event categories is shown in Fig.~\ref{fig:ttbbsyst}. 
The renormalisation scale choice and the shower recoil scheme 
have a large effect on the modelling of \ttbb. They provide 
large shape variations of the NN discriminants resulting in the 
fourth and sixth leading uncertainties in this analysis.

\begin{figure*}[ht!]
\begin{center}
\subfigure[]{\includegraphics[width=0.49\textwidth]{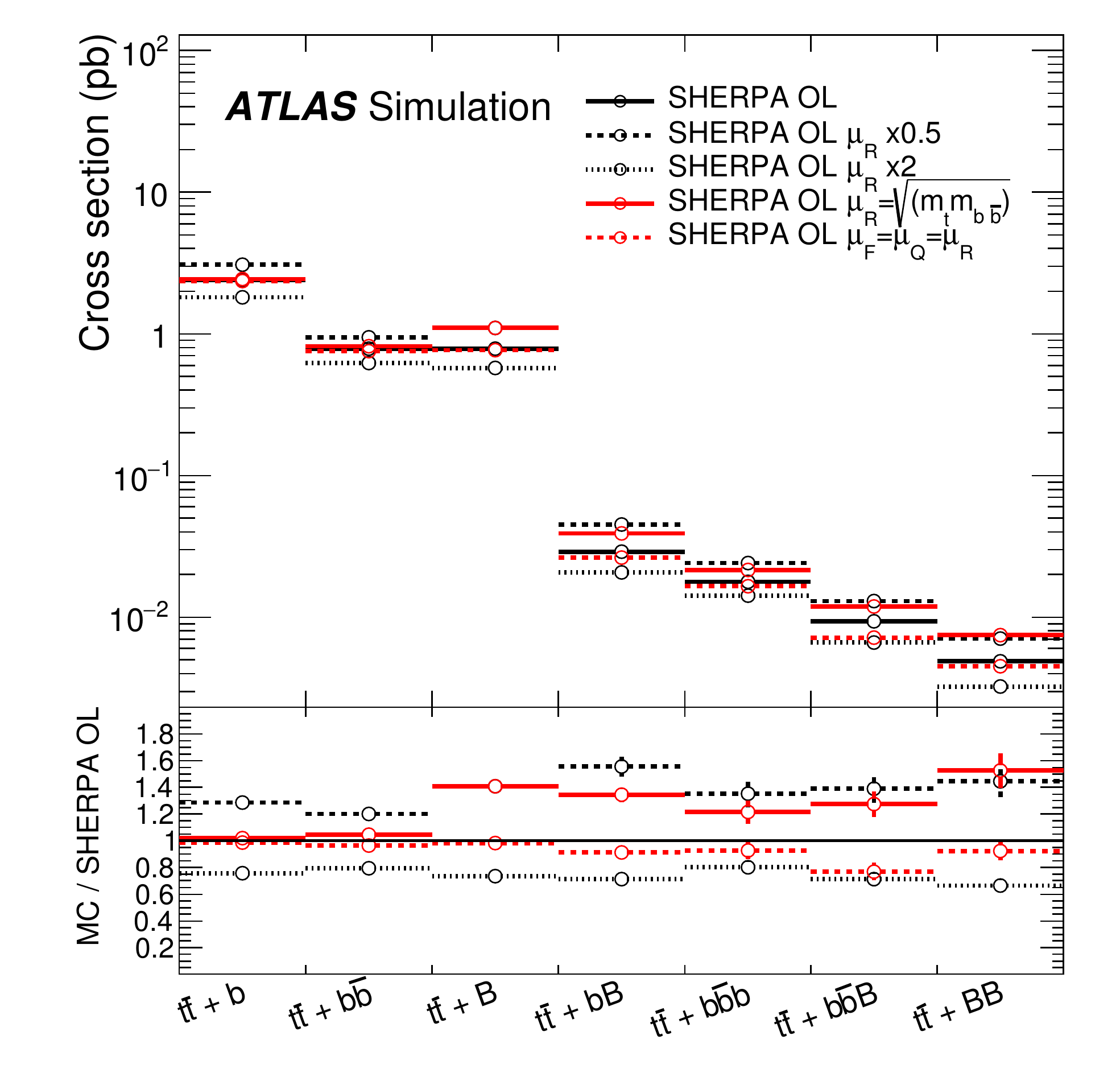}}\label{fig:ttbbsyst_a}
\subfigure[]{\includegraphics[width=0.49\textwidth]{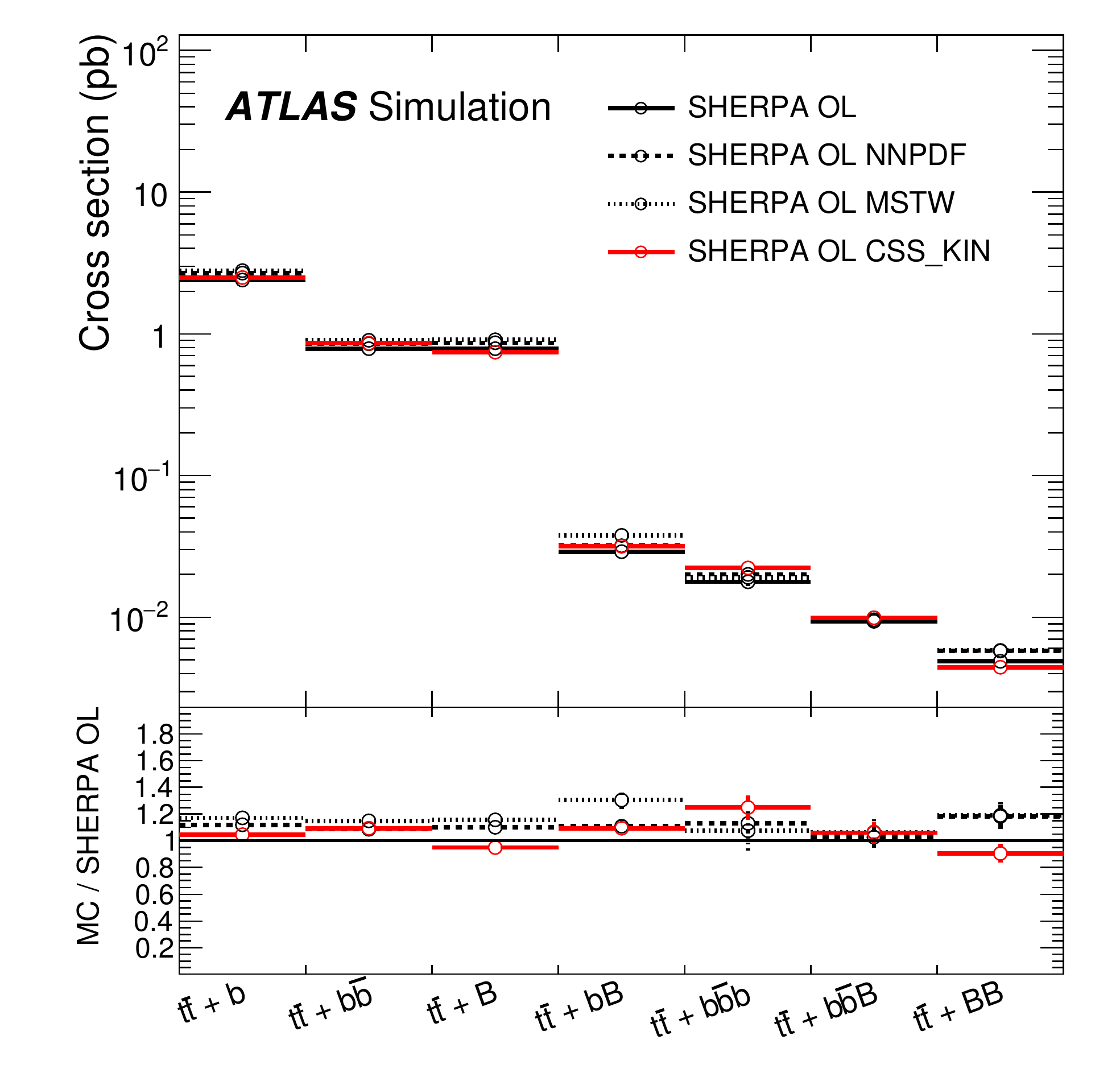}}\label{fig:ttbbsyst_b}	
\caption{Systematic uncertainties on the \ttbar+\bbbar\ contribution
based on (a) scale variations and (b) PDF choice and shower recoil model 
of the \ShOL\ simulation. The effect
of a given systematic uncertainty is shown across the different \ttbar+\bbbar\ categories.   
The effect of migration between categories is covered by variations of these systematic
uncertainties.}
\label{fig:ttbbsyst} 
\end{center}
\end{figure*}

Finally, two uncertainties due to
\ttbar+\bbbar\ production via multiparton interaction and final-state 
radiation which are not present in the \ShOL\ NLO calculation are applied.  
Overall, the uncertainties on  
\ttbb\ normalisation and modelling result in about a 55\% total uncertainty on 
the \ttbb\ background contribution in the most sensitive \sixfour\ and 
\fourfourdi\ regions.

\subsubsection{The $W/Z$+jets modelling}
\label{sec:syst_vjetsnorm}
As discussed in Sect.~\ref{sec:SimulatedSamples}, the $W/Z$+jets contributions 
are  obtained from the simulation and normalised to the inclusive theoretical cross 
sections, and a reweighting is applied to improve the modelling of the $W/Z$ boson $\pt$ spectrum.   
The full difference between applying and not applying the $W/Z$ boson $\pt$
reweighting is taken as a systematic uncertainty, which is then assumed to be 
symmetric with respect to the central value.
Additional uncertainties are assigned due to the extrapolation of the $W/Z$+jets 
estimate to high jet multiplicity.  

\subsubsection{Misidentified lepton background modelling}
\label{sec:qcd_norm}
Systematic uncertainties on the misidentified lepton background 
estimated via the matrix method~\cite{ttbar_3pb} 
in the single-lepton channel receive contributions from the limited 
number of data events, 
particularly at high jet and $b$-tag multiplicities, from the subtraction of the prompt-lepton contribution 
as well as from the uncertainty on 
the lepton misidentification rates, estimated in different control regions. The statistical uncertainty is
uncorrelated among the different jet and $b$-tag multiplicity bins. An uncertainty 
of 50\% associated with the lepton misidentification rate measurements is 
assumed, which is taken 
as correlated across jet and $b$-tag multiplicity bins, but uncorrelated between electron 
and muon channels. Uncertainty on the shape of the misidentified lepton 
background arises from the 
prompt-lepton background subtraction and the misidentified lepton rate 
measurement.   

In the dilepton channel, since the misidentified lepton background is estimated
using both the simulation and same-sign dilepton events in data, a 50\% normalisation
uncertainty is assigned to cover the maximum difference between the two 
methods. It is 
taken as correlated among the different jet and $b$-tag multiplicity bins. 
An additional uncertainty is applied to cover the difference in shape between the
predictions derived from the simulation and from same-sign dilepton events in data.

\subsubsection{Electroweak background modelling}
\label{sec:ew_norm}
Uncertainties of +5\%/--4\% and $\pm 6.8$\% are 
used for the theoretical cross sections of single
top production in the single-lepton and dilepton 
channels~\cite{stopxs,stopxs_2}, respectively. The former corresponds to the weighted average 
of the theoretical uncertainties on $s$-, $t$- and $Wt$-channel production, 
while the latter corresponds to the theoretical uncertainty on 
$Wt$-channel production, the only single top process contributing to the dilepton final 
state. 

The uncertainty on the diboson background rates includes an uncertainty 
on the inclusive diboson NLO cross section of $\pm 5\%$~\cite{dibosonxs} and 
uncertainties to account for the extrapolation to high jet multiplicity. 

Finally, an uncertainty of $\pm 30\%$ is assumed for the theoretical cross sections of the 
$t\bar{t}+V$~\cite{ttbarVxs1,ttbarVxs2} background. An additional uncertainty 
on $t\bar{t}+V$
modelling arises from variations in the amount of initial-state radiation.  
The $t\bar{t}+Z$ background with $Z$ boson decaying into a $\bbbar$ pair is 
an irreducible background to the \tth, \htobb\ signal, and as such, has   
kinematics and an NN discriminant shape similar to those of the signal.   
The uncertainty on the $t\bar{t}+V$ background normalisation is the fifth 
leading uncertainty in the analysis. 

\subsection{Uncertainties on signal modelling}
\label{sec:ttH_PS}
Dedicated NLO {\sc PowHel} samples are used to evaluate the impact of the
choice of factorisation and renormalisation scales on the $t\bar{t}H$ 
signal kinematics. In these samples the default scale is varied  
by a factor of two up and down. The effect of the variations on 
$t\bar{t}H$ distributions was studied at particle level and the nominal  
{\sc PowHel} $t\bar{t}H$ sample was reweighted to reproduce these 
variations.  In a similar way, the nominal sample is reweighted to 
reproduce the 
effect of changing the functional form of the scale. 
Additional uncertainties on the \tth\ signal due to the choice of PDF, parton
shower and fragmentation model and NLO generator are also considered. 
The effect of the PDF uncertainty on the \tth\ signal is evaluated following the 
recommendation of the PDF4LHC.  
The uncertainty in the parton shower and 
fragmentation is evaluated by comparing {\sc Powhel}+{\sc Pythia8} and {\sc Powhel}+{\sc Herwig} 
samples, while the uncertainty due to a generator choice is evaluated by 
comparing {\sc Powhel}+{\sc Pythia8} with \\ {\sc {Madgraph5\_aMC@NLO}}~\cite{Alwall:2014hca} interfaced with \\ 
{\sc Herwig++}~\cite{Bahr:2008pv,Bellm:2013lba}.

\section{Statistical methods}
\label{sec:statmethods}

The distributions of the discriminants from each of the channels and regions considered are combined to
test for the presence of a signal, assuming a Higgs boson mass of $\mH= 125 \gev$.
The statistical analysis is based on a binned likelihood function ${\cal L}(\mu,\theta)$ constructed as
a product of Poisson probability terms over all bins considered in the 
analysis. The likelihood function depends
on the signal-strength parameter $\mu$, defined as the ratio of the observed/expected 
cross section to the SM cross section,   
and $\theta$, denoting the set of nuisance parameters that encode the
effects of systematic uncertainties
on the signal and background expectations. They 
are implemented in the likelihood function as Gaussian or 
log-normal priors. Therefore, the total number of expected events in a
given bin depends on $\mu$ and $\theta$. The nuisance parameters
$\theta$ adjust the expectations for signal and background 
according to the corresponding systematic uncertainties, 
and their fitted values correspond to the amount that best fits the data. 
This procedure allows the impact of systematic uncertainties on the 
search sensitivity to be reduced by taking advantage of the highly 
populated background-dominated control regions 
included in the likelihood fit.  
It requires a good understanding of the systematic effects affecting 
the shapes of the discriminant distributions. 
The test statistic $q_\mu$ is defined as the profile likelihood ratio: 
$q_\mu = -2\ln({\cal L}(\mu,\hat{\hat{\theta}}_\mu)/{\cal L}(\hat{\mu},\hat{\theta}))$,
where $\hat{\mu}$ and $\hat{\theta}$ are the values of the parameters
that maximise the likelihood function (with the constraints 
$0\leq \hat{\mu} \leq \mu$), and $\hat{\hat{\theta}}_\mu$ are the values of the 
nuisance parameters that maximise the 
likelihood function for a given value of $\mu$. This test statistic is used to measure the 
compatibility of the observed data 
with the background-only hypothesis (i.e. for $\mu=0$), 
and to make statistical inferences about $\mu$, such as upper limits using the 
CL$_{\rm{s}}$ method~\cite{cls,cls_2,cls_3} as implemented in the {\sc RooFit} 
package~\cite{RooFit,RooFitManual}. 

To obtain the final result, a simultaneous fit to the data 
is performed on the distributions of the discriminants 
in 15 regions: nine analysis regions in the single-lepton channel and six 
regions in the dilepton channel. Fits are performed under 
the signal-plus-background hypothesis, where the 
signal-strength parameter $\mu$ is the parameter of interest in the fit and 
is allowed to float freely, but is required to be the same in all 15 fit regions.  
The normalisation of each background is determined from the fit simultaneously 
with $\mu$. Contributions from \ttbar, $W/Z$+jets production, single 
top, diboson and $t\bar{t}V$ backgrounds are constrained by the uncertainties 
of the respective theoretical calculations, the uncertainty on the 
luminosity, and the data themselves. 
Statistical uncertainties in each bin of the discriminant 
distributions are taken into account by dedicated parameters in the fit.    
The performance of the fit is tested using simulated events by injecting \tth\  
signal with a variable signal strength and comparing it to the fitted value. 
Good agreement between the injected and measured signal strength is observed. 

\section{Results}
\label{sec:Results}
   
The results of the binned likelihood fit to data described in 
Sect.~\ref{sec:statmethods} are presented in this section.
Figure~\ref{fig:PRplot_lj} shows the yields after the fit in
all analysis regions in the single-lepton and dilepton channels.
The post-fit event yields and the corresponding $S/B$ and $S/\sqrt{B}$ ratios 
are summarised in Appendix~\ref{sec:postfit_tables}. 

\begin{figure*}[!ht]
\begin{center}
\subfigure[]{\includegraphics[width=0.485\textwidth]{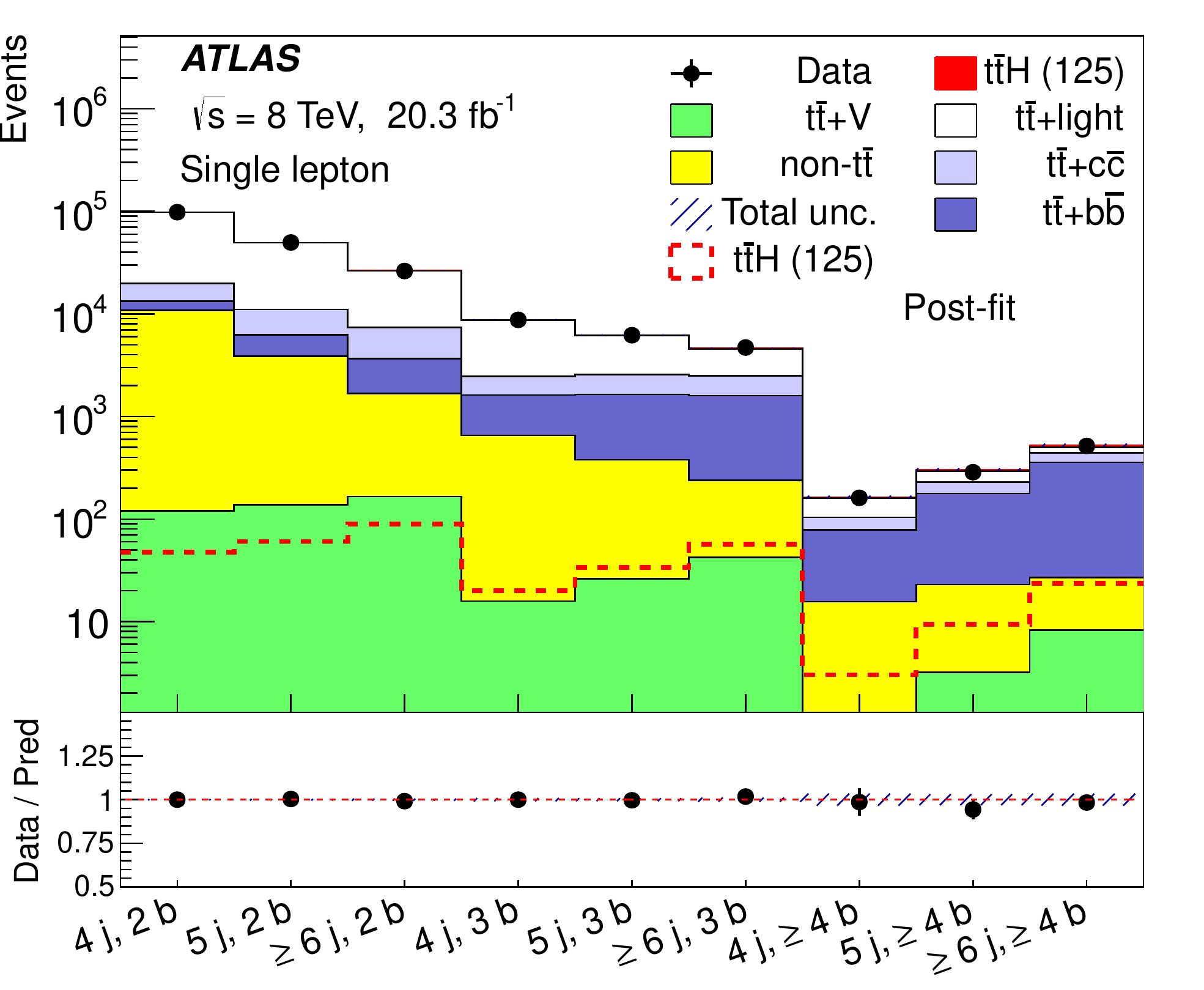}}\label{fig:PRplot_lj_a}
\subfigure[]{\includegraphics[width=0.485\textwidth]{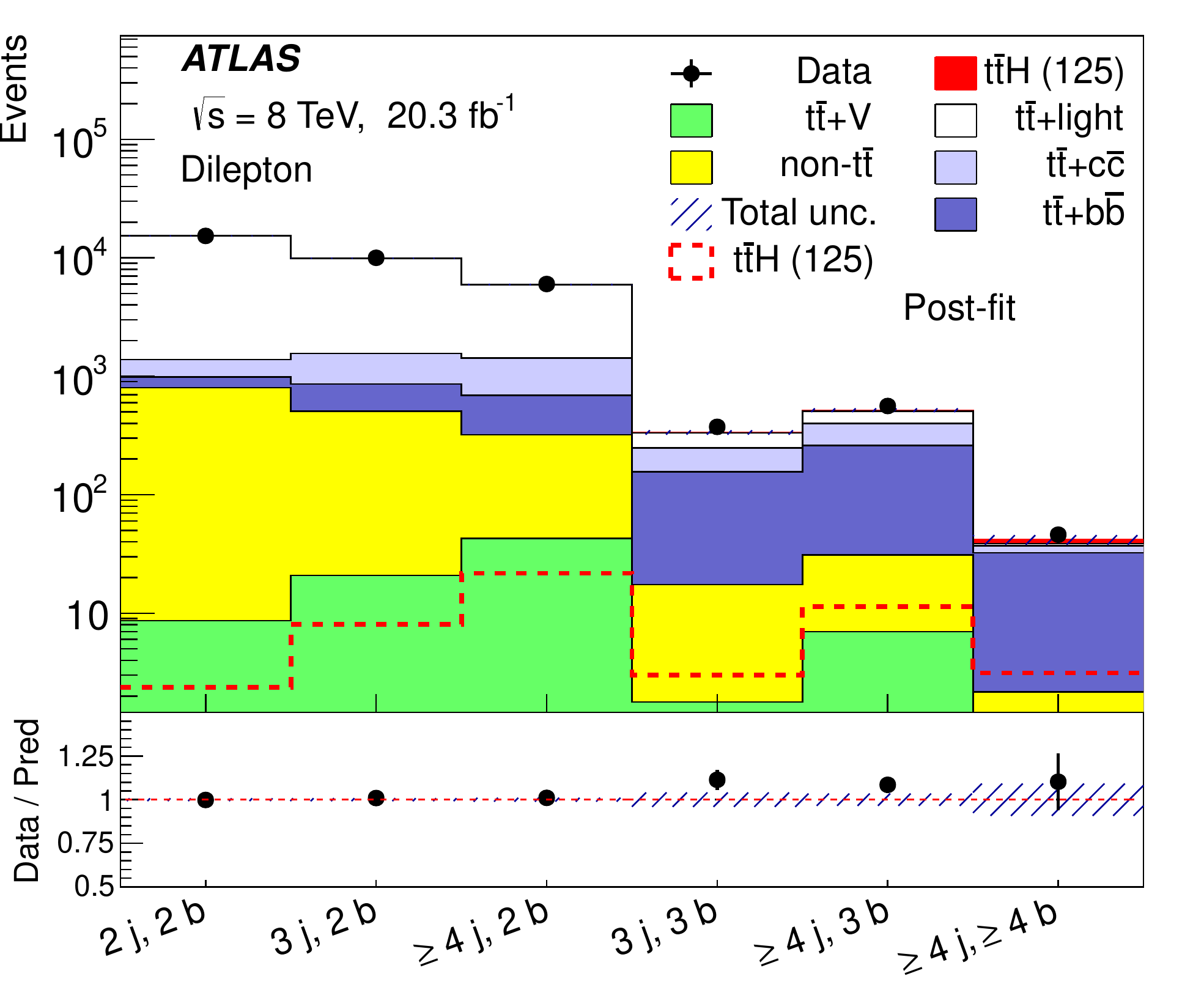}}\label{fig:PRplot_lj_b}
\caption{Event yields in all analysis regions in (a) the single-lepton channel and (b) the dilepton channel after the combined fit 
to data under the signal-plus-background
hypothesis. The signal, normalised to the fitted $\mu$, is shown both as a
filled area stacked on the other backgrounds and separately as
a dashed line. The hashed area represents the 
total uncertainty on the yields. }
\label{fig:PRplot_lj} 
\end{center}
\end{figure*}

Figures~\ref{fig:prepost_lj_1}--\ref{fig:prepost_lj_3} and~\ref{fig:prepost_dil_1}--\ref{fig:prepost_dil_2} 
show a comparison of data and prediction for the discriminating variables 
(either \hthad, \htlep, or NN discriminants) for
each of the regions considered in the single-lepton and dilepton channels, respectively, 
both pre- and post-fit to data. The uncertainties decrease 
significantly in all regions due to constraints provided by data and correlations between 
different sources of uncertainty introduced by the fit to the data. In Appendix~\ref{sec:postfitinput}, the 
most highly discriminating variables in the NN are shown post-fit compared to data.

\begin{figure*}[!ht]
\begin{center}
\subfigure[]{\includegraphics[width=0.34\textwidth]{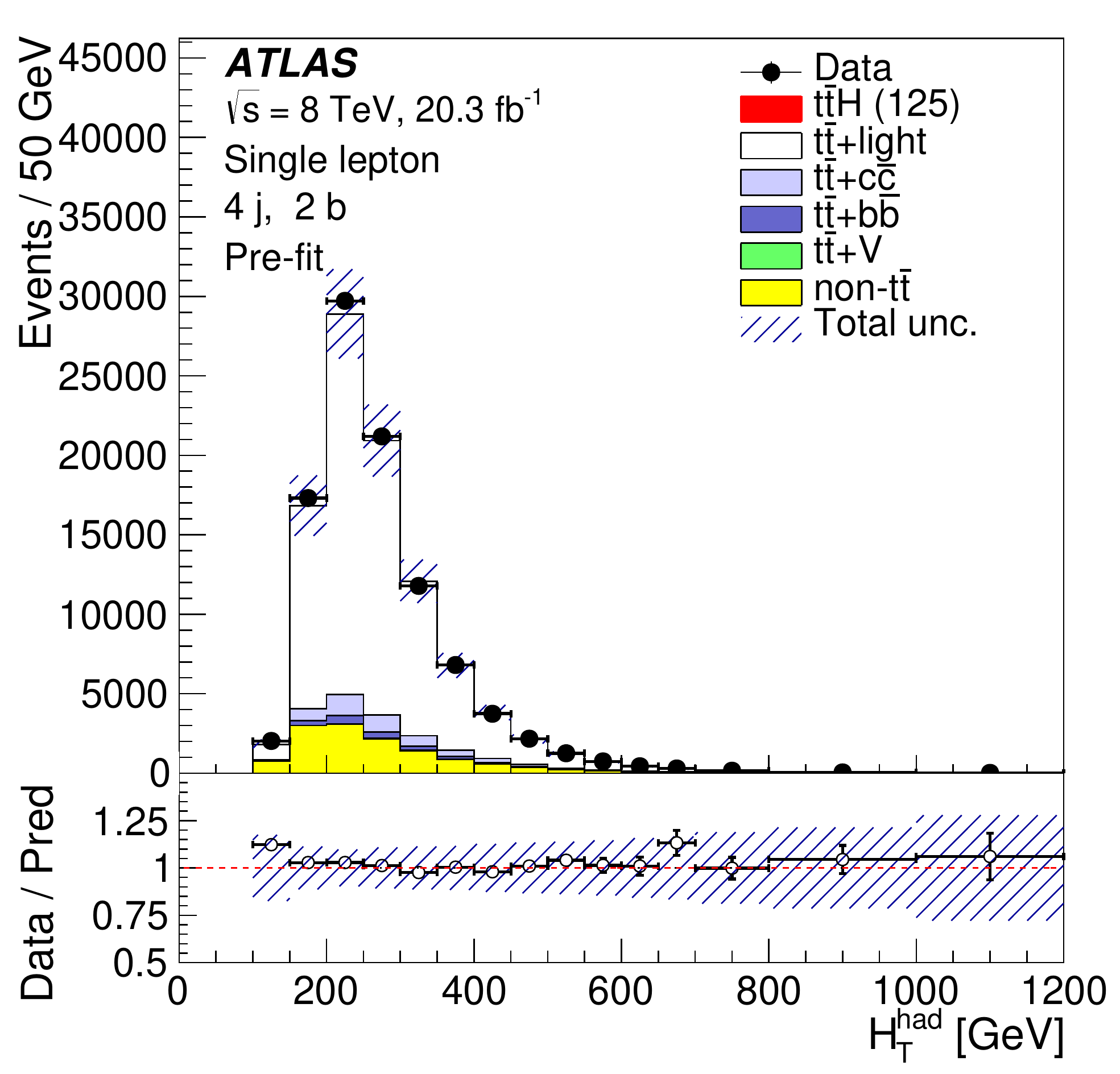}}\label{fig:prepost_lj_1_a}
\subfigure[]{\includegraphics[width=0.34\textwidth]{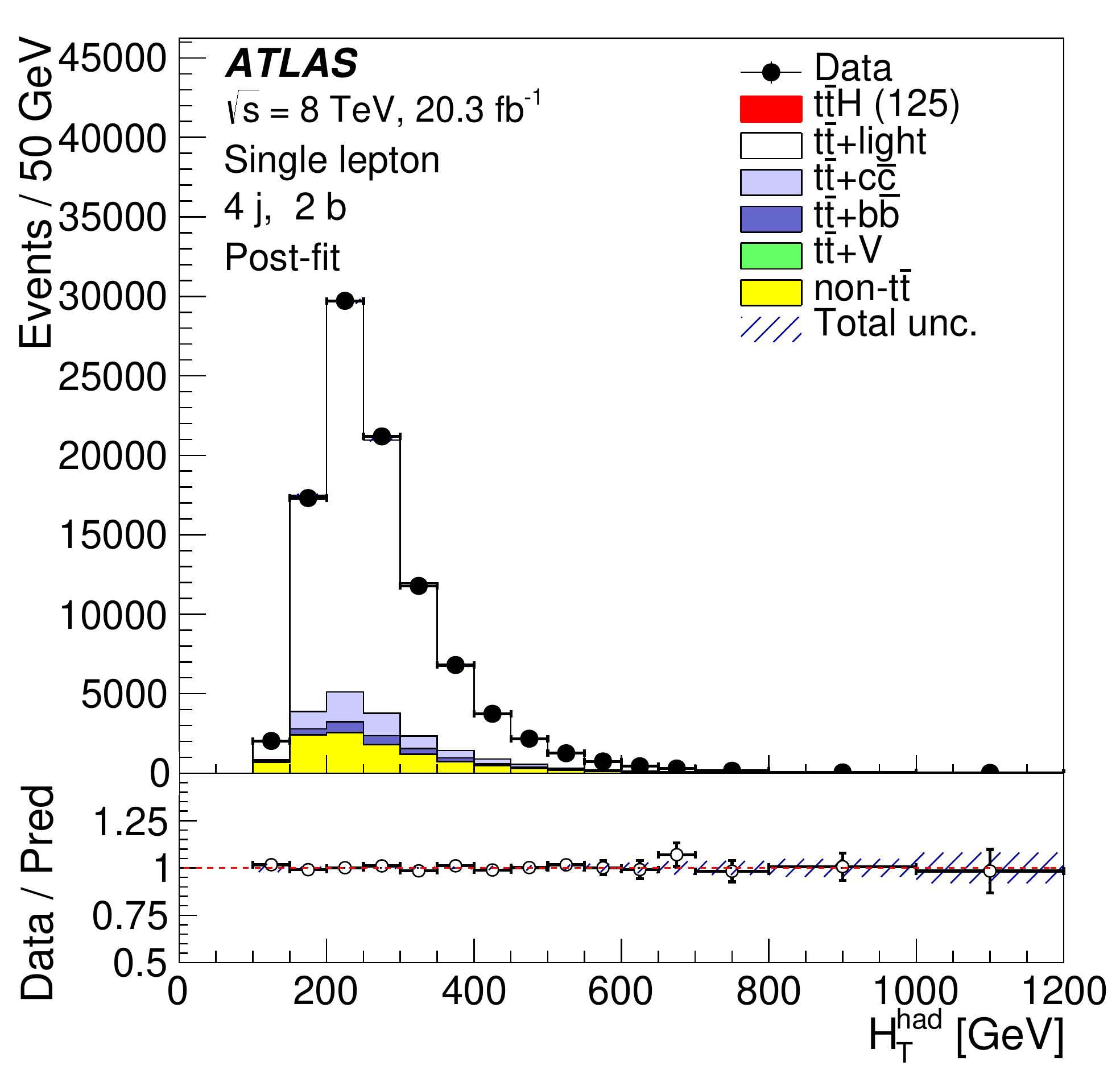}}\label{fig:prepost_lj_1_b}\\
\subfigure[]{\includegraphics[width=0.34\textwidth]{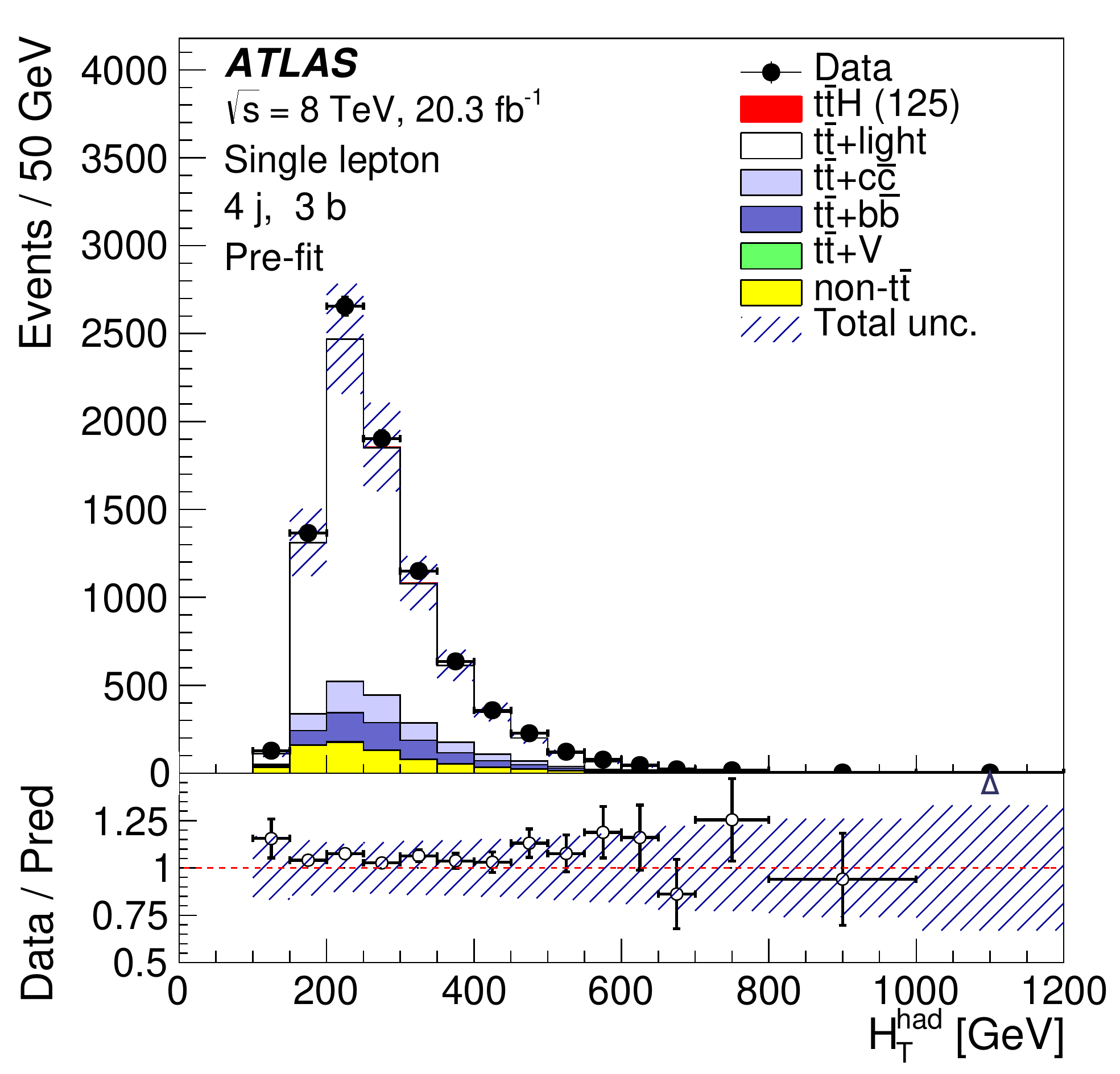}}\label{fig:prepost_lj_1_c}
\subfigure[]{\includegraphics[width=0.34\textwidth]{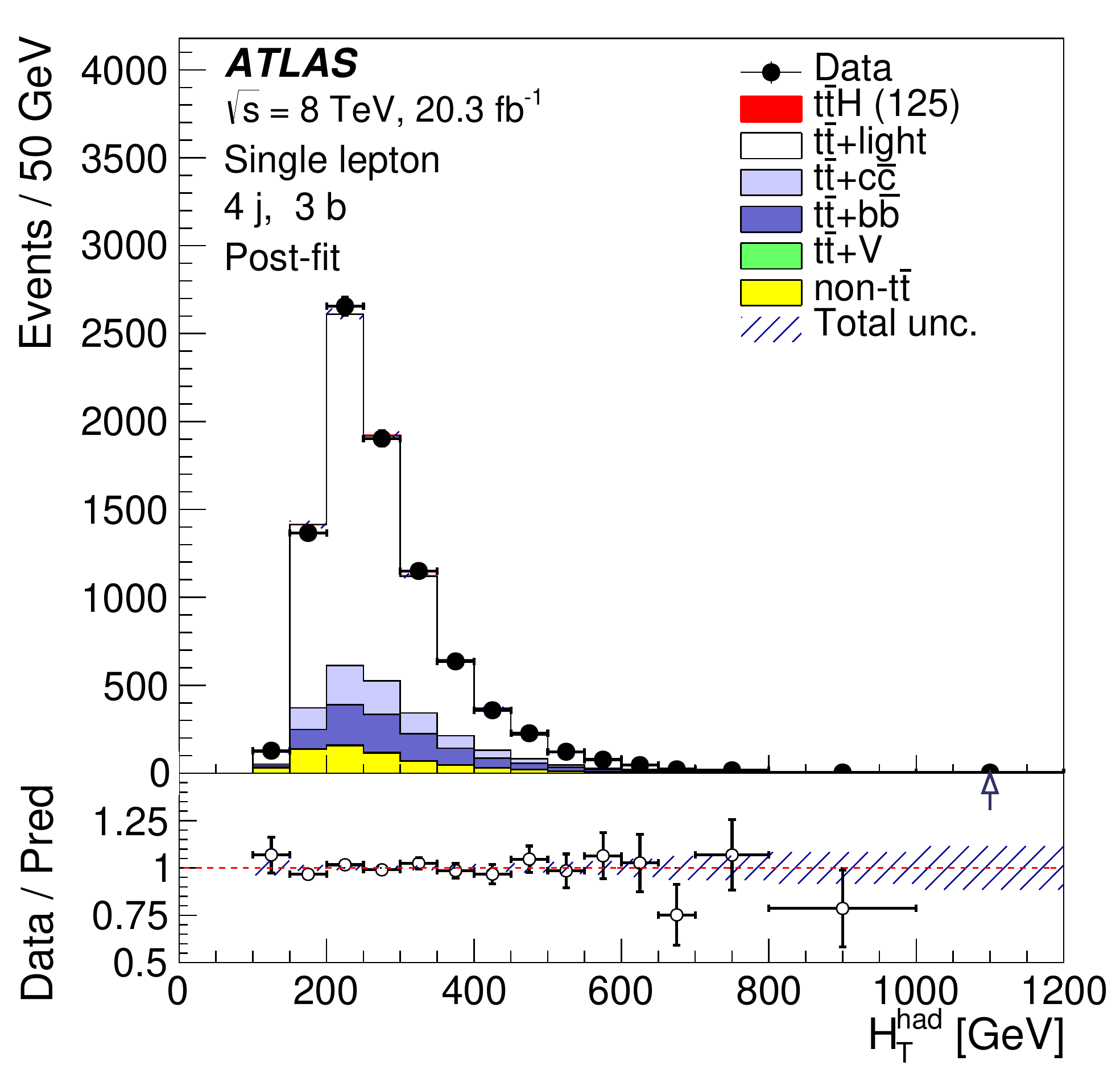}}\label{fig:prepost_lj_1_d}\\
\subfigure[]{\includegraphics[width=0.34\textwidth]{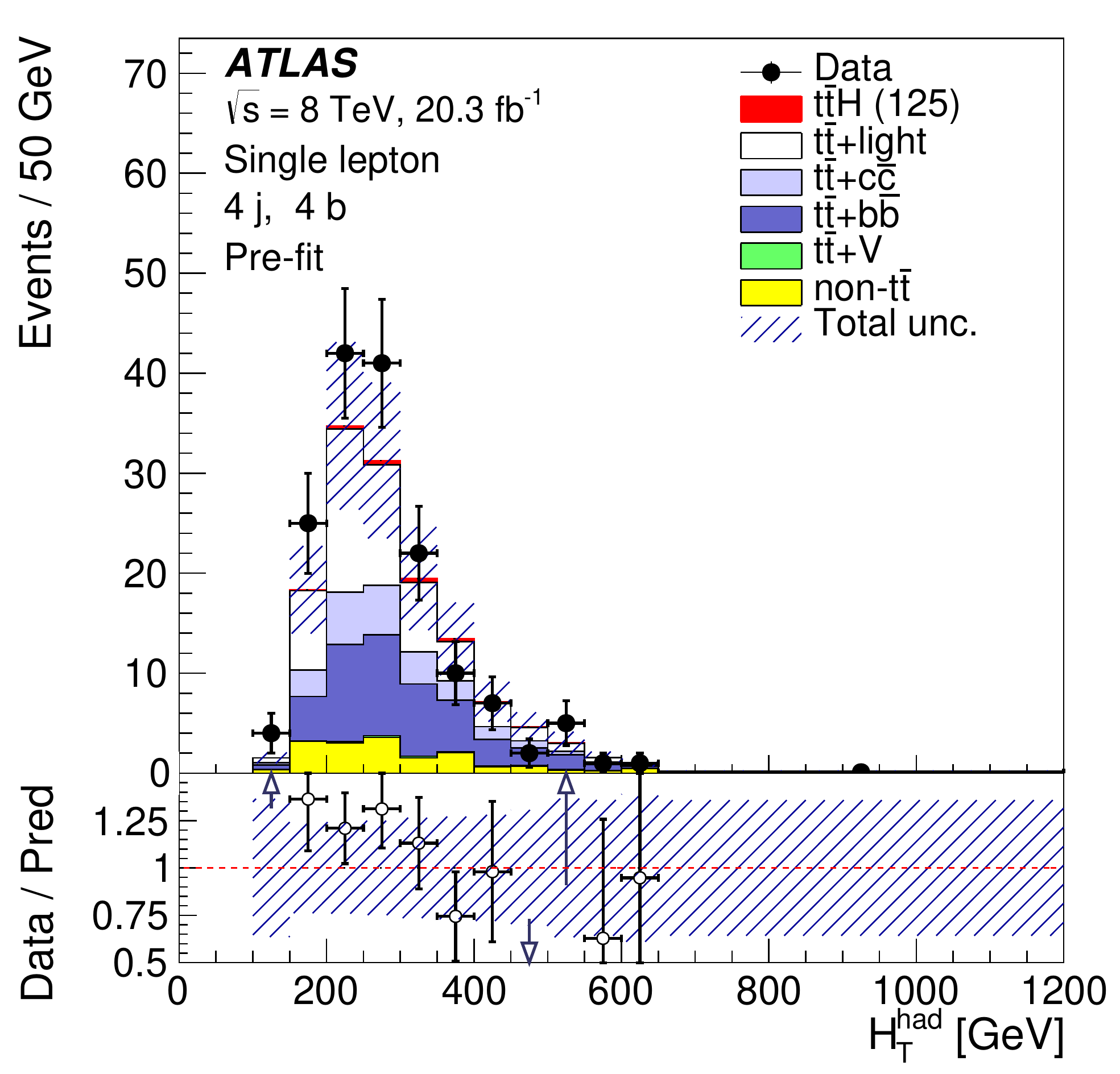}}\label{fig:prepost_lj_1_e}
\subfigure[]{\includegraphics[width=0.34\textwidth]{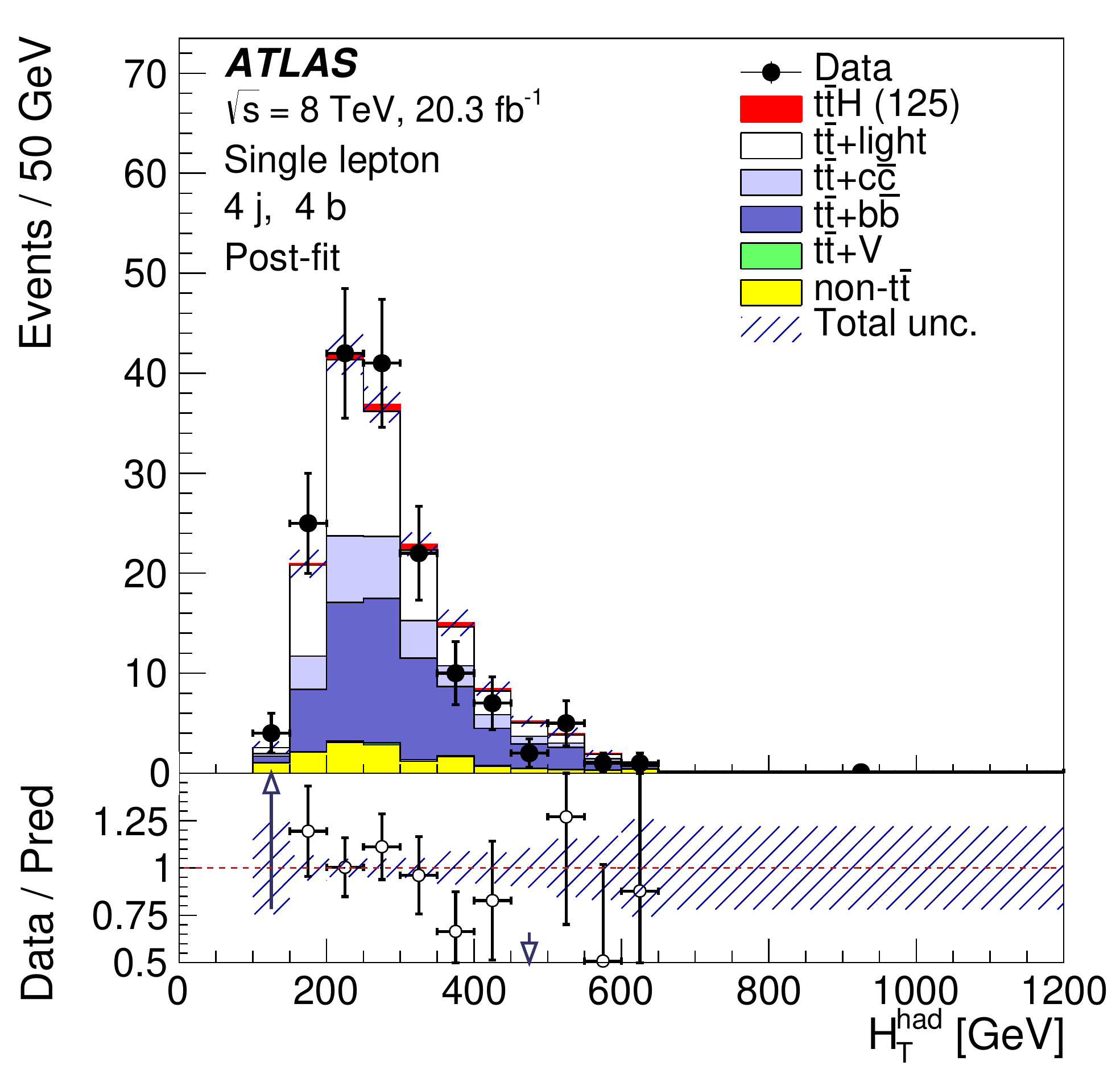}}\label{fig:prepost_lj_1_f}\\
\caption{Single-lepton channel: comparison between data and prediction for the 
discriminant variable used in the \fourtwolj\ region (a) before the fit and (b) after the fit, 
in the \fourthreelj\ region (c) before the fit and (d) after the fit, 
in the \fourfourlj\ region (e) before the fit and (f) after the fit. 
The fit is performed on data under the signal-plus-background 
hypothesis. The last bin in all figures contains the overflow. 
The bottom panel displays the ratio of data to the total prediction. An arrow indicates that the point is off-scale. 
The hashed area represents the uncertainty on the background.  
The \tth\ signal yield (solid) is normalised to the SM cross section before the fit and to the fitted $\mu$ 
after the fit.  In several regions, predominantly the control regions, the \tth\ signal yield is not visible on top 
of the large background. }
\label{fig:prepost_lj_1} 
\end{center}
\end{figure*}

\begin{figure*}[!ht]
\begin{center}
\subfigure[]{\includegraphics[width=0.34\textwidth]{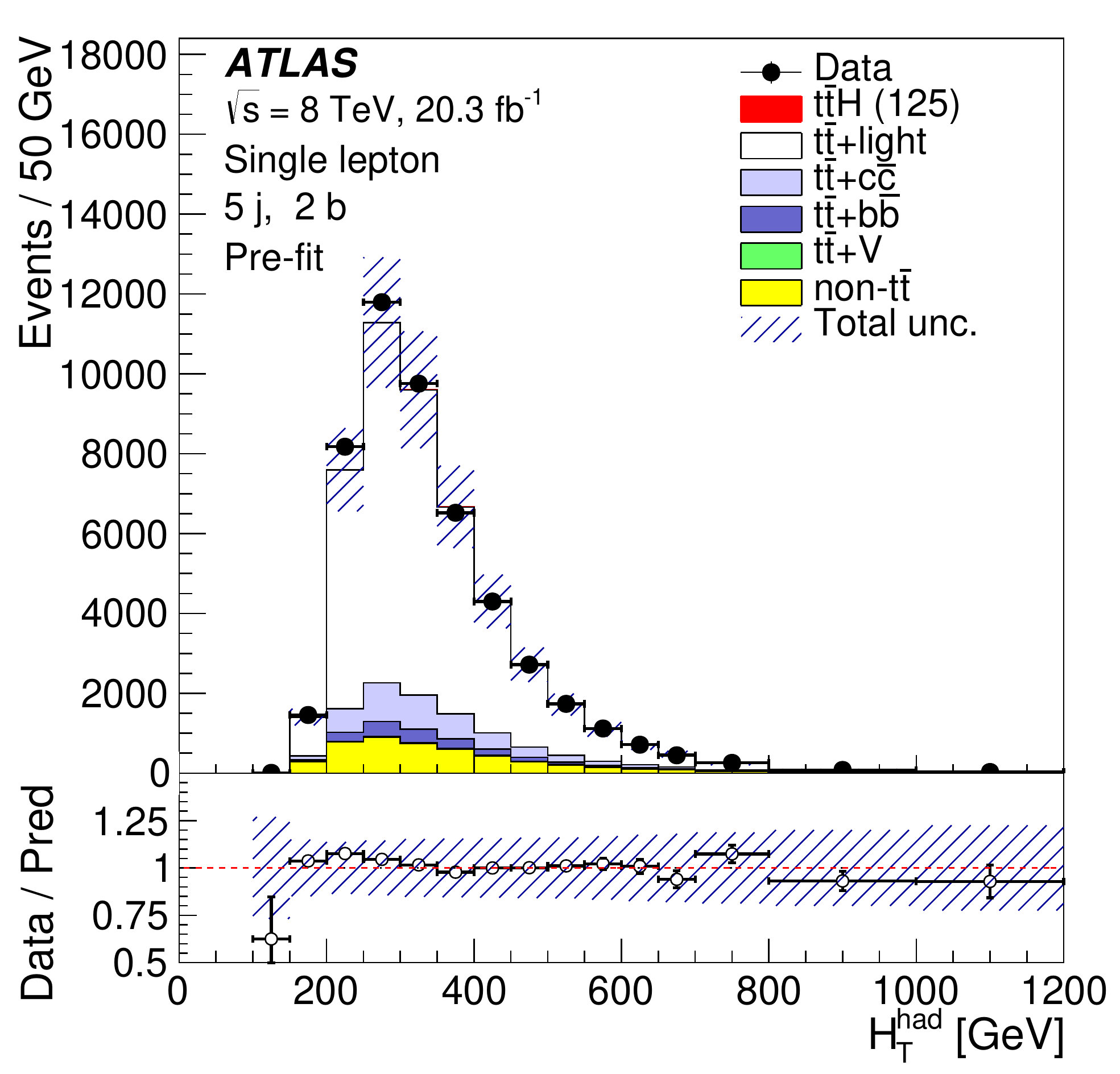}}\label{fig:prepost_lj_2_a} 
\subfigure[]{\includegraphics[width=0.34\textwidth]{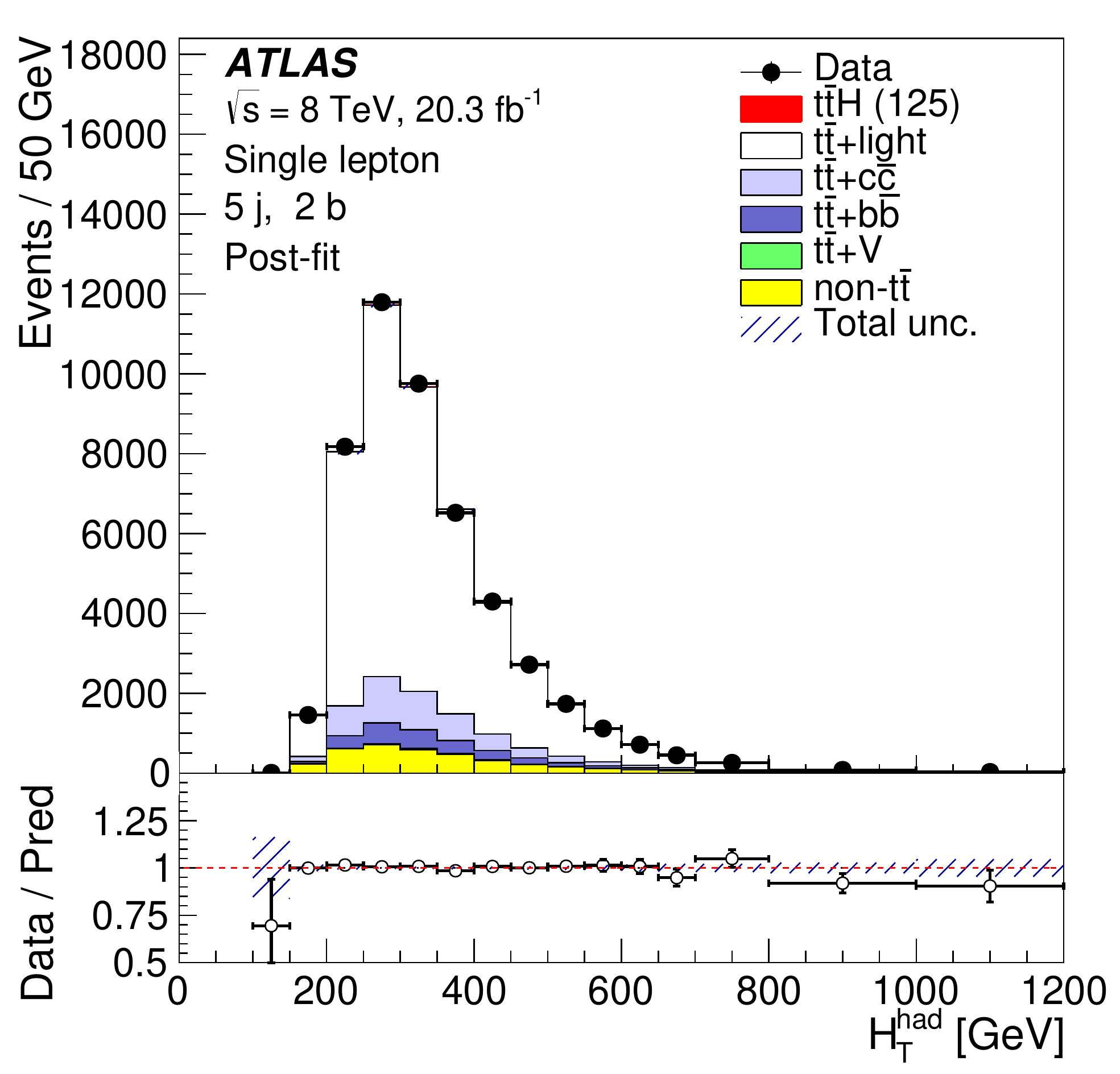}}\label{fig:prepost_lj_2_b} \\
\subfigure[]{\includegraphics[width=0.34\textwidth]{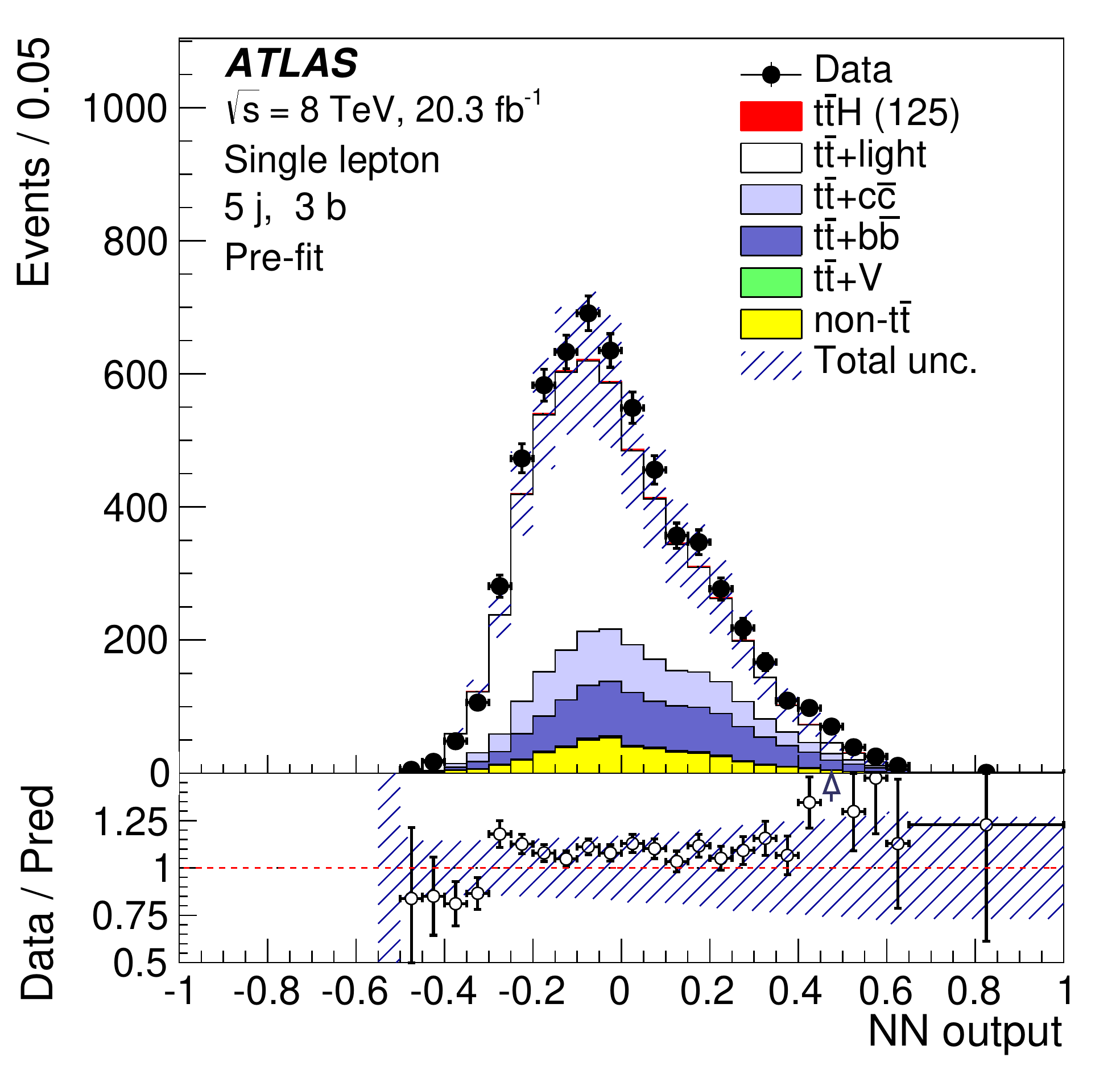}}\label{fig:prepost_lj_2_c} 
\subfigure[]{\includegraphics[width=0.34\textwidth]{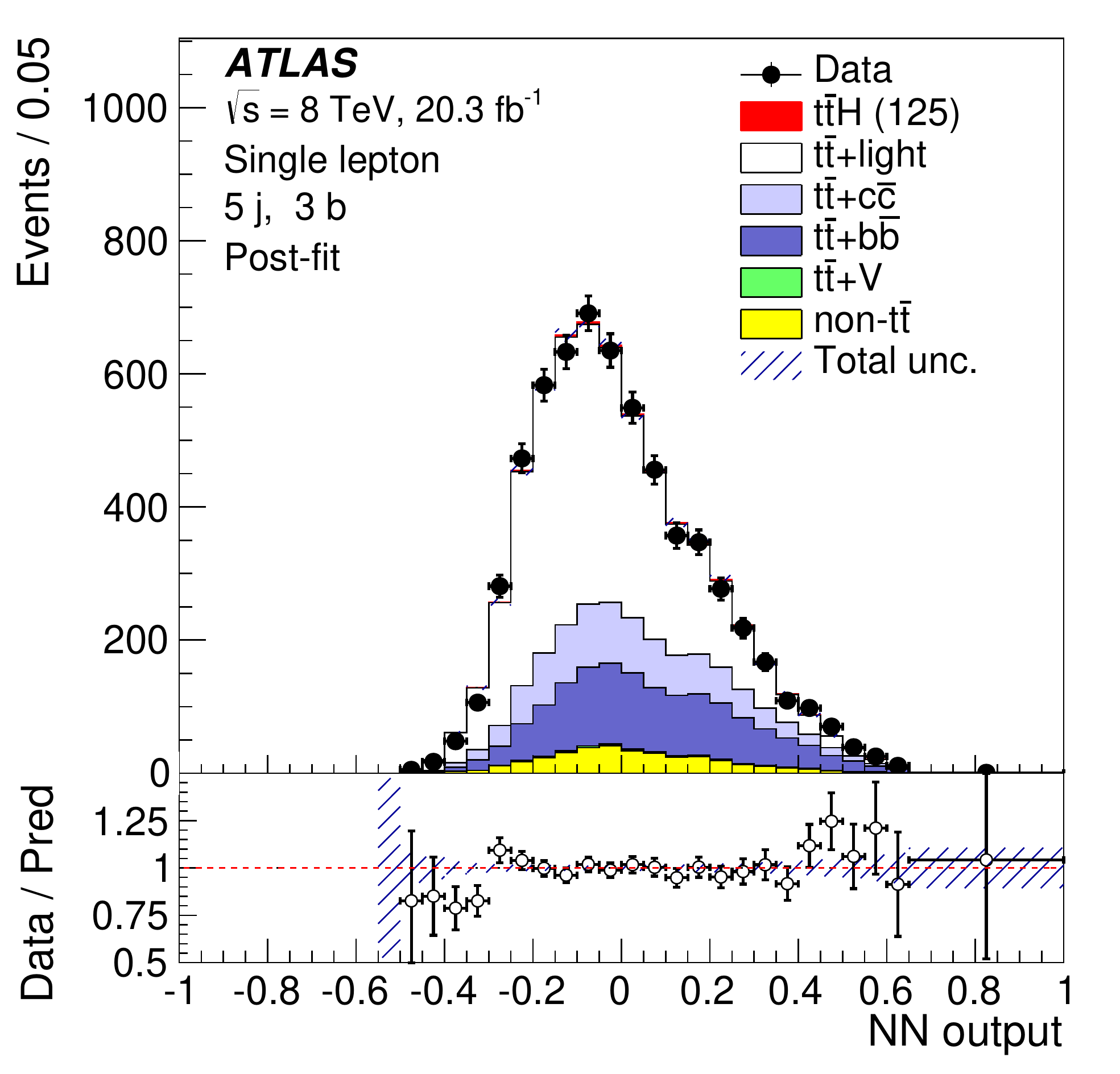}}\label{fig:prepost_lj_2_d} \\
\subfigure[]{\includegraphics[width=0.34\textwidth]{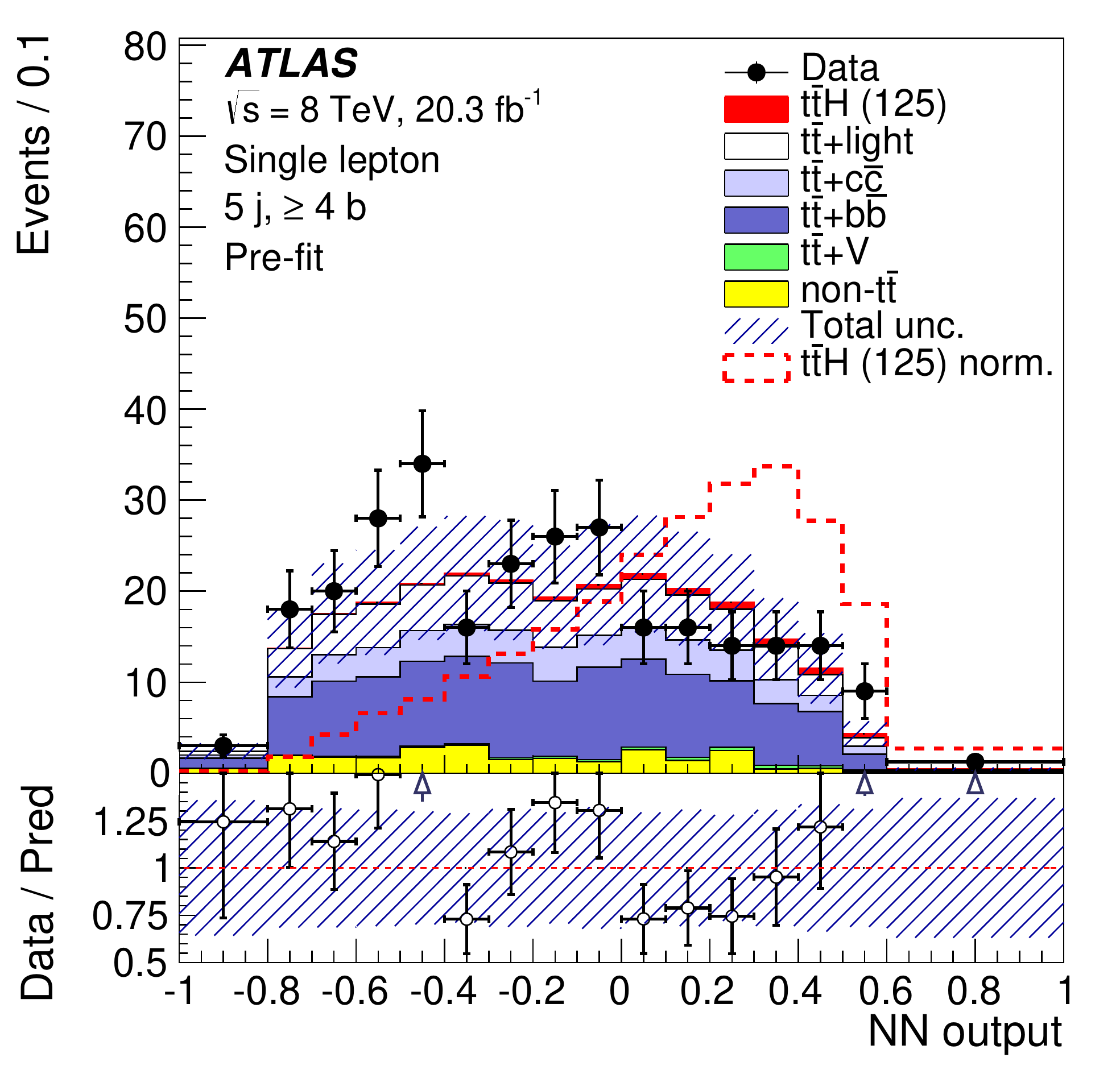}}\label{fig:prepost_lj_2_e}  
\subfigure[]{\includegraphics[width=0.34\textwidth]{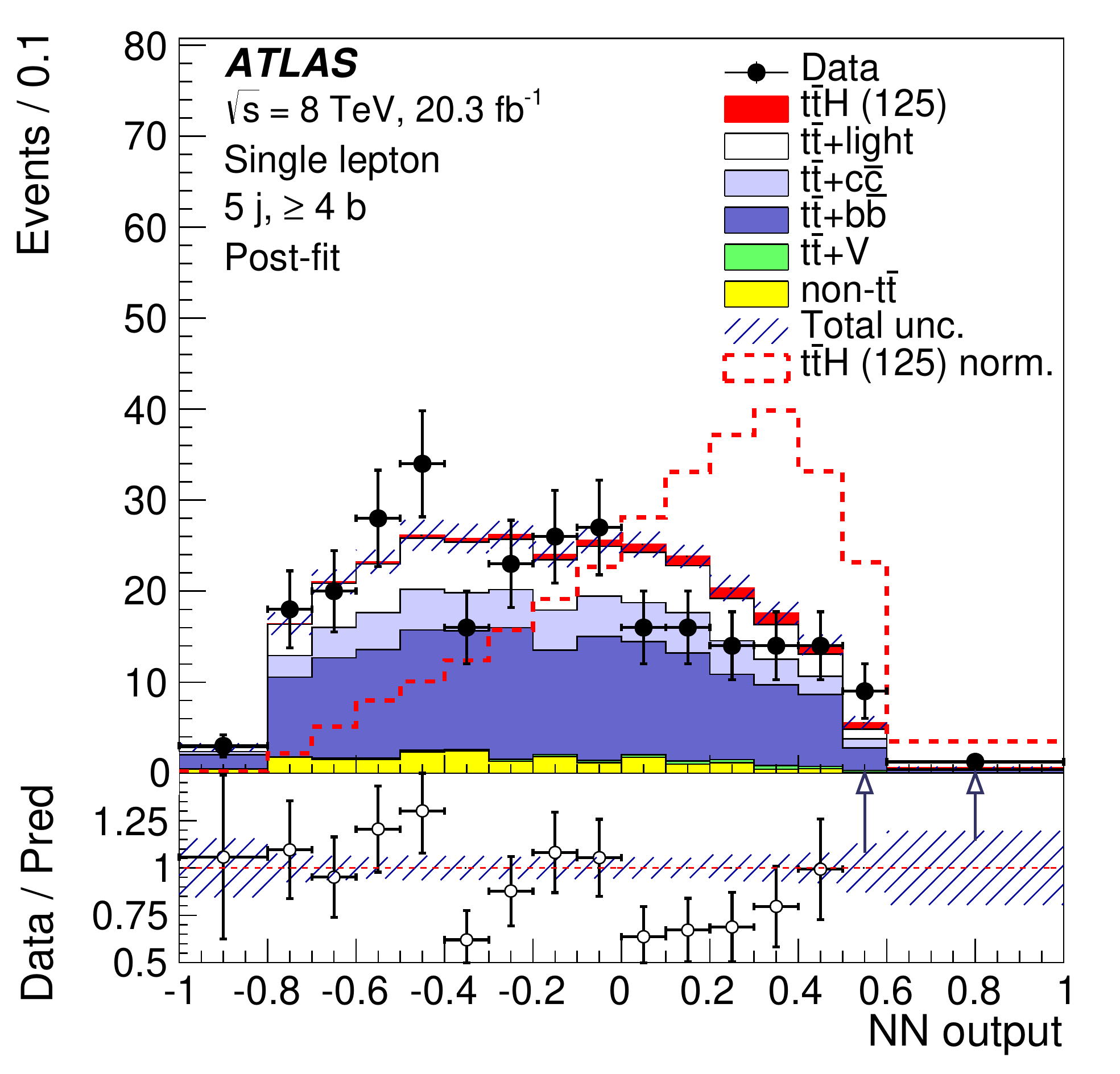}}\label{fig:prepost_lj_2_f}  \\
\caption{Single-lepton channel: comparison of data and prediction for the discriminant variable used in the 
\fivetwo\ region (a) before the fit and (b) after the fit, 
in the \fivethree\ region (c) before the fit and (d) after the fit, 
in the \fivefour\ region (e) before the fit and (f) after the fit. 
The fit is peformed on data under the signal-plus-background
hypothesis. The last bin in all figures contains the overflow. The bottom 
panel displays the ratio of 
data to the total prediction. An arrow indicates that the point is off-scale. The hashed area represents the uncertainty on the background.
The dashed line shows \tth\ signal
distribution normalised to background yield. The \tth\ signal yield (solid) 
is normalised to the SM cross section before the fit and to the fitted $\mu$ after the fit.  
In several regions, predominantly the control regions, the \tth\ signal yield is not visible on top 
of the large background. }
\label{fig:prepost_lj_2} 
\end{center}
\end{figure*}

\begin{figure*}[!ht]
\begin{center}
\subfigure[]{\includegraphics[width=0.34\textwidth]{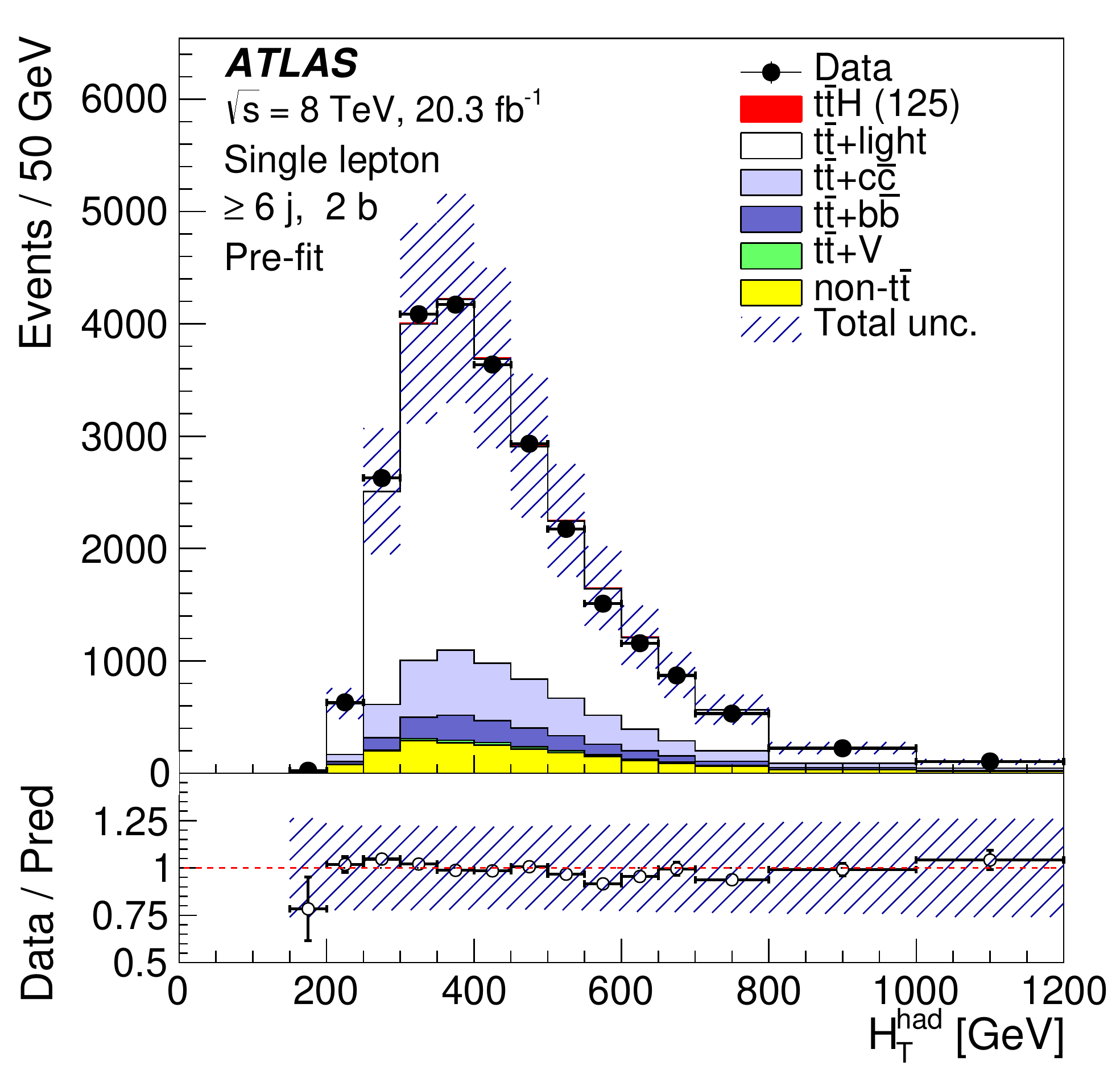}}\label{fig:prepost_lj_3_a}  
\subfigure[]{\includegraphics[width=0.34\textwidth]{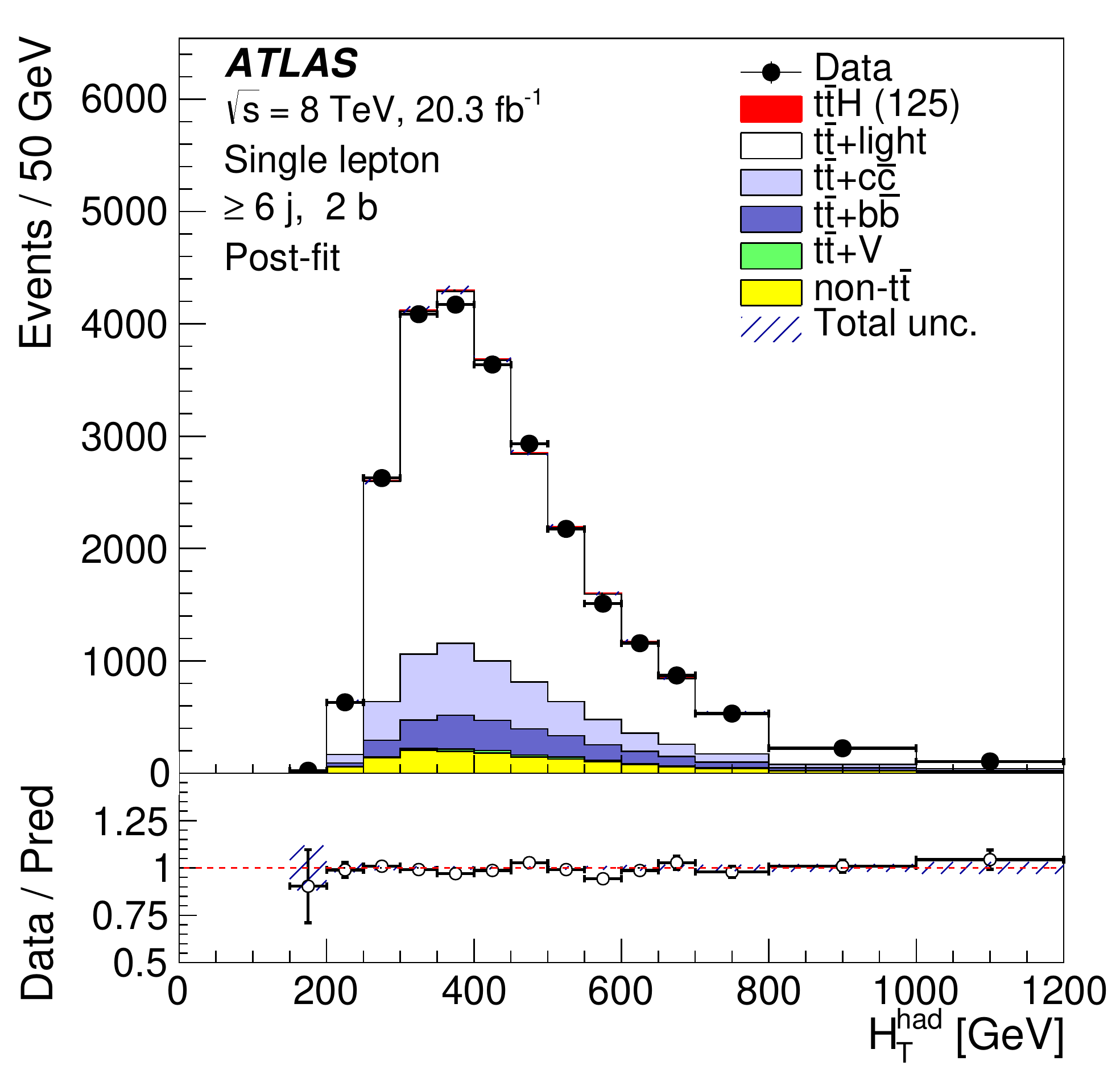}}\label{fig:prepost_lj_3_b} \\ 
\subfigure[]{\includegraphics[width=0.34\textwidth]{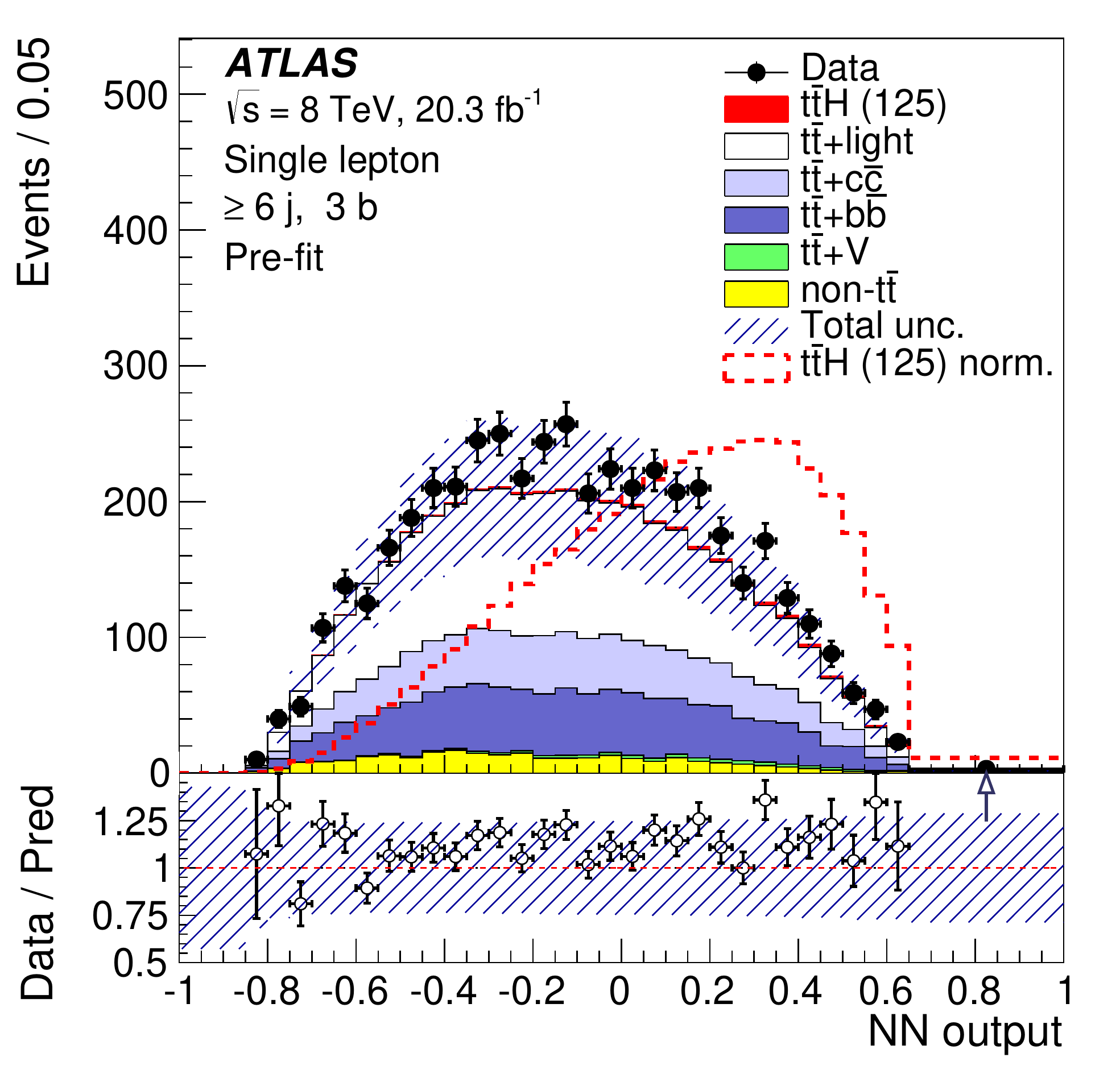}}\label{fig:prepost_lj_3_c} 
\subfigure[]{\includegraphics[width=0.34\textwidth]{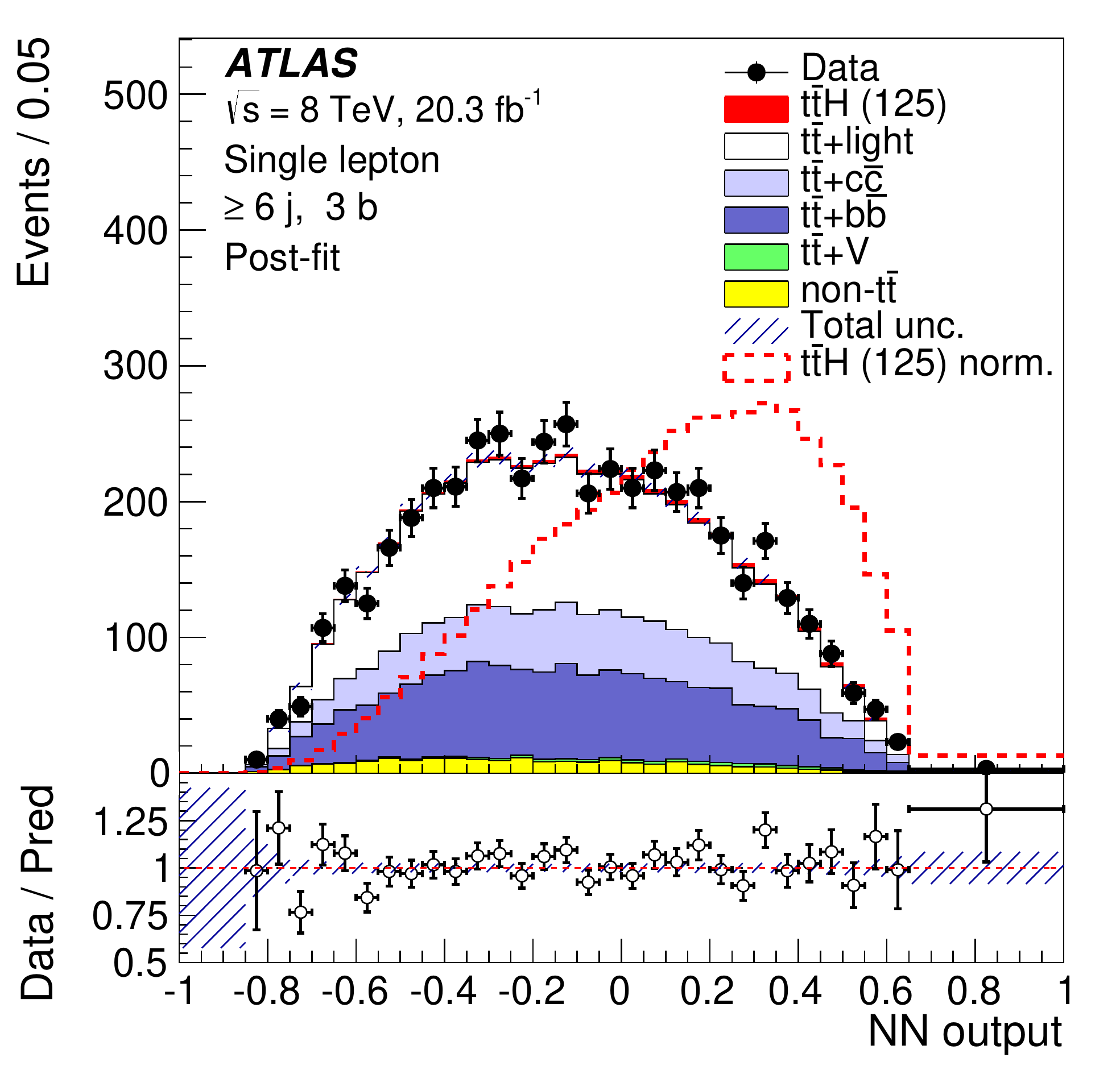}}\label{fig:prepost_lj_3_d} \\
\subfigure[]{\includegraphics[width=0.34\textwidth]{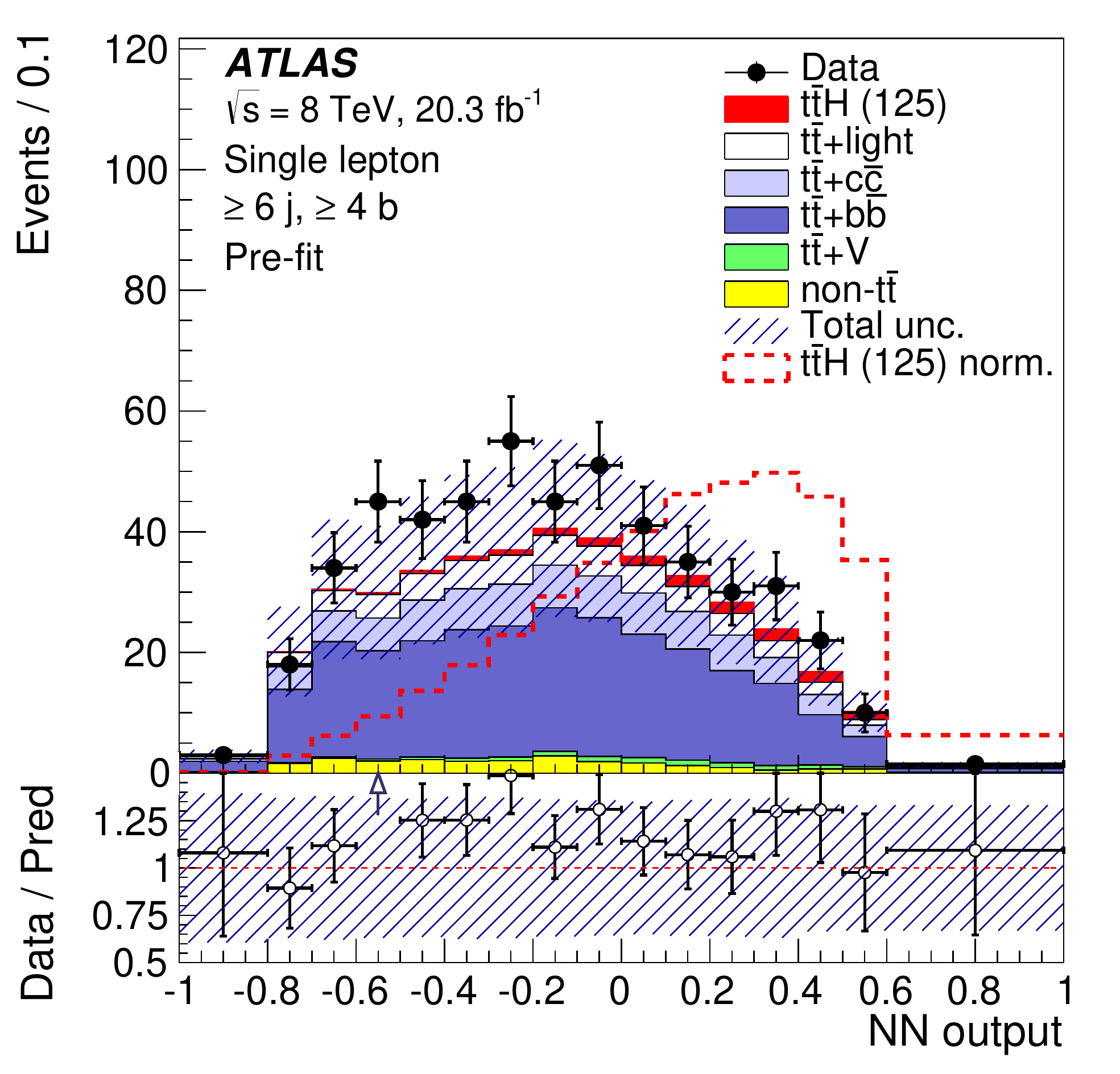}}\label{fig:prepost_lj_3_e} 
\subfigure[]{\includegraphics[width=0.34\textwidth]{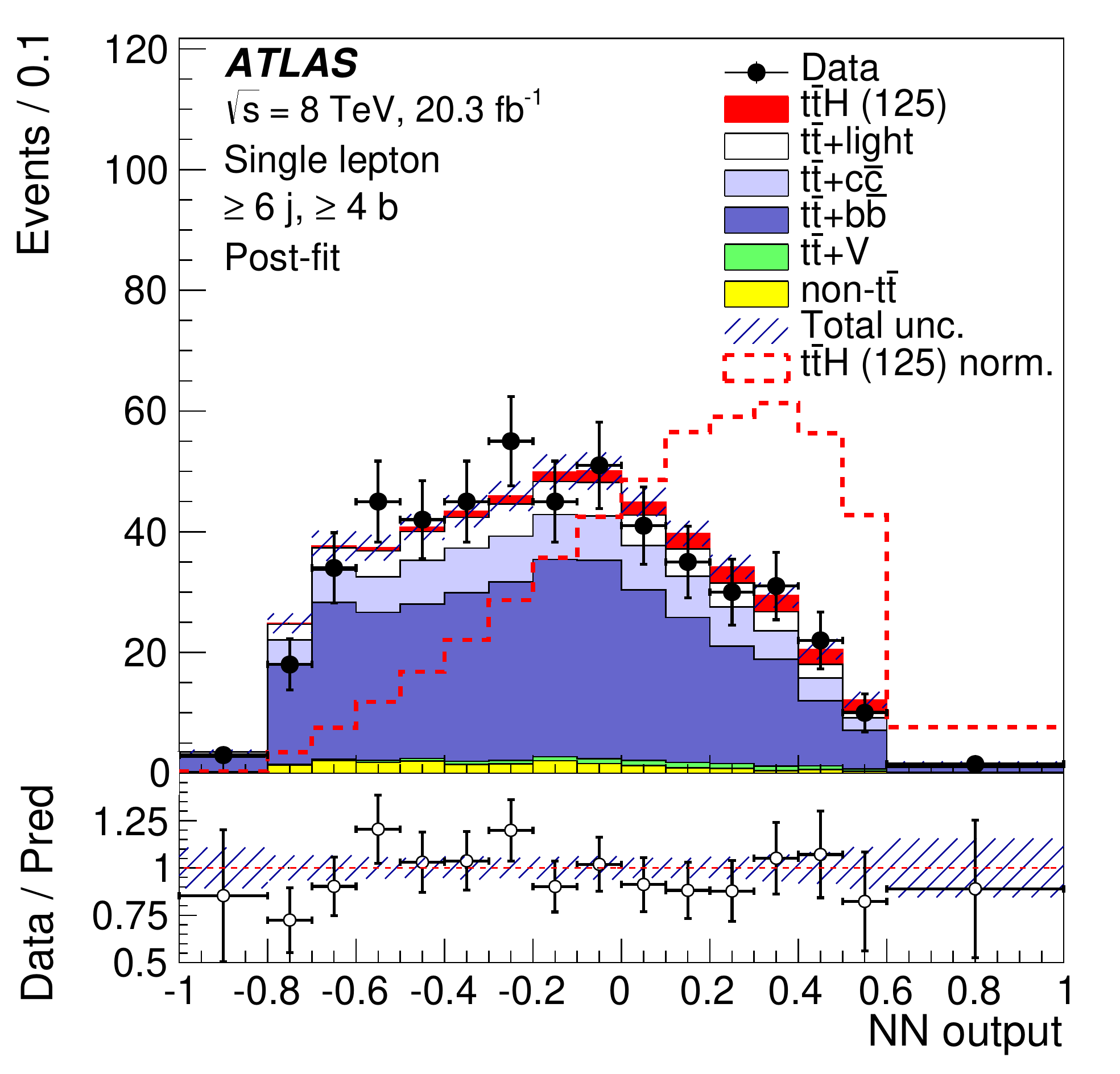}}\label{fig:prepost_lj_3_f} \\
\caption{Single-lepton channel: comparison of data and prediction for the discriminant variable used in the 
\sixtwo\ region (a) before the fit and (b) after the fit, 
in the \sixthree\ region (c) before the fit and (d) after the fit, 
in the \sixfour\ region (e) before the fit and (f) after the fit. 
The fit is performed on data under the signal-plus-background hypothesis. The last bin in all figures contains the overflow. 
The bottom panel displays the ratio of data to the total prediction. An arrow indicates that the point is off-scale.
The hashed area represents the uncertainty on the background. 
The dashed line shows \tth\ signal distribution normalised to background yield. The \tth\ signal 
yield (solid) 
is normalised to the SM cross section before the fit and to the fitted $\mu$ after the fit.
In several regions, predominantly the control regions, the \tth\ signal yield is not visible on top 
of the large background. }
\label{fig:prepost_lj_3} 
\end{center}
\end{figure*}

\begin{figure*}[!ht]
\begin{center}
\subfigure[]{\includegraphics[width=0.34\textwidth]{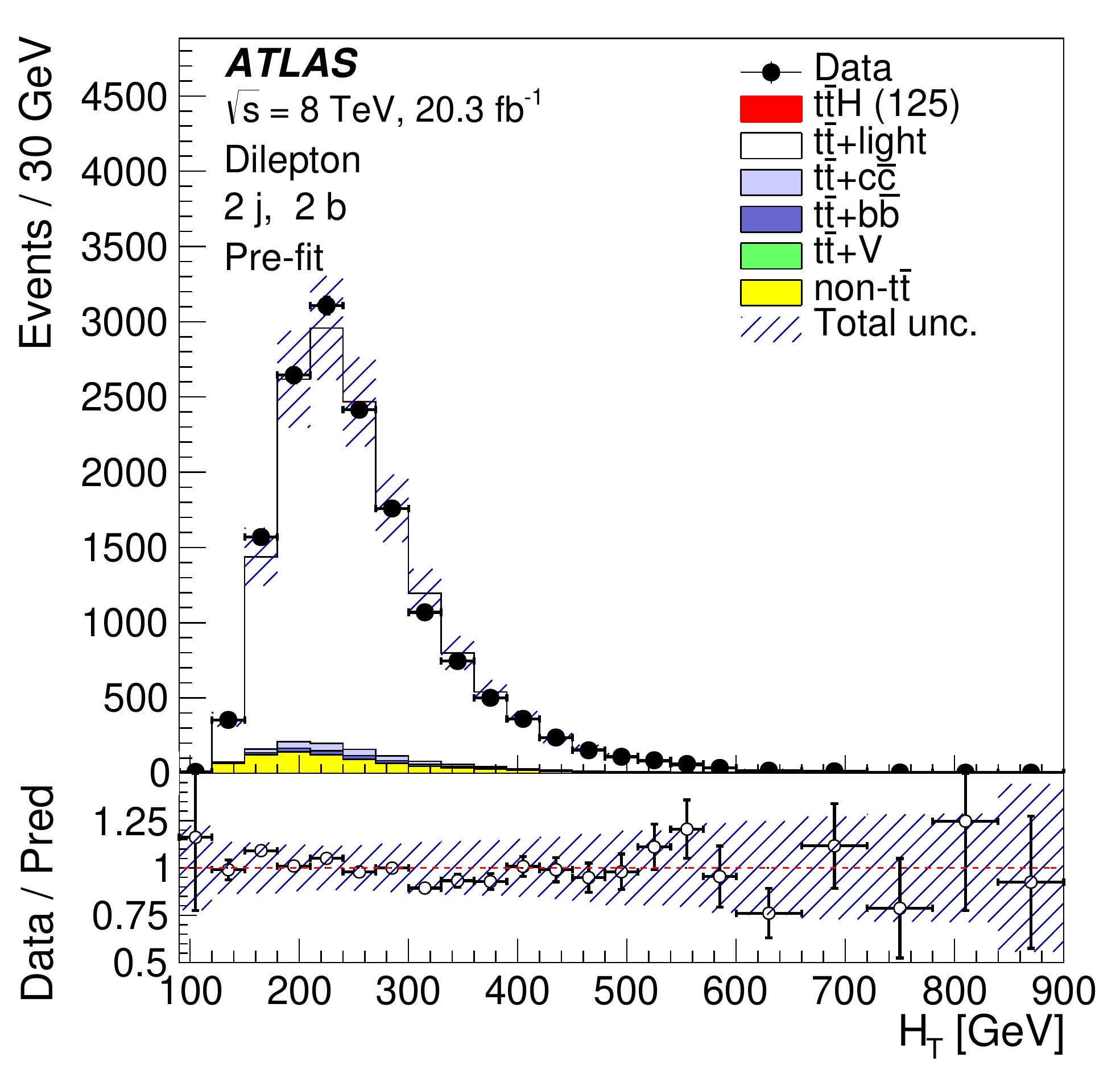}}\label{fig:prepost_dil_1_a} 
\subfigure[]{\includegraphics[width=0.34\textwidth]{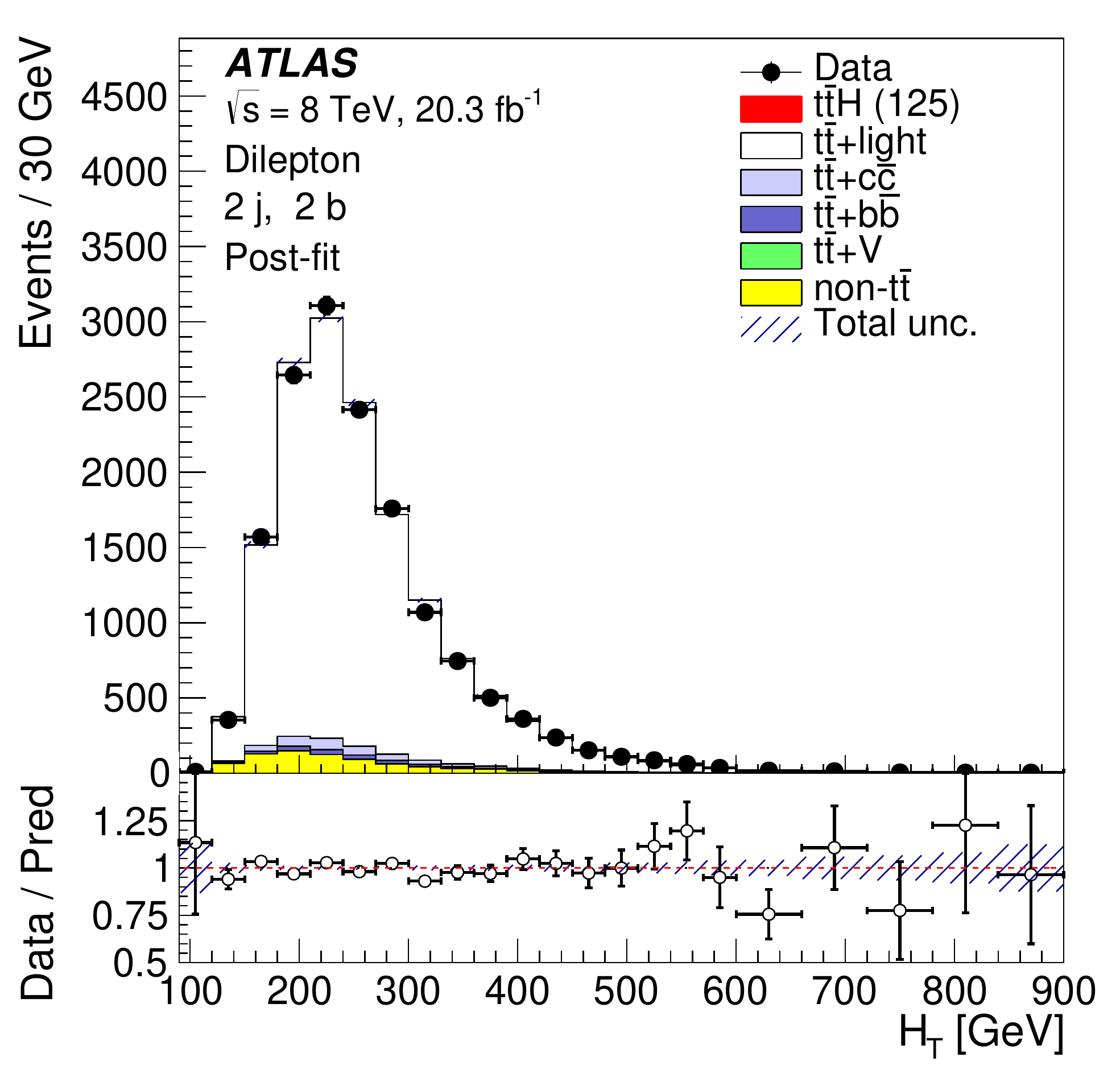}}\label{fig:prepost_dil_1_b}\\
\subfigure[]{\includegraphics[width=0.34\textwidth]{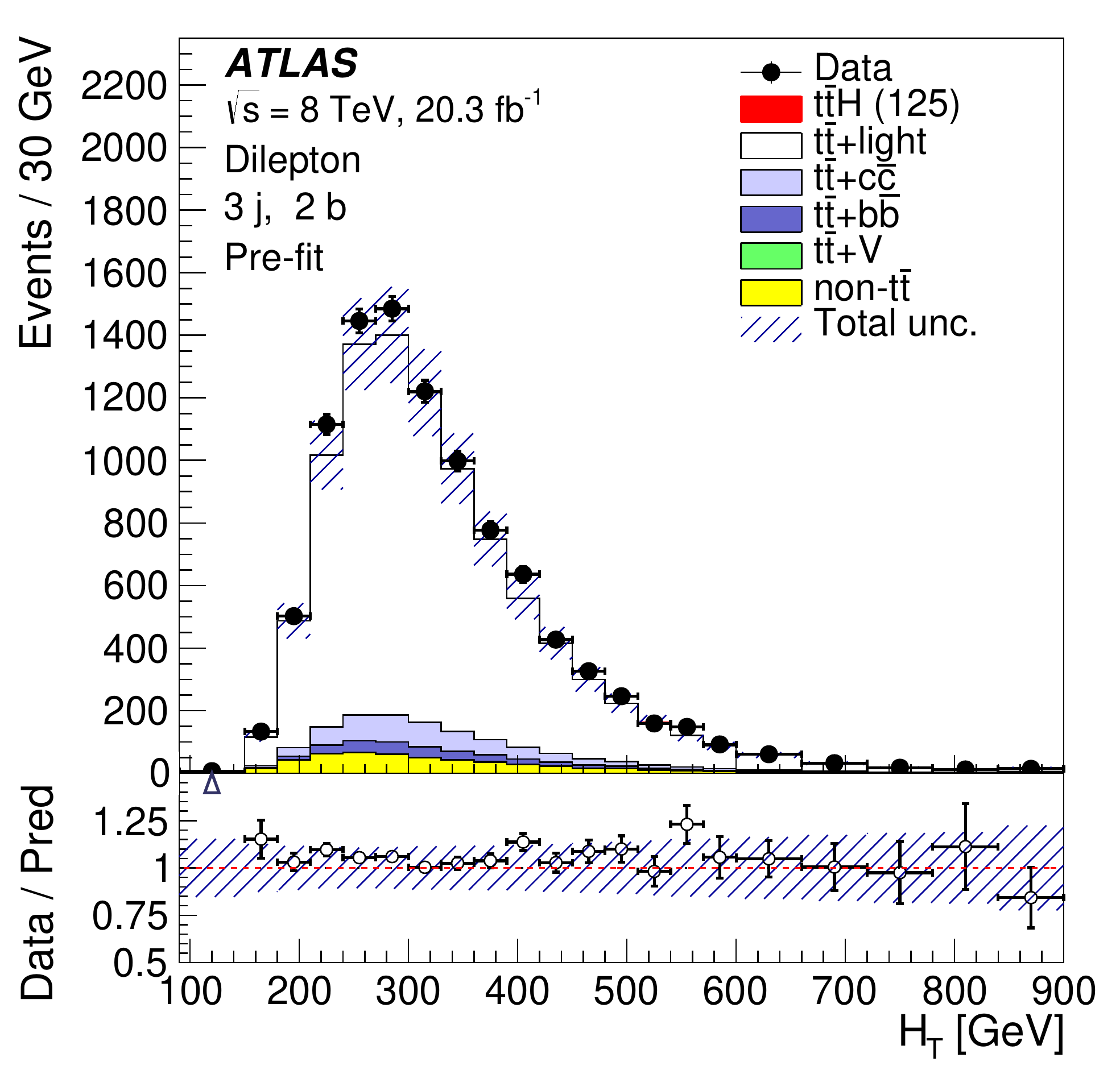}}\label{fig:prepost_dil_1_c} 
\subfigure[]{\includegraphics[width=0.34\textwidth]{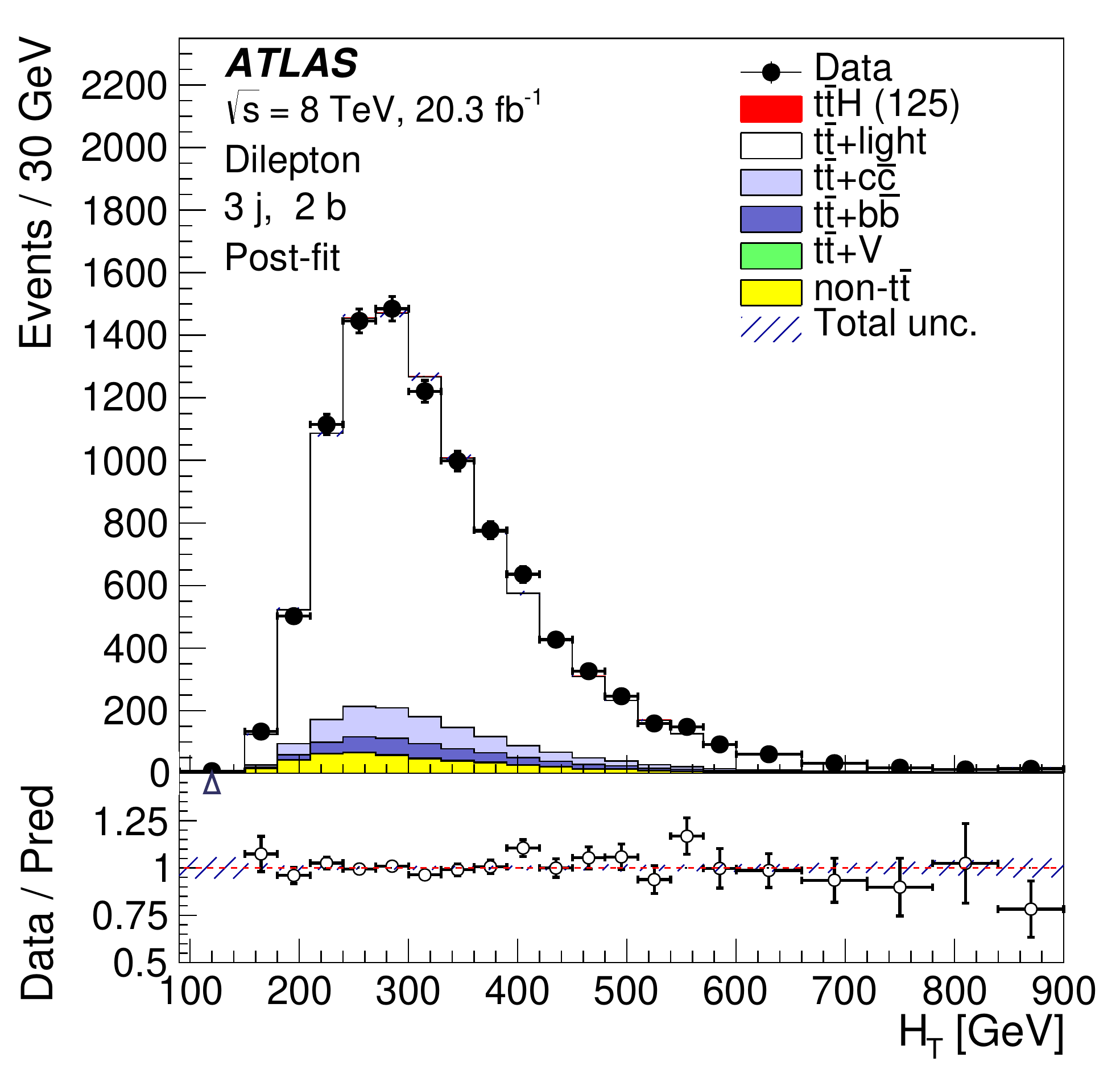}}\label{fig:prepost_dil_1_d} \\
\subfigure[]{\includegraphics[width=0.34\textwidth]{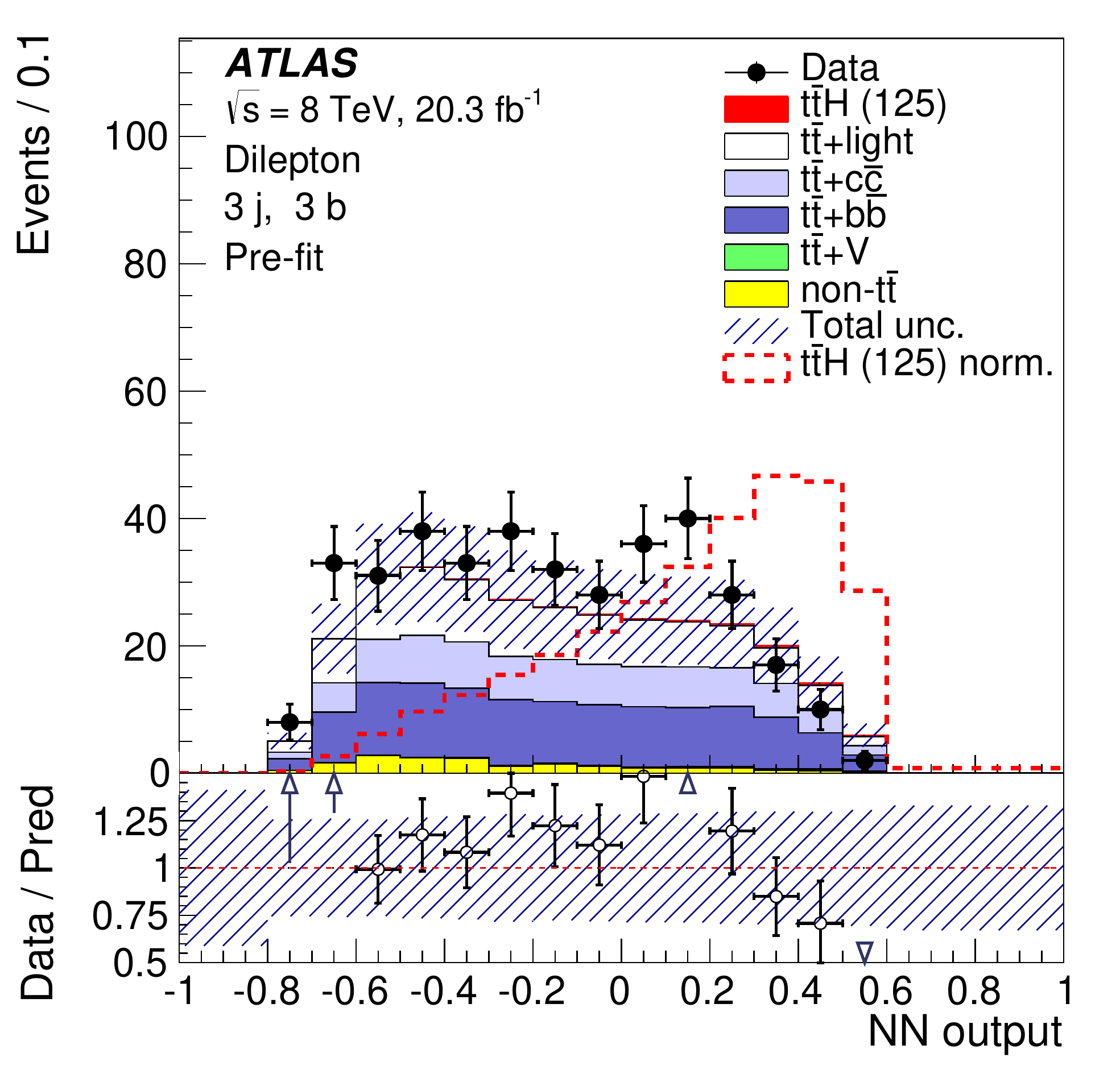}}\label{fig:prepost_dil_1_e} 
\subfigure[]{\includegraphics[width=0.34\textwidth]{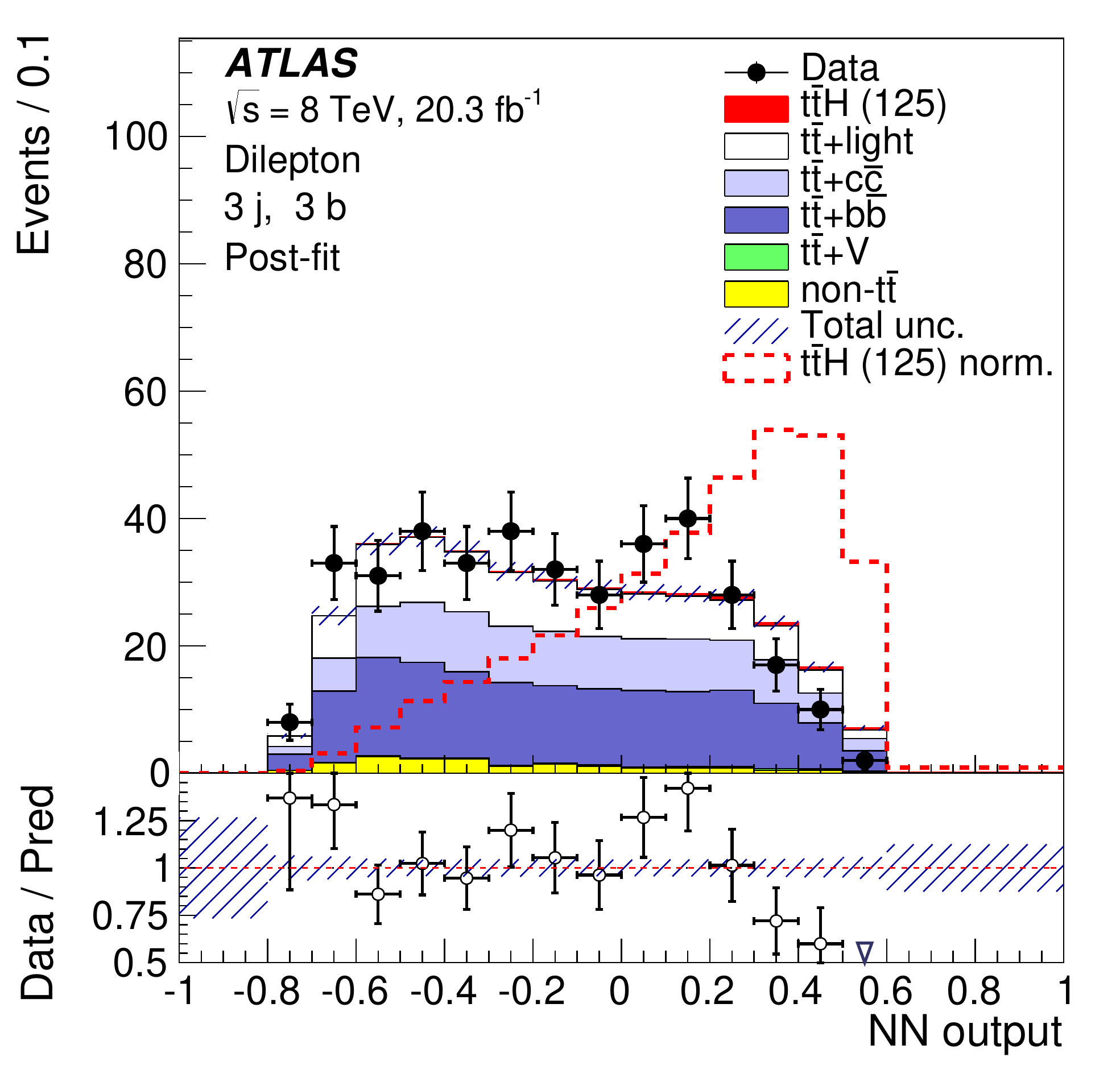}}\label{fig:prepost_dil_1_f} \\ 
\caption{Dilepton channel: comparison of data and prediction for the discriminant variable used in the 
\twotwo\ region (a) before the fit and (b) after the fit, 
in the \threetwo\ region (c) before the fit and (d) after the fit, 
in the \threethree\ region (e) before the fit and (f) after the fit.  The fit is performed on data under the signal-plus-background
hypothesis. The last bin in all figures contains the overflow. 
The bottom panel displays the ratio of data to the total prediction. An arrow indicates that the point is off-scale.
The hashed area represents the uncertainty on the background. 
The dashed line shows \tth\ signal 
distribution normalised to background yield. The \tth\ signal yield (solid ) 
is normalised to the SM cross section before the fit and to the fitted $\mu$ after the fit.
In several regions, predominantly the control regions, the \tth\ signal yield is not visible on top 
of the large background. }
\label{fig:prepost_dil_1} 
\end{center}
\end{figure*}

\begin{figure*}[!ht]
\begin{center}
\subfigure[]{\includegraphics[width=0.34\textwidth]{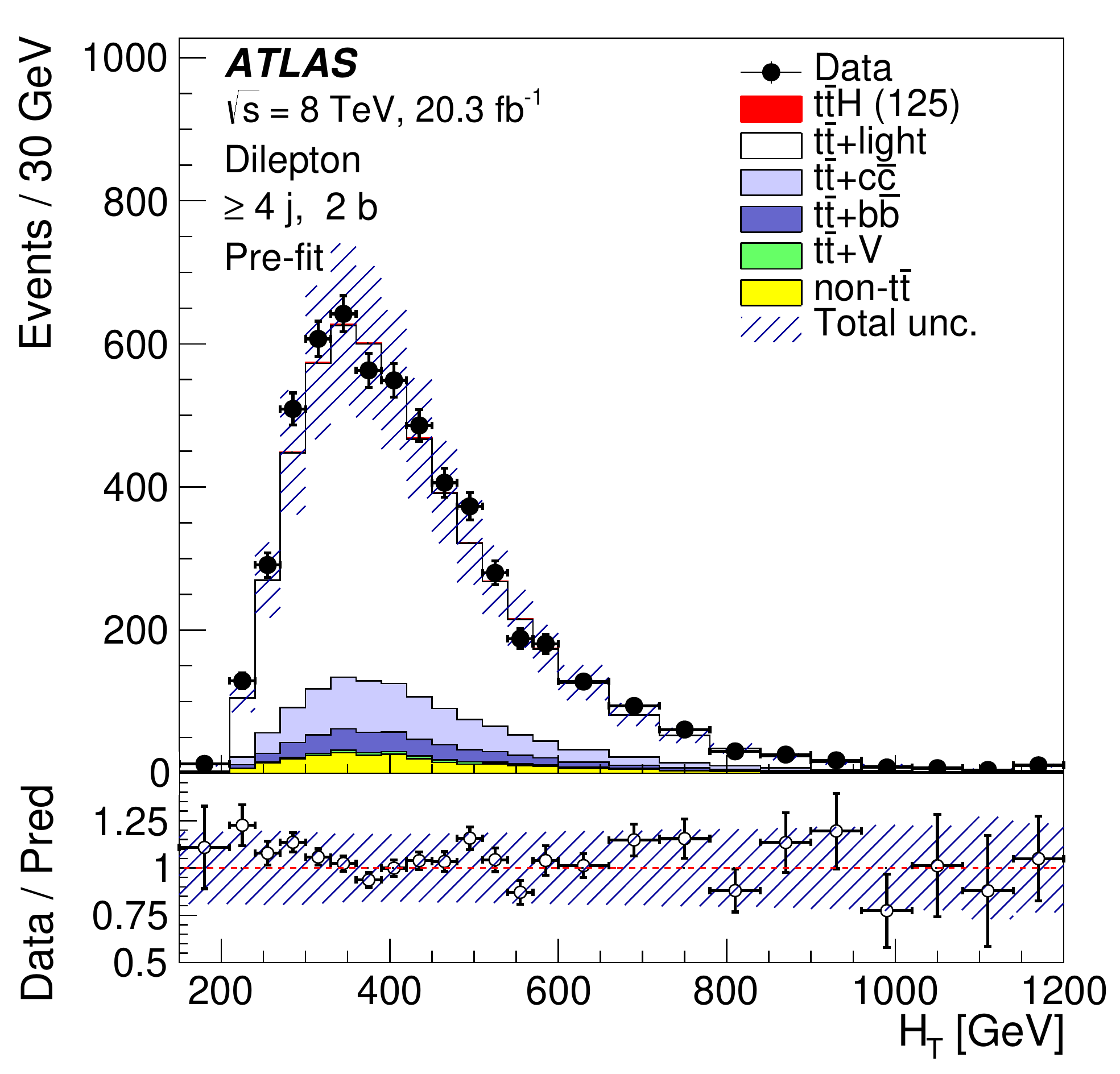}}\label{fig:prepost_dil_2_a}
\subfigure[]{\includegraphics[width=0.34\textwidth]{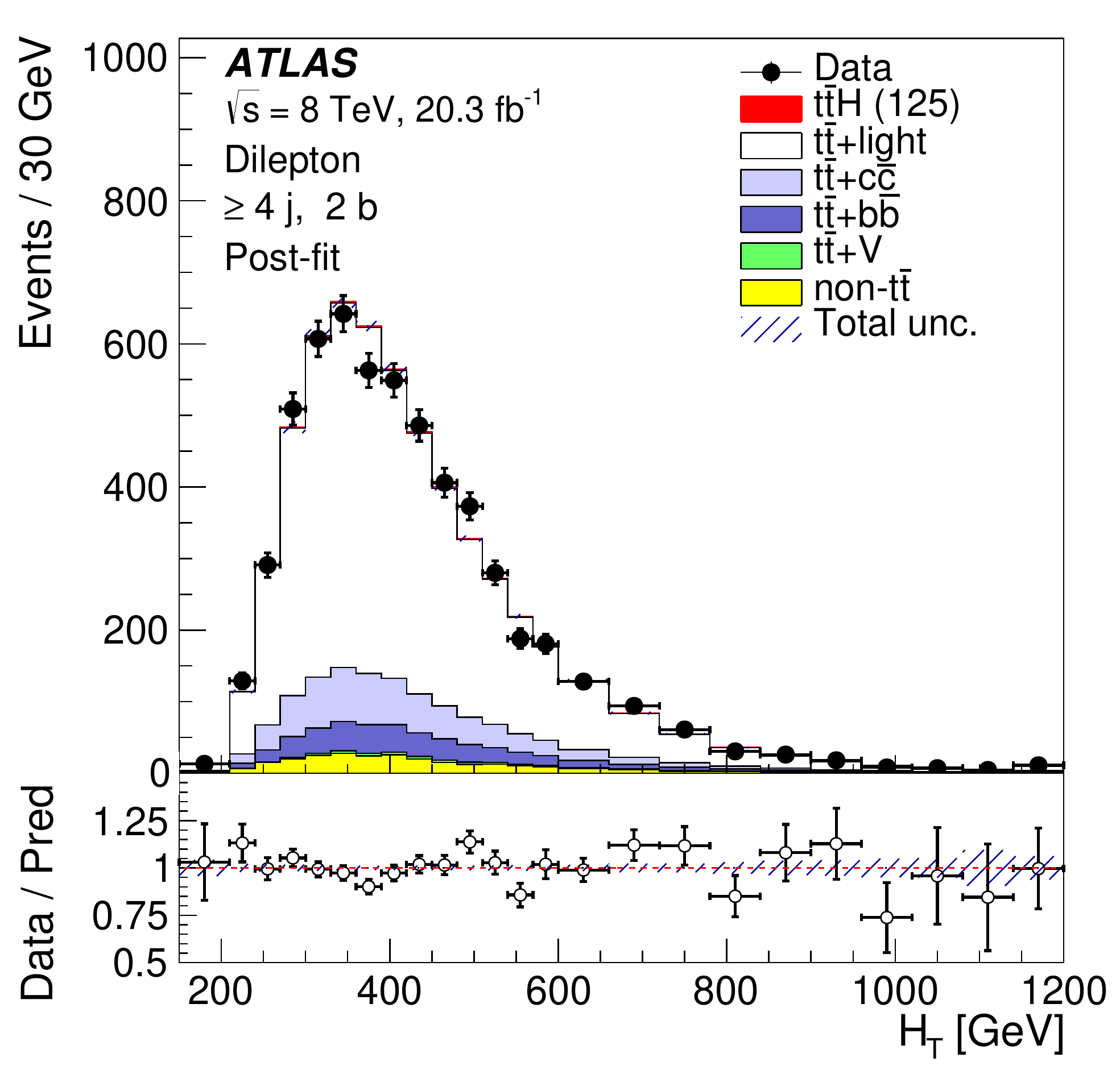}}\label{fig:prepost_dil_2_b} \\ 
\subfigure[]{\includegraphics[width=0.34\textwidth]{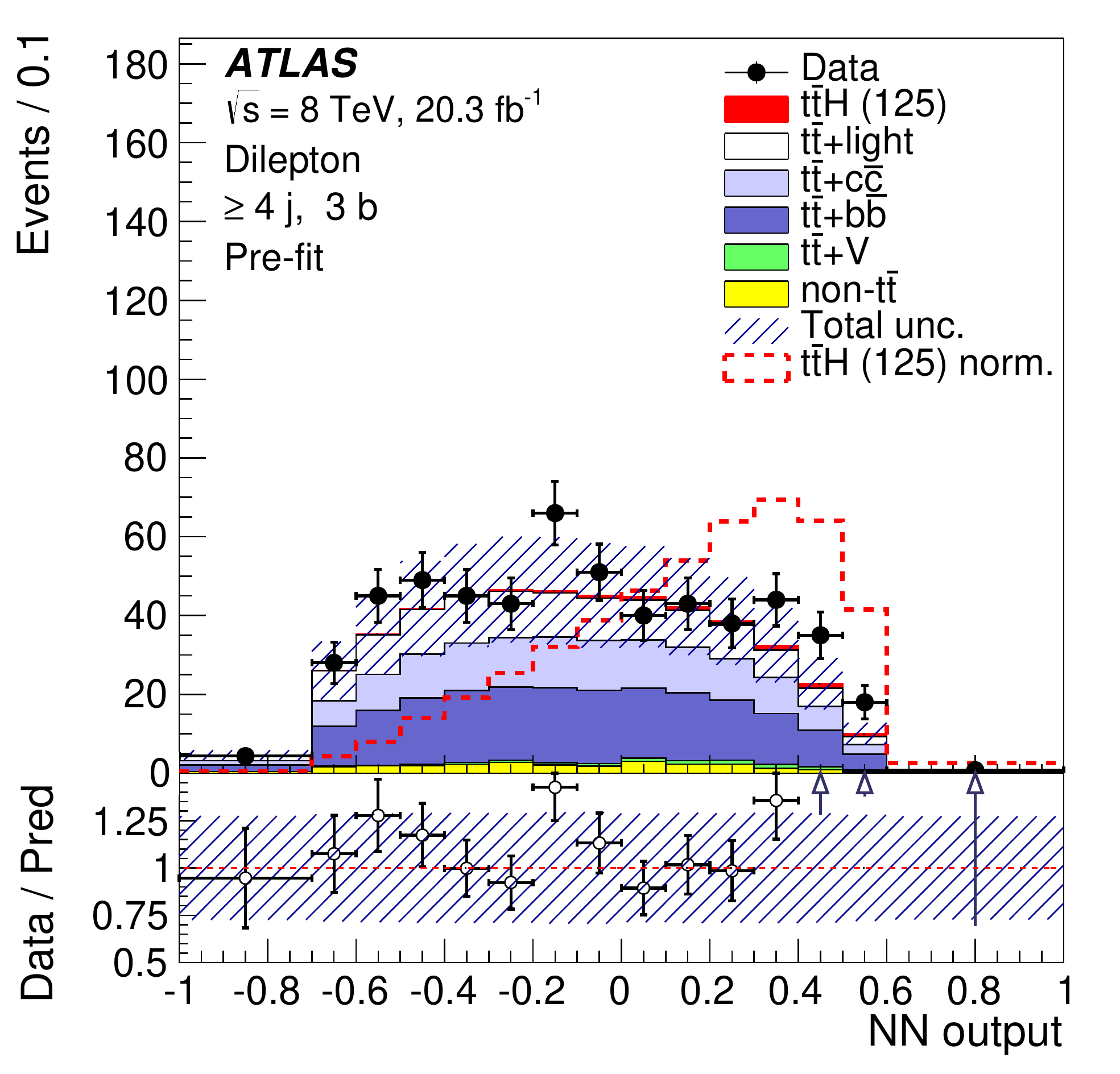}}\label{fig:prepost_dil_2_c} 
\subfigure[]{\includegraphics[width=0.34\textwidth]{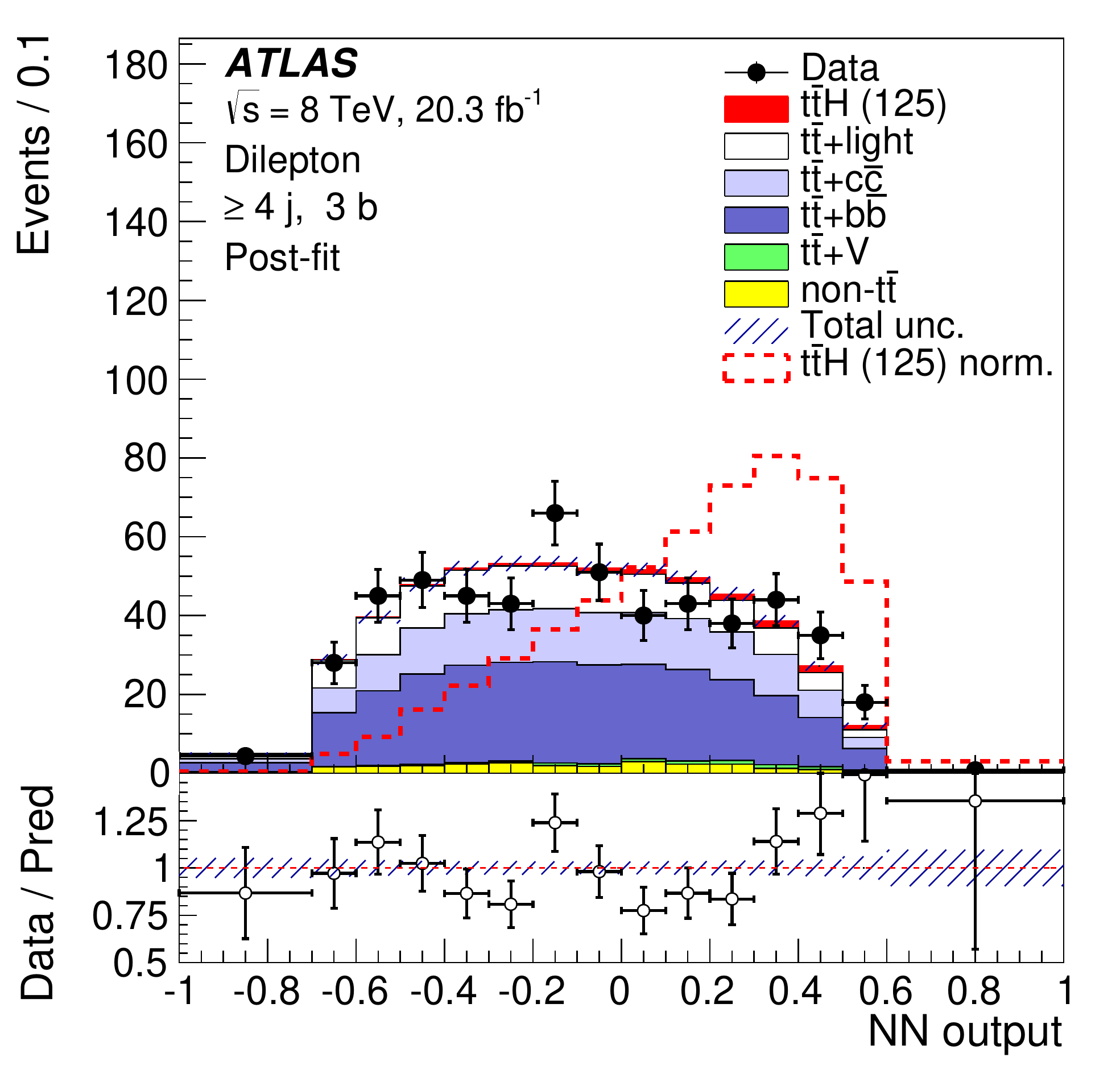}}\label{fig:prepost_dil_2_d} \\ 
\subfigure[]{\includegraphics[width=0.34\textwidth]{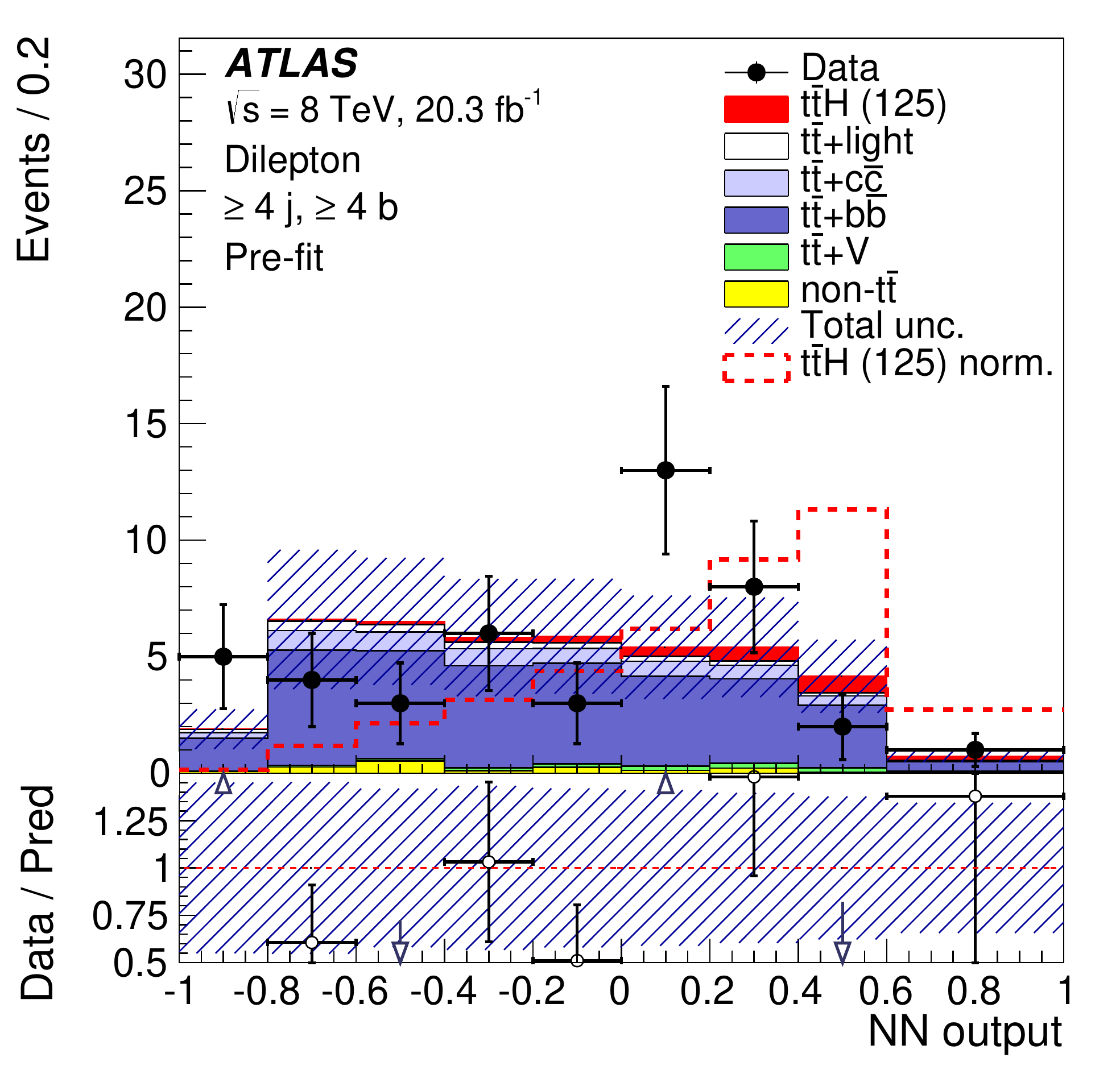}}\label{fig:prepost_dil_2_e} 
\subfigure[]{\includegraphics[width=0.34\textwidth]{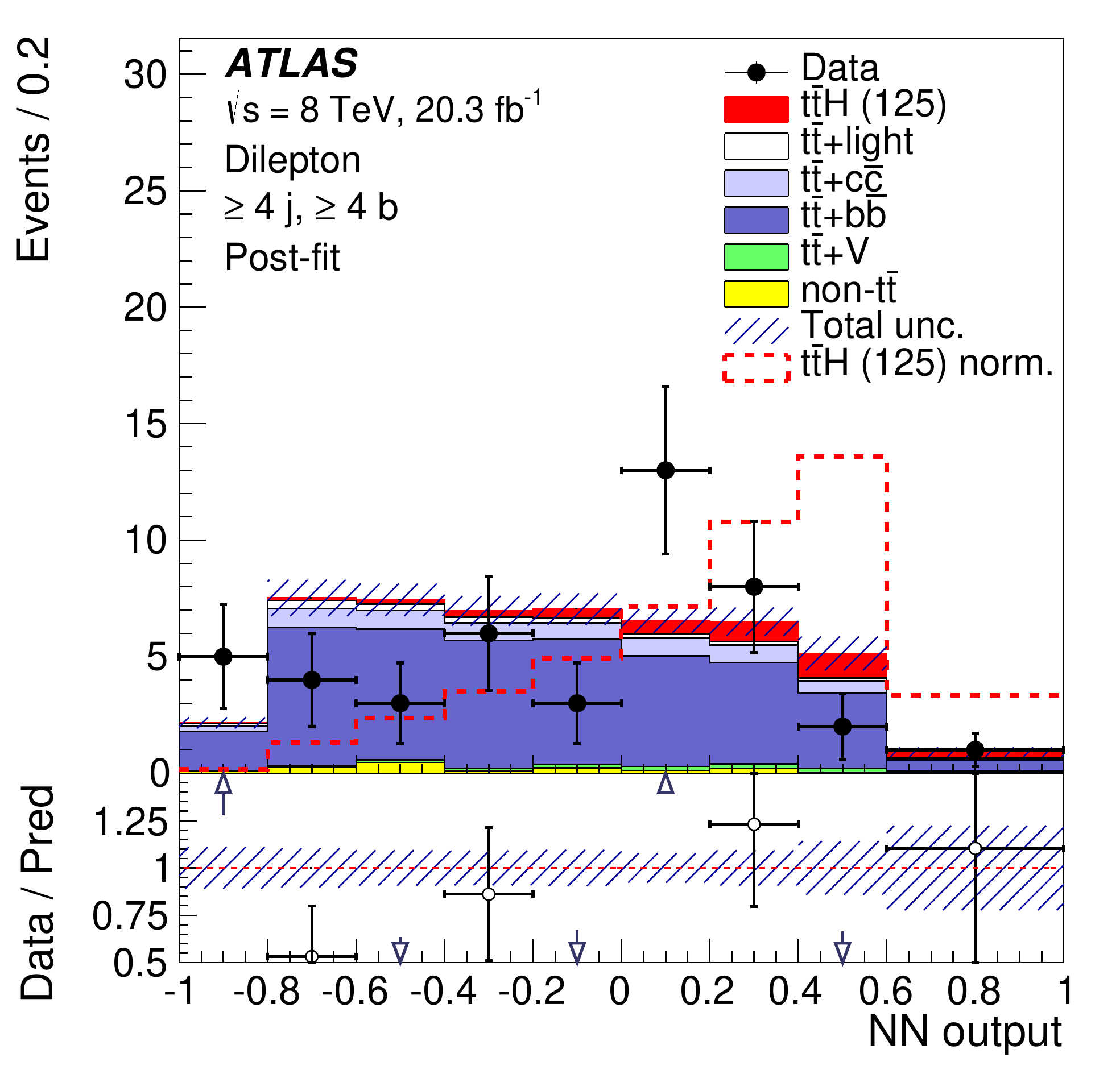}}\label{fig:prepost_dil_2_f} \\
\caption{Dilepton channel: comparison of data and prediction for the discriminant variable used in the 
\fourtwodi\ region (a) before the fit and (b) after the fit, 
in the \fourthreedi\ region (c) before the fit and (d) after the fit, 
in the \fourfourdi\ region (e) before the fit and (f) after the fit.  The fit is performed on data under the signal-plus-background
hypothesis. The last bin in all figures contains the overflow. 
The bottom panel displays the ratio of data to the total prediction. An arrow indicates that the point is off-scale.
The hashed area represents the uncertainty on the background. 
The dashed line shows \tth\ signal 
distribution normalised to background yield. The \tth\ signal yield (solid) 
is normalised to the SM cross section before the fit and to the fitted $\mu$ after the fit.
In several regions, predominantly the control regions, the \tth\ signal yield is not visible on top 
of the large background. }
\label{fig:prepost_dil_2} 
\end{center}
\end{figure*}

Table~\ref{tab:MuVSAnalysis} shows the observed $\mu$ values obtained from the individual 
fits in the single-lepton and dilepton channels, and their combination.  
The signal strength from the combined fit for $m_H=125\gev$ is:
\begin{equation}
\mu (m_H=125\gev) = 1.5 \pm 1.1.
\end{equation} 
The expected uncertainty for the signal strength ($\mu~=~1$) is $\pm 1.1$.  
The observed (expected) significance of the signal is 1.4 (1.1)
standard deviations, which corresponds to an observed (expected) $p$-value of 8\% (15\%).   
The probability, $p$, to obtain a result
at least as signal-like as observed if no signal is present is calculated 
using  $q_0 = -2{\rm ln} (\mathcal{L}(0,\hat{\hat{\theta_{\mu}}})/\mathcal{L}(\hat{\mu},
\hat{\theta}))$ as a
test statistic. 

\begin{table}[!ht]
\begin{center}
\begin{tabular}{l|cc}
\hline\hline
Signal strength    & $\mu$ & Uncertainty \\

\hline
Single lepton      &  1.2   &   1.3  \\
Dilepton           &  2.8   &   2.0   \\
\hline 
Combination        &  1.5   &   1.1   \\
\hline\hline
\end{tabular}

\caption{\label{tab:MuVSAnalysis} The fitted values of signal strength and their 
uncertainties for the individual channels as well as their combination, 
assuming $m_H=125\gev$. Total uncertainties are shown. }
\end{center}
\end{table}

The fitted values of the signal strength and their  
uncertainties for the individual channels and their combination 
are shown in Fig.~\ref{fig:fittedmu}.   

The observed limits, those expected with and without assuming a SM 
Higgs boson with $m_H=125\gev$, for each channel 
and their combination are shown in Fig.~\ref{fig:limits}. A signal 3.4 times larger than 
predicted by the SM is excluded at 95\% CL using the CL$_{\rm{s}}$ 
method. A signal 2.2 times larger than for the SM 
Higgs boson is expected to be excluded in the case of no SM Higgs 
boson, and 3.1 times larger in the case of a SM Higgs boson.  This 
is also summarised in Table~\ref{tab:LimitsVSAnalysis}.

\begin{table*}[!ht]
\small
\vspace{0.2cm}
\begin{center}
\begin{tabular}{l|c|ccccc|c}
\hline\hline
95\% CL upper limit     & Observed & $-2\sigma$  & $-1\sigma$  & Median    & $+1\sigma$  & $+2\sigma$ & Median ($\mu = 1$) \\

\hline
Single lepton      &  3.6   &  1.4    & 1.9  &   2.6   & 3.7  & 4.9  & 3.6 \\
Dilepton           &  6.7   &  2.2    & 3.0  &   4.1   & 5.8  & 7.7  & 4.7 \\
\hline
Combination        &  3.4   &  1.2    & 1.6  &   2.2   & 3.0  & 4.1  & 3.1 \\
\hline\hline
\end{tabular}
\caption{\label{tab:LimitsVSAnalysis} Observed and expected
(median, for the background-only hypothesis) 95\% CL upper limits on
$\sigma(t\bar{t}H)$ relative to the SM prediction, for the individual channels as well as their
combination, assuming $m_H=125\gev$. The 68\% and 95\% confidence intervals around the expected limits under the background-only hypothesis are also provided, denoted by $\pm 1 \sigma$ and $\pm 2 \sigma$, respectively.  The expected (median) 95\% CL upper limits assuming the SM prediction for $\sigma(t\bar{t}H)$ are shown in the last column. }
\end{center}
\end{table*}

Figure~\ref{fig:logSBplot} summarises post-fit event yields as a function of $\log_{10}(S/B)$, 
for all bins of the distributions used in the combined fit of the single-lepton and dilepton channels. 
The value of $\log_{10}(S/B)$ is calculated according to the post-fit yields in each bin of the fitted distributions, 
either \hthad, \htlep, or NN. The total number of background and signal events is displayed 
in bins of $\log_{10}(S/B)$. In particular, the last bin of 
Fig.~\ref{fig:logSBplot} includes the 
two last bins from the most signal-rich region of the NN 
distribution in \sixfour\ and the two last bins from the most 
signal-rich region of the NN in \fourfourdi\ from the fit.
The signal is normalised to the fitted value of the signal strength ($\mu = 1.5$) and the background is obtained
from the global fit. A signal strength 3.4 times larger than predicted by the SM, which is excluded at 95\% CL by this analysis, is also shown.	

Figure~\ref{fig:ranking} demonstrates the effect of various systematic uncertainties on
the fitted value of $\mu$ and the constraints provided by the data. 
The post-fit effect on $\mu$ is calculated by fixing the corresponding 
nuisance parameter at $\hat{\rm{\theta}} \pm \sigma_{\rm{\theta}}$, where  
$\hat{\rm{\theta}}$ is the fitted value of the nuisance parameter and 
$\sigma_{\rm{\theta}}$ is its post-fit uncertainty, and 
performing the fit again. The difference between the default and the modified $\mu$, $\Delta\mu$, 
represents the effect on $\mu$ of this particular systematic uncertainty.
The largest effect arises
from the uncertainty in normalisation of the irreducible \ttbb\ 
background. 
This uncertainty is reduced by more than one half from 
the initial 50\%.
The \ttbb\ background normalisation
is pulled up by about 40\% in the fit, resulting in an increase in the observed 
\ttbb\ yield with respect to the {\sc Powheg}+{\sc Pythia} prediction.
Most of the reduction in uncertainty on the \ttbb\ normalisation
is the result of the significant number of data events in the signal-rich regions 
dominated by \ttbb\ background.  With no Gaussian prior considered on the \ttbb\ normalisation, as described in 
Sect.~\ref{sec:SystematicUncertainties}, the fit still prefers an increase in the amount of \ttbb\ background by about 40\%. 

\begin{figure}[!ht]
\begin{center}
\includegraphics[width=0.47\textwidth]{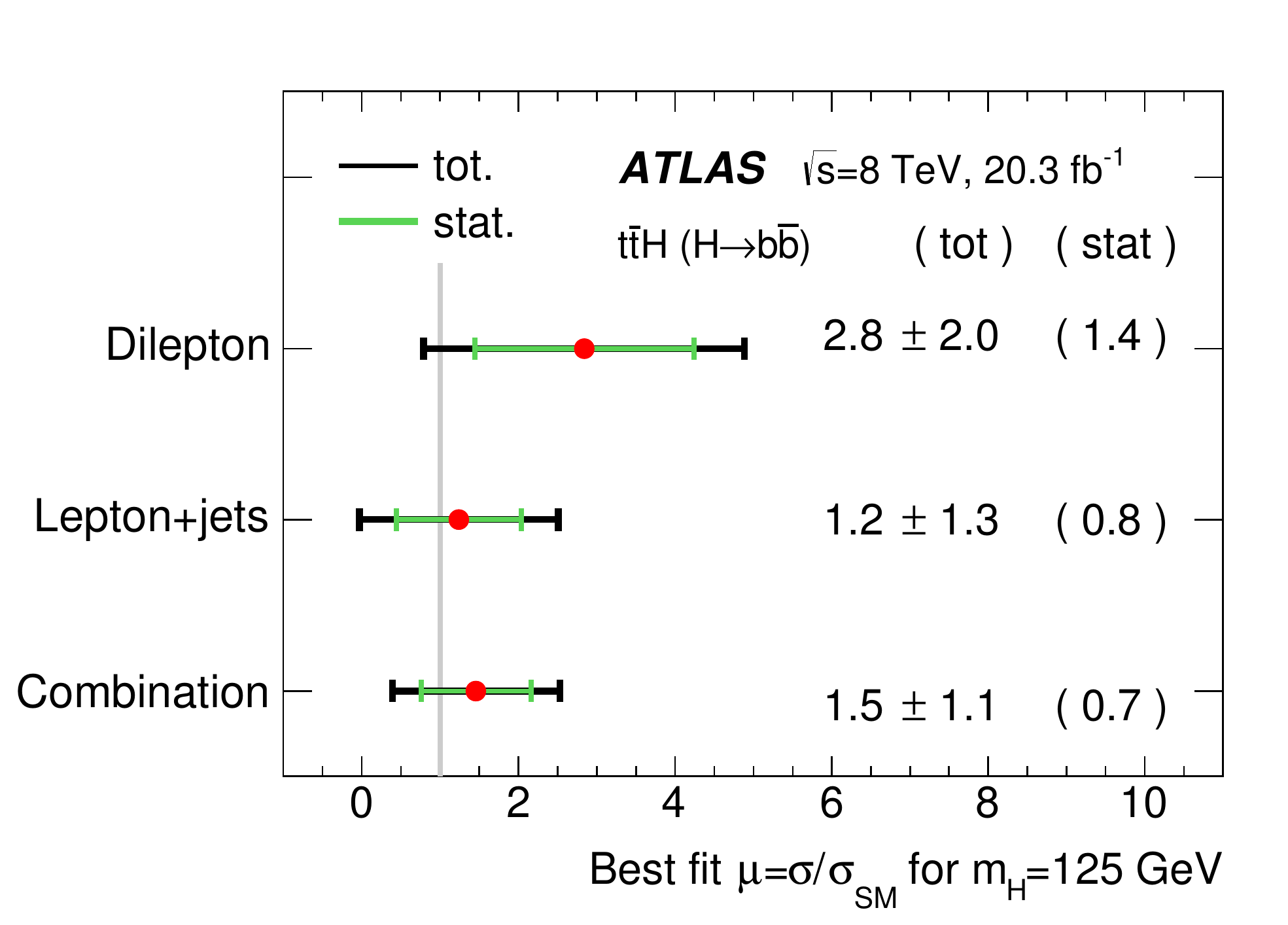}
\caption{The fitted values of the signal strength and their uncertainties for 
the individual channels and their 
combination.  The green line shows the statistical
uncertainty on the signal strength.}
\label{fig:fittedmu} 
\end{center}
\end{figure}

\begin{figure}[!htb]
\begin{center}
\includegraphics[width=0.47\textwidth]{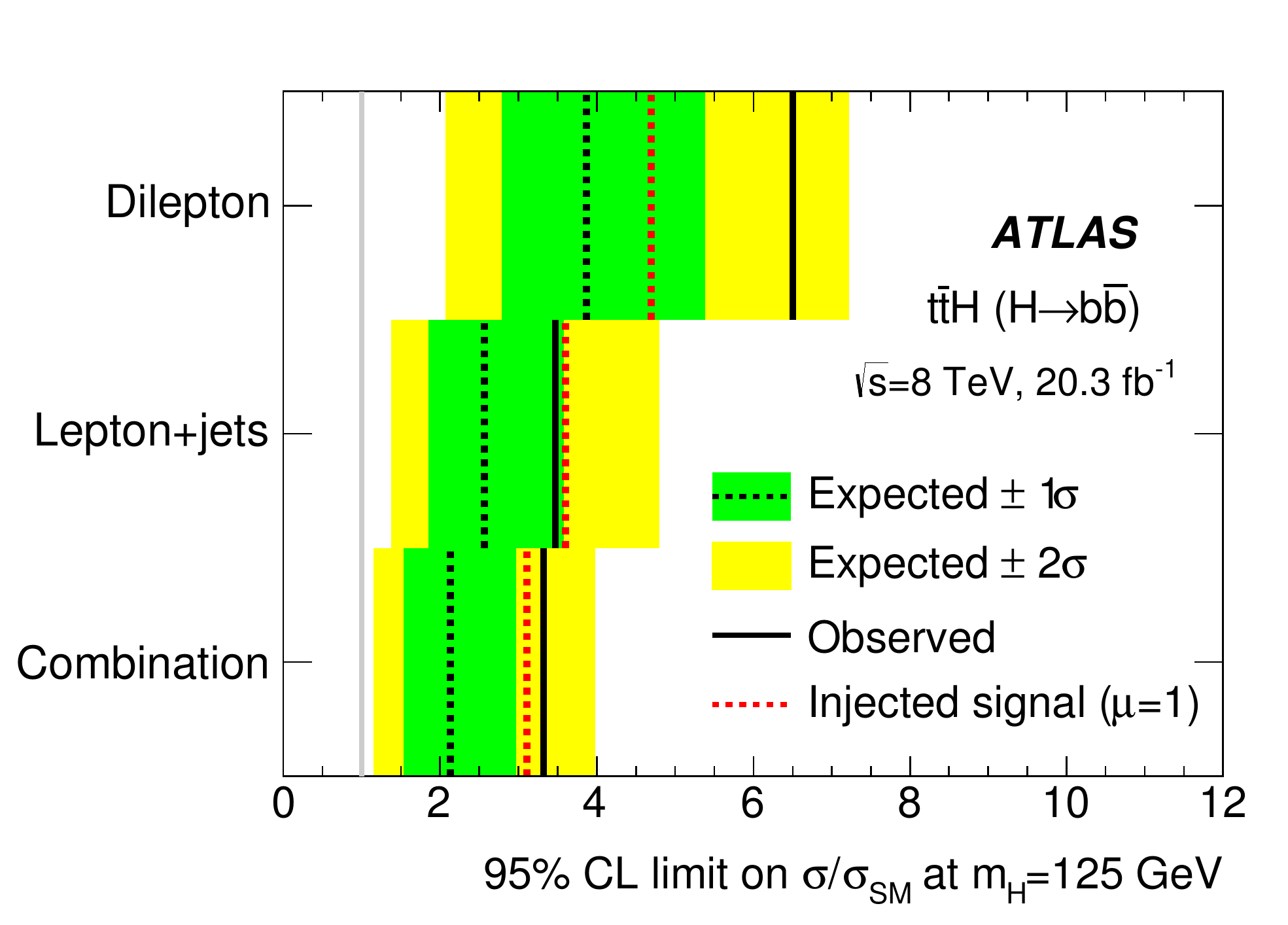}
\caption{95\% CL upper limits on $\sigma(t\bar{t}H)$ relative to the SM prediction, $\sigma/\sigma_{\mathrm{SM}}$, for the individual 
channels as well as their combination. The observed 
limits (solid lines) are compared
to the expected (median) limits under the background-only hypothesis and under
the signal-plus-background hypothesis assuming the SM prediction 
for $\sigma(t\bar{t}H)$ and pre-fit prediction for the background.
The surrounding shaded bands correspond to the 68\% and 95\% confidence 
intervals around the expected limits under the background-only hypothesis, 
denoted by $\pm 1 \sigma$ and $\pm 2 \sigma$, respectively. }
\label{fig:limits} 
\end{center}
\end{figure}

\begin{figure}[!ht]
\begin{center}
\includegraphics[width=0.47\textwidth]{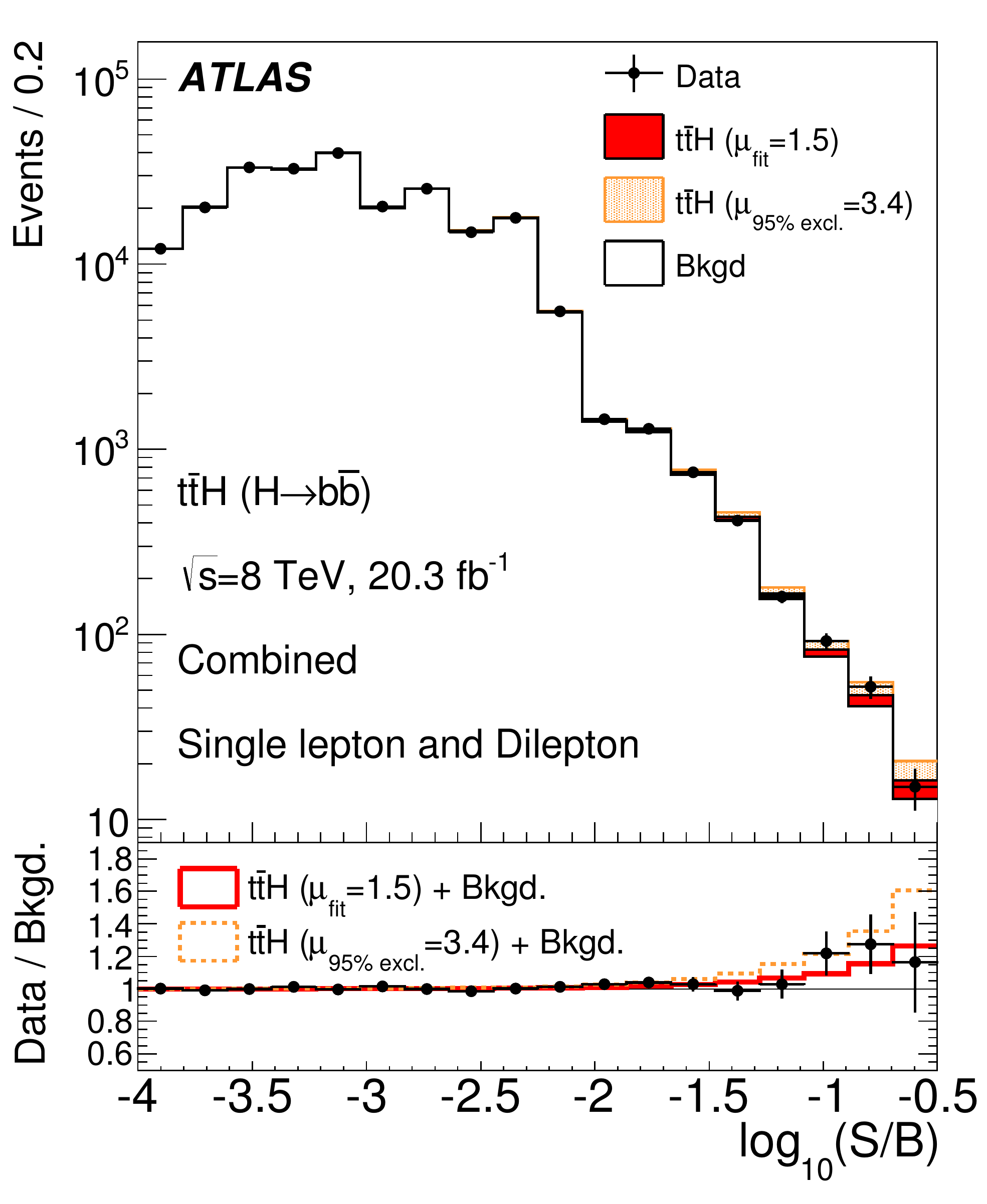}
\caption{Event yields as a function of $\log_{10}(S/B)$, where $S$ 
(signal yield) and $B$ (background yield) are taken from the \hthad, 
\htlep, and NN output bin of each event.  Events in all fitted regions are included. 
The predicted background is obtained from the global signal-plus-background fit.  
The \tth\ signal is shown both for the best fit value ($\mu = 1.5$) and for the upper limit at 95\% CL ($\mu=3.4$).}
\label{fig:logSBplot}
\end{center}
\end{figure}

The \ttbb\ modelling uncertainties affecting the 
shape of this background also have a significant effect on $\mu$. 
These systematic uncertainties affect only the \ttbb\ modelling and
are not correlated with the other \ttbar+jets backgrounds.  The largest
of the uncertainties is given by the renormalisation scale choice.  The
uncertainty drastically changes the shape of the NN for the \ttbb\ background,
making it appear more signal-like.

\begin{figure*}[!htb]
\begin{center}
\includegraphics[width=0.67\textwidth]{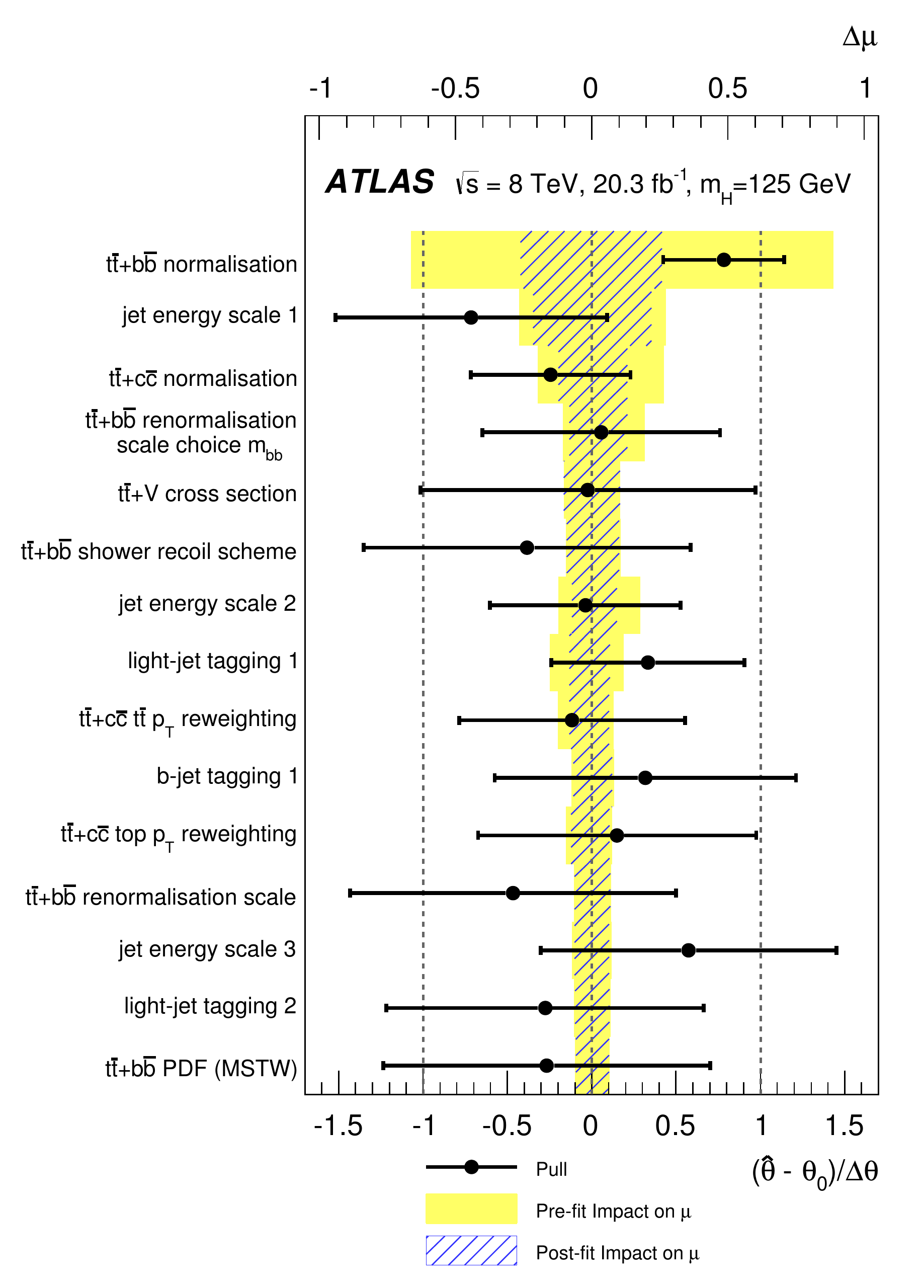}\hspace{0.15cm}
\begin{minipage}[b]{12pc}\caption{\label{fig:ranking} The fitted values of the 
nuisance parameters with the largest impact on the measured signal strength. The 
points, which are drawn conforming to the scale of the bottom axis, show the 
deviation of each of the fitted nuisance parameters, $\hat{\rm{\theta}}$, from 
$\rm{\theta_{0}}$, which is the nominal value of that nuisance parameter, in units 
of the pre-fit standard deviation $\Delta\theta$. The error bars show the 
post-fit uncertainties, $\sigma_{\theta}$, which are close to 1 if the data do not provide 
any further constraint on that uncertainty. 
Conversely, a value of $\sigma_{\theta}$ much smaller than 1 indicates a significant 
reduction with respect to the original uncertainty. The nuisance parameters are 
sorted according to the post-fit effect of each on $\mu$ (hashed blue 
area) conforming to the 
scale of the top axis, with those with the largest impact at the top. }
\end{minipage}
\end{center}
\end{figure*}

The \ttcc\ normalisation uncertainty is ranked third (Fig.~\ref{fig:ranking}) and its pull is slightly negative, while
the post-fit yields for \ttcc\ increase significantly in the four- and 
five-jet regions in the single-lepton channel 
and in the two- and three-jet regions of the dilepton channel 
(see Tables~\ref{tab:Postfit_EventsTable_lj} 
and~\ref{tab:Postfit_EventsTable_dil} of Appendix~\ref{sec:postfit_tables}). 
It was verified that this effect is caused by the interplay between 
the \ttcc\ normalisation uncertainty and 
several other systematic uncertainties affecting the \ttcc\ background yield. 

The noticeable effect of the light-jet tagging (mistag) systematic uncertainty 
is explained by the relatively large fraction of the \ttbar+light 
background in the signal 
region with four $b$-jets in the single-lepton channel. 
The \ttbar+light events enter the 4-$b$-tag region through a mistag as 
opposed to the 3-$b$-tag region where tagging a $c$-jet from a $W$ boson decay is 
more likely. Since the amount of data in the 4-$b$-tag regions is not 
large this uncertainty cannot be constrained significantly. 

The $t\bar{t}+Z$ background with $Z \rightarrow b\bar{b}$ is an irreducible 
background to the \tth\ signal as it has the same number of $b$-jets in the final 
state and similar event kinematics. Its normalisation has a notable effect on $\mu$ 
(${\rm d}\mu /{\rm d}\sigma({t\bar{t}V}) = 0.3$) and the uncertainty arising from   
the $t\bar{t}+V$ normalisation cannot be significantly constrained by the fit.    
Other leading uncertainties include $b$-tagging and some components of the 
JES uncertainty. 

Uncertainties arising from jet energy resolution, jet vertex 
fraction, jet reconstruction and JES that affect primarily low \pt\ jets 
as well as the $\ttbar$+light-jet background modelling uncertainties 
are constrained mainly in the signal-depleted regions. These uncertainties 
do not have a significant effect on the fitted value of $\mu$. 

\FloatBarrier
\section{Summary}
\label{sec:Summary}
A search has been performed for the Standard Model Higgs boson produced in 
association with a top-quark pair ($t\bar{t}H$) using 20.3~\ifb~of 
$pp$ collision data at $\sqrt{s}=8\tev$ collected with the ATLAS detector 
during the first run of the Large Hadron Collider. The search focuses on
\htobb\ decays, and is performed in events with either one or 
two charged leptons.  

To improve sensitivity, the search employs a likelihood fit to data 
in several jet and $b$-tagged jet multiplicity regions. Systematic 
uncertainties included in the fit are significantly constrained by the data. 
Discrimination between signal and background is obtained in both final 
states by employing neural networks in the signal-rich regions.  
In the single-lepton channel, discriminating variables are calculated using 
the matrix element technique. They are used in addition to kinematic 
variables as input to the neural network. 
No significant excess of events above the background expectation is 
found for a Standard Model Higgs boson with a mass of 125~\gev. 
An observed (expected)
95\% confidence-level upper limit of 3.4 (2.2) times the Standard Model cross 
section is obtained. By performing a fit under the signal-plus-background 
hypothesis, the ratio of the measured signal strength 
to the Standard Model expectation is found to be $\mu = 1.5 \pm 1.1$.

\section*{Acknowledgements}

We honour the memory of our colleague Richard St.Denis, who 
was a driving force of the work described
here for a long time and died shortly before its completion. 

We thank CERN for the very successful operation of the LHC, as well as the
support staff from our institutions without whom ATLAS could not be
operated efficiently.

We acknowledge the support of ANPCyT, Argentina; YerPhI, Armenia; ARC,
Australia; BMWFW and FWF, Austria; ANAS, Azerbaijan; SSTC, Belarus; CNPq and FAPESP,
Brazil; NSERC, NRC and CFI, Canada; CERN; CONICYT, Chile; CAS, MOST and NSFC,
China; COLCIENCIAS, Colombia; MSMT CR, MPO CR and VSC CR, Czech Republic;
DNRF, DNSRC and Lundbeck Foundation, Denmark; EPLANET, ERC and NSRF, European Union;
IN2P3-CNRS, CEA-DSM/IRFU, France; GNSF, Georgia; BMBF, DFG, HGF, MPG and AvH
Foundation, Germany; GSRT and NSRF, Greece; RGC, Hong Kong SAR, China; ISF, MINERVA, GIF, I-CORE and Benoziyo Center, Israel; INFN, Italy; MEXT and JSPS, Japan; CNRST, Morocco; FOM and NWO, Netherlands; BRF and RCN, Norway; MNiSW and NCN, Poland; GRICES and FCT, Portugal; MNE/IFA, Romania; MES of Russia and ROSATOM, Russian Federation; JINR; MSTD,
Serbia; MSSR, Slovakia; ARRS and MIZ\v{S}, Slovenia; DST/NRF, South Africa;
MINECO, Spain; SRC and Wallenberg Foundation, Sweden; SER, SNSF and Cantons of
Bern and Geneva, Switzerland; NSC, Taiwan; TAEK, Turkey; STFC, the Royal
Society and Leverhulme Trust, United Kingdom; DOE and NSF, United States of
America.

The crucial computing support from all WLCG partners is acknowledged
gratefully, in particular from CERN and the ATLAS Tier-1 facilities at
TRIUMF (Canada), NDGF (Denmark, Norway, Sweden), CC-IN2P3 (France),
KIT/GridKA (Germany), INFN-CNAF (Italy), NL-T1 (Netherlands), PIC (Spain),
ASGC (Taiwan), RAL (UK) and BNL (USA) and in the Tier-2 facilities
worldwide.

\bibliographystyle{atlasBibStyleWithTitle}
\bibliography{paper.bib}

\appendix
\clearpage
\section{Higgs boson decay modes}
\label{sec:SignalPie}

Figure~\ref{fig:modes} shows the contributions of different Higgs boson decay modes in each of the analysis 
regions in the single-lepton and dilepton channels. The $\htobb$ decay is the dominant contribution  
in the signal-rich regions. 

\begin{figure*}[hpb]
\centering
\subfigure[]{\includegraphics[width=0.47\textwidth]{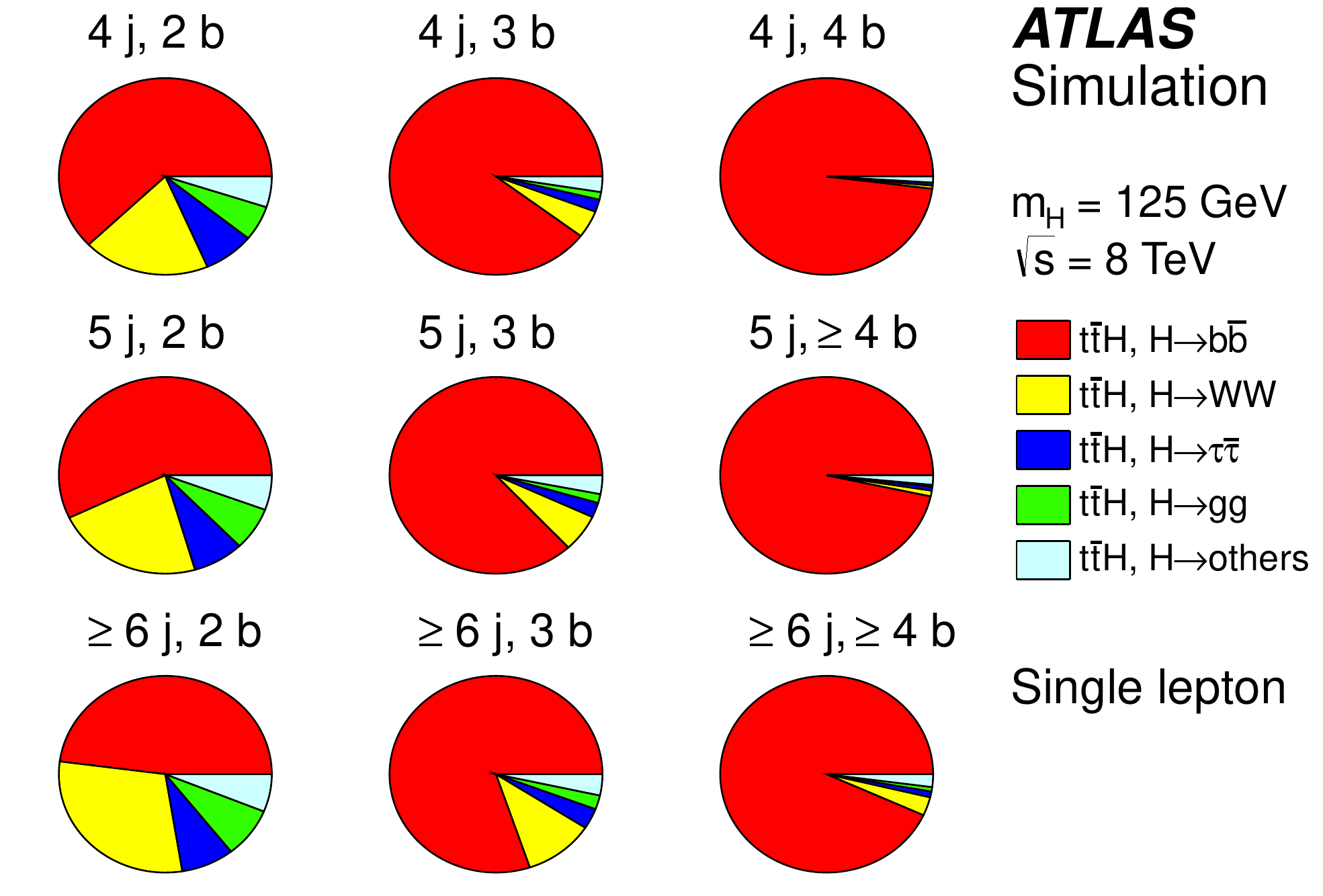}}\label{fig:modes_a}
\subfigure[]{\includegraphics[width=0.47\textwidth]{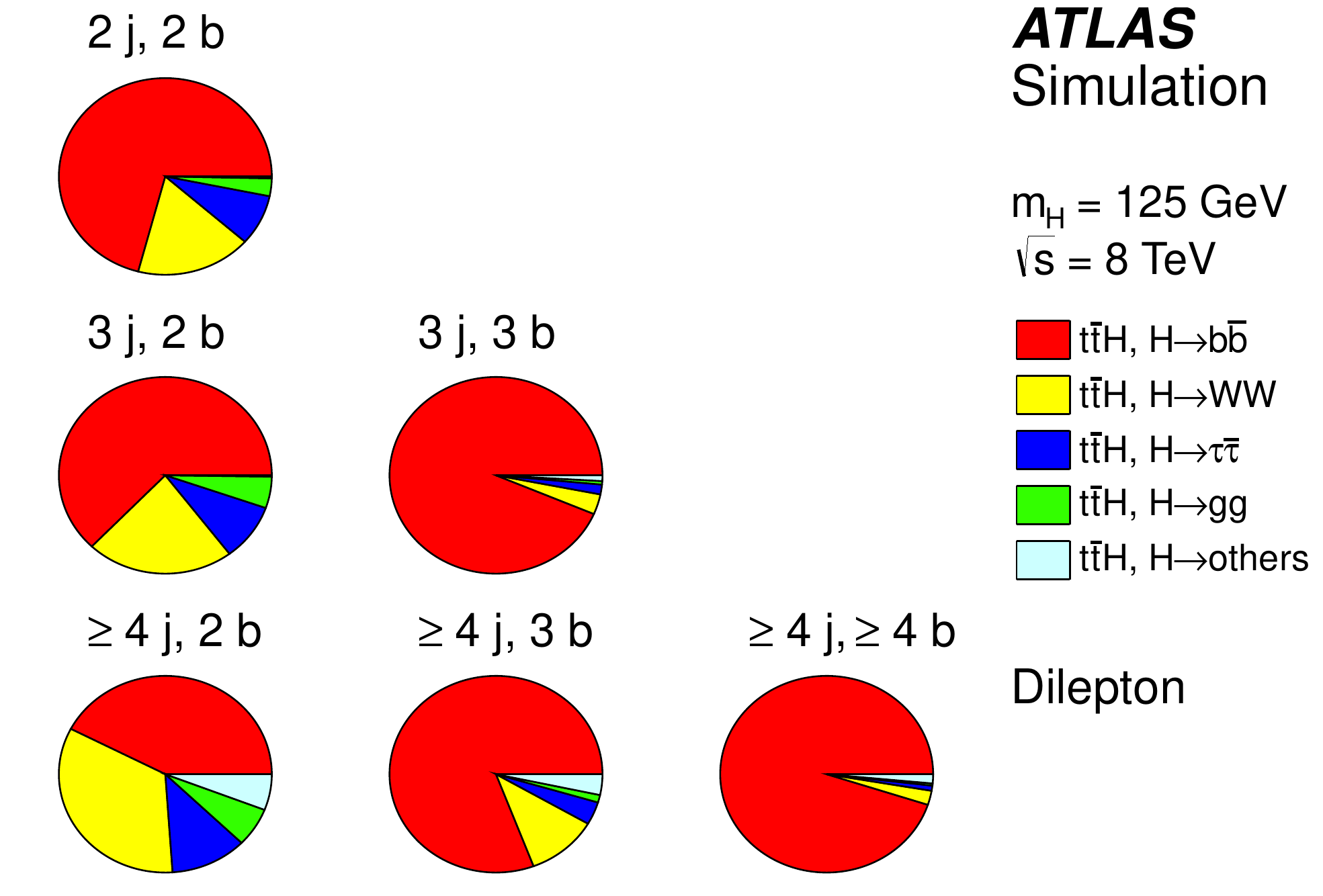}}\label{fig:modes_b}
\caption{Contribution of various Higgs boson decay modes to the 
 analysis regions in (a) the single-lepton channel and (b) the dilepton channel.}
\label{fig:modes}
\end{figure*}

\section{Event yields prior to the fit}
\label{sec:prefit_tables}
The event yields prior to the fit for the combined $e$+jets and $\mu$+jets 
samples for the different regions considered in the analysis are summarised 
in Table~\ref{tab:Prefit_EventsTable_lj}. 

The event yields prior to the fit for the combined $ee$+jets, 
$\mu\mu$+jets and $e\mu$+jets 
samples for the different regions considered in the dilepton channel are 
summarised in Table~\ref{tab:Prefit_EventsTable_dil}. 

\begin{table*}[ht!]
\begin{center}
\begin{tabular}{l*{3}{r@{$\,\pm\,$}r}}
\hline\hline
 & \multicolumn{2}{c}{4 j, 2 b} & \multicolumn{2}{c}{4 j, 3 b} & \multicolumn{2}{c}{4 j, 4 b}\\
\hline
$t\bar{t}H$ (125) & \numRF{30.66}{2} & \numRF{2.83}{1} & \numRF{12.85}{2} & \numRF{1.50}{1} & \numRF{1.95}{2} & \numRF{0.29}{1}\\
$t\bar{t}+$ light & \numRF{76659.58}{2} & \numRF{7528.04}{2} & \numRF{6165.66}{2} & \numRF{745.14}{2} & \numRF{53.15}{2} & \numRF{12.13}{2}\\
$t\bar{t}+c\bar{c}$ & \numRF{4868.29}{2} & \numRF{2964.75}{2} & \numRF{682.12}{2} & \numRF{390.52}{2} & \numRF{21.21}{2} & \numRF{11.95}{2}\\
$t\bar{t}+b\bar{b}$ & \numRF{1844.41}{2} & \numRF{1071.64}{2} & \numRF{679.72}{2} & \numRF{378.59}{2} & \numRF{44.24}{2} & \numRF{24.75}{2}\\
$W$+jets & \numRF{5118.04}{2} & \numRF{2981.90}{2} & \numRF{224.65}{2} & \numRF{133.00}{2} & \numRF{5.52}{2} & \numRF{3.31}{2}\\
$Z$+jets & \numRF{1125.82}{2} & \numRF{597.54}{2} & \numRF{50.23}{2} & \numRF{27.25}{2} & \numRF{0.90}{1} & \numRF{0.56}{1}\\
Single top & \numRF{4932.12}{2} & \numRF{644.49}{2} & \numRF{337.36}{2} & \numRF{60.37}{2} & \numRF{6.78}{2} & \numRF{1.58}{2}\\
Diboson & \numRF{216.76}{2} & \numRF{70.50}{2} & \numRF{11.49}{2} & \numRF{4.06}{2} & \numRF{0.24}{1} & \numRF{0.12}{1}\\
$t\bar{t}+V$ & \numRF{122.18}{2} & \numRF{40.45}{2} & \numRF{15.46}{2} & \numRF{5.10}{2} & \numRF{0.89}{1} & \numRF{0.30}{1}\\
Lepton misID & \numRF{1564.61}{2} & \numRF{619.84}{2} & \numRF{102.21}{2} & \numRF{37.13}{2} & \numRF{3.52}{2} & \numRF{1.28}{2}\\
\hline
Total & \numRF{96482.46}{2} & \numRF{9509.42}{2}   & \numRF{8281.77}{2} & \numRF{1073.67}{2}   & \numRF{138.40}{2} & \numRF{33.85}{2}  \\
\hline
Data & \multicolumn{2}{l}{\num{98049}}  & \multicolumn{2}{l}{\num{8752}}  & \multicolumn{2}{l}{\num{161}} \\
\hline
S/B & \multicolumn{2}{c}{\num{<0.001}}  & \multicolumn{2}{c}{\num{0.002}}  & \multicolumn{2}{c}{\num{0.014}}  \\
S/$\sqrt{\rm{B}}$ &  \multicolumn{2}{c}{\num{0.099}} &  \multicolumn{2}{c}{\num{0.141}} & \multicolumn{2}{c}{\num{0.167}} \\
\hline\hline     \\
\end{tabular}
\vspace{0.1cm}

\begin{tabular}{l*{3}{r@{$\,\pm\,$}r}}
\hline\hline
 & \multicolumn{2}{c}{5 j, 2 b} & \multicolumn{2}{c}{5 j, 3 b} & \multicolumn{2}{c}{5 j, $\geq$ 4 b}\\
\hline
$t\bar{t}H$ (125) & \numRF{40.86}{2} & \numRF{2.05}{1} & \numRF{22.65}{2} & \numRF{1.77}{1} & \numRF{6.22}{2} & \numRF{0.80}{1}\\
$t\bar{t}+$ light & \numRF{37606.38}{2} & \numRF{5512.23}{2} & \numRF{3484.78}{2} & \numRF{524.44}{2} & \numRF{60.84}{2} & \numRF{14.73}{2}\\
$t\bar{t}+c\bar{c}$ & \numRF{4298.98}{2} & \numRF{2380.58}{2} & \numRF{809.59}{2} & \numRF{455.46}{2} & \numRF{42.80}{2} & \numRF{24.68}{2}\\
$t\bar{t}+b\bar{b}$ & \numRF{1665.02}{2} & \numRF{876.06}{2} & \numRF{886.03}{2} & \numRF{477.26}{2} & \numRF{114.90}{2} & \numRF{63.25}{2}\\
$W$+jets & \numRF{1938.28}{2} & \numRF{1232.45}{2} & \numRF{135.32}{2} & \numRF{86.95}{2} & \numRF{5.89}{2} & \numRF{3.85}{2}\\
$Z$+jets & \numRF{405.34}{2} & \numRF{237.67}{2} & \numRF{28.91}{2} & \numRF{17.07}{2} & \numRF{1.47}{2} & \numRF{0.90}{1}\\
Single top & \numRF{1880.73}{2} & \numRF{364.26}{2} & \numRF{194.65}{2} & \numRF{41.44}{2} & \numRF{8.32}{2} & \numRF{1.32}{2}\\
Diboson & \numRF{96.52}{2} & \numRF{38.51}{2} & \numRF{8.02}{2} & \numRF{3.43}{2} & \numRF{0.40}{1} & \numRF{0.20}{1}\\
$t\bar{t}+V$ & \numRF{145.41}{2} & \numRF{47.67}{2} & \numRF{26.47}{2} & \numRF{8.58}{1} & \numRF{3.10}{2} & \numRF{1.02}{2}\\
Lepton misID & \numRF{460.79}{2} & \numRF{165.45}{2} & \numRF{69.93}{2} & \numRF{27.57}{2} & \numRF{8.31}{2} & \numRF{3.70}{2}\\
\hline
Total & \numRF{48538.31}{2} & \numRF{6957.37}{2}   & \numRF{5666.35}{2} & \numRF{982.65}{2}   & \numRF{252.25}{2} & \numRF{75.03}{2}  \\
\hline
Data & \multicolumn{2}{l}{\num{49699}}  & \multicolumn{2}{l}{\num{6199}}  & \multicolumn{2}{l}{\num{286}} \\
\hline
S/B &  \multicolumn{2}{c}{\num{0.001}} &  \multicolumn{2}{c}{\num{0.004}} & \multicolumn{2}{c}{\num{0.025}} \\
S/$\sqrt{\rm{B}}$ &  \multicolumn{2}{c}{\num{0.186}} &  \multicolumn{2}{c}{\num{0.301}} & \multicolumn{2}{c}{\num{0.397}} \\
\hline\hline     \\
\end{tabular}
\vspace{0.1cm}

\begin{tabular}{l*{3}{r@{$\,\pm\,$}r}}
\hline\hline
 & \multicolumn{2}{c}{$\geq$ 6 j, 2 b} & \multicolumn{2}{c}{$\geq$ 6 j, 3 b} & \multicolumn{2}{c}{$\geq$ 6 j, $\geq$ 4 b}\\
\hline
$t\bar{t}H$ (125) & \numRF{63.73}{2} & \numRF{4.99}{1} & \numRF{40.23}{2} & \numRF{3.46}{1} & \numRF{16.49}{2} & \numRF{2.04}{1}\\
$t\bar{t}+$ light & \numRF{18849.67}{2} & \numRF{4402.80}{2} & \numRF{2008.43}{2} & \numRF{463.33}{2} & \numRF{52.40}{2} & \numRF{16.73}{2}\\
$t\bar{t}+c\bar{c}$ & \numRF{3725.70}{2} & \numRF{2086.67}{2} & \numRF{845.95}{2} & \numRF{482.52}{2} & \numRF{79.07}{2} & \numRF{46.46}{2}\\
$t\bar{t}+b\bar{b}$ & \numRF{1424.18}{2} & \numRF{767.00}{2} & \numRF{973.67}{2} & \numRF{526.66}{2} & \numRF{245.44}{2} & \numRF{132.77}{2}\\
$W$+jets & \numRF{911.56}{2} & \numRF{624.38}{2} & \numRF{96.70}{2} & \numRF{66.00}{2} & \numRF{8.60}{2} & \numRF{6.18}{2}\\
$Z$+jets & \numRF{183.25}{2} & \numRF{118.21}{2} & \numRF{19.00}{2} & \numRF{12.44}{2} & \numRF{1.54}{2} & \numRF{1.01}{2}\\
Single top & \numRF{836.37}{2} & \numRF{220.14}{2} & \numRF{121.73}{2} & \numRF{35.35}{2} & \numRF{11.91}{2} & \numRF{3.68}{2}\\
Diboson & \numRF{50.45}{2} & \numRF{24.31}{2} & \numRF{5.98}{2} & \numRF{3.01}{2} & \numRF{0.54}{1} & \numRF{0.27}{1}\\
$t\bar{t}+V$ & \numRF{182.33}{2} & \numRF{59.31}{2} & \numRF{44.58}{2} & \numRF{14.25}{2} & \numRF{8.45}{2} & \numRF{2.77}{2}\\
Lepton misID & \numRF{181.42}{2} & \numRF{66.00}{2} & \numRF{21.31}{2} & \numRF{7.63}{1} & \numRF{1.09}{2} & \numRF{0.52}{1}\\
\hline
Total & \numRF{26408.66}{2} & \numRF{5760.23}{2}   & \numRF{4177.59}{2} & \numRF{1018.85}{2}   & \numRF{425.53}{2} & \numRF{151.98}{2}  \\
\hline
Data & \multicolumn{2}{l}{\num{26185}}  & \multicolumn{2}{l}{\num{4701}}  & \multicolumn{2}{l}{\num{516}} \\
\hline
S/B & \multicolumn{2}{c}{\num{0.002}} & \multicolumn{2}{c}{\num{0.01}} &  \multicolumn{2}{c}{\num{0.04}} \\
S/$\sqrt{\rm{B}}$ &  \multicolumn{2}{c}{\num{0.393}} &  \multicolumn{2}{c}{\num{0.63}} & \multicolumn{2}{c}{\num{0.815}} \\
\hline\hline     \\
\end{tabular}
\vspace{0.1cm}

\end{center}
\vspace{-0.5cm}
\caption{Single lepton channel:  
pre-fit event yields 
for signal, backgrounds and data in each of the analysis regions. The
quoted uncertainties are the sum in quadrature of the statistical and 
systematic uncertainties on the yields.  
}
\label{tab:Prefit_EventsTable_lj}
\end{table*}

\begin{table*}
\begin{center}
\begin{tabular}{l*{3}{r@{$\,\pm\,$}r}}
\hline\hline
 & \multicolumn{2}{c}{2 j, 2 b} & \multicolumn{2}{c}{3 j, 2 b} & \multicolumn{2}{c}{3 j, 3 b}\\
\hline
$t\bar{t}H$ (125) & \numRF{1.54}{2} & \numRF{0.21}{1} & \numRF{5.34}{2} & \numRF{0.47}{1} & \numRF{2.20}{2} & \numRF{0.27}{1}\\
$t\bar{t}+$ light & \numRF{14054.89}{2} & \numRF{1784.19}{2} & \numRF{8091.81}{2} & \numRF{877.65}{2} & \numRF{95.93}{2} & \numRF{21.48}{2}\\
$t\bar{t}+c\bar{c}$ & \numRF{267.69}{2} & \numRF{172.87}{2} & \numRF{599.56}{2} & \numRF{322.46}{2} & \numRF{76.26}{2} & \numRF{44.36}{2}\\
$t\bar{t}+b\bar{b}$ & \numRF{146.67}{2} & \numRF{87.29}{2} & \numRF{261.92}{2} & \numRF{134.50}{2} & \numRF{117.85}{2} & \numRF{64.71}{2}\\
$Z$+jets & \numRF{331.32}{2} & \numRF{29.59}{2} & \numRF{192.24}{2} & \numRF{49.45}{2} & \numRF{8.16}{2} & \numRF{3.13}{2}\\
Single top & \numRF{429.84}{2} & \numRF{70.52}{2} & \numRF{266.09}{2} & \numRF{30.05}{2} & \numRF{7.56}{2} & \numRF{3.50}{2}\\
Diboson & \numRF{6.76}{2} & \numRF{2.18}{2} & \numRF{4.17}{2} & \numRF{1.45}{2} & $\leq$ \numRF{0.1}{1} & \numRF{0.1}{1}\\
$t\bar{t}+V$ & \numRF{8.40}{2} & \numRF{2.70}{2} & \numRF{20.64}{2} & \numRF{6.45}{1} & \numRF{1.90}{2} & \numRF{0.61}{1}\\
Lepton misID & \numRF{20.81}{2} & \numRF{10.40}{2} & \numRF{33.35}{2} & \numRF{16.67}{2} & \numRF{0.82}{1} & \numRF{0.41}{1}\\
\hline
Total & \numRF{15267.91}{2} & \numRF{1860.26}{2}   & \numRF{9475.12}{2} & \numRF{1037.11}{2}   & \numRF{310.76}{2} & \numRF{85.45}{2}  \\
\hline
Data & \multicolumn{2}{l}{\num{15296}}  & \multicolumn{2}{l}{\num{9996}}  & \multicolumn{2}{l}{\hphantom{00}\num{374}} \\
\hline
$S/B$ & \multicolumn{2}{c}{\num{<0.001}}  & \multicolumn{2}{c}{\num{0.001}}  & \multicolumn{2}{c}{\num{0.006}}  \\
$S/\sqrt{\rm{B}}$ &  \multicolumn{2}{c}{\num{0.012}} &  \multicolumn{2}{c}{\num{0.053}} & \multicolumn{2}{c}{\num{0.114}} \\
\hline\hline     \\
\end{tabular}
\vspace{0.1cm}

\begin{tabular}{l*{3}{r@{$\,\pm\,$}r}}
\hline\hline
 & \multicolumn{2}{c}{$\geq$ 4 j, 2 b} & \multicolumn{2}{c}{$\geq$ 4 j, 3 b} & \multicolumn{2}{c}{$\geq$ 4 j, $\geq$ 4 b}\\
\hline
$t\bar{t}H$ (125) & \numRF{15.29}{2} & \numRF{1.00}{1} & \numRF{8.64}{2} & \numRF{0.60}{1} & \numRF{2.72}{2} & \numRF{0.33}{1}\\
$t\bar{t}+$ light & \numRF{4430.30}{2} & \numRF{810.50}{2} & \numRF{120.31}{2} & \numRF{30.57}{2} & \numRF{1.93}{2} & \numRF{0.77}{1}\\
$t\bar{t}+c\bar{c}$ & \numRF{711.64}{2} & \numRF{382.24}{2} & \numRF{132.50}{2} & \numRF{74.09}{2} & \numRF{5.01}{2} & \numRF{2.95}{2}\\
$t\bar{t}+b\bar{b}$ & \numRF{290.63}{2} & \numRF{150.73}{2} & \numRF{197.50}{2} & \numRF{103.97}{2} & \numRF{30.66}{2} & \numRF{16.96}{2}\\
$Z$+jets & \numRF{103.65}{2} & \numRF{38.86}{2} & \numRF{10.18}{2} & \numRF{3.91}{1} & \numRF{0.55}{1} & \numRF{0.24}{1}\\
Single top & \numRF{135.76}{2} & \numRF{54.50}{2} & \numRF{10.96}{2} & \numRF{4.96}{1} & \numRF{0.77}{1} & \numRF{0.21}{1}\\
Diboson & \numRF{3.98}{2} & \numRF{1.27}{2} & \numRF{0.37}{1} & \numRF{0.13}{1} & $\leq$ \numRF{0.1}{1} & \numRF{0.1}{1}\\
$t\bar{t}+V$ & \numRF{45.00}{2} & \numRF{13.86}{2} & \numRF{7.76}{2} & \numRF{2.42}{2} & \numRF{1.07}{2} & \numRF{0.36}{1}\\
Lepton misID & \numRF{38.05}{2} & \numRF{19.03}{2} & \numRF{4.30}{2} & \numRF{2.15}{2} & \numRF{0.35}{1} & \numRF{0.18}{1}\\
\hline
Total & \numRF{5774.30}{2} & \numRF{1034.54}{2}   & \numRF{492.51}{2} & \numRF{141.69}{2}   & \numRF{43.11}{2} & \numRF{17.73}{2}  \\
\hline
Data & \multicolumn{2}{l}{\num{6006}}  & \multicolumn{2}{l}{\num{561}}  & \multicolumn{2}{l}{\hphantom{.00}\num{46}} \\
\hline
$S/B$ & \multicolumn{2}{c}{\num{0.003}}  & \multicolumn{2}{c}{\num{0.015}}  & \multicolumn{2}{c}{\num{0.059}}  \\
$S/\sqrt{\rm{B}}$ &  \multicolumn{2}{c}{\num{0.197}} &  \multicolumn{2}{c}{\num{0.365}} & \multicolumn{2}{c}{\num{0.401}} \\
\hline\hline     \\
\end{tabular}
\vspace{0.1cm}

\end{center}
\vspace{-0.5cm}
\caption{Dilepton channel:  
pre-fit event yields 
for signal, backgrounds and data in each of the analysis regions. The
quoted uncertainties are the sum in quadrature of the statistical and 
systematic uncertainties on the yields.
}
\label{tab:Prefit_EventsTable_dil}                                                               
\end{table*}

\section{Discrimination power of input variables}
\label{sec:separation}

Figures~\ref{fig:sepinput_lj_0}--\ref{fig:sepinput_lj_3} and 
~\ref{fig:sepinput_dil_1}--\ref{fig:sepinput_dil_3} 
show the discrimination between 
signal and background for the top four input variables in each region 
where NN is used in the single-lepton and dilepton channels, respectively. 
In Fig.~\ref{fig:sepinput_lj_0}, the NN is designed to separate $t\bar{t}$+HF from $t\bar{t}$+light.

\begin{figure*}[ht!]
\begin{center}
\subfigure[]{\includegraphics[width=0.24\textwidth]{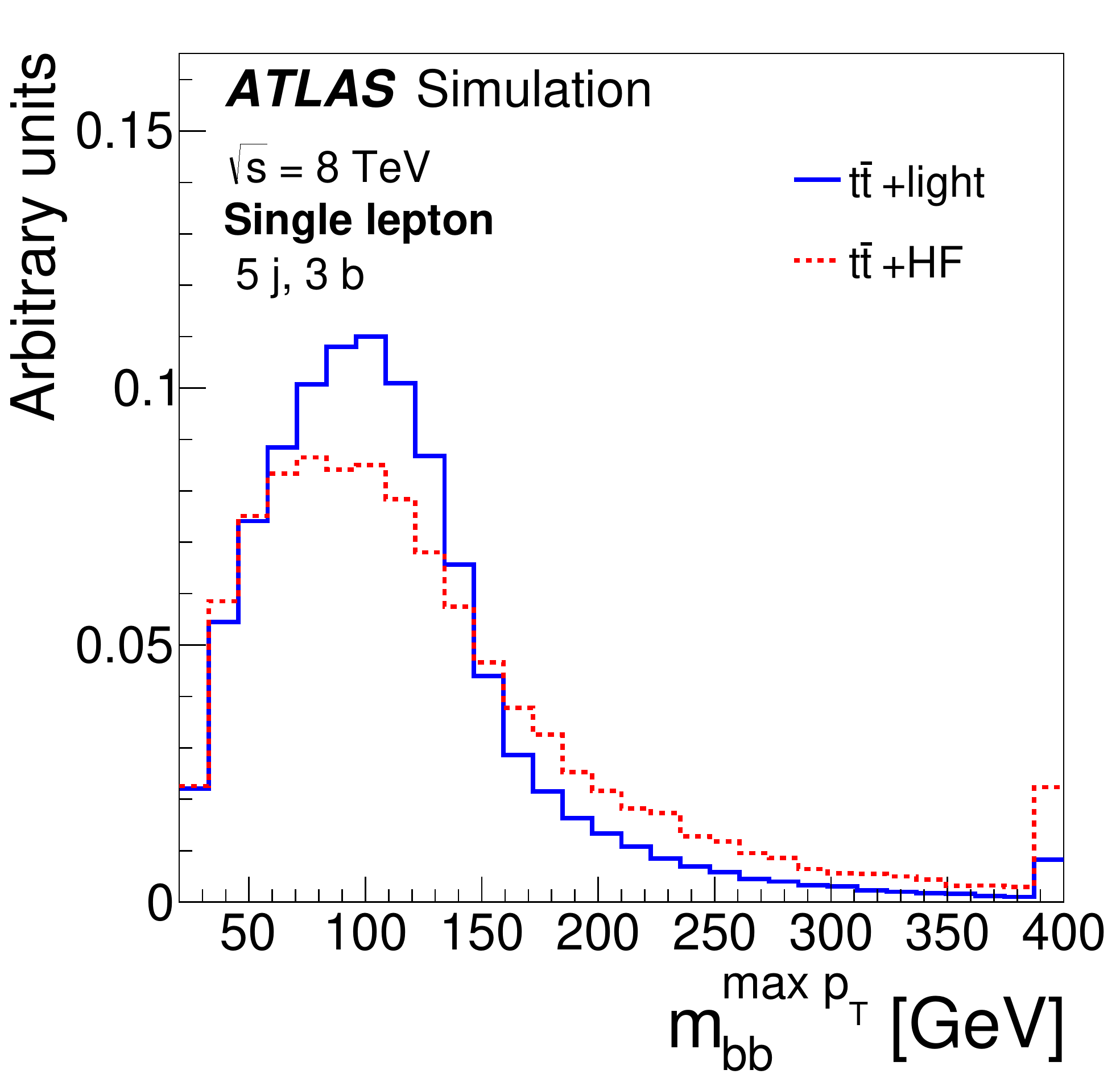}}\label{fig:sepinput_lj_0_a}
\subfigure[]{\includegraphics[width=0.24\textwidth]{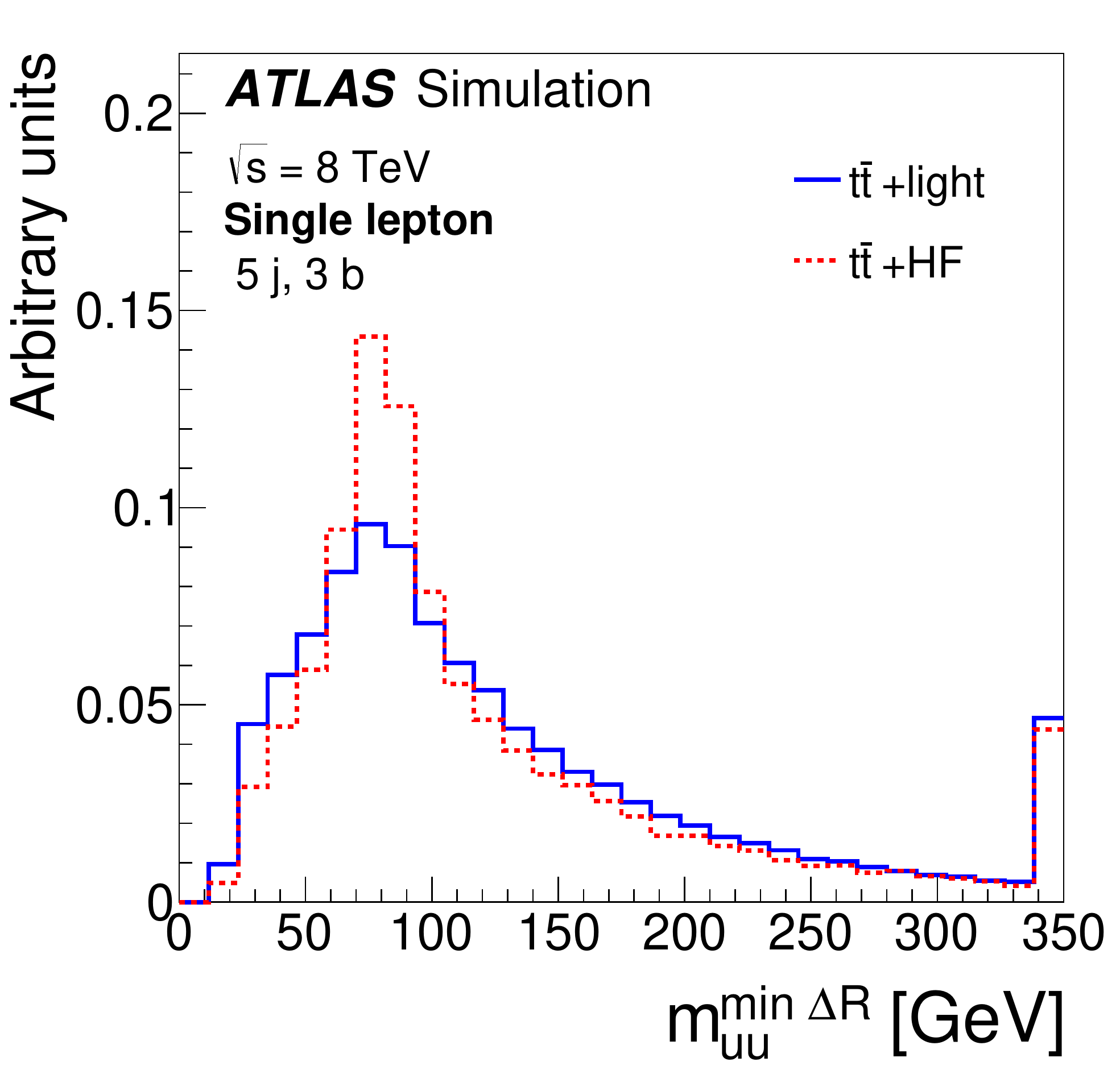}}\label{fig:sepinput_lj_0_b} 
\subfigure[]{\includegraphics[width=0.24\textwidth]{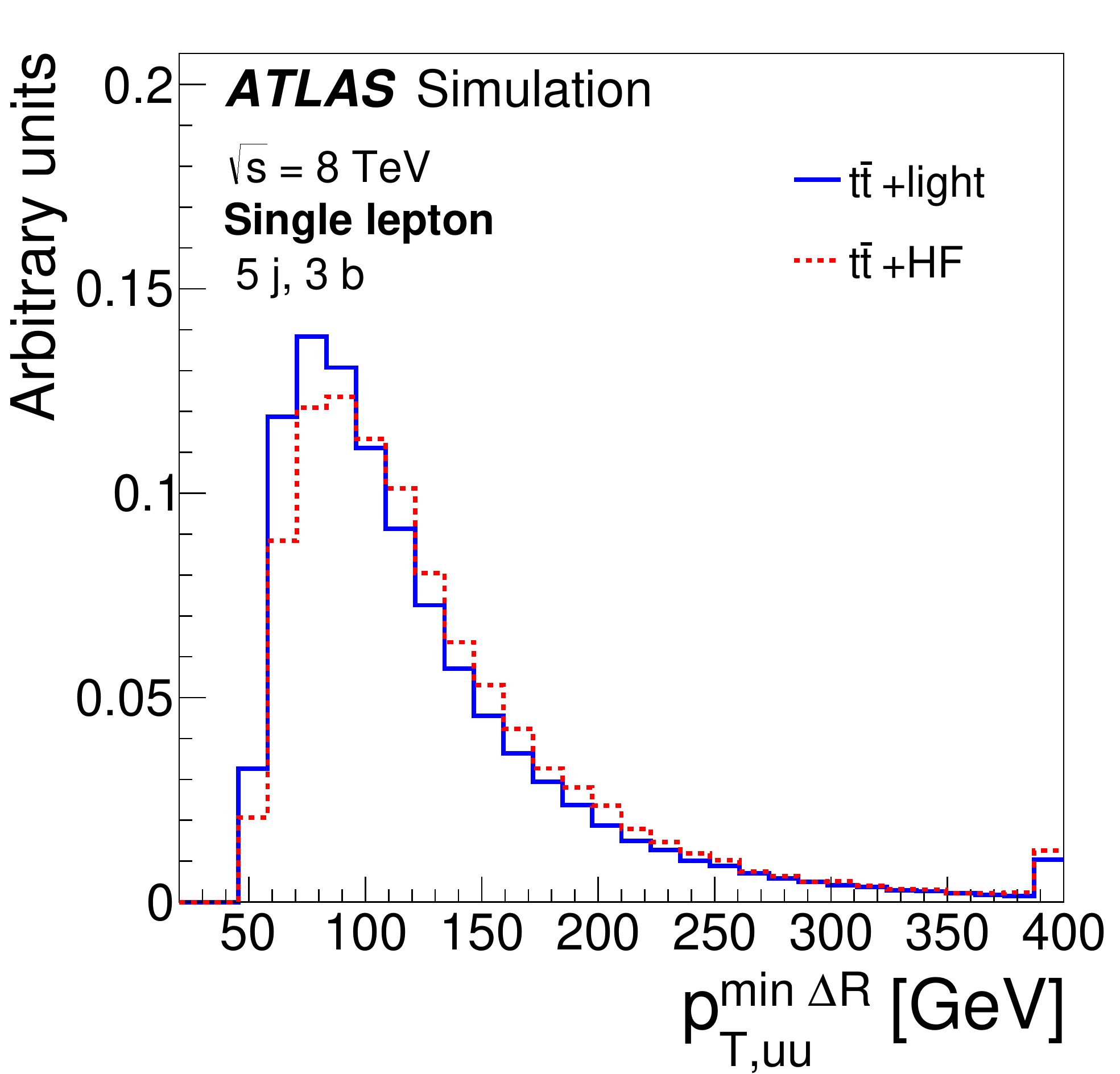}}\label{fig:sepinput_lj_0_c}
\subfigure[]{\includegraphics[width=0.24\textwidth]{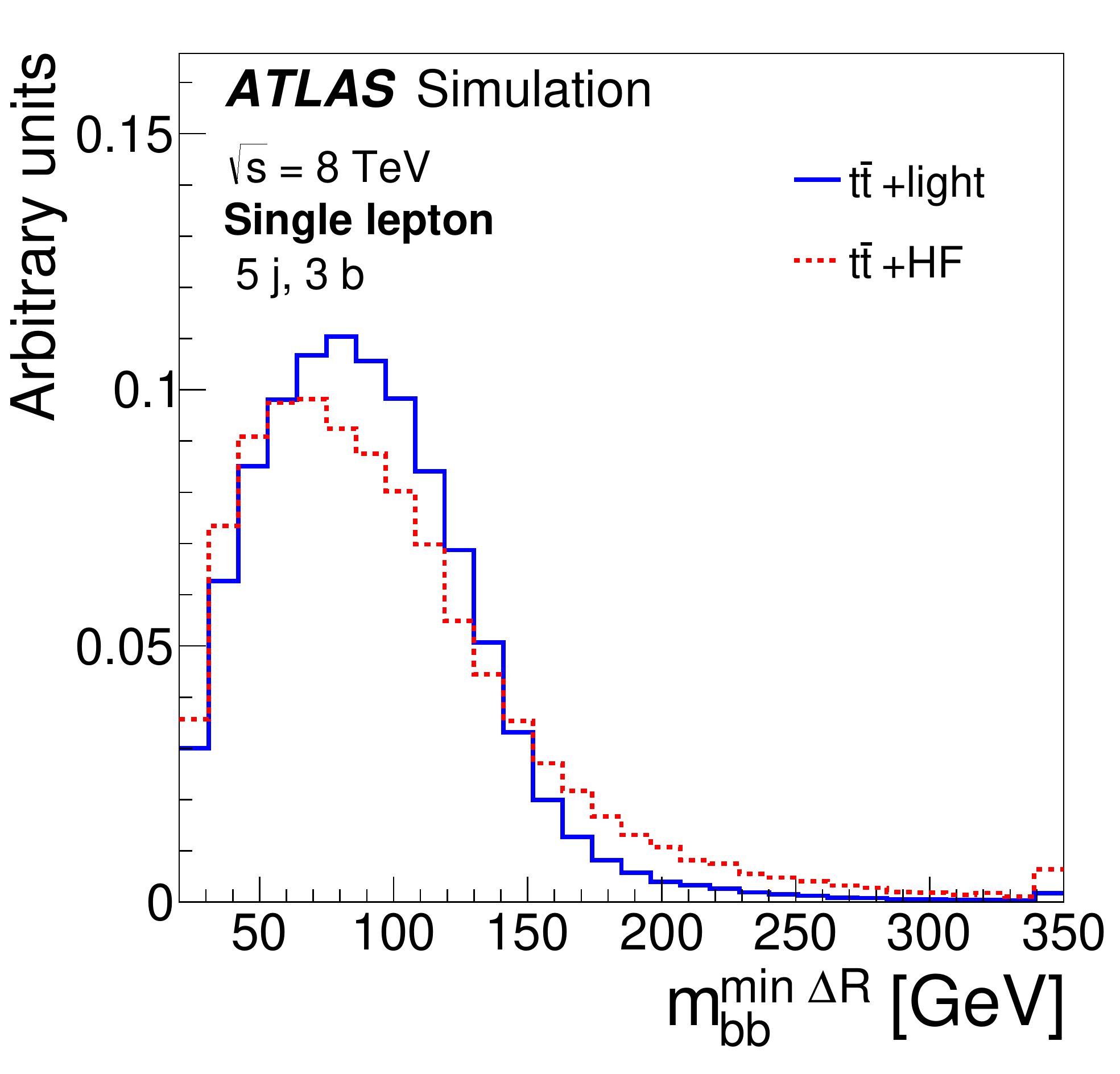}}\label{fig:sepinput_lj_0_d}
\caption{Single-lepton channel: comparison of $t\bar{t}$+HF (dashed) and $t\bar{t}$+light (solid) background for the four top-ranked 
input variables in the \fivethree\ region where the NN is designed to separate these two backgrounds. The plots 
include (a) \mbbmaxpt, (b) \whadmass, (c)  \whadpt and (d) \mbbmindr.
}
\label{fig:sepinput_lj_0} 
\end{center}
\end{figure*}

\begin{figure*}[ht!]
\begin{center}
\subfigure[]{\includegraphics[width=0.24\textwidth]{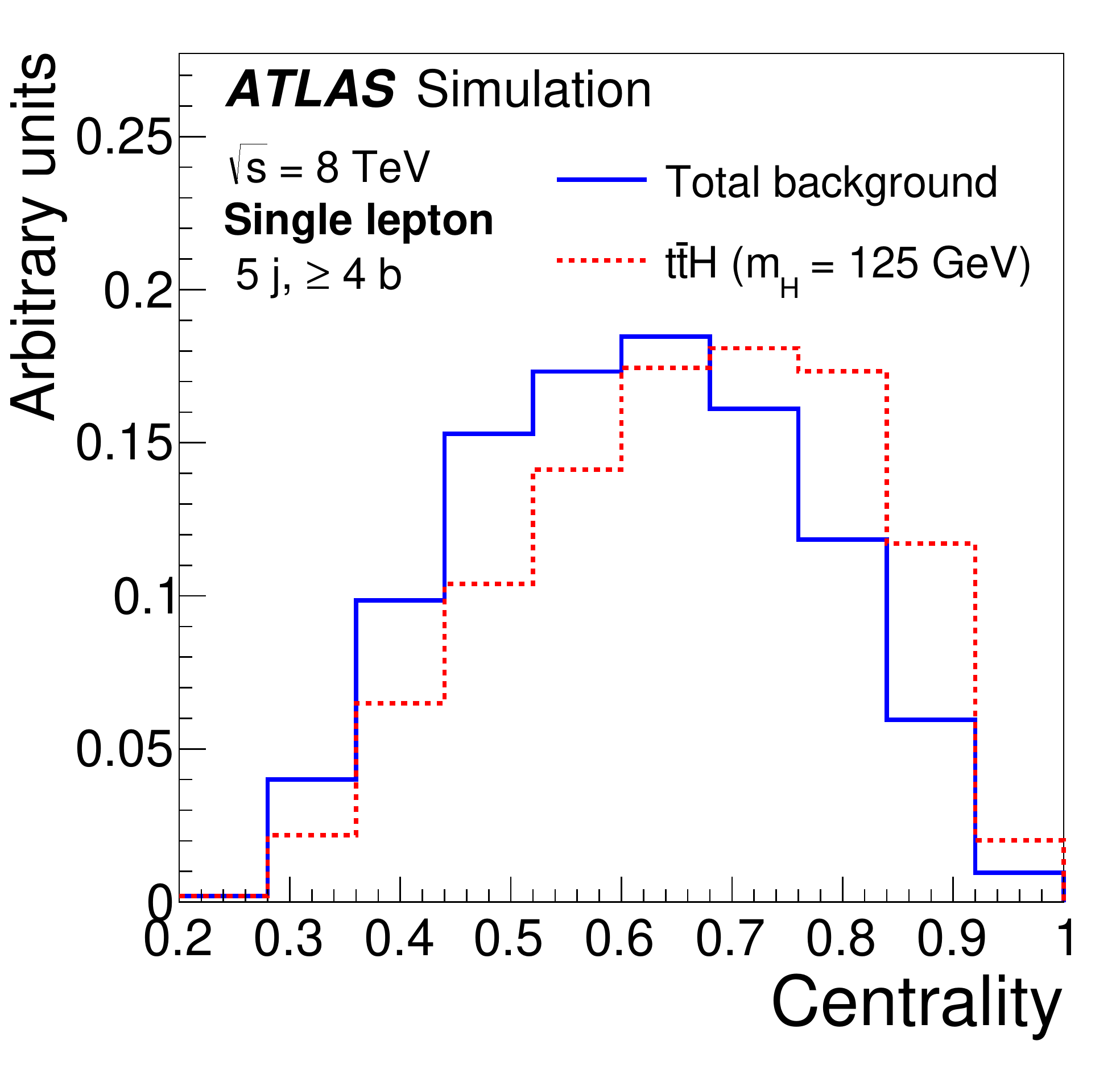}}\label{fig:sepinput_lj_1_a}
\subfigure[]{\includegraphics[width=0.24\textwidth]{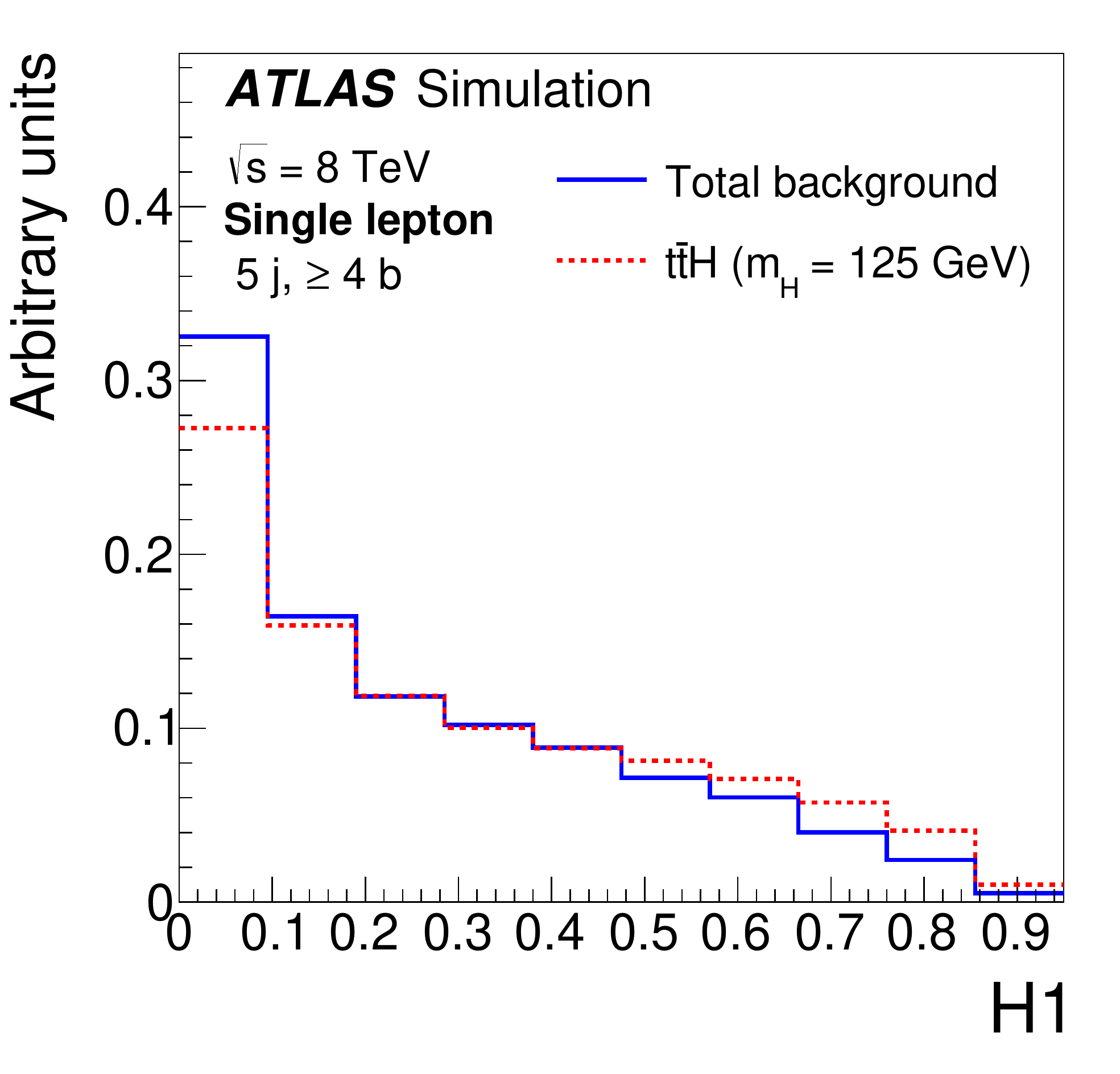}}\label{fig:sepinput_lj_1_b} 
\subfigure[]{\includegraphics[width=0.24\textwidth]{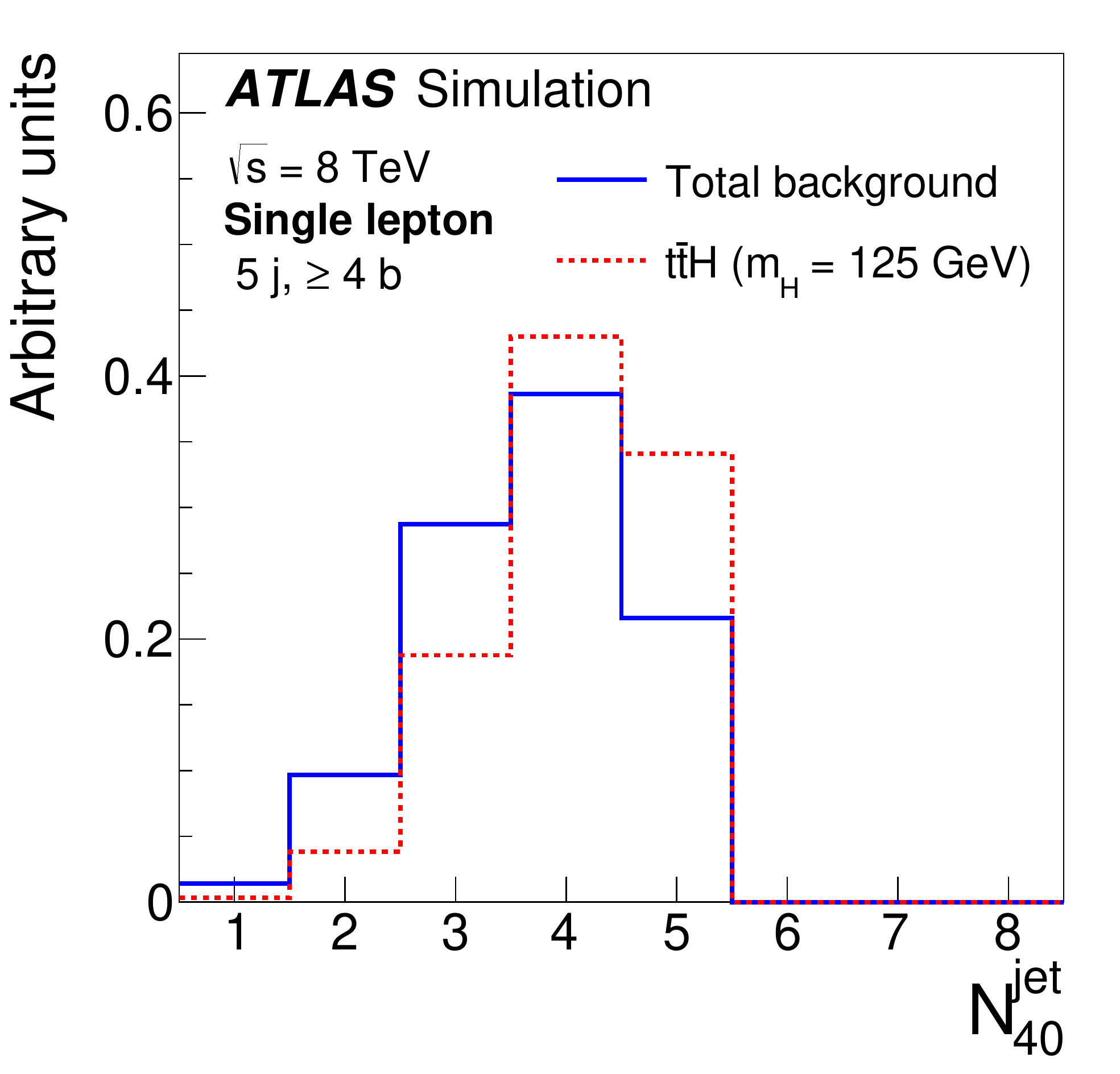}}\label{fig:sepinput_lj_1_c}
\subfigure[]{\includegraphics[width=0.24\textwidth]{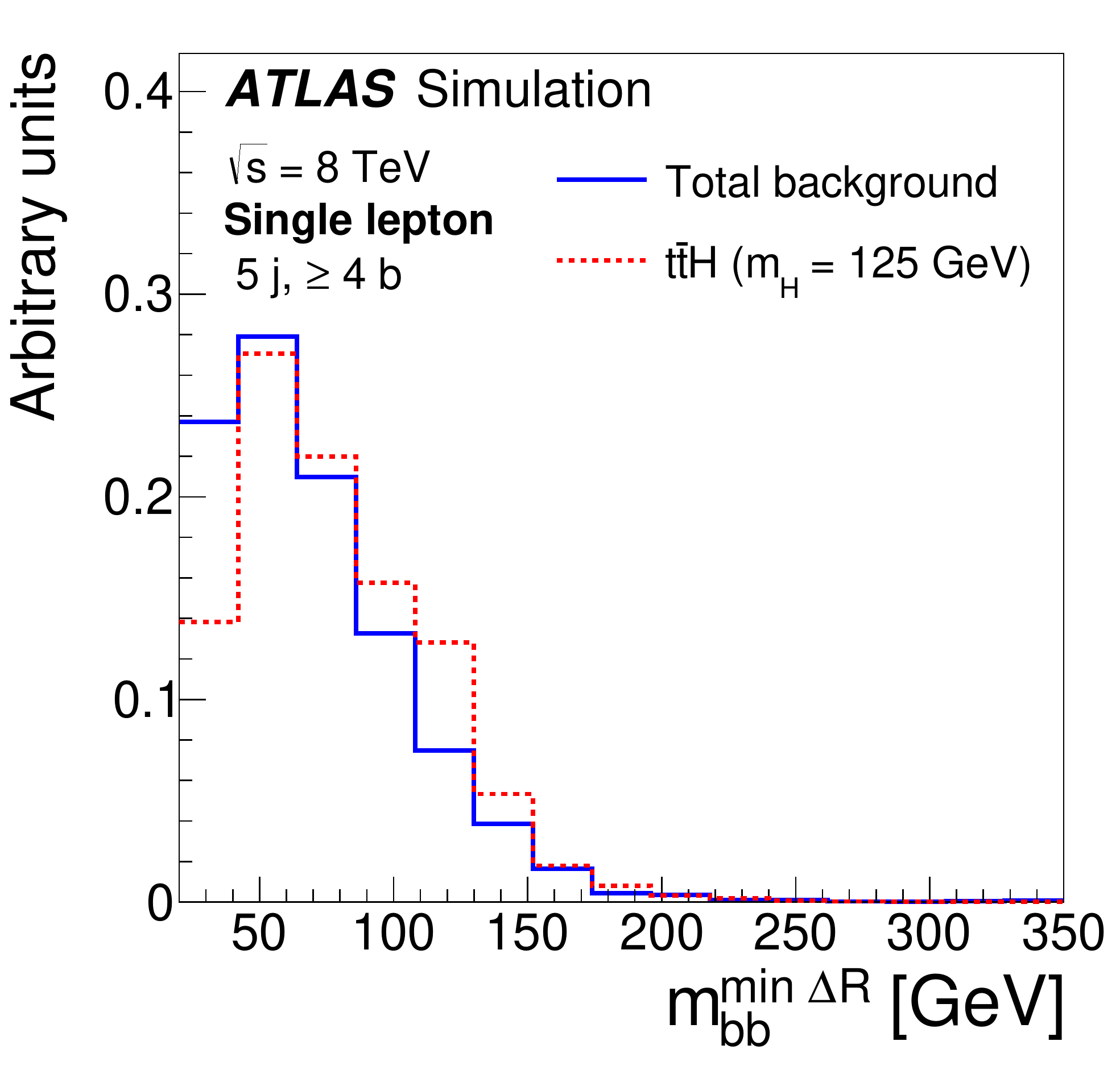}}\label{fig:sepinput_lj_1_d}
\caption{Single-lepton channel: comparison of \tth\ signal (dashed) and background (solid) for the four top-ranked input variables in the 
\fivefour\ region.  The plots include (a) \cent, (b) $H1$, (c)  \numjetforty and (d) \mbbmindr.
}
\label{fig:sepinput_lj_1} 
\end{center}
\end{figure*}

\begin{figure*}[ht!]
\begin{center}
\subfigure[]{\includegraphics[width=0.24\textwidth]{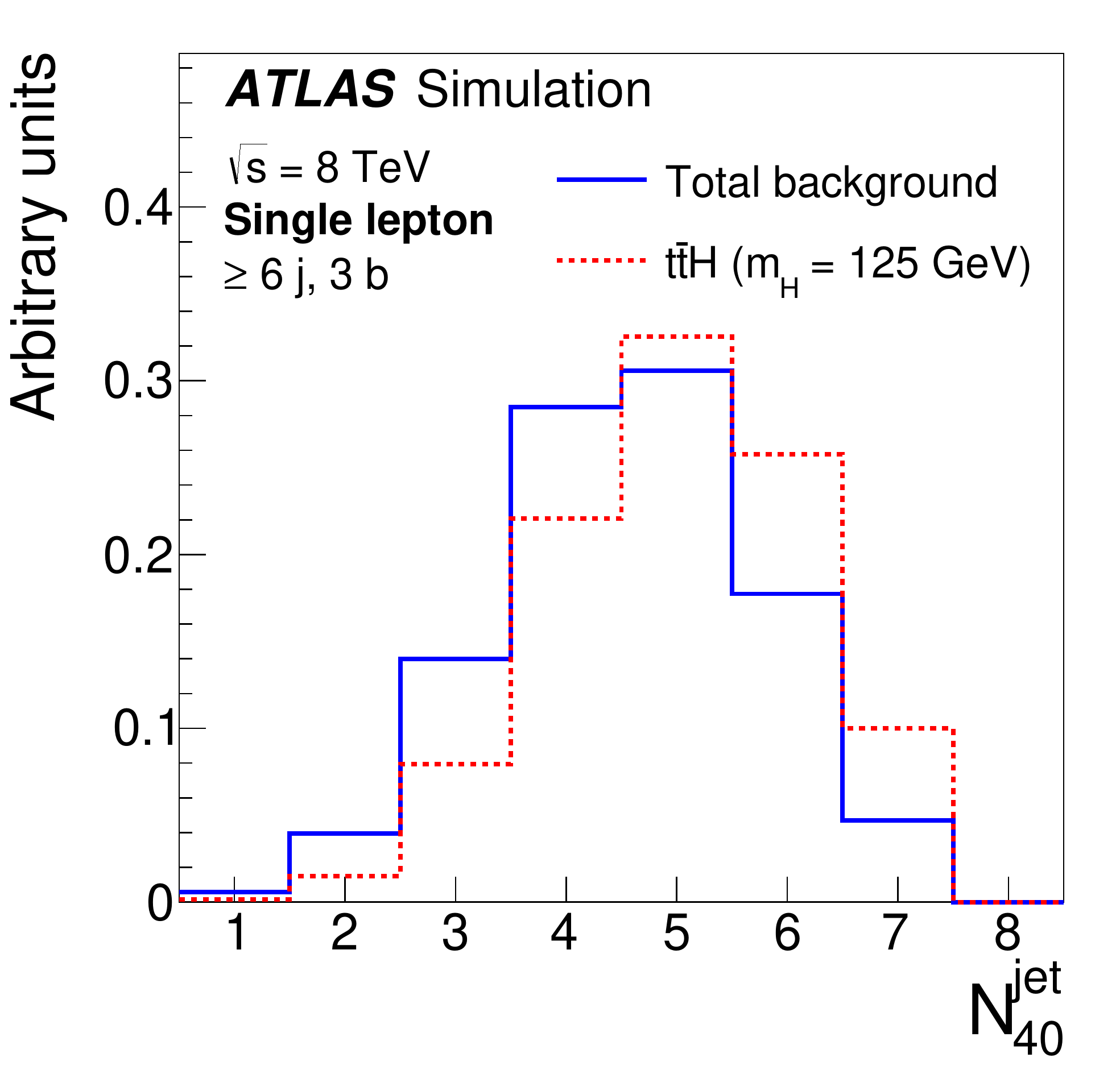}}\label{fig:sepinput_lj_2_a}
\subfigure[]{\includegraphics[width=0.24\textwidth]{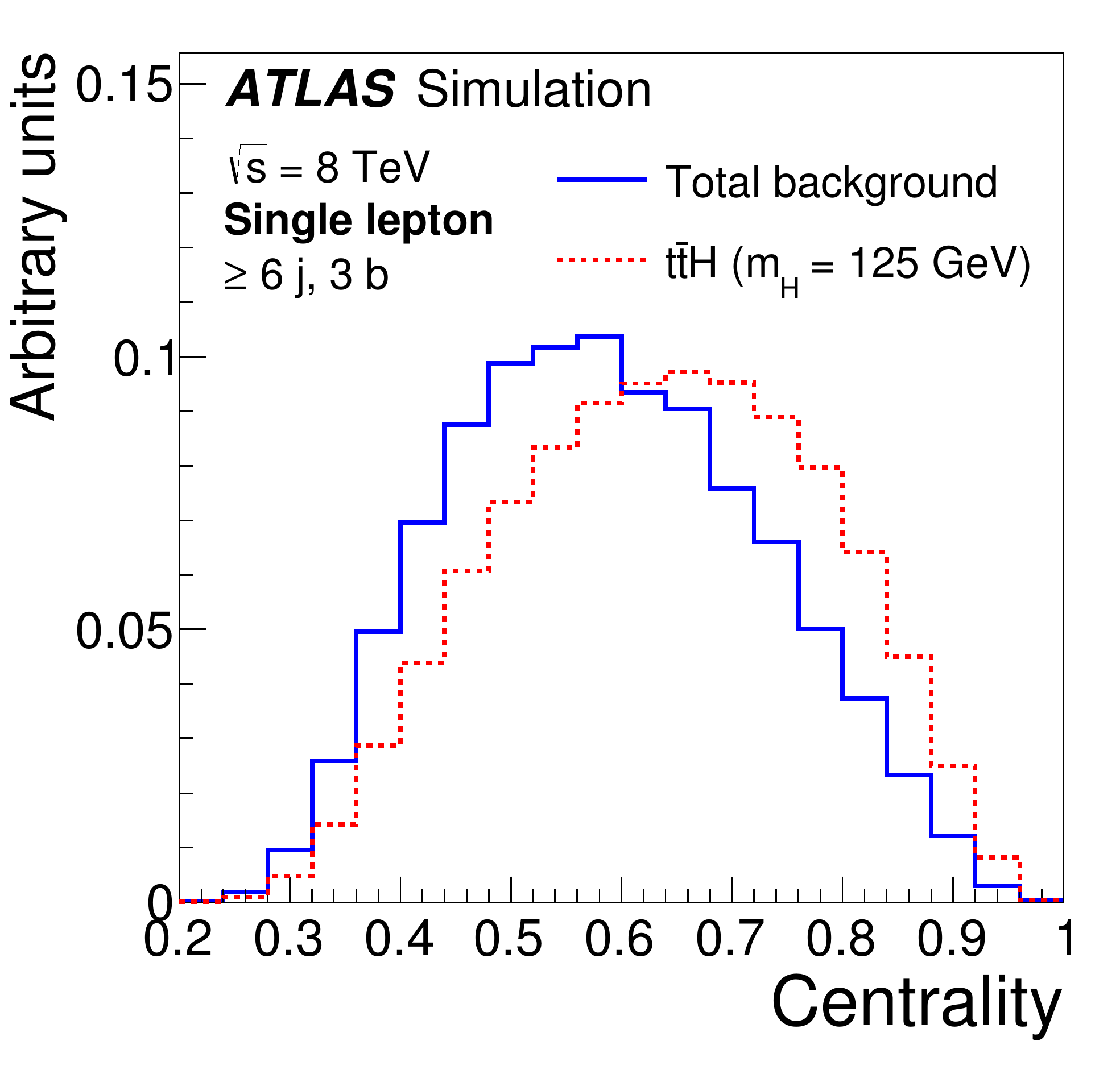}}\label{fig:sepinput_lj_2_b} 
\subfigure[]{\includegraphics[width=0.24\textwidth]{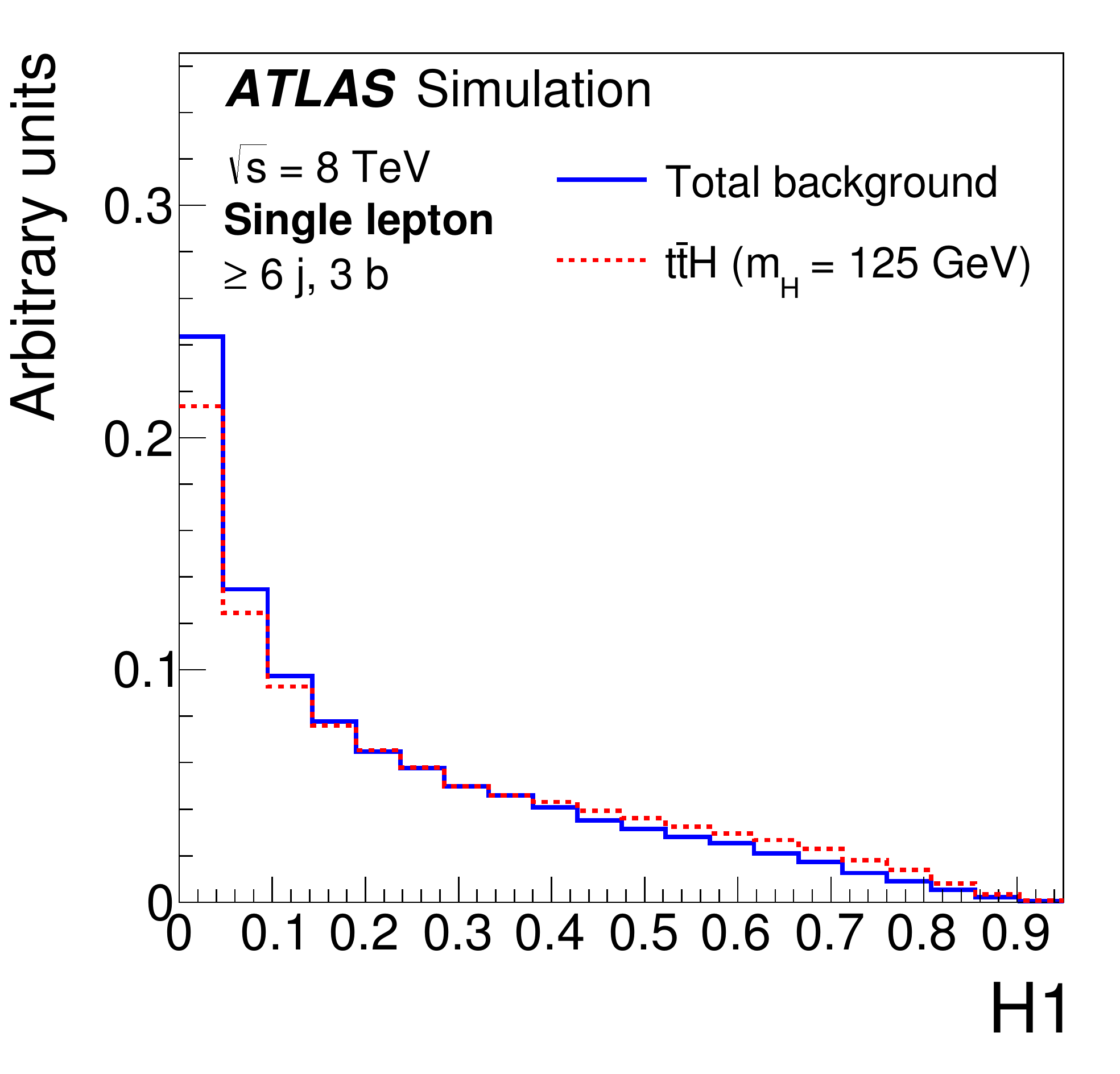}} \label{fig:sepinput_lj_2_c}
\subfigure[]{\includegraphics[width=0.24\textwidth]{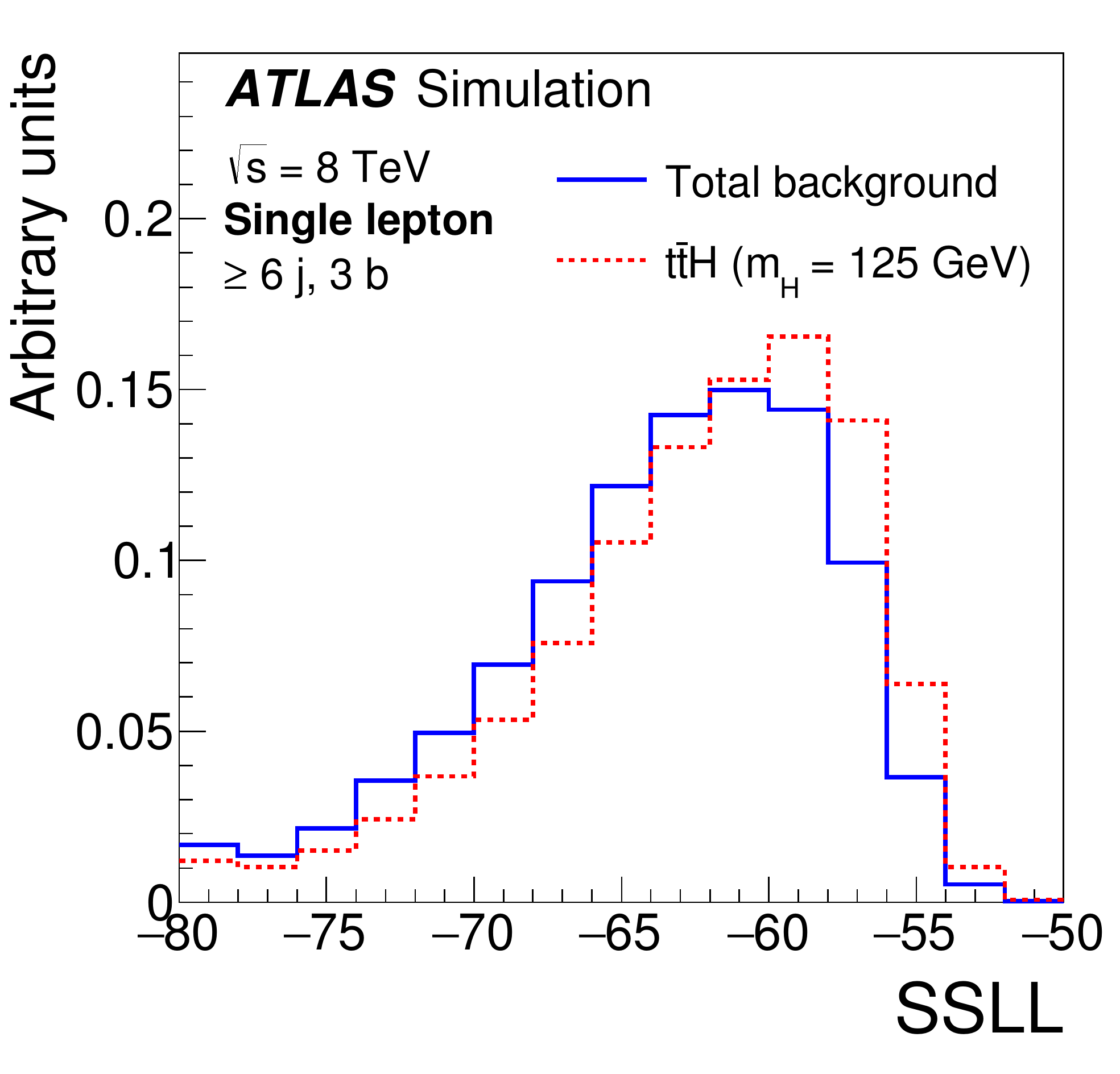}}\label{fig:sepinput_lj_2_d}
\caption{Single-lepton channel: comparison of \tth\ signal (dashed) and background (solid) for the four top-ranked input variables 
in the \sixthree\ region.  The plots include (a) \numjetforty, (b) \cent, (c) $H1$, and (d) SSLL.
}
\label{fig:sepinput_lj_2} 
\end{center}
\end{figure*}

\begin{figure*}[ht!]
\begin{center}
\subfigure[]{\includegraphics[width=0.24\textwidth]{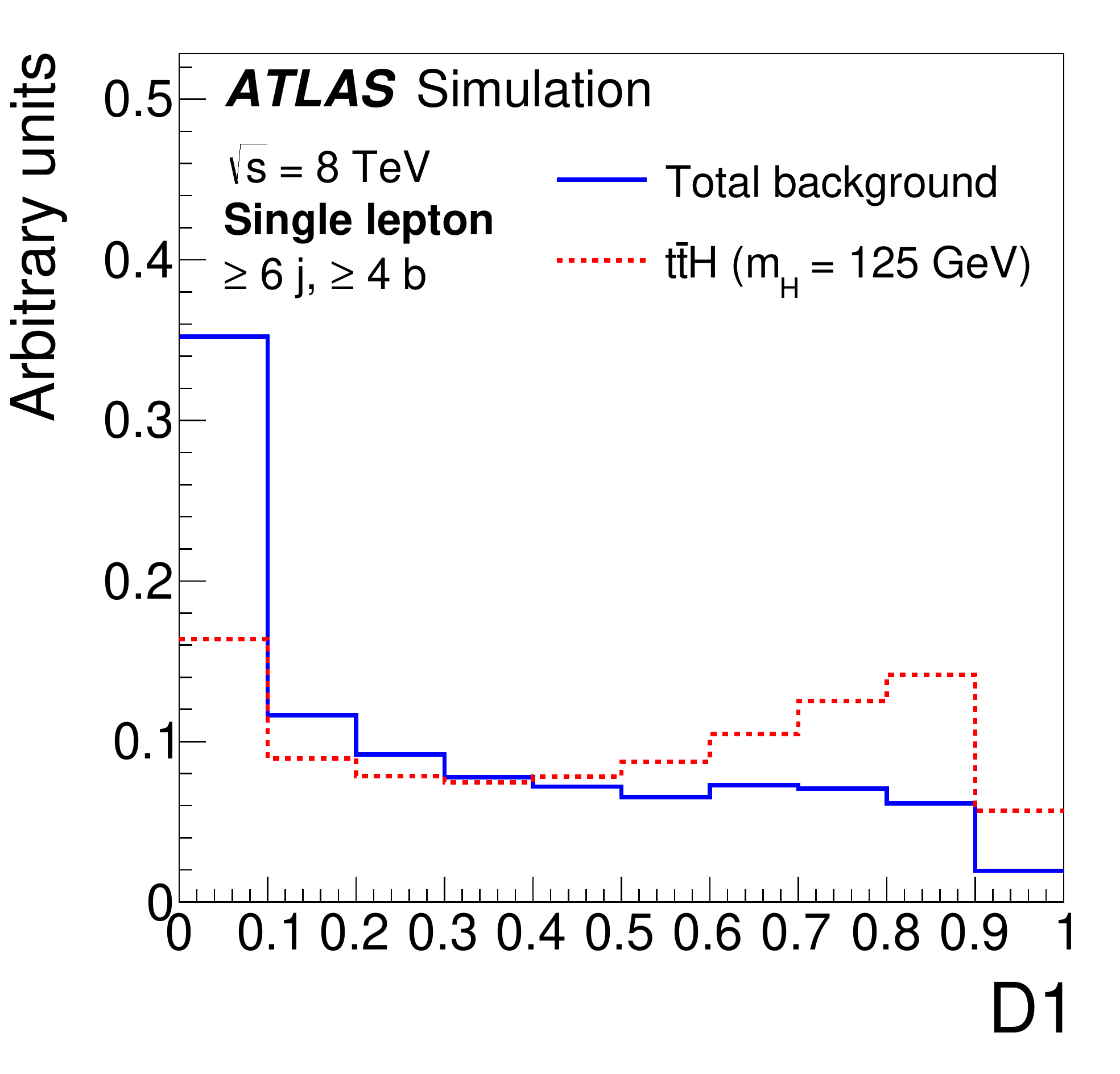}}\label{fig:sepinput_lj_3_a}
\subfigure[]{\includegraphics[width=0.24\textwidth]{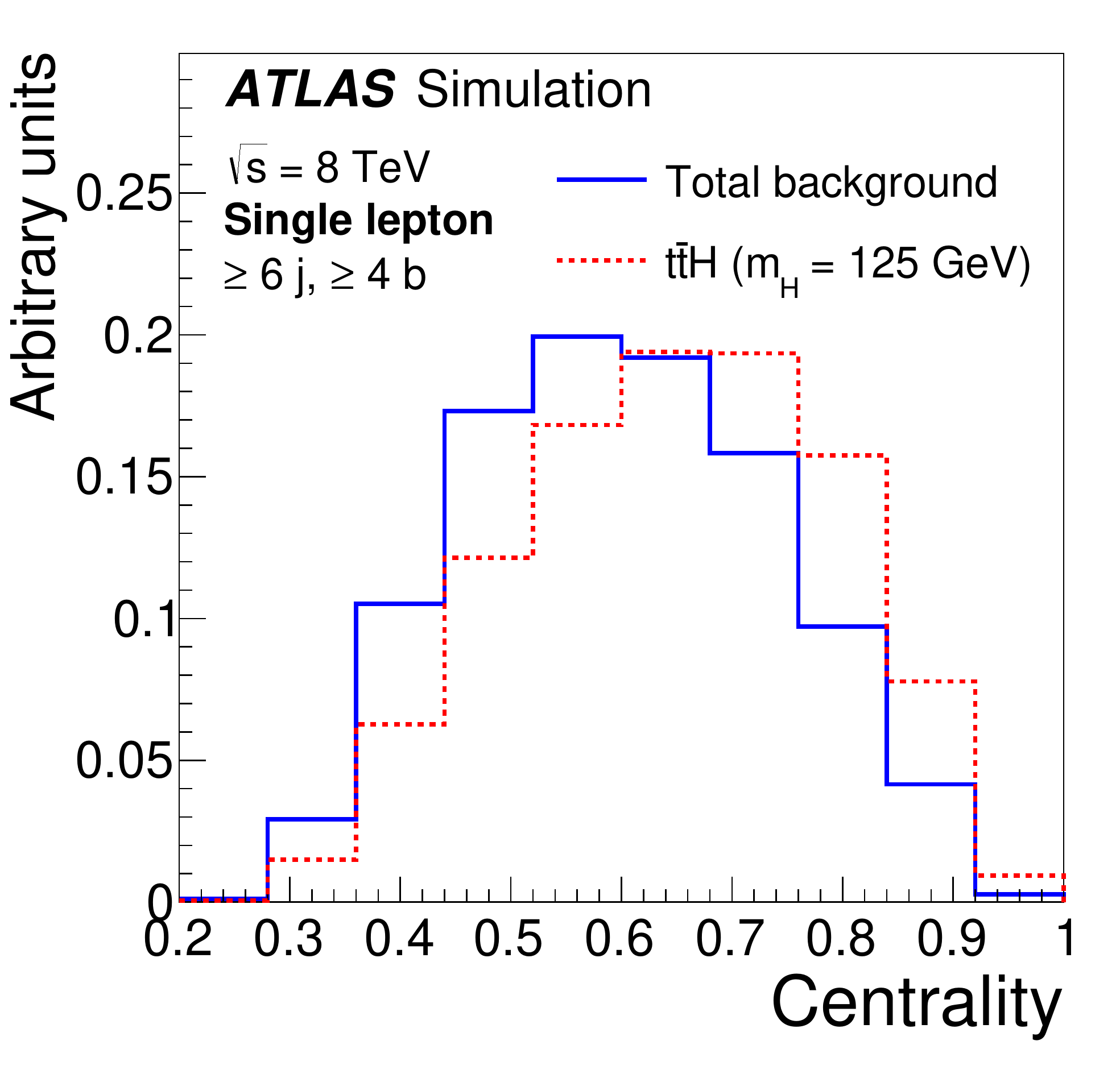}}\label{fig:sepinput_lj_3_b} 
\subfigure[]{\includegraphics[width=0.24\textwidth]{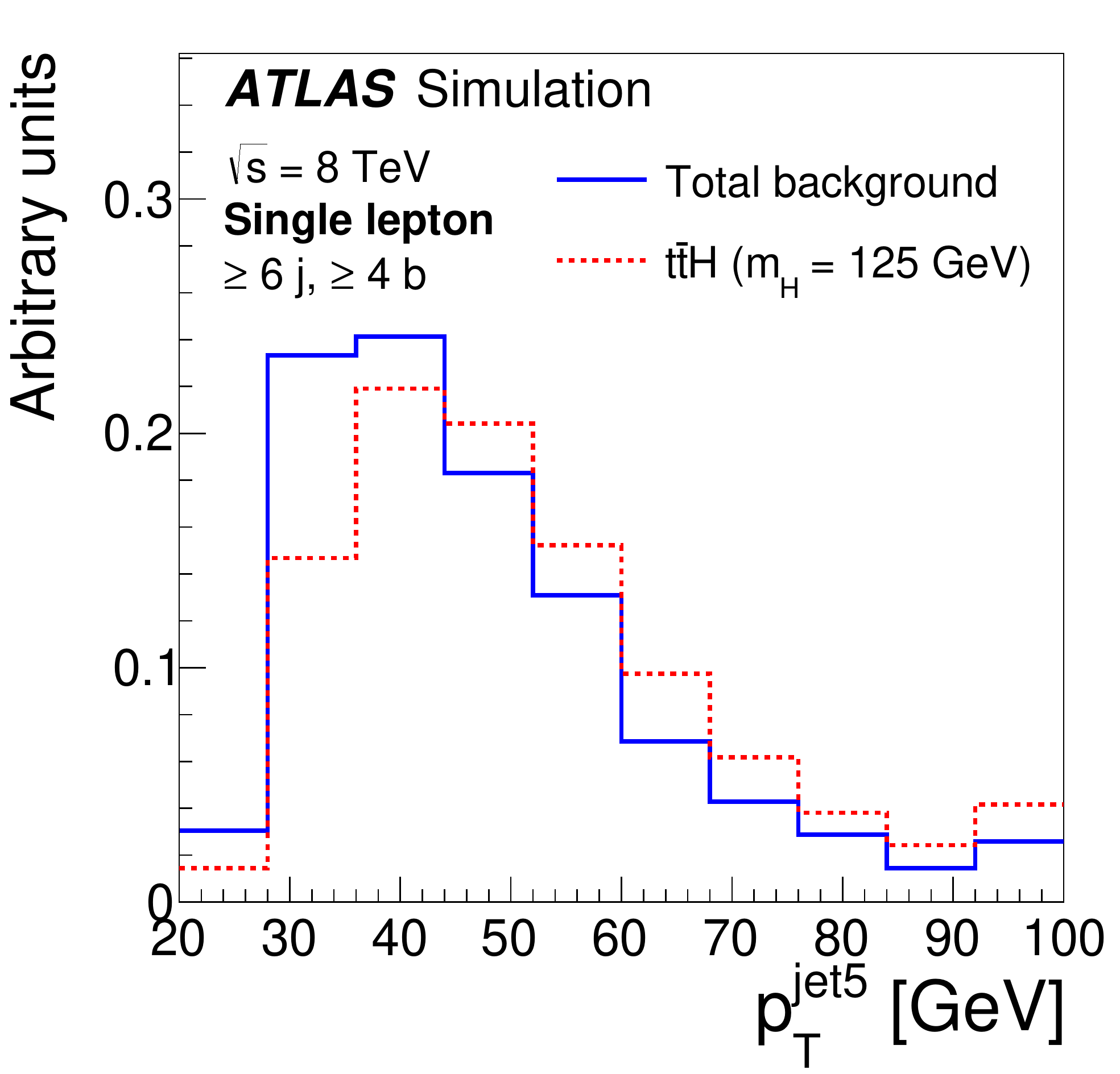}}\label{fig:sepinput_lj_3_c}
\subfigure[]{\includegraphics[width=0.24\textwidth]{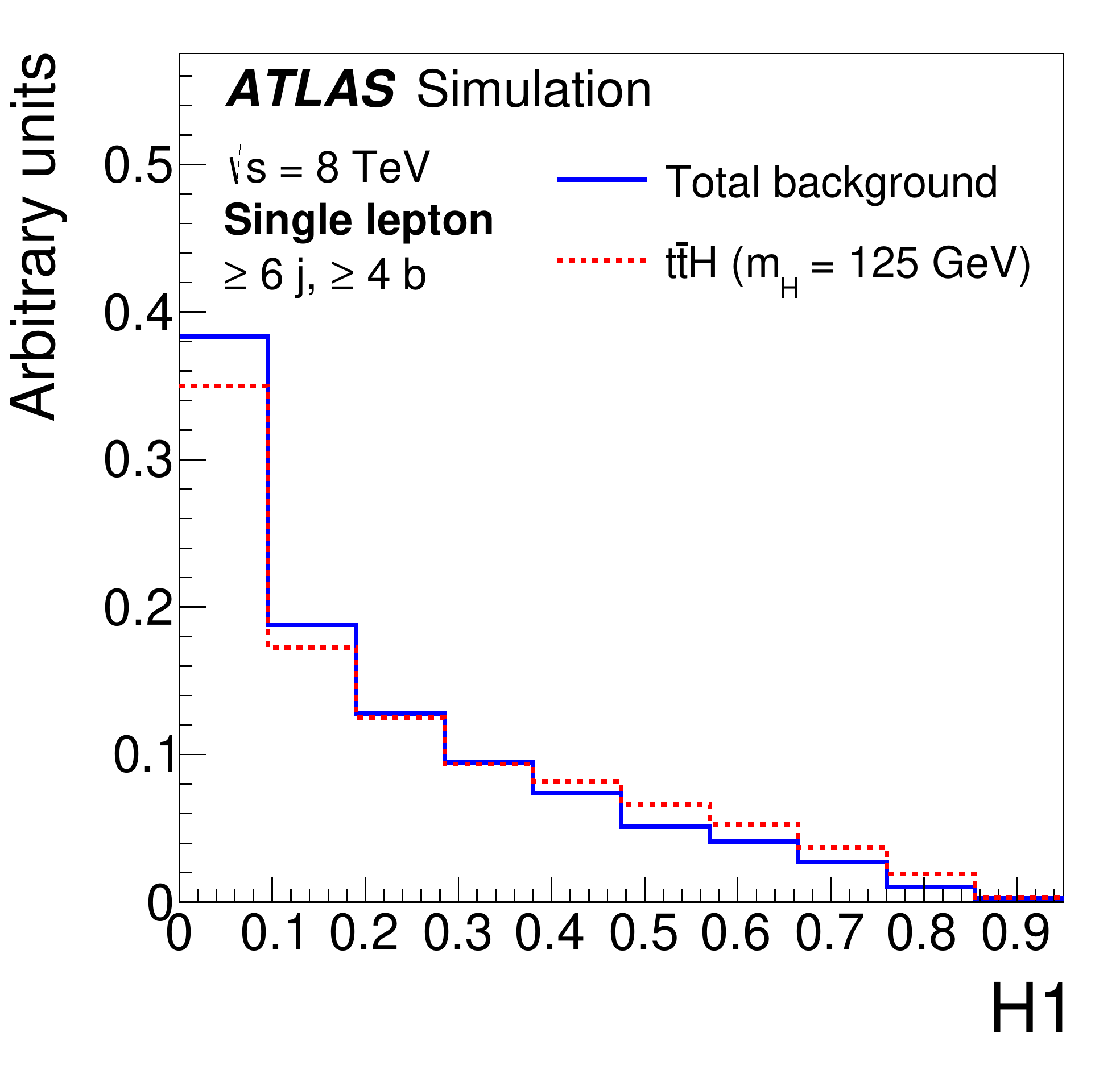}} \label{fig:sepinput_lj_3_d}
\caption{Single-lepton channel: comparison of \tth\ signal (dashed) and background (solid) for the four top-ranked input variables 
in the \sixfour\ region.  The plots include (a) $D1$, (b) \cent, (c) \ptjetfive, and (d) $H1$.
}
\label{fig:sepinput_lj_3} 
\end{center}
\end{figure*}

\begin{figure*}[ht!]
\begin{center}
\subfigure[]{\includegraphics[width=0.24\textwidth]{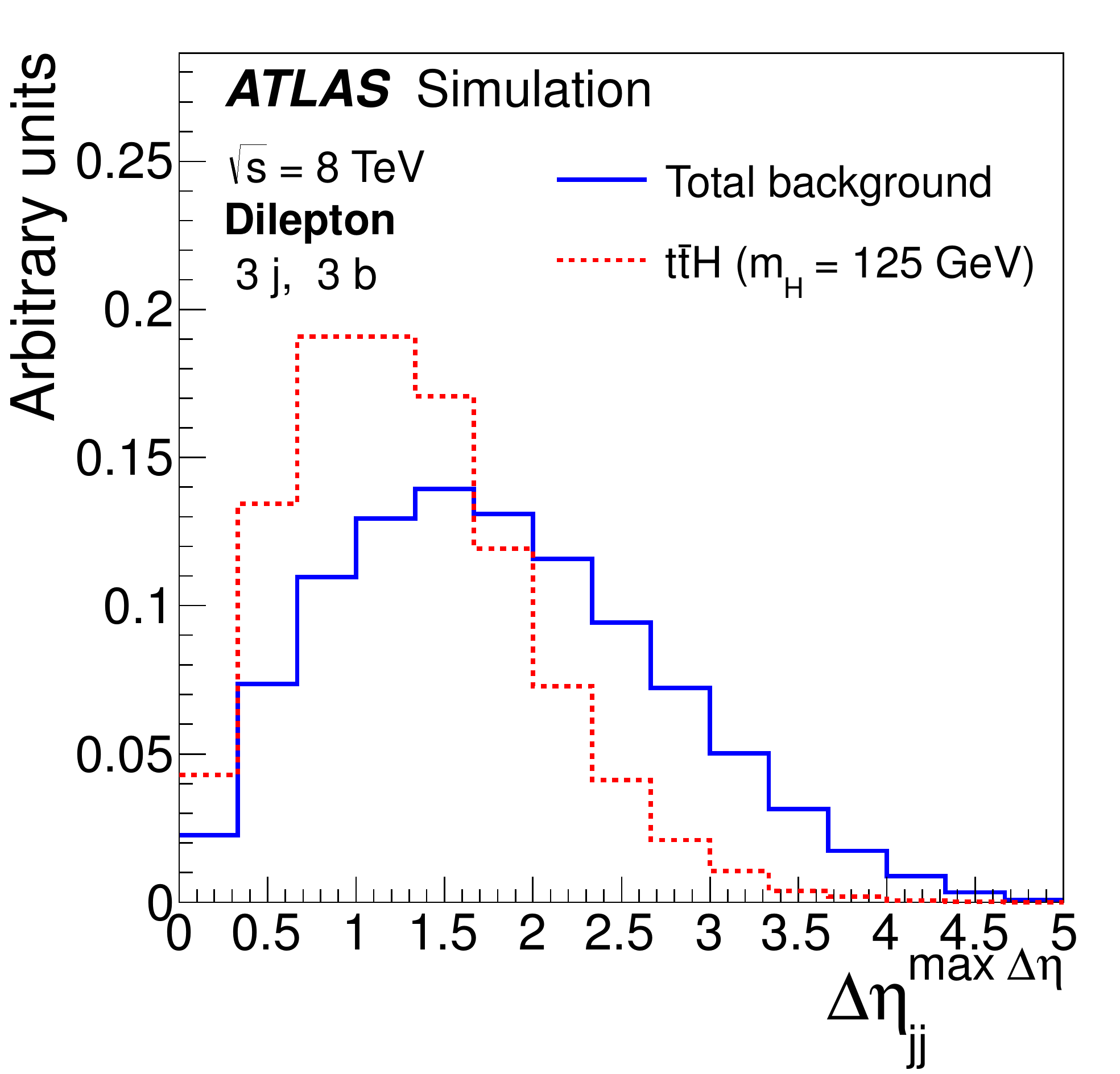}}\label{fig:sepinput_dil_1_a}
\subfigure[]{\includegraphics[width=0.24\textwidth]{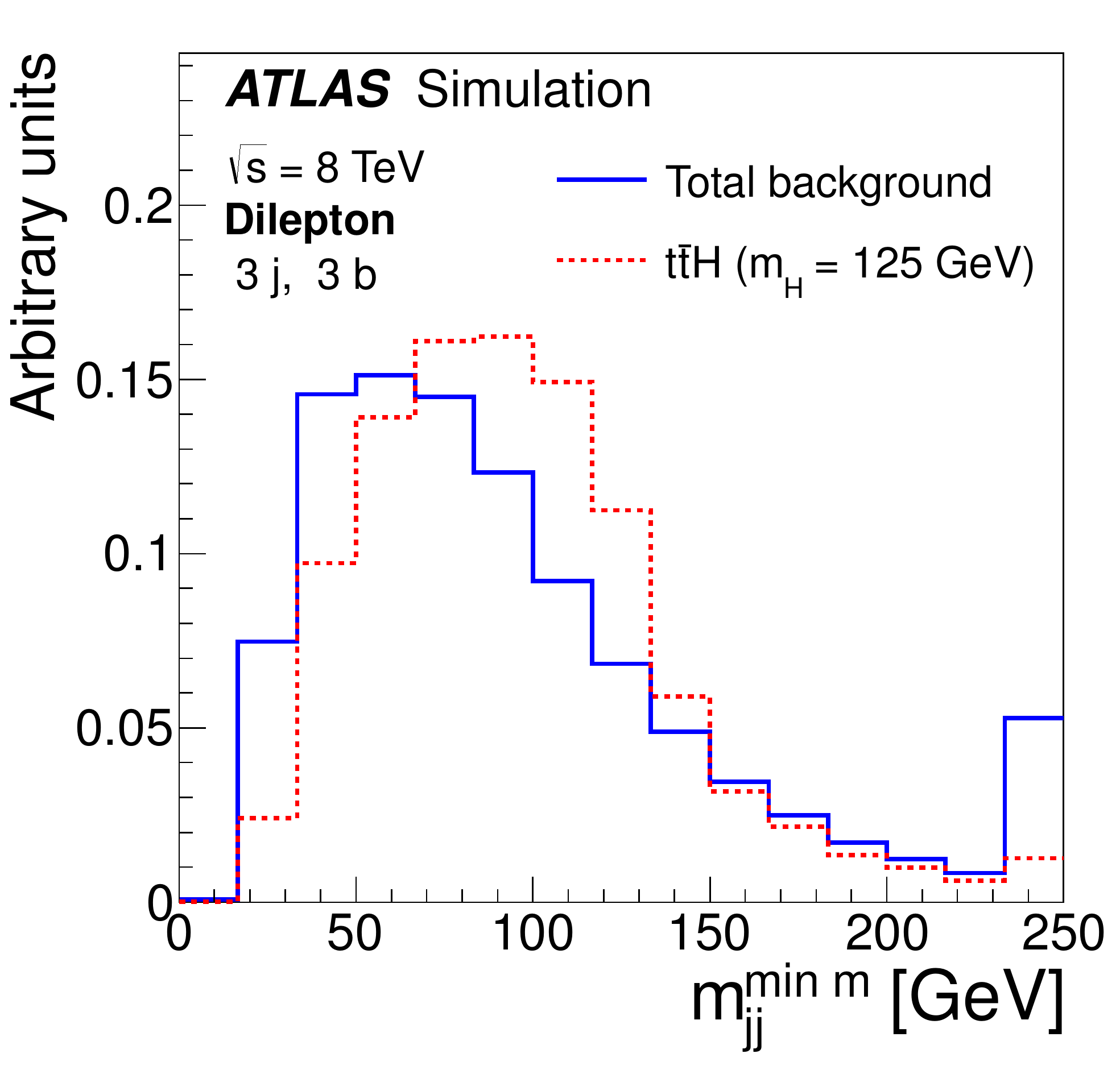}}\label{fig:sepinput_dil_1_b}
\subfigure[]{\includegraphics[width=0.24\textwidth]{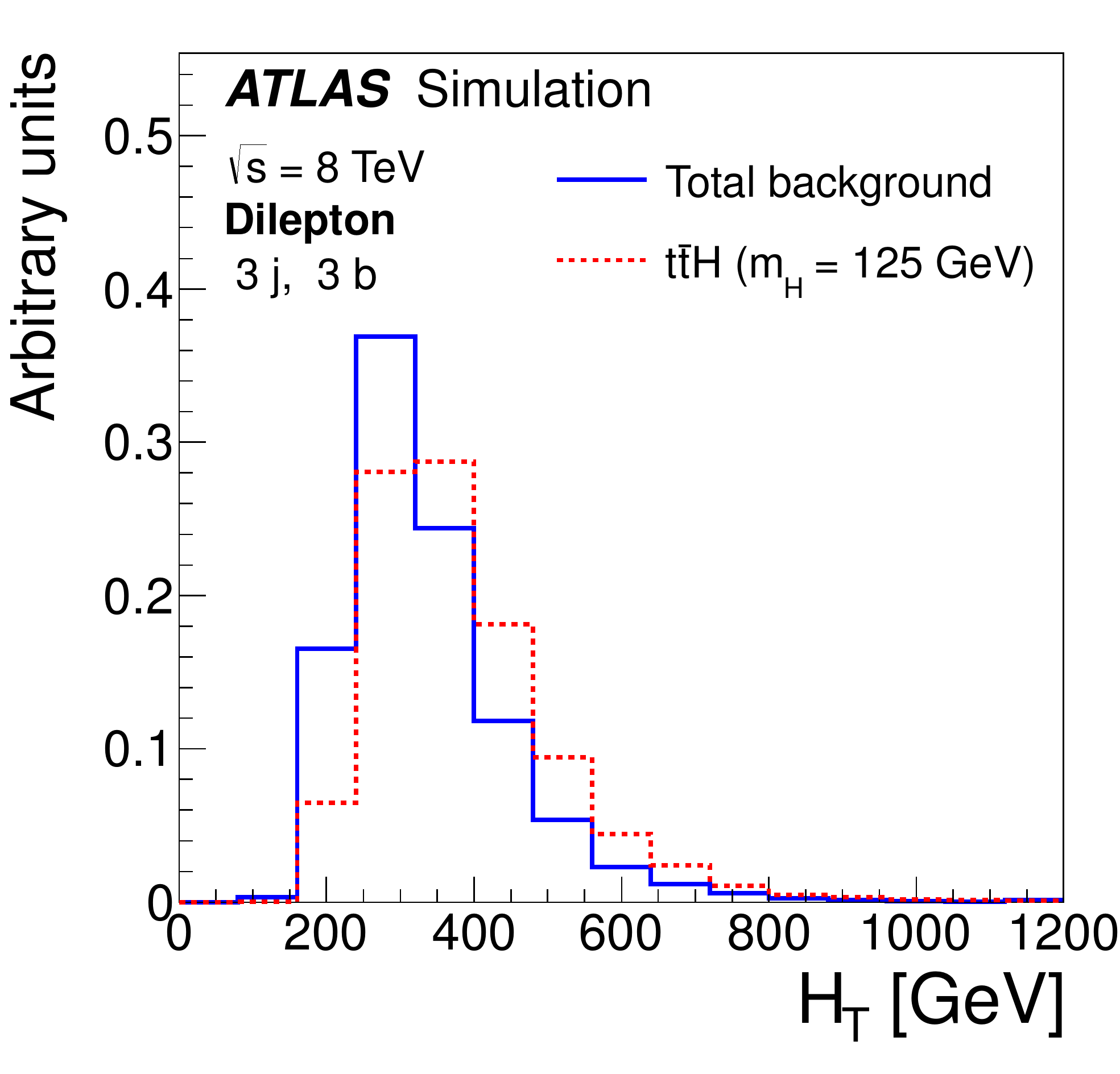}}\label{fig:sepinput_dil_1_c}
\subfigure[]{\includegraphics[width=0.24\textwidth]{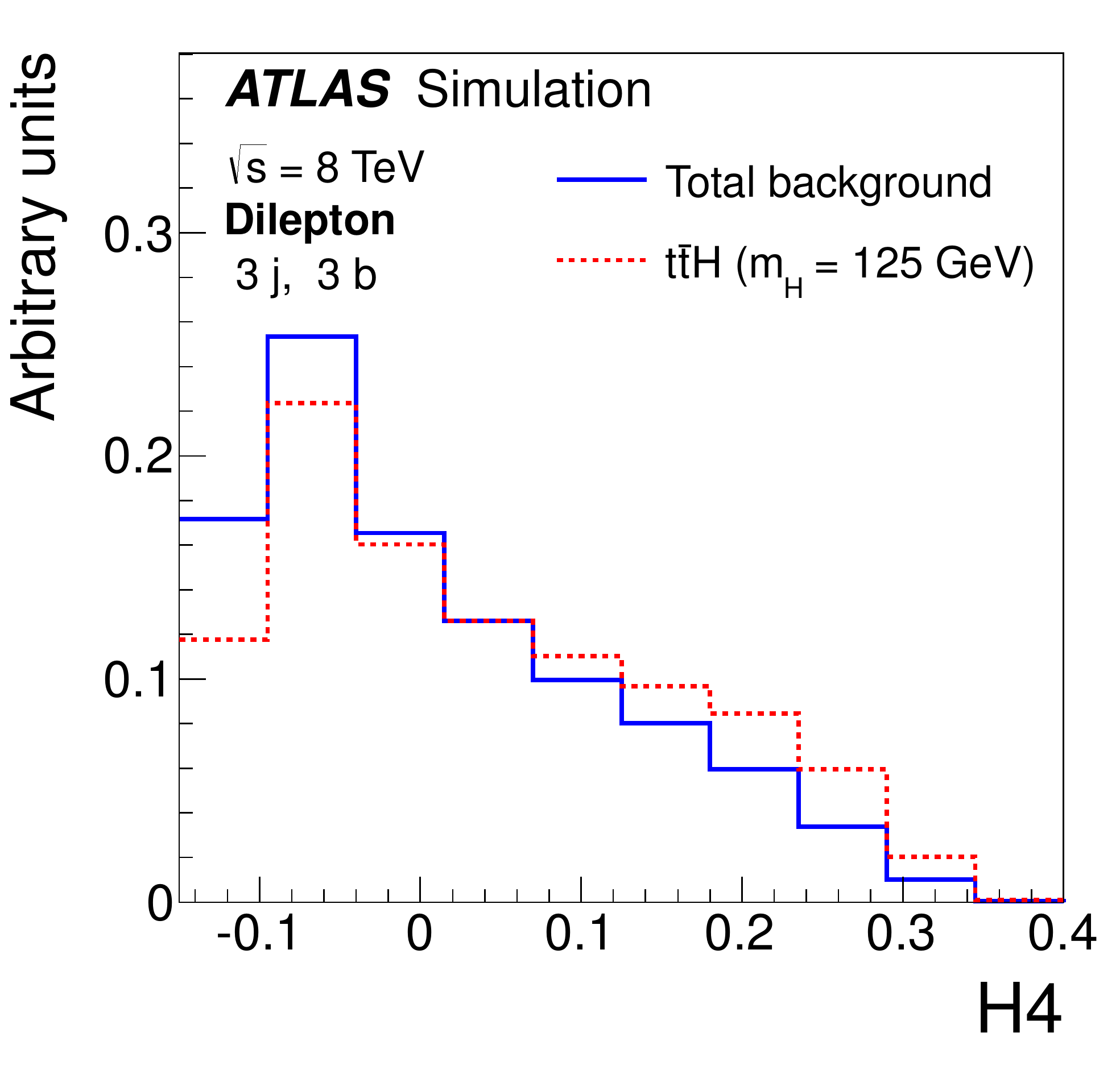}}\label{fig:sepinput_dil_1_d}
\caption{Dilepton channel: comparison of \tth\ signal (dashed) and background (solid) for the four top-ranked input variables 
in the \threethree\ region.  The plots include (a) \maxdeta, (b) \mindijetmass, (c) \htlep, and (d) $H4$.
}
\label{fig:sepinput_dil_1} 
\end{center}
\end{figure*}

\begin{figure*}[ht!]
\begin{center}
\subfigure[]{\includegraphics[width=0.24\textwidth]{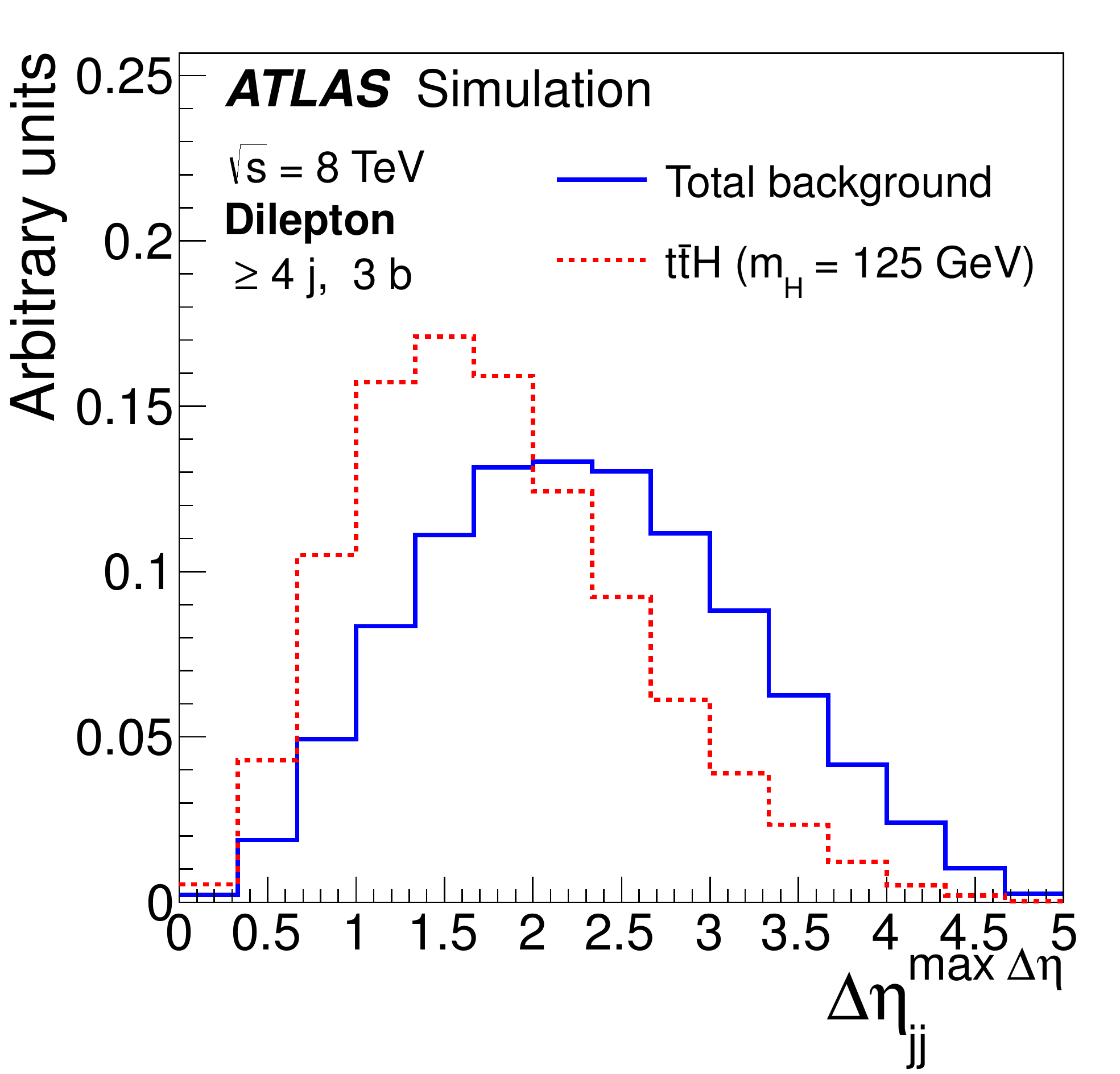}}\label{fig:sepinput_dil_2_a}
\subfigure[]{\includegraphics[width=0.24\textwidth]{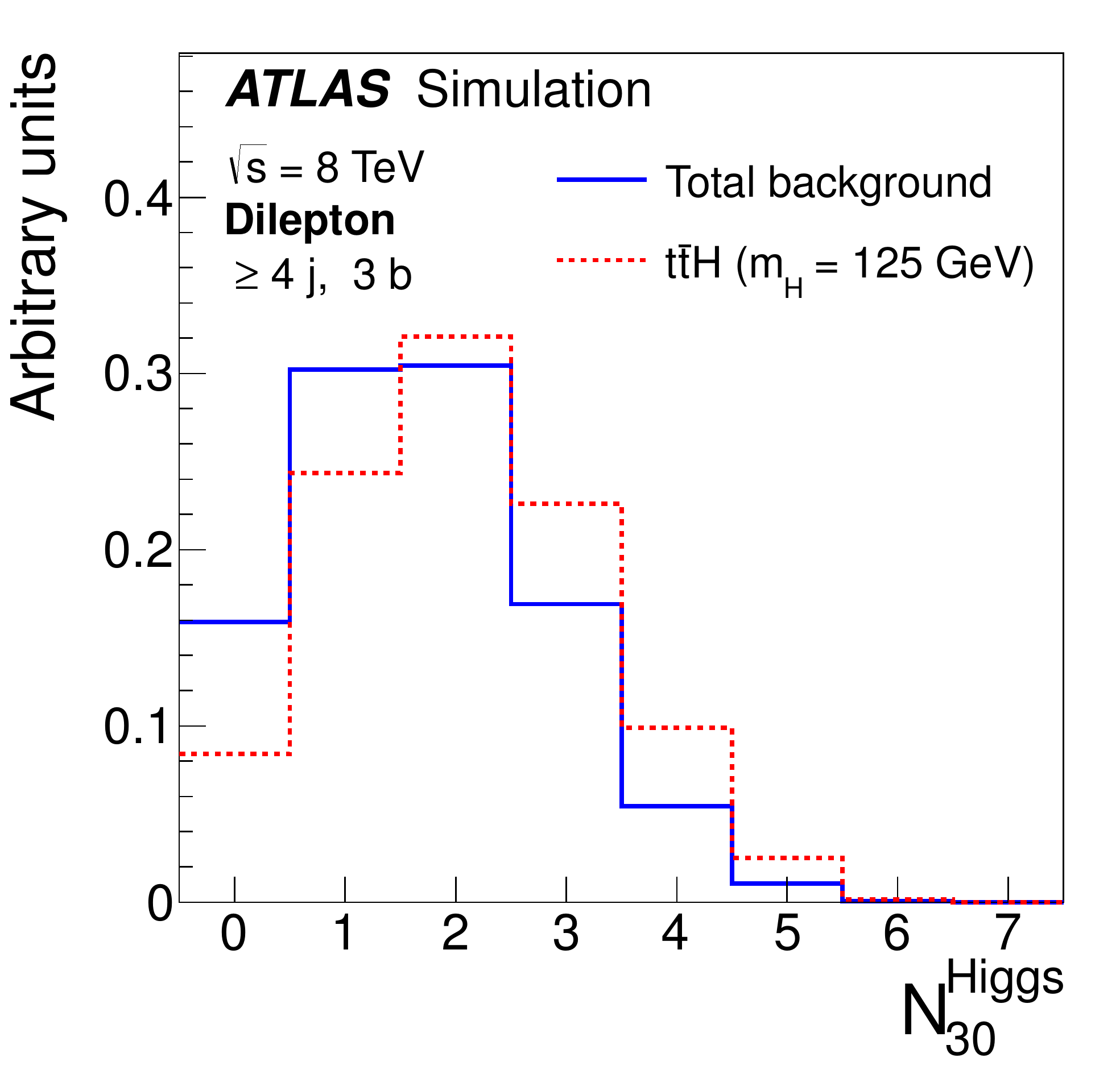}}\label{fig:sepinput_dil_2_b} 
\subfigure[]{\includegraphics[width=0.24\textwidth]{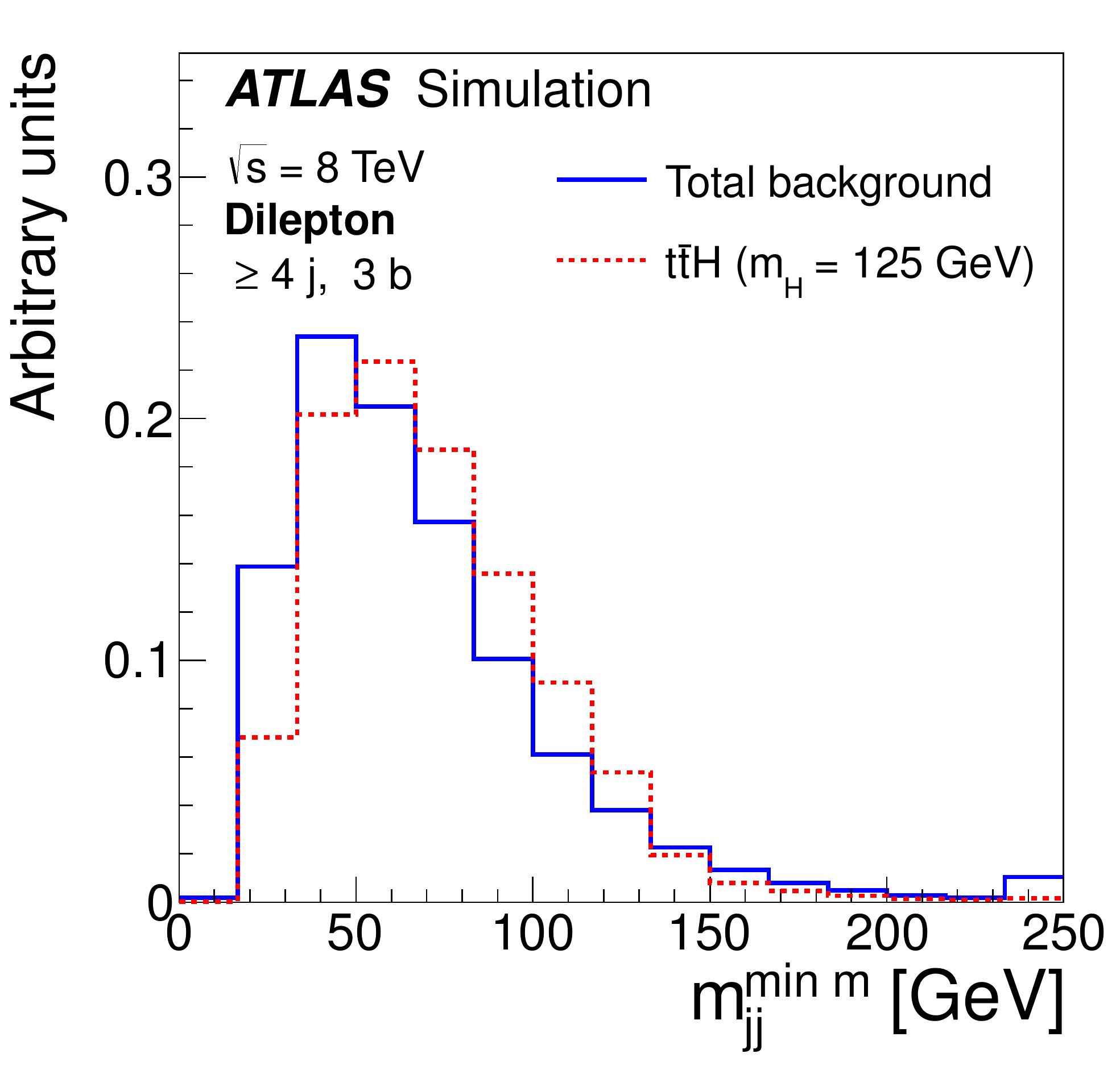}}\label{fig:sepinput_dil_2_c}
\subfigure[]{\includegraphics[width=0.24\textwidth]{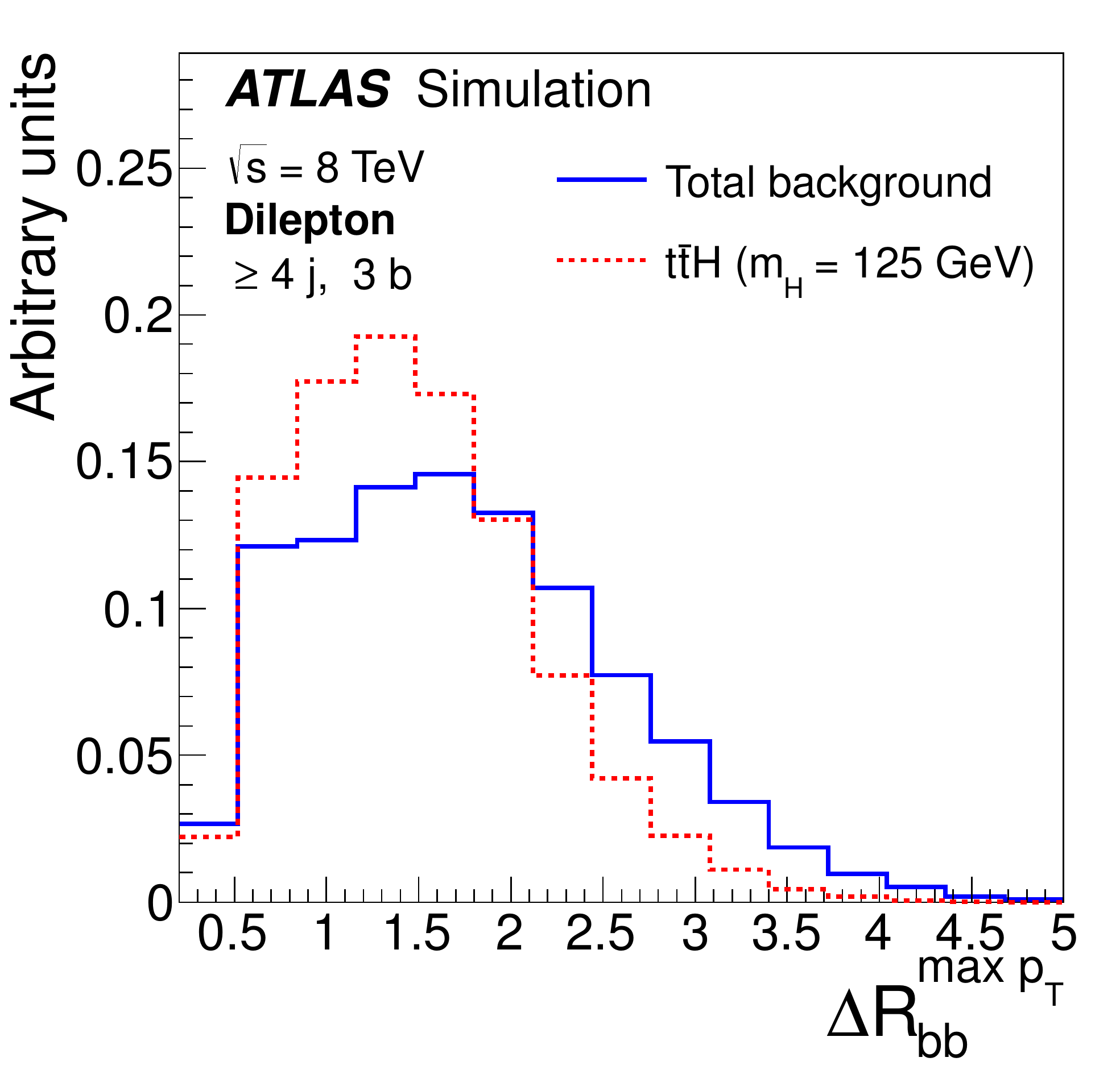}}\label{fig:sepinput_dil_2_d}
\caption{Dilepton channel: comparison of \tth\ signal (dashed) and background (solid) for the four top-ranked input variables 
in the \fourthreedi\ region. The plots include (a) \maxdeta, (b) \nhiggsthirty, (c) \mindijetmass, and (d) \drbbmaxpt.
}
\label{fig:sepinput_dil_2} 
\end{center}
\end{figure*}

\begin{figure*}[ht!]
\begin{center}
\subfigure[]{\includegraphics[width=0.24\textwidth]{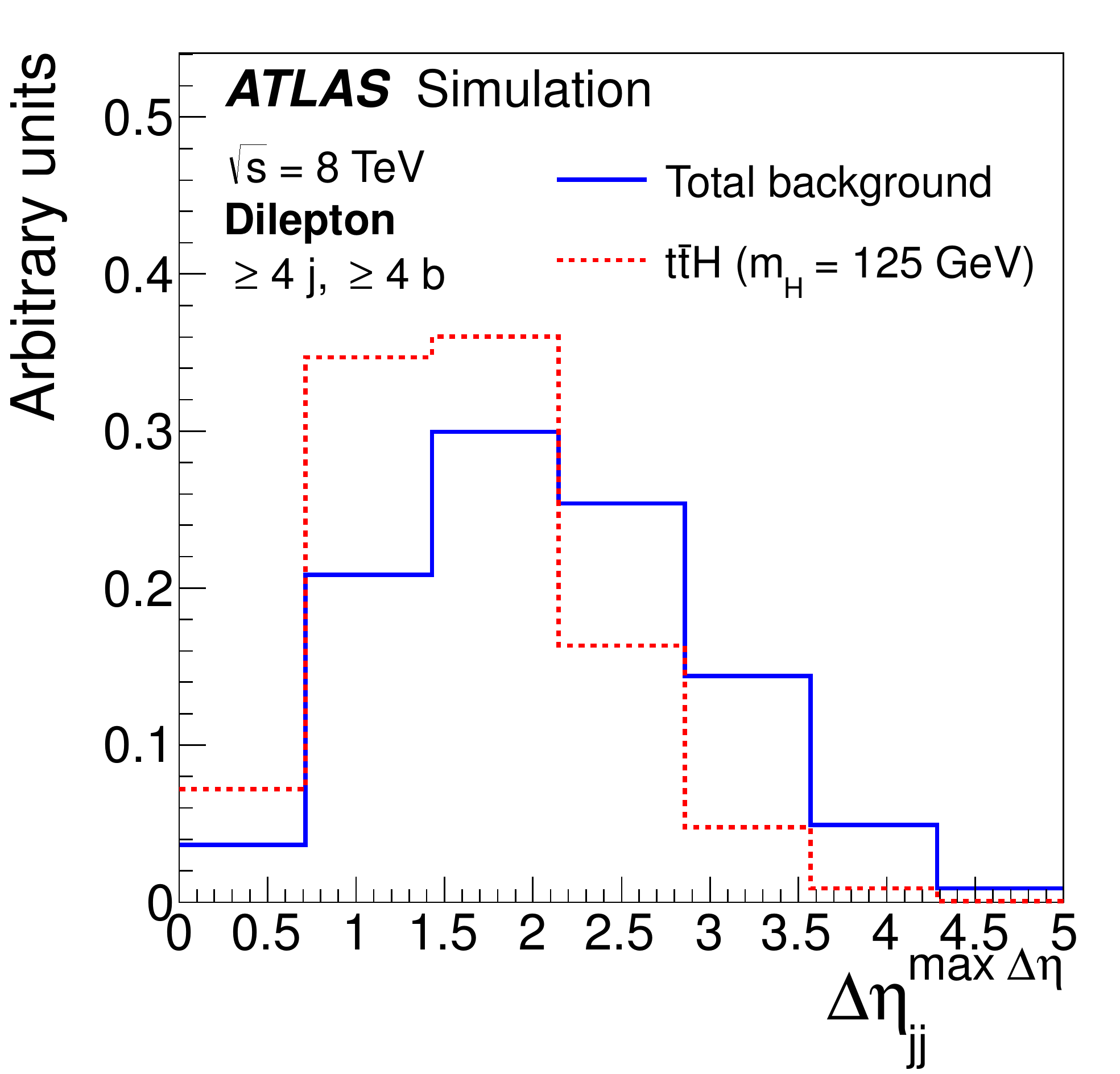}}\label{fig:sepinput_dil_3_a}
\subfigure[]{\includegraphics[width=0.24\textwidth]{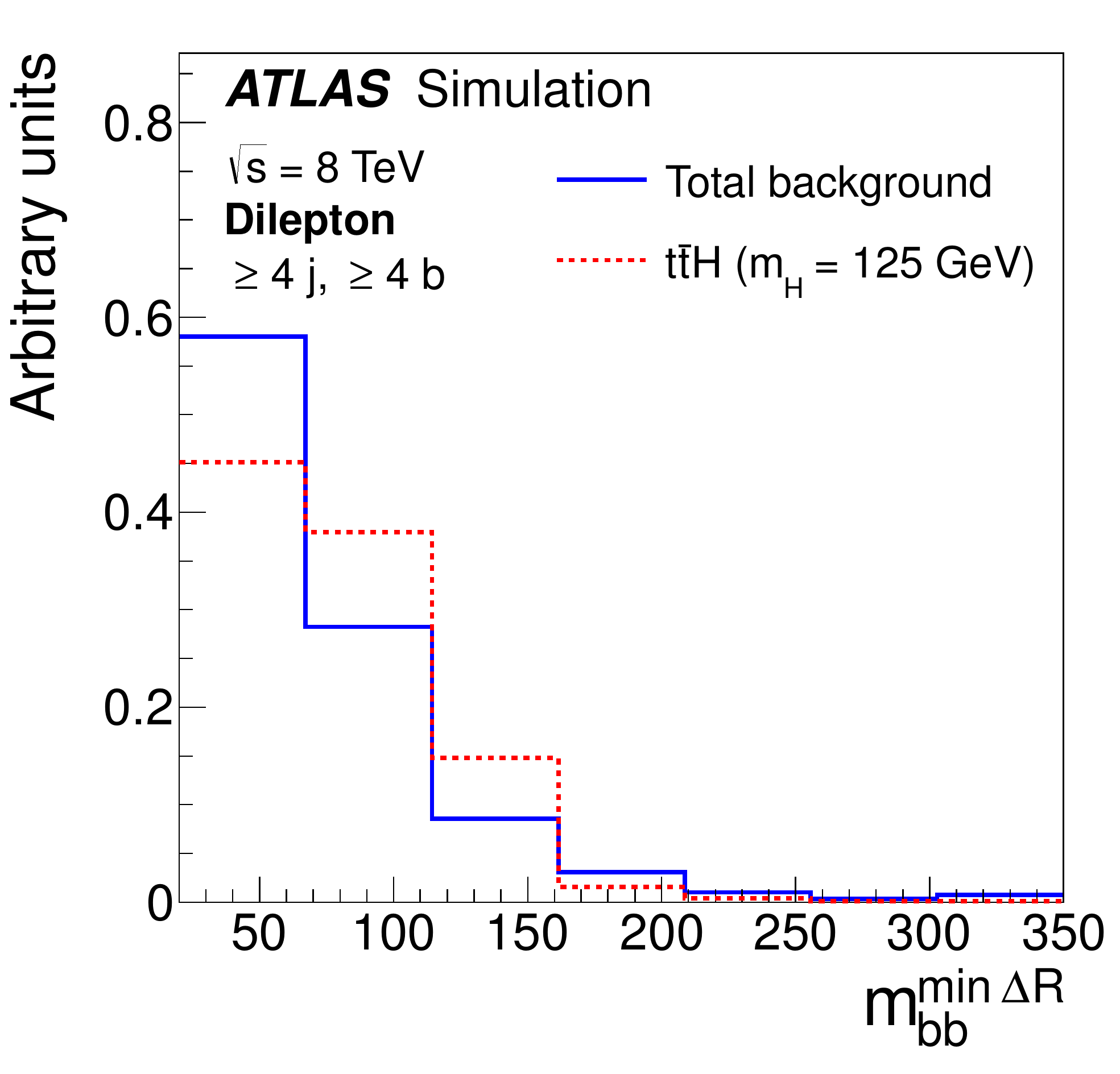}}\label{fig:sepinput_dil_3_b} 
\subfigure[]{\includegraphics[width=0.24\textwidth]{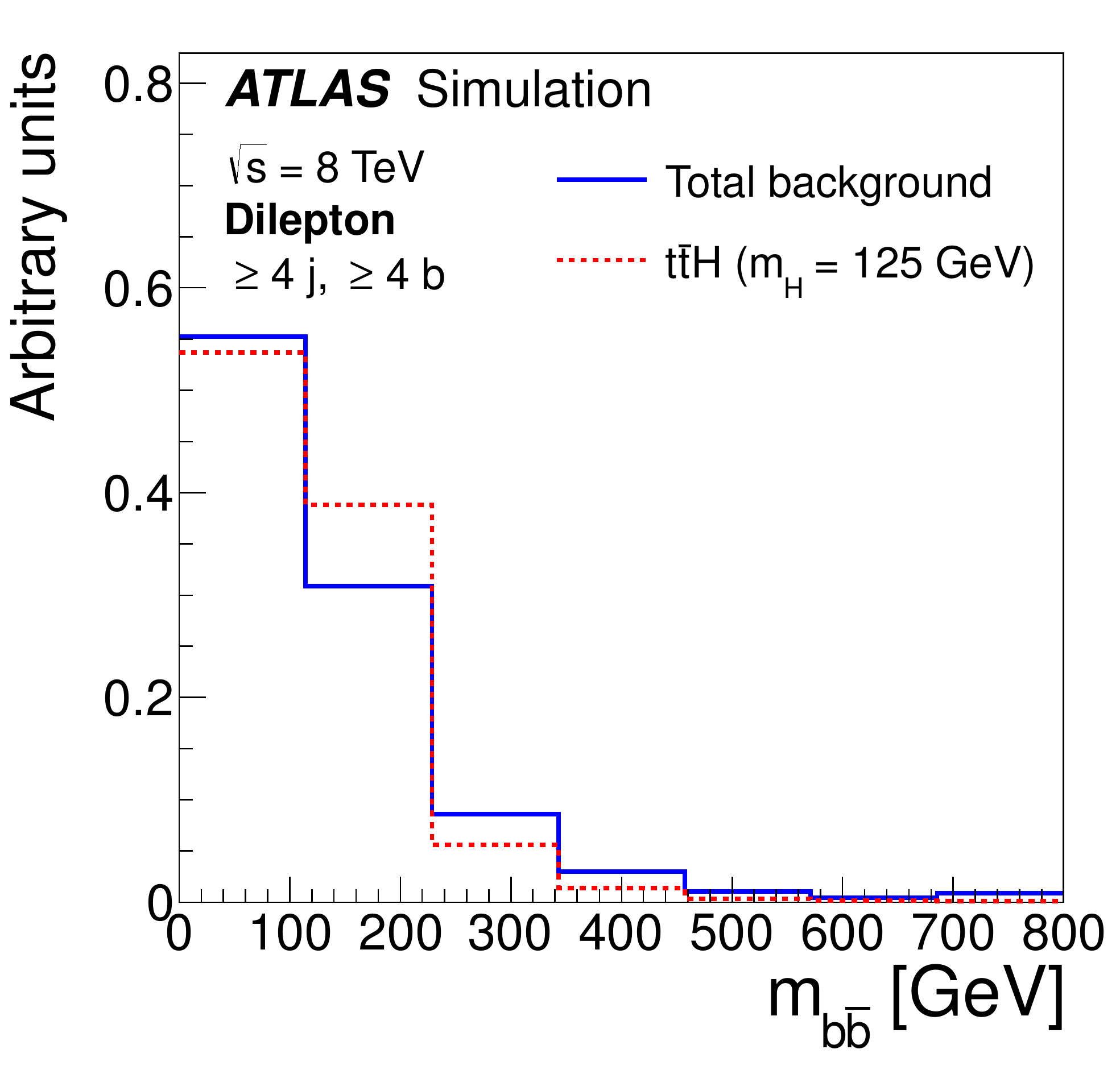}}\label{fig:sepinput_dil_3_c}
\subfigure[]{\includegraphics[width=0.24\textwidth]{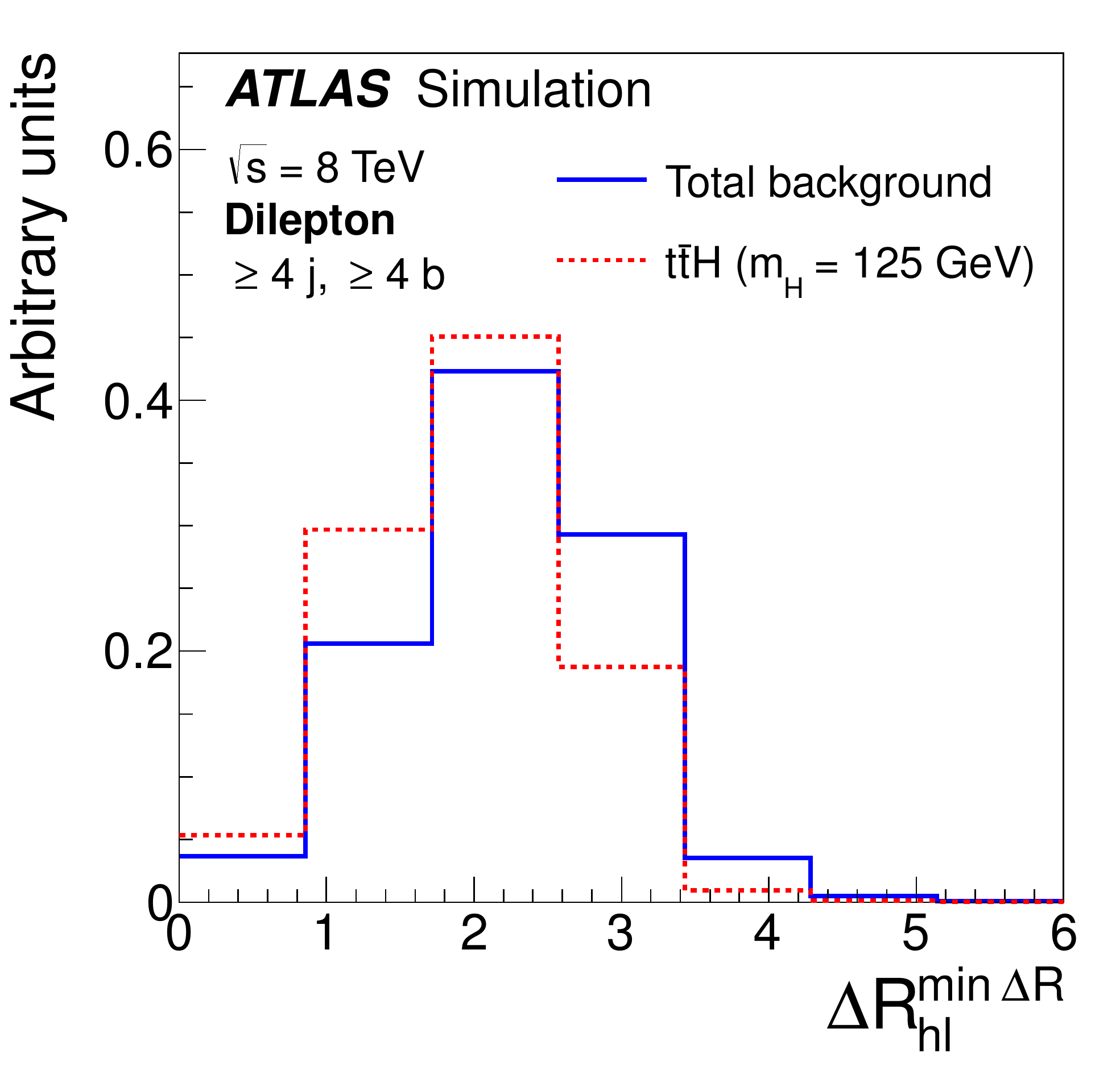}}\label{fig:sepinput_dil_3_d}
\caption{Dilepton channel: comparison of \tth\ signal (dashed) and background (solid) for the four top-ranked input variables 
in the \fourfourdi\ region.  The plots include (a) \maxdeta, (b) \mbbmindr, (c) \mbb, and (d) \mindrhl.
}
\label{fig:sepinput_dil_3} 
\end{center}
\end{figure*}

\section{Tables of systematic uncertainties in the signal region}
\label{sec:norm_syst}

Tables~\ref{tab:t6jetin4btagin8TeV_Postfit_test_SystTable} 
and~\ref{tab:t4jetin4btagin8TeV_Postfit_test_SystTable} show pre-fit and post-fit 
contributions of the different categories of uncertainties (expressed in \%) 
for the \tth\ signal and main background processes in the \sixfour\ 
region of the single-lepton channel and the \fourfourdi\ region of the dilepton channel, respectively. 

The ``Lepton efficiency'' category includes sysmematic uncertanties on electrons and muons 
listed in Table~\ref{tab:SystSummary}. The ``Jet efficiency'' category includes
uncertainties on the jet vertex fraction and jet reconstruction. The ``\ttbar\ heavy-flavour
modelling'' category includes uncertainties on the \ttbar+\bbbar\ NLO shape and on the     
$t\bar{t}$+$c\bar{c}$ \pt\ reweighting and generator. The ``Theoretical cross sections''
category includes uncertainties on the single top, diboson, $V$+jets and $\ttbar+V$
theoretical cross sections. The ``\tth\ modelling'' category includes contributions from
\tth\ scale, generator, hadronisation model and PDF choice. The details of the 
evaluation of the uncertainties can be found in Sect.~\ref{sec:SystematicUncertainties}. 

\begin{table*}[ht!]
	\centering
	\begin{tabular}{l | l l l l | l l l l}
\multicolumn{9}{c}{$\geq$ 6 j, $\geq$ 4 b}\\
\hline\hline
 & \multicolumn{4}{c|}{Pre-fit} & \multicolumn{4}{c}{Post-fit} \\ 
 & $t\bar{t}H$ (125) & $t\bar{t}$ + light & $t\bar{t}+c\bar{c}$ & $t\bar{t}+b\bar{b}$ & $t\bar{t}H$ (125) & $t\bar{t}$ + light & $t\bar{t}+c\bar{c}$ & $t\bar{t}+b\bar{b}$ \\
\hline

Luminosity  & $\pm 2.8 $  & $\pm 2.8 $  & $\pm 2.8 $  & $\pm 2.8 $  & $\pm 2.6 $  & $\pm 2.6 $  & $\pm 2.6 $  & $\pm 2.6 $ \\ 
Lepton efficiencies  & $\pm 1.4 $  & $\pm 1.4 $  & $\pm 1.4 $  & $\pm 1.5 $  & $\pm 1.3 $  & $\pm 1.3 $  & $\pm 1.3 $  & $\pm 1.3 $ \\ 
Jet energy scale  & $\pm 6.4 $  & $\pm 13 $  & $\pm 11 $  & $\pm 9.2 $  & $\pm 2.3 $  & $\pm 5.3 $  & $\pm 4.7 $  & $\pm 3.6 $ \\ 
Jet efficiencies  & $\pm 1.7 $  & $\pm 5.2 $  & $\pm 2.7 $  & $\pm 2.5 $  & $\pm 0.7 $  & $\pm 2.3 $  & $\pm 1.2 $  & $\pm 1.1 $ \\ 
Jet energy resolution  & $\pm 0.1 $  & $\pm 4.4 $  & $\pm 2.5 $  & $\pm 1.6 $  & $\pm 0.1 $  & $\pm 2.3 $  & $\pm 1.3 $  & $\pm 0.8 $ \\ 
$b$-tagging efficiency  & $\pm 9.2 $  & $\pm 5.6 $  & $\pm 5.1 $  & $\pm 9.3 $  & $\pm 5.0 $  & $\pm 3.1 $  & $\pm 2.9 $  & $\pm 5.0 $ \\ 
$c$-tagging efficiency  & $\pm 1.7 $  & $\pm 6.0 $  & $\pm 12 $  & $\pm 2.4 $  & $\pm 1.4 $  & $\pm 5.1 $  & $\pm 10 $  & $\pm 2.1 $ \\ 
$l$-tagging efficiency  & $\pm 1.0 $  & $\pm 19 $  & $\pm 5.2 $  & $\pm 2.1 $  & $\pm 0.6 $  & $\pm 11 $  & $\pm 3.0 $  & $\pm 1.1 $ \\ 
High \pt\ tagging efficiency  & $\pm 0.6 $  & --  & $\pm 0.7 $  & $\pm 0.6 $  & $\pm 0.3 $  & --  & $\pm 0.4 $  & $\pm 0.3 $ \\ 
$t\bar{t}$: \pt\ reweighting  & --  & $\pm 5.4 $  & $\pm 6.1 $  & --  & --  & $\pm 4.7 $  & $\pm 5.4 $  & -- \\ 
$t\bar{t}$: parton shower  & --  & $\pm 13 $  & $\pm 16 $  & $\pm 11 $  & --  & $\pm 3.6 $  & $\pm 10 $  & $\pm 6.0 $ \\ 
$t\bar{t}$+HF: normalisation  & --  & --  & $\pm 50 $  & $\pm 50 $  & --  & --  & $\pm 28 $  & $\pm 14 $ \\ 
$t\bar{t}$+HF: modelling  & --  & $\pm 11 $  & $\pm 16 $  & $\pm 8.3 $  & --  & $\pm 3.6 $  & $\pm 9.1 $  & $\pm 7.1 $ \\ 
Theoretical cross sections  & --  & $\pm 6.3 $  & $\pm 6.3 $  & $\pm 6.3 $  & --  & $\pm 4.1 $  & $\pm 4.1 $  & $\pm 4.1 $ \\ 
$t\bar{t}H$ modelling  & $\pm 2.7 $  & --  & --  & --  & $\pm 2.6 $  & --  & --  & -- \\ 
\hline
Total   & $\pm 12 $  & $\pm 32 $  & $\pm 59 $  & $\pm 54 $  & $\pm 6.9 $  & $\pm 9.2 $  & $\pm 23 $  & $\pm 12 $ \\ 
\hline\hline
\end{tabular}
\parbox{18cm}{\caption{\label{tab:t6jetin4btagin8TeV_Postfit_test_SystTable}
Single lepton channel: normalisation uncertainties (expressed in
\% ) on signal and main background processes
for the systematic uncertainties considered, before and after the fit to data in
\sixfour\ region of the single lepton channel.
The total uncertainty can be different from the sum in quadrature 
of individual sources due to the anti-correlations between them.
}}

\end{table*}

\begin{table*}[h!]
	\centering
	\begin{tabular}{l | l l l l | l l l l}
\multicolumn{9}{c}{$\geq$ 4 j, $\geq$ 4 b} \\
\hline\hline
 & \multicolumn{4}{c|}{Pre-fit} & \multicolumn{4}{c}{Post-fit} \\ 
 & $t\bar{t}H$ (125) & $t\bar{t}$ + light & $t\bar{t}+c\bar{c}$ & $t\bar{t}+b\bar{b}$ & $t\bar{t}H$ (125) & $t\bar{t}$ + light & $t\bar{t}+c\bar{c}$ & $t\bar{t}+b\bar{b}$ \\
\hline
Luminosity  & $\pm 2.8 $  & $\pm 2.8 $  & $\pm 2.8 $  & $\pm 2.8 $  & $\pm 2.6 $  & $\pm 2.6 $  & $\pm 2.6 $  & $\pm 2.6 $ \\ 
Lepton efficiencies  & $\pm 2.5 $  & $\pm 2.5 $  & $\pm 2.5 $  & $\pm 2.5 $  & $\pm 1.8 $  & $\pm 1.8 $  & $\pm 1.8 $  & $\pm 1.8 $ \\ 
Jet energy scale  & $\pm 4.5 $  & $\pm 12 $  & $\pm 9.4 $  & $\pm 7.0 $  & $\pm 2.0 $  & $\pm 5.5 $  & $\pm 4.5 $  & $\pm 3.3 $ \\ 
Jet efficiencies  & --  & $\pm 5.9 $  & $\pm 1.6 $  & $\pm 0.9 $  & --  & $\pm 2.6 $  & $\pm 0.7 $  & $\pm 0.4 $ \\ 
Jet energy resolution  & $\pm 0.1 $  & $\pm 4.5 $  & $\pm 1.1 $  & --  & $\pm 0.1 $  & $\pm 2.3 $  & $\pm 0.6 $  & -- \\ 
$b$-tagging efficiency  & $\pm 10 $  & $\pm 5.5 $  & $\pm 5.4 $  & $\pm 11 $  & $\pm 5.6 $  & $\pm 3.1 $  & $\pm 3.0 $  & $\pm 5.8 $ \\ 
$c$-tagging efficiency  & $\pm 0.5 $  & --  & $\pm 12 $  & $\pm 0.6 $  & $\pm 0.3 $  & --  & $\pm 10 $  & $\pm 0.3 $ \\ 
$l$-tagging efficiency  & $\pm 0.7 $  & $\pm 34 $  & $\pm 7.0 $  & $\pm 1.6 $  & $\pm 0.4 $  & $\pm 21 $  & $\pm 4.2 $  & $\pm 0.9 $ \\ 
High \pt\ tagging efficiency  & --  & --  & $\pm 0.6 $  & --  & --  & --  & $\pm 0.3 $  & -- \\ 
$t\bar{t}$: \pt\ reweighting  & --  & $\pm 5.8 $  & $\pm 6.2 $  & --  & --  & $\pm 5.0 $  & $\pm 5.4 $  & -- \\ 
$t\bar{t}$: parton shower  & --  & $\pm 14$  & $\pm 18 $  & $\pm 14 $  & --  & $\pm 4.8$  & $\pm 11 $  & $\pm 8.1 $ \\ 
$t\bar{t}$+HF: normalisation  & --  & --  & $\pm 50 $  & $\pm 50 $  & --  & --  & $\pm 28 $  & $\pm 14 $ \\ 
$t\bar{t}$+HF: modelling  & --  & $\pm 11 $  & $\pm 16 $  & $\pm 12 $  & --  & $\pm 3.8 $  & $\pm 10 $  & $\pm 10 $ \\ 
Theoretical cross sections  & --  & $\pm 6.3 $  & $\pm 6.3 $  & $\pm 6.2 $  & --  & $\pm 4.1 $  & $\pm 4.1 $  & $\pm 4.1 $ \\ 
$t\bar{t}H$ modelling  & $\pm 1.9 $  & --  & --  & --  & $\pm 1.8 $  & --  & --  & -- \\ 
\hline
Total   & $\pm 12 $  & $\pm 40 $  & $\pm 59 $  & $\pm 55 $  & $\pm 6.7 $  & $\pm 22 $  & $\pm 22 $  & $\pm 13 $ \\ 
\hline\hline
\end{tabular}
\parbox{18cm}{
\caption{\label{tab:t4jetin4btagin8TeV_Postfit_test_SystTable}
Dilepton channel: normalisation uncertainties (expressed in
\% ) on signal and main background processes
for the systematic uncertainties considered, before and after the fit to data in
\fourfourdi\ region of the dilepton channel.
The total uncertainty can be different from the sum in quadrature 
of individual sources due to the anti-correlations between them.
}}

\end{table*}

\section{Post-fit event yields}
\label{sec:postfit_tables}
The post-fit event yields for the combined single-lepton channel for 
the different regions considered in the analysis are summarised 
in Table~\ref{tab:Postfit_EventsTable_lj}.  Similarly, the post-fit 
event yields for the combined dilepton channels for the different 
regions are summarised in Table~\ref{tab:Postfit_EventsTable_dil}.

\begin{table*}
\begin{center}
\begin{tabular}{l*{3}{r@{$\,\pm\,$}r}}
\hline\hline
 & \multicolumn{2}{c}{4 j, 2 b} & \multicolumn{2}{c}{4 j, 3 b} & \multicolumn{2}{c}{4 j, 4 b}\\
\hline
$t\bar{t}H$ (125) & \numRF{47.58}{2} & \numRF{34.91}{2} & \numRF{20.03}{2} & \numRF{14.69}{2} & \numRF{3.03}{2} & \numRF{2.23}{2}\\
$t\bar{t}+$ light & \numRF{78240.75}{2} & \numRF{1574.18}{2} & \numRF{6263.72}{2} & \numRF{161.34}{2} & \numRF{56.45}{2} & \numRF{4.71}{1}\\
$t\bar{t}+c\bar{c}$ & \numRF{6433.38}{2} & \numRF{1796.53}{2} & \numRF{845.05}{2} & \numRF{220.43}{2} & \numRF{25.54}{2} & \numRF{6.54}{1}\\
$t\bar{t}+b\bar{b}$ & \numRF{2475.84}{2} & \numRF{487.98}{2} & \numRF{969.03}{2} & \numRF{148.87}{2} & \numRF{62.51}{2} & \numRF{8.48}{1}\\
$W$+jets & \numRF{3654.81}{2} & \numRF{1116.58}{2} & \numRF{165.88}{2} & \numRF{51.34}{2} & \numRF{4.00}{2} & \numRF{1.24}{2}\\
$Z$+jets & \numRF{1058.08}{2} & \numRF{535.13}{2} & \numRF{49.12}{2} & \numRF{25.09}{2} & \numRF{1.06}{2} & \numRF{0.57}{1}\\
Single top & \numRF{4712.43}{2} & \numRF{322.17}{2} & \numRF{332.61}{2} & \numRF{28.10}{2} & \numRF{6.81}{2} & \numRF{0.73}{1}\\
Diboson & \numRF{215.89}{2} & \numRF{64.89}{2} & \numRF{11.33}{2} & \numRF{3.65}{1} & \numRF{0.28}{1} & \numRF{0.12}{1}\\
$t\bar{t}+V$ & \numRF{120.03}{2} & \numRF{37.66}{2} & \numRF{15.84}{2} & \numRF{4.92}{1} & \numRF{0.94}{1} & \numRF{0.29}{1}\\
Lepton misID & \numRF{1082.01}{2} & \numRF{367.87}{2} & \numRF{78.36}{2} & \numRF{26.20}{2} & \numRF{2.59}{2} & \numRF{0.99}{1}\\
\hline
Total & \numRF{98040.79}{2} & \numRF{336.60}{2}   & \numRF{8750.98}{2} & \numRF{81.58}{2}   & \numRF{163.21}{2} & \numRF{5.63}{1}  \\
\hline
Data & \multicolumn{2}{l}{\num{98049}}  & \multicolumn{2}{l}{\num{8752}}  & \multicolumn{2}{l}{\num{161}} \\
\hline\hline     \\
\end{tabular}
\vspace{0.1cm}

\begin{tabular}{l*{3}{r@{$\,\pm\,$}r}}
\hline\hline
 & \multicolumn{2}{c}{5 j, 2 b} & \multicolumn{2}{c}{5 j, 3 b} & \multicolumn{2}{c}{5 j, $\geq$ 4 b}\\
\hline
$t\bar{t}H$ (125) & \numRF{60.39}{2} & \numRF{44.18}{2} & \numRF{33.72}{2} & \numRF{24.66}{2} & \numRF{9.42}{2} & \numRF{6.89}{2}\\
$t\bar{t}+$ light & \numRF{38387.98}{2} & \numRF{1046.42}{2} & \numRF{3614.11}{2} & \numRF{116.75}{2} & \numRF{65.28}{2} & \numRF{5.59}{1}\\
$t\bar{t}+c\bar{c}$ & \numRF{4801.21}{2} & \numRF{1203.63}{2} & \numRF{934.58}{2} & \numRF{230.20}{2} & \numRF{50.70}{2} & \numRF{12.47}{2}\\
$t\bar{t}+b\bar{b}$ & \numRF{2376.48}{2} & \numRF{364.28}{2} & \numRF{1260.77}{2} & \numRF{176.75}{2} & \numRF{154.69}{2} & \numRF{19.93}{2}\\
$W$+jets & \numRF{1208.16}{2} & \numRF{424.84}{2} & \numRF{86.55}{2} & \numRF{30.69}{2} & \numRF{4.04}{2} & \numRF{1.45}{2}\\
$Z$+jets & \numRF{367.88}{2} & \numRF{204.43}{2} & \numRF{27.88}{2} & \numRF{15.59}{2} & \numRF{1.40}{2} & \numRF{0.80}{1}\\
Single top & \numRF{1726.09}{2} & \numRF{152.89}{2} & \numRF{185.02}{2} & \numRF{17.78}{2} & \numRF{8.17}{2} & \numRF{0.66}{1}\\
Diboson & \numRF{93.76}{2} & \numRF{35.15}{2} & \numRF{7.96}{2} & \numRF{3.14}{2} & \numRF{0.45}{1} & \numRF{0.19}{1}\\
$t\bar{t}+V$ & \numRF{137.94}{2} & \numRF{43.15}{2} & \numRF{26.11}{2} & \numRF{8.06}{1} & \numRF{3.19}{2} & \numRF{0.98}{1}\\
Lepton misID & \numRF{342.54}{2} & \numRF{112.21}{2} & \numRF{43.52}{2} & \numRF{15.83}{2} & \numRF{5.73}{2} & \numRF{2.20}{2}\\
\hline
Total & \numRF{49502.44}{2} & \numRF{221.76}{2}   & \numRF{6220.20}{2} & \numRF{53.55}{2}   & \numRF{303.08}{2} & \numRF{9.54}{1}  \\
\hline
Data & \multicolumn{2}{l}{\num{49699}}  & \multicolumn{2}{l}{\num{6199}}  & \multicolumn{2}{l}{\num{286}} \\
\hline\hline     \\
\end{tabular}
\vspace{0.1cm}

\begin{tabular}{l*{3}{r@{$\,\pm\,$}r}}
\hline\hline
 & \multicolumn{2}{c}{$\geq$ 6 j, 2 b} & \multicolumn{2}{c}{$\geq$ 6 j, 3 b} & \multicolumn{2}{c}{$\geq$ 6 j, $\geq$ 4 b}\\
\hline
$t\bar{t}H$ (125) & \numRF{88.96}{2} & \numRF{65.04}{2} & \numRF{57.00}{2} & \numRF{41.64}{2} & \numRF{23.56}{2} & \numRF{17.19}{2}\\
$t\bar{t}+$ light & \numRF{18938.70}{2} & \numRF{704.70}{2} & \numRF{2077.28}{2} & \numRF{87.41}{2} & \numRF{57.88}{2} & \numRF{5.26}{1}\\
$t\bar{t}+c\bar{c}$ & \numRF{3733.31}{2} & \numRF{889.24}{2} & \numRF{888.02}{2} & \numRF{210.88}{2} & \numRF{85.40}{2} & \numRF{20.57}{2}\\
$t\bar{t}+b\bar{b}$ & \numRF{1980.03}{2} & \numRF{311.13}{2} & \numRF{1357.44}{2} & \numRF{187.71}{2} & \numRF{330.81}{2} & \numRF{37.31}{2}\\
$W$+jets & \numRF{454.72}{2} & \numRF{173.62}{2} & \numRF{50.73}{2} & \numRF{19.42}{2} & \numRF{4.43}{2} & \numRF{1.85}{2}\\
$Z$+jets & \numRF{151.74}{2} & \numRF{86.28}{2} & \numRF{15.55}{2} & \numRF{8.89}{1} & \numRF{1.22}{2} & \numRF{0.70}{1}\\
Single top & \numRF{734.34}{2} & \numRF{83.37}{2} & \numRF{110.67}{2} & \numRF{13.91}{2} & \numRF{11.44}{2} & \numRF{1.55}{1}\\
Diboson & \numRF{44.67}{2} & \numRF{19.76}{2} & \numRF{5.58}{2} & \numRF{2.57}{2} & \numRF{0.53}{1} & \numRF{0.23}{1}\\
$t\bar{t}+V$ & \numRF{165.88}{2} & \numRF{51.54}{2} & \numRF{42.28}{2} & \numRF{12.86}{2} & \numRF{8.23}{2} & \numRF{2.51}{2}\\
Lepton misID & \numRF{116.90}{2} & \numRF{40.67}{2} & \numRF{13.78}{2} & \numRF{5.25}{1} & \numRF{1.13}{2} & \numRF{0.51}{1}\\
\hline
Total & \numRF{26409.24}{2} & \numRF{160.08}{2}   & \numRF{4618.33}{2} & \numRF{54.65}{2}   & \numRF{524.62}{2} & \numRF{17.91}{2}  \\
\hline
Data & \multicolumn{2}{l}{\num{26185}}  & \multicolumn{2}{l}{\num{4701}}  & \multicolumn{2}{l}{\num{516}} \\
\hline\hline     \\
\end{tabular}
\vspace{0.1cm}

\end{center}
\vspace{-0.5cm}
\caption{Single lepton channel: post-fit event yields under the 
signal-plus-background hypothesis
for signal, backgrounds and data in each of the analysis regions.
The
quoted uncertainties are the sum in quadrature of statistical and
systematic uncertainties on the yields, computed taking into
account correlations among nuisance parameters and among processes.
}
\label{tab:Postfit_EventsTable_lj}
\end{table*}

\begin{table*}
\begin{center}
\begin{tabular}{l*{3}{r@{$\,\pm\,$}r}}
\hline\hline
 & \multicolumn{2}{c}{2 j, 2 b} & \multicolumn{2}{c}{3 j, 2 b} & \multicolumn{2}{c}{3 j, 3 b}\\
\hline
$t\bar{t}H$ (125) & \numRF{2.38}{2} & \numRF{1.75}{2} & \numRF{8.06}{2} & \numRF{5.91}{2} & \numRF{2.99}{2} & \numRF{2.20}{2}\\
$t\bar{t}+$ light & \numRF{13935.91}{2} & \numRF{155.98}{2} & \numRF{8337.41}{2} & \numRF{172.39}{2} & \numRF{83.58}{2} & \numRF{9.58}{2}\\
$t\bar{t}+c\bar{c}$ & \numRF{396.73}{2} & \numRF{112.47}{2} & \numRF{696.70}{2} & \numRF{164.29}{2} & \numRF{92.47}{2} & \numRF{21.75}{2}\\
$t\bar{t}+b\bar{b}$ & \numRF{187.18}{2} & \numRF{36.17}{2} & \numRF{352.43}{2} & \numRF{49.37}{2} & \numRF{139.29}{2} & \numRF{19.27}{2}\\
$Z$+jets & \numRF{329.99}{2} & \numRF{21.81}{2} & \numRF{195.69}{2} & \numRF{43.41}{2} & \numRF{7.31}{2} & \numRF{2.44}{2}\\
Single top & \numRF{434.89}{2} & \numRF{34.65}{2} & \numRF{257.55}{2} & \numRF{20.65}{2} & \numRF{7.58}{2} & \numRF{1.46}{2}\\
Diboson & \numRF{6.83}{2} & \numRF{2.09}{2} & \numRF{4.51}{2} & \numRF{1.38}{2} & $\leq$ \numRF{0.1}{1} & \numRF{0.1}{1}\\
$t\bar{t}+V$ & \numRF{8.65}{2} & \numRF{2.67}{2} & \numRF{20.86}{2} & \numRF{6.35}{1} & \numRF{1.78}{2} & \numRF{0.55}{1}\\
Lepton misID & \numRF{18.78}{2} & \numRF{9.67}{1} & \numRF{29.92}{2} & \numRF{15.41}{2} & \numRF{0.68}{1} & \numRF{0.35}{1}\\
\hline
Total & \numRF{15321.34}{2} & \numRF{115.89}{2}   & \numRF{9903.13}{2} & \numRF{81.83}{2}   & \numRF{335.74}{2} & \numRF{13.79}{2}  \\
\hline
Data & \multicolumn{2}{l}{\num{15296}}  & \multicolumn{2}{l}{\num{9996}}  & \multicolumn{2}{l}{\hphantom{00}\num{374}} \\
\hline\hline     \\
\end{tabular}

\vspace{0.1cm}

\begin{tabular}{l*{3}{r@{$\,\pm\,$}r}}
\hline\hline
 & \multicolumn{2}{c}{$\geq$ 4 j, 2 b} & \multicolumn{2}{c}{$\geq$ 4 j, 3 b} & \multicolumn{2}{c}{$\geq$ 4 j, $\geq$ 4 b}\\
\hline
$t\bar{t}H$ (125) & \numRF{21.71}{2} & \numRF{15.82}{2} & \numRF{11.34}{2} & \numRF{8.27}{1} & \numRF{3.13}{2} & \numRF{2.29}{2}\\
$t\bar{t}+$ light & \numRF{4494.93}{2} & \numRF{145.34}{2} & \numRF{104.85}{2} & \numRF{12.35}{2} & \numRF{1.38}{2} & \numRF{0.30}{1}\\
$t\bar{t}+c\bar{c}$ & \numRF{735.45}{2} & \numRF{167.91}{2} & \numRF{139.60}{2} & \numRF{30.28}{2} & \numRF{4.80}{2} & \numRF{1.05}{2}\\
$t\bar{t}+b\bar{b}$ & \numRF{371.29}{2} & \numRF{58.60}{2} & \numRF{230.13}{2} & \numRF{31.19}{2} & \numRF{30.24}{2} & \numRF{4.44}{1}\\
$Z$+jets & \numRF{102.61}{2} & \numRF{33.22}{2} & \numRF{9.47}{2} & \numRF{3.10}{2} & \numRF{0.43}{1} & \numRF{0.15}{1}\\
Single top & \numRF{136.93}{2} & \numRF{23.13}{2} & \numRF{10.77}{2} & \numRF{1.99}{1} & \numRF{0.61}{1} & \numRF{0.10}{1}\\
Diboson & \numRF{4.23}{2} & \numRF{1.27}{2} & \numRF{0.33}{1} & \numRF{0.11}{1} & $\leq$ \numRF{0.1}{1} & \numRF{0.1}{1}\\
$t\bar{t}+V$ & \numRF{42.68}{2} & \numRF{12.85}{2} & \numRF{6.97}{2} & \numRF{2.11}{2} & \numRF{0.86}{1} & \numRF{0.27}{1}\\
Lepton misID & \numRF{33.98}{2} & \numRF{17.50}{2} & \numRF{3.53}{2} & \numRF{1.82}{2} & \numRF{0.24}{1} & \numRF{0.13}{1}\\
\hline
Total & \numRF{5943.81}{2} & \numRF{65.18}{2}   & \numRF{516.99}{2} & \numRF{18.10}{2}   & \numRF{41.71}{2} & \numRF{3.85}{1}  \\
\hline
Data & \multicolumn{2}{l}{\num{6006}}  & \multicolumn{2}{l}{\num{561}}  & \multicolumn{2}{l}{\hphantom{..00}\num{46}} \\
\hline\hline     \\
\end{tabular}

\vspace{0.1cm}

\end{center}
\vspace{-0.5cm}
\caption{Dilepton channel: 
post-fit event yields under the signal-plus-background hypothesis
for signal, backgrounds and data in each of the analysis regions. The
quoted uncertainties are the sum in quadrature of statistical and
systematic uncertainties on the yields, computed taking into
account correlations among nuisance parameters and among processes.
}
\label{tab:Postfit_EventsTable_dil}
\end{table*}

\section{Post-fit input variables}
\label{sec:postfitinput}

Figures~\ref{fig:postinput_lj_0} --\ref{fig:postinput_lj_3} and 
~\ref{fig:postinput_dil_1} --\ref{fig:postinput_dil_3} 
show a comparison of data and prediction for the top four input variables in each region with a neural network in the single-lepton channel and dilepton channel, respectively. 
All of the plots are made using post-fit predictions.

\begin{figure*}[ht!]
\begin{center}
\subfigure[]{\includegraphics[width=0.24\textwidth]{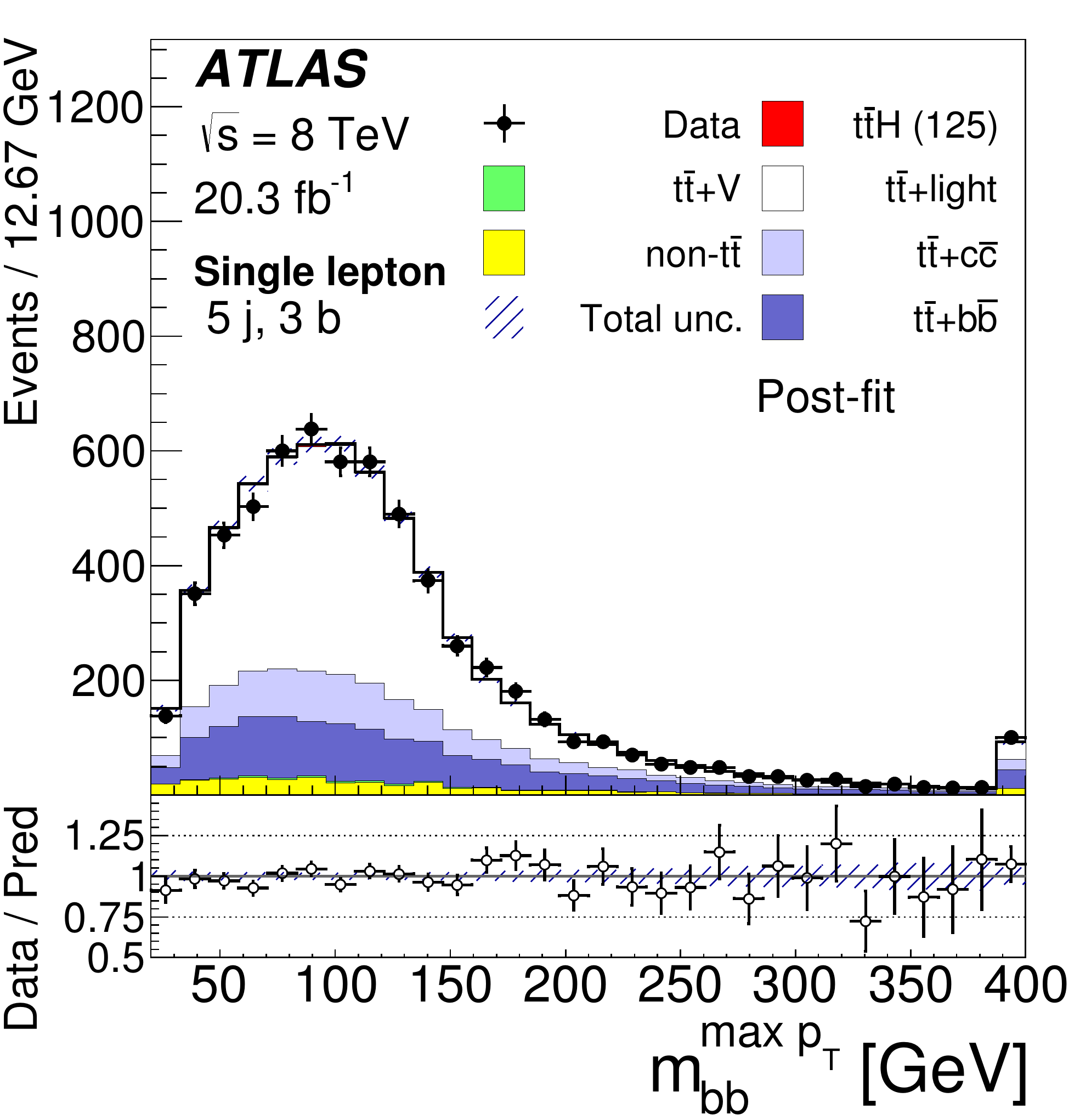}}\label{fig:postinput_lj_0a}
\subfigure[]{\includegraphics[width=0.24\textwidth]{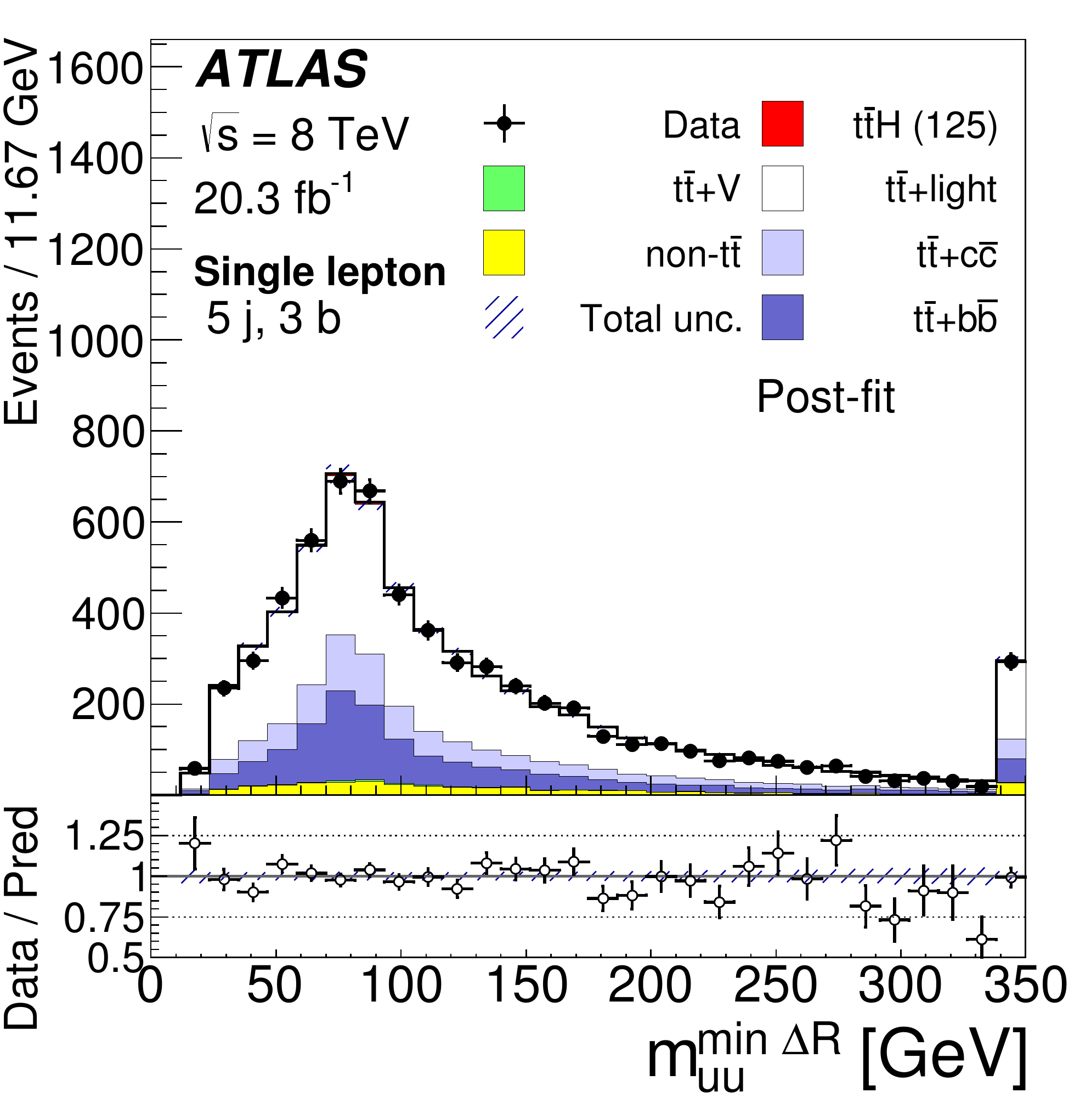}}\label{fig:postinput_lj_0b} 
\subfigure[]{\includegraphics[width=0.24\textwidth]{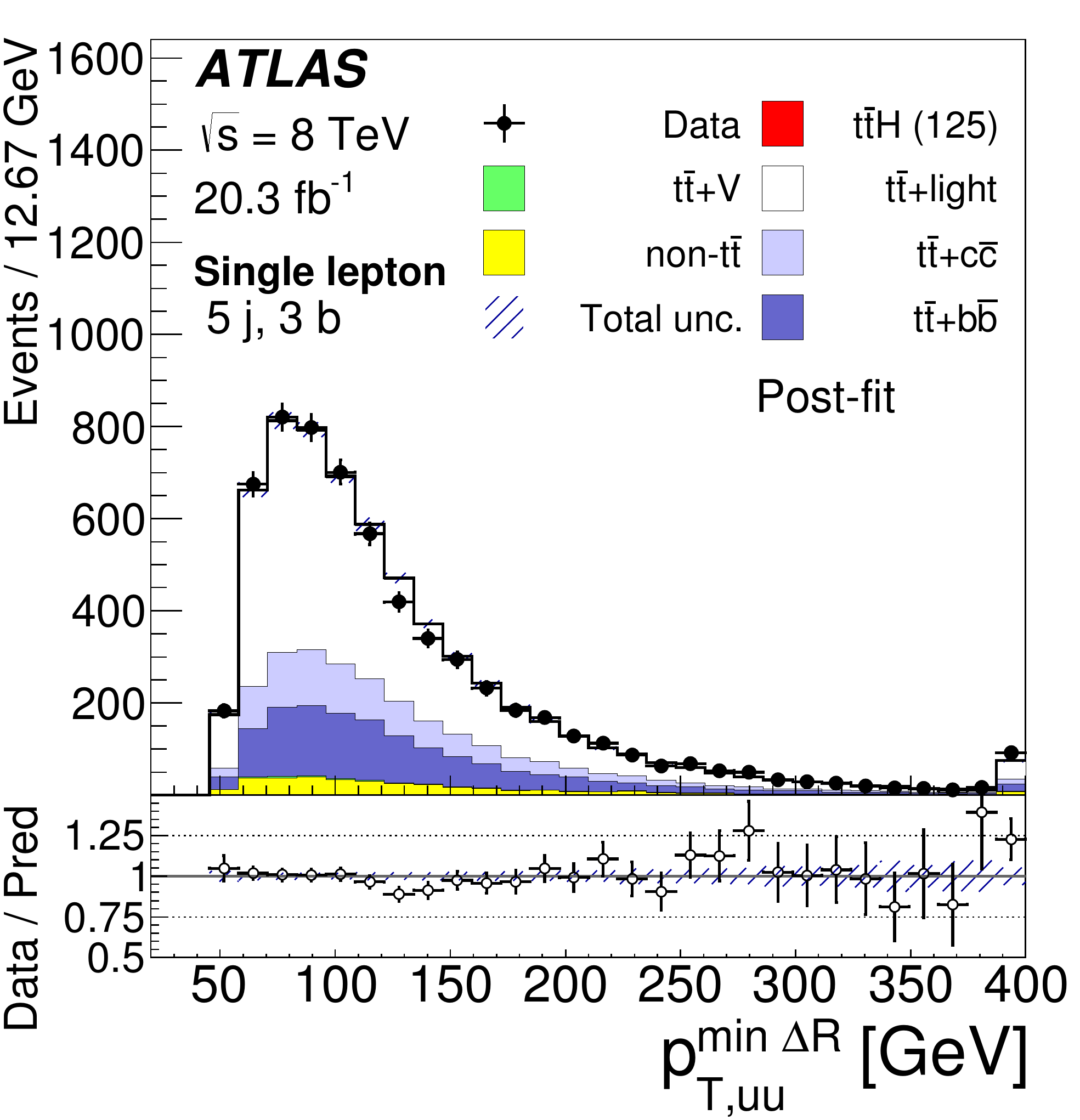}}\label{fig:postinput_lj_0c}
\subfigure[]{\includegraphics[width=0.24\textwidth]{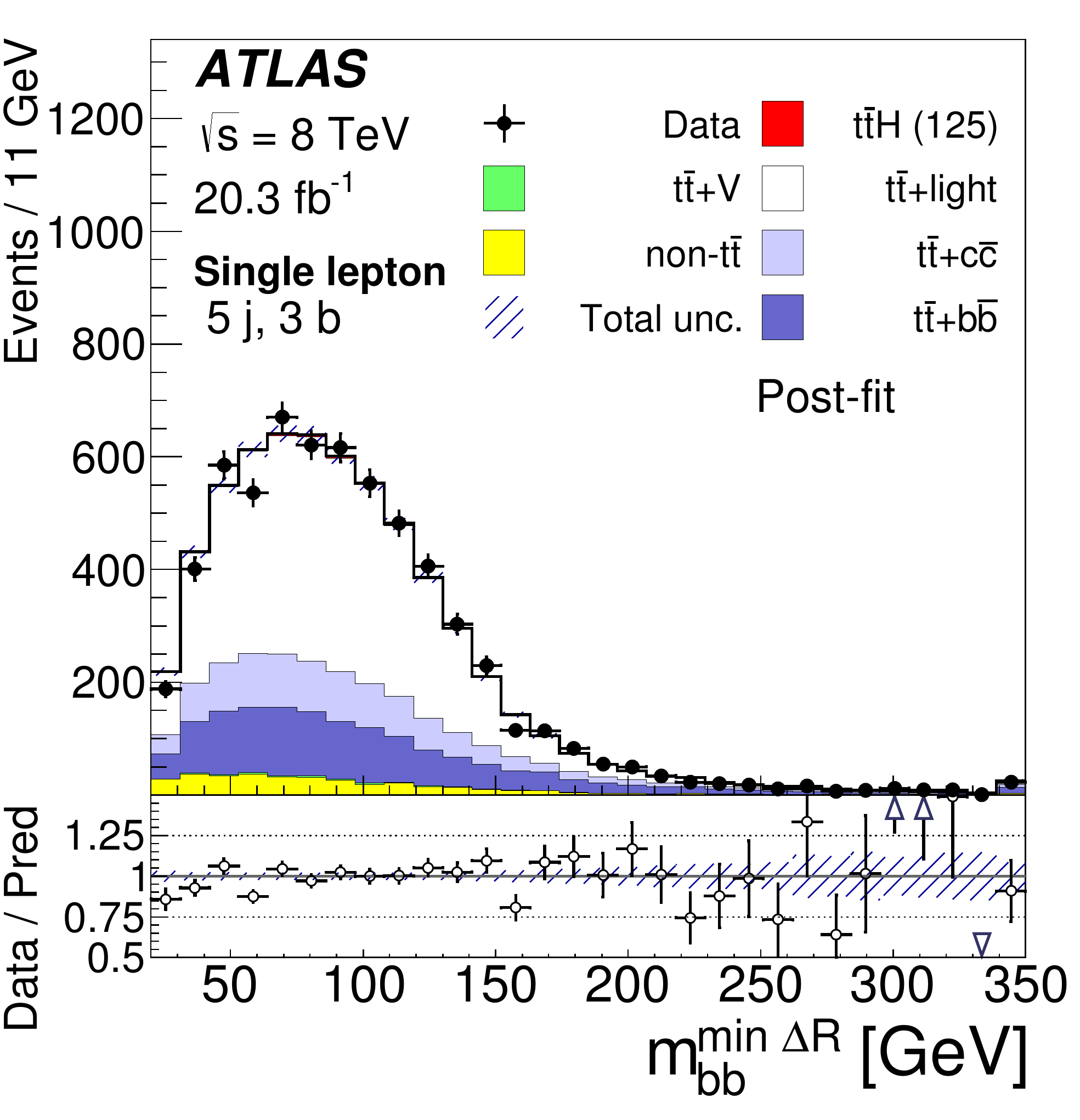}}\label{fig:postinput_lj_0d}
\caption{Single-lepton channel: post-fit comparison of data and prediction for the four top-ranked input variables in the 
\fivethree\ region. The plots include (a) \mbbmaxpt, (b) \whadmass, (c)  \whadpt and (d) \mbbmindr.
The first and last bins in all figures contain the underflow and 
overflow, respectively. The bottom panel displays the ratio of 
data to the total prediction. An arrow indicates that the point is off-scale. The hashed area represents the uncertainty on the background. }
\label{fig:postinput_lj_0} 
\end{center}
\end{figure*}

\clearpage

\begin{figure*}[ht!]
\begin{center}
\subfigure[]{\includegraphics[width=0.24\textwidth]{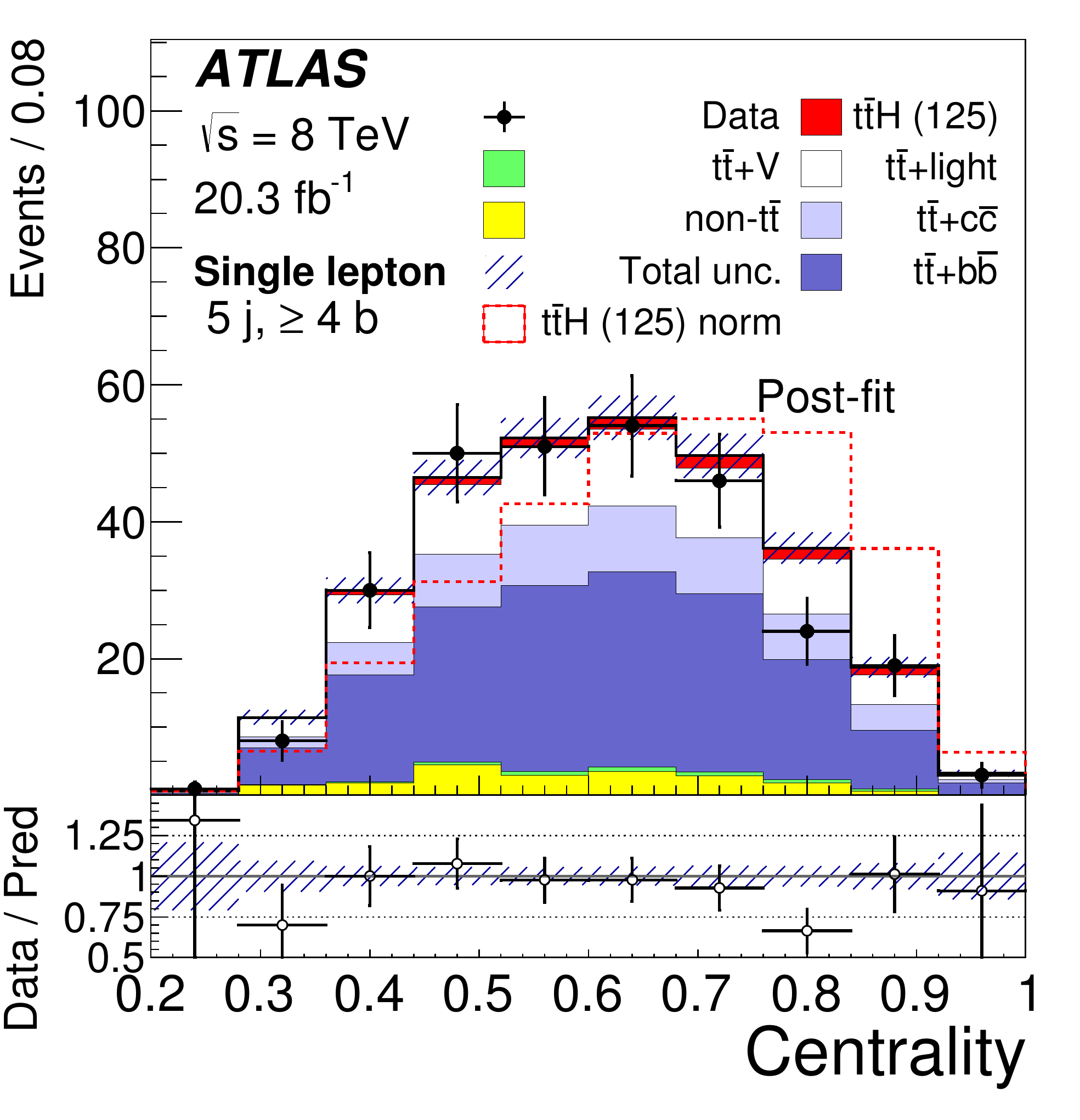}}\label{fig:postinput_lj_1a}
\subfigure[]{\includegraphics[width=0.24\textwidth]{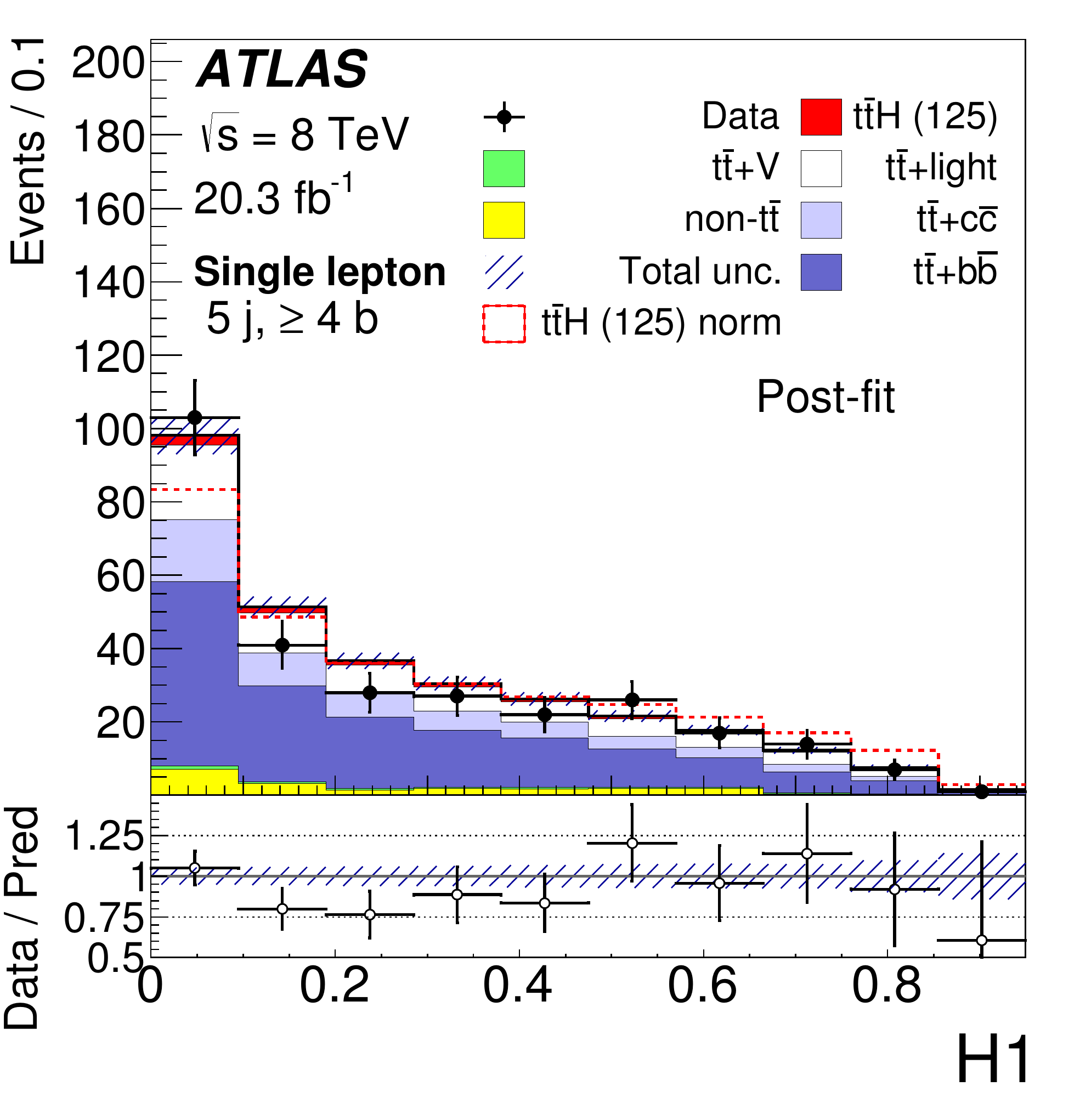}}\label{fig:postinput_lj_1b} 
\subfigure[]{\includegraphics[width=0.24\textwidth]{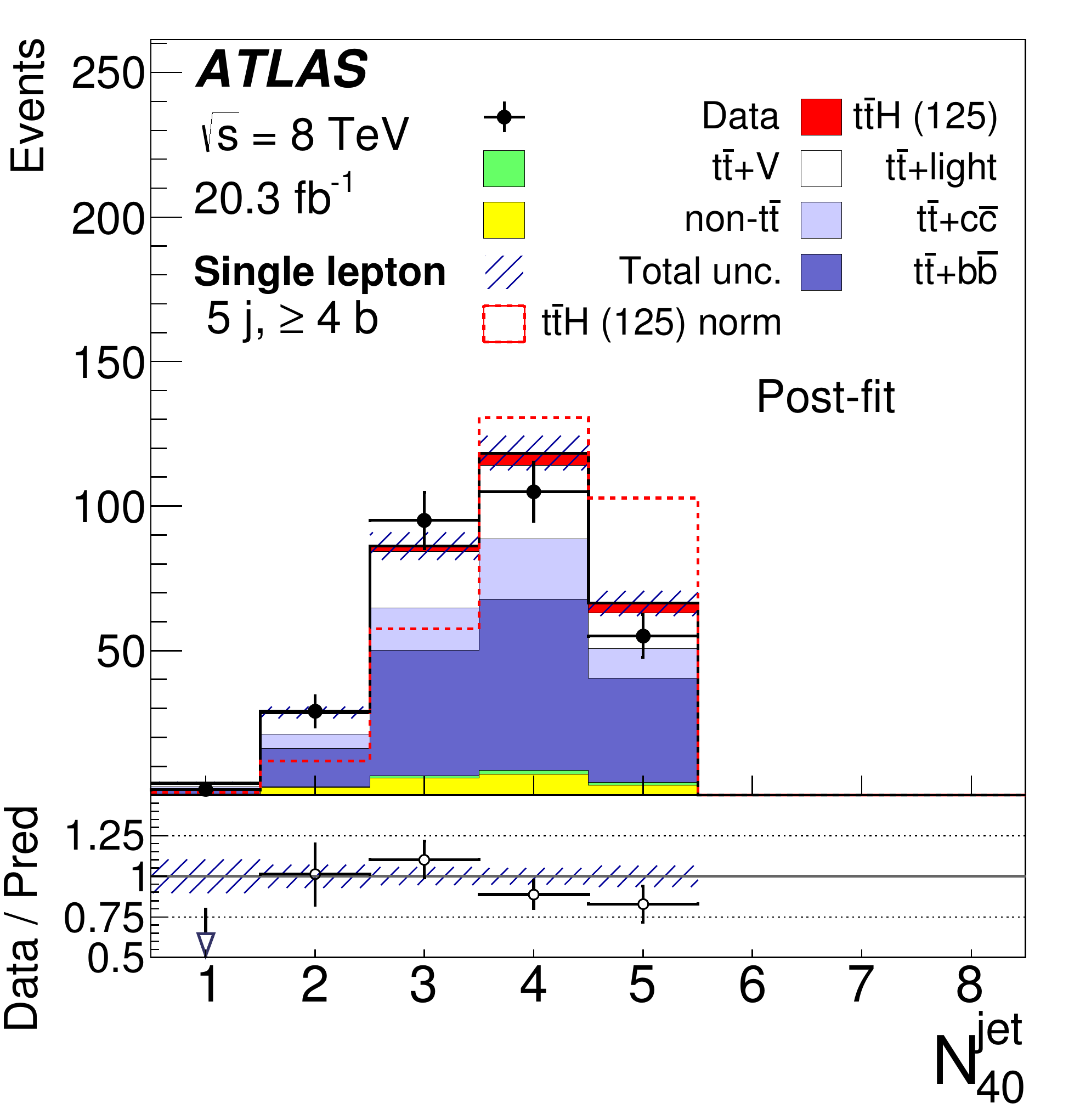}}\label{fig:postinput_lj_1c}
\subfigure[]{\includegraphics[width=0.24\textwidth]{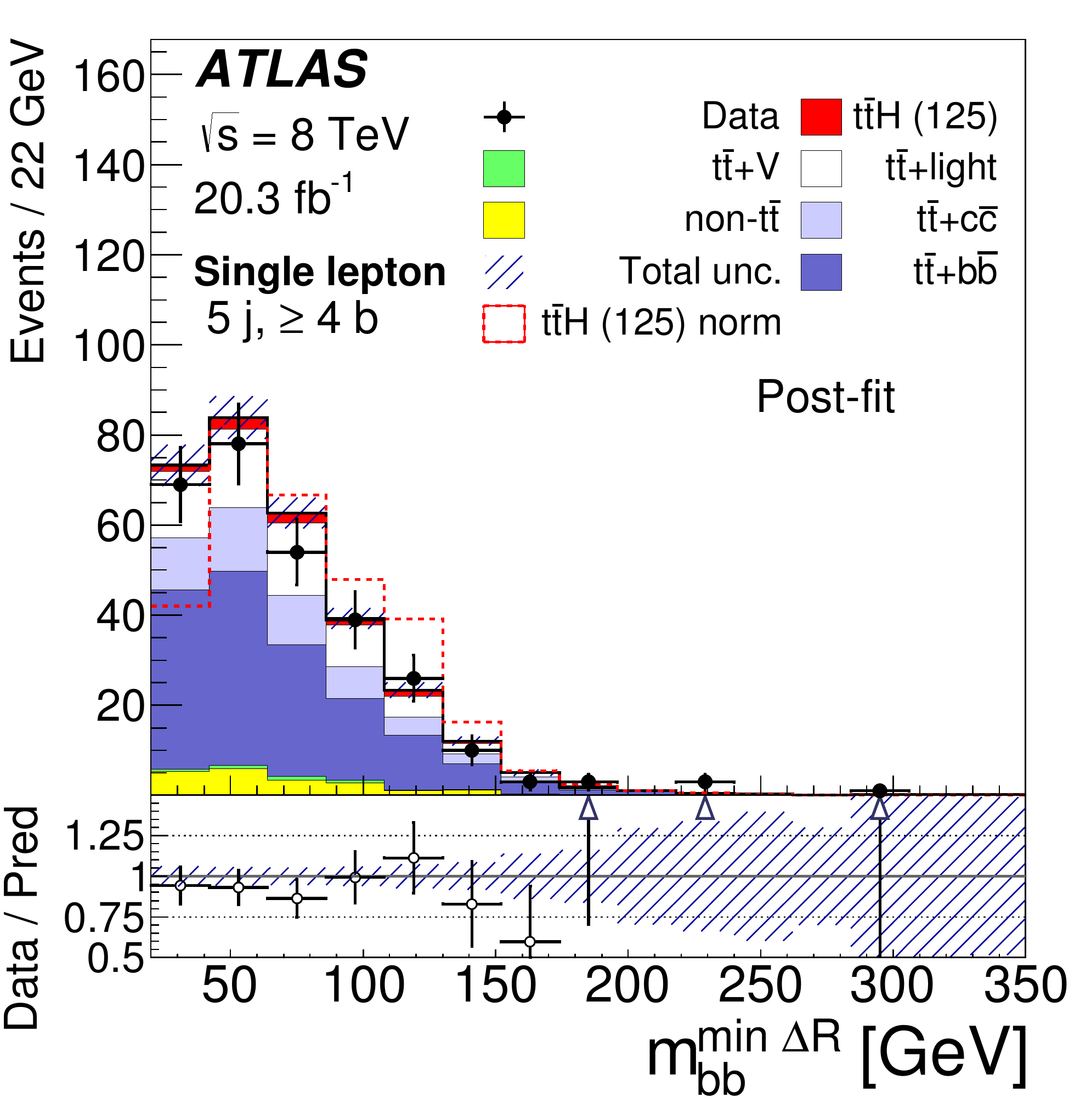}}\label{fig:postinput_lj_1d}
\caption{Single-lepton channel: post-fit comparison of data and prediction for the four top-ranked input variables in the 
\fivefour\ region. The plots 
include (a) \cent, (b) $H1$, (c)  \numjetforty and (d) \mbbmindr.
The first and last bins in all figures contain the underflow and
overflow, respectively. The bottom panel displays the ratio of 
data to the total prediction. An arrow indicates that the point is off-scale. The hashed area represents the uncertainty on the background.
The dashed line shows \tth\ signal
distribution normalised to background yield. The \tth\ signal yield (solid) 
is normalised to the fitted $\mu$.}
\label{fig:postinput_lj_1} 
\end{center}
\end{figure*}

\begin{figure*}[ht!]
\begin{center}
\subfigure[]{\includegraphics[width=0.24\textwidth]{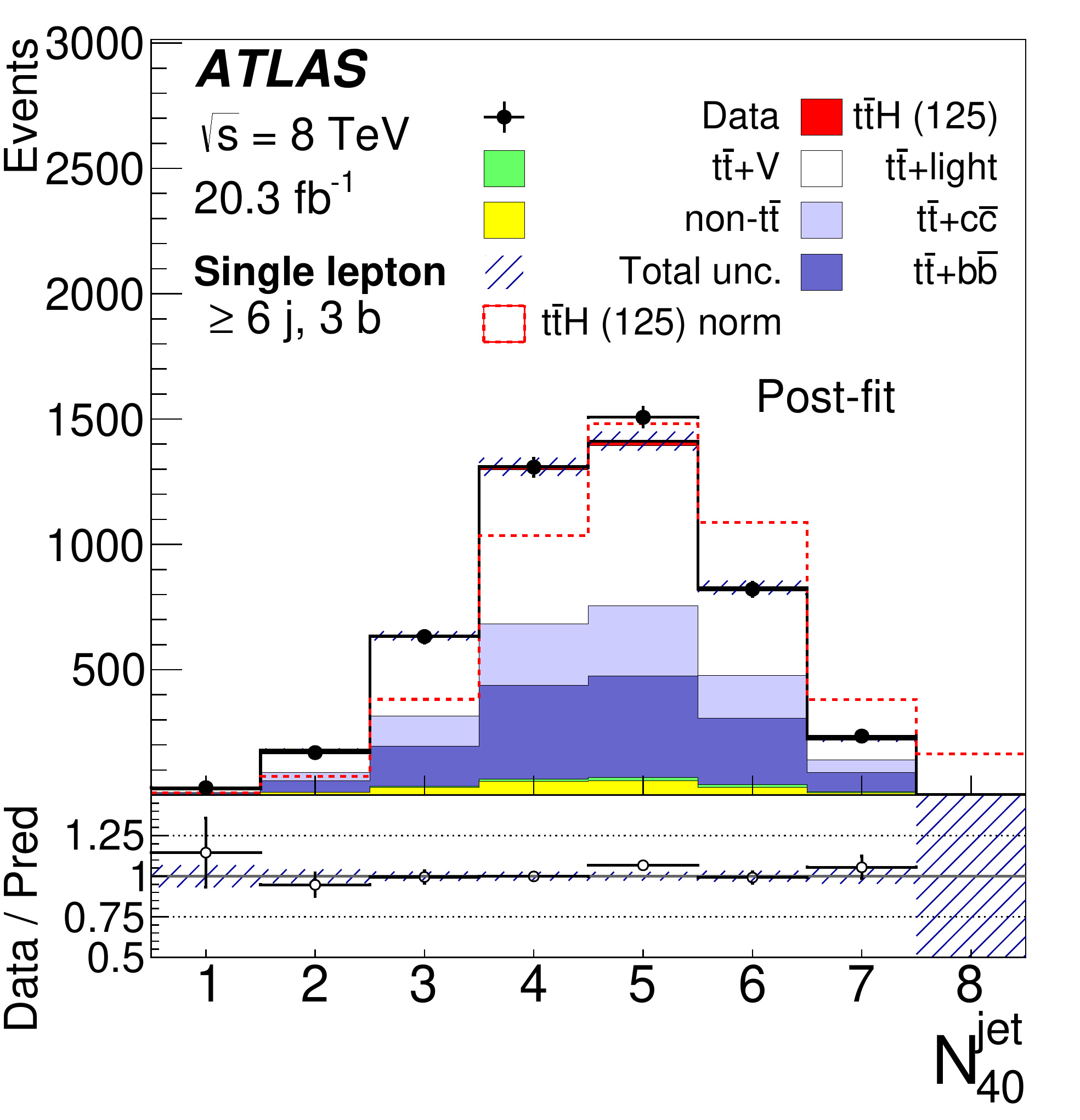}}\label{fig:postinput_lj_2a}
\subfigure[]{\includegraphics[width=0.24\textwidth]{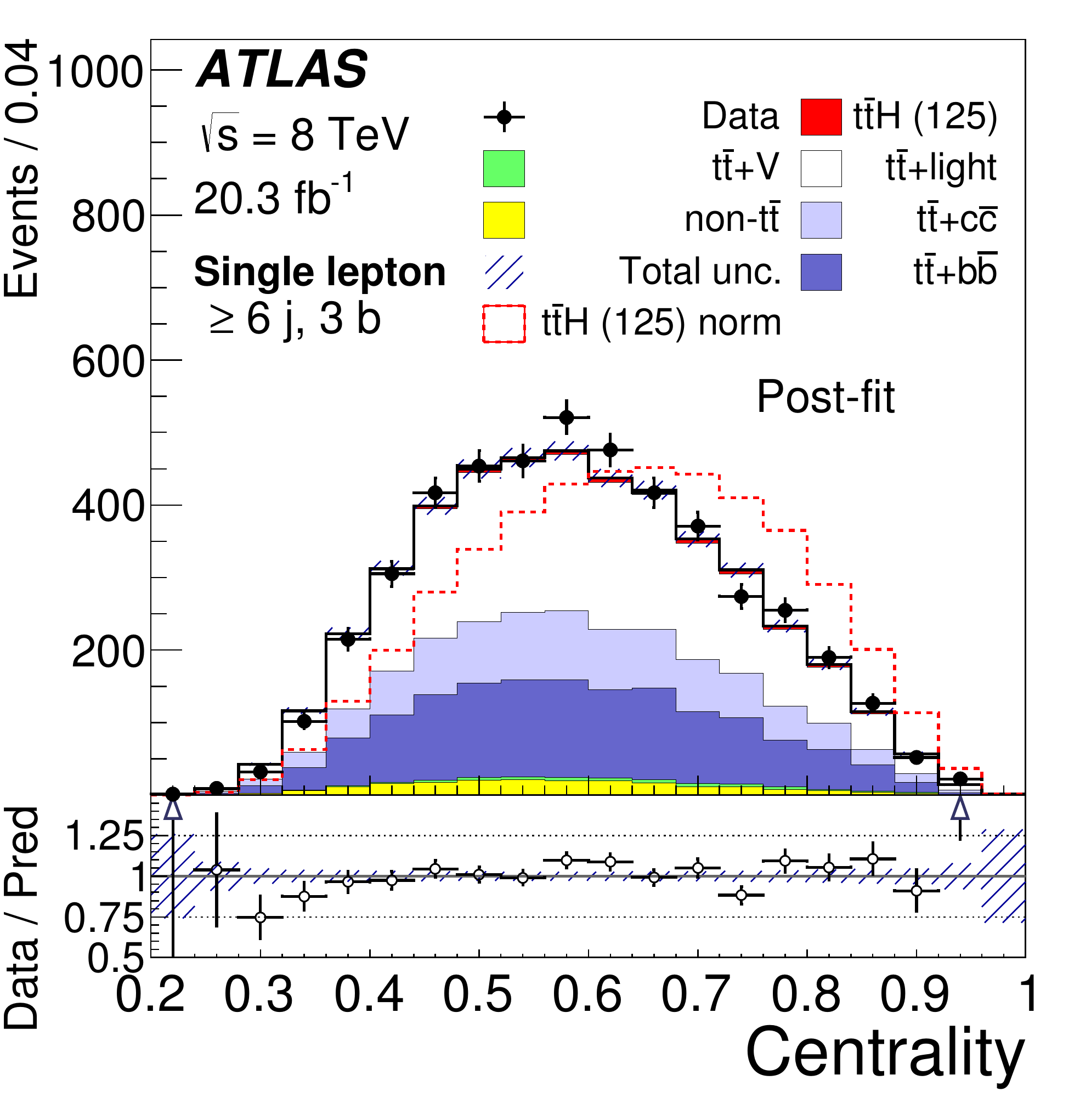}}\label{fig:postinput_lj_2b} 
\subfigure[]{\includegraphics[width=0.24\textwidth]{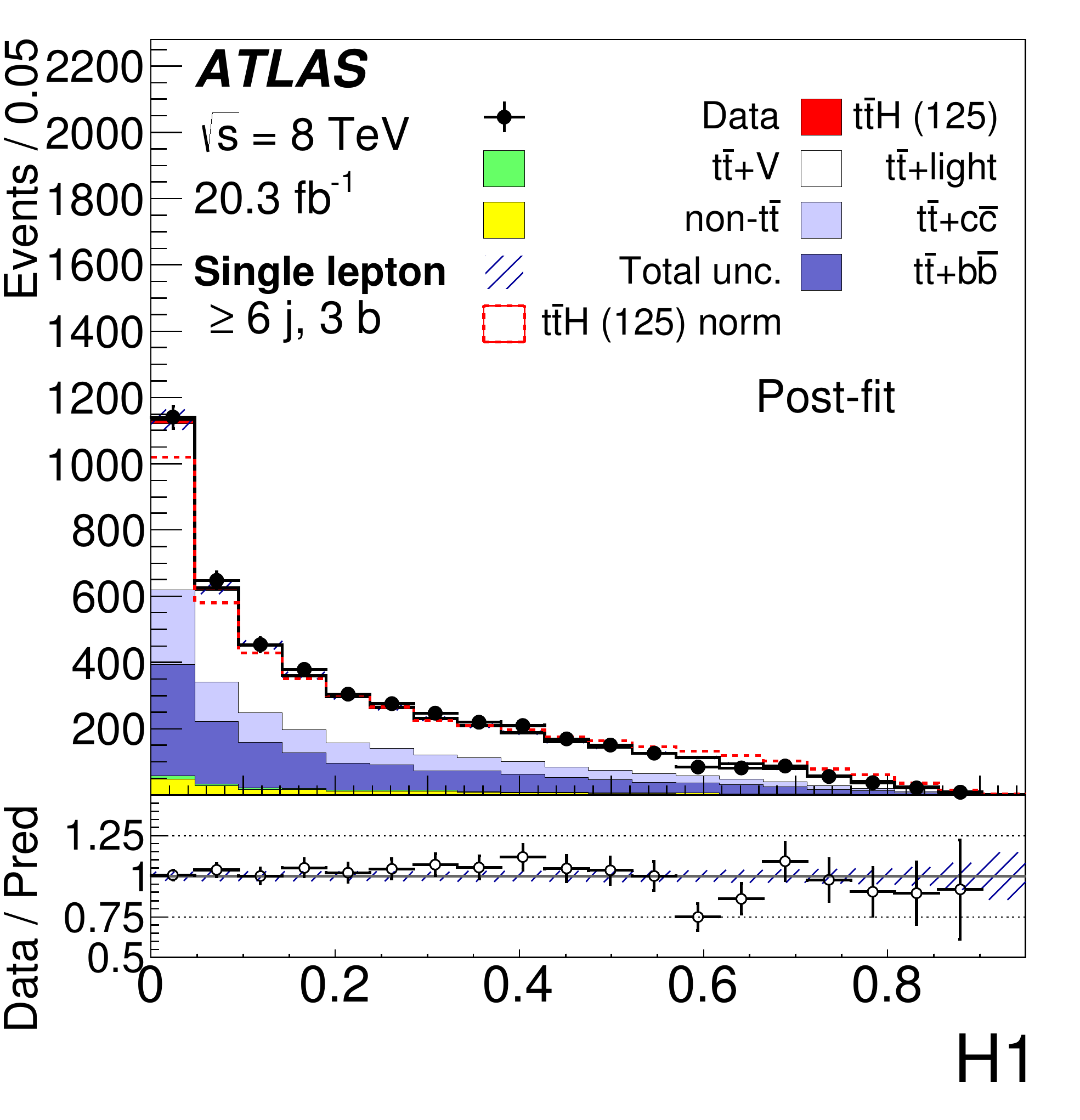}}\label{fig:postinput_lj_2c} 
\subfigure[]{\includegraphics[width=0.24\textwidth]{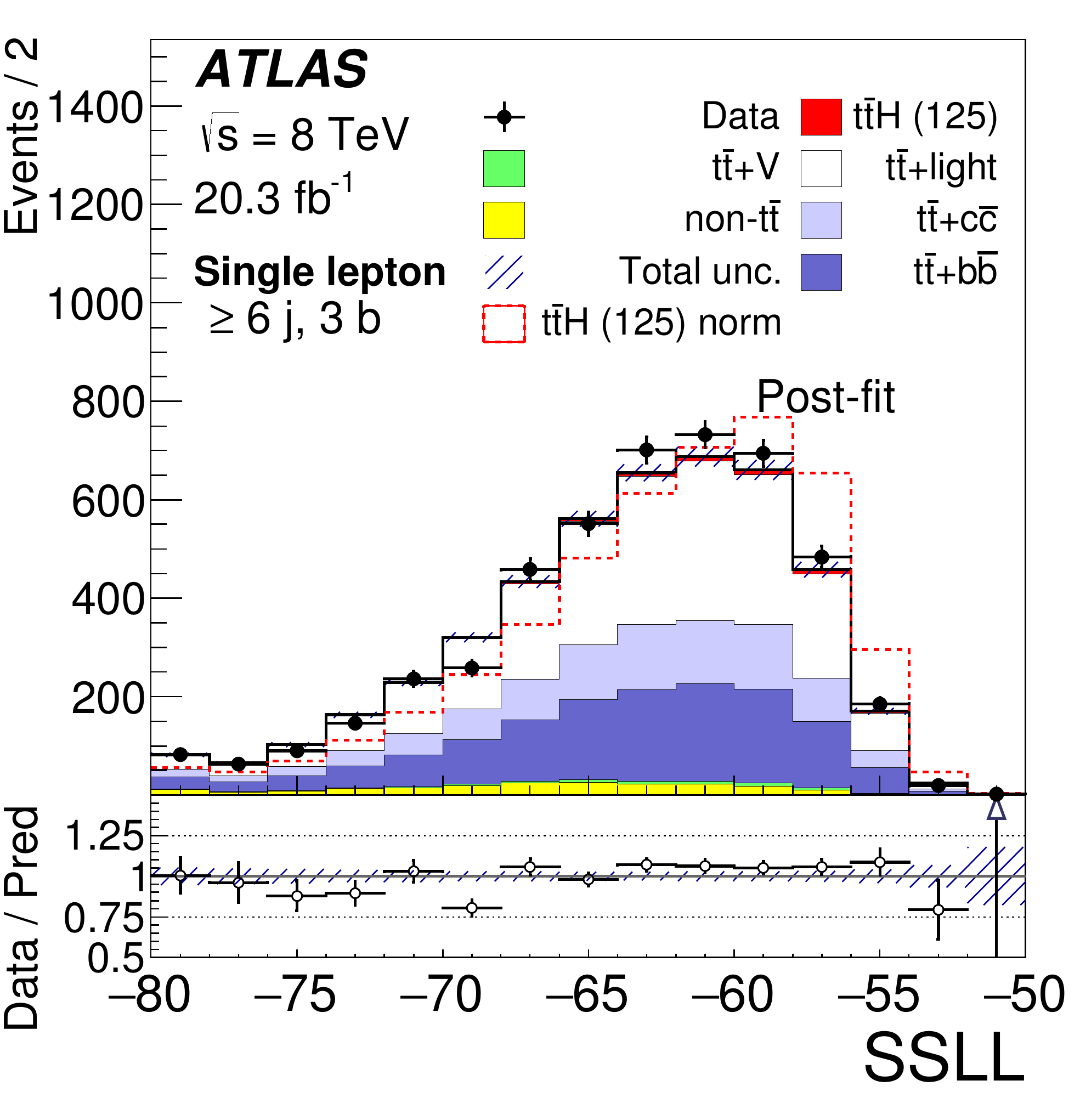}}\label{fig:postinput_lj_2d}
\caption{Single-lepton channel: post-fit comparison of data and prediction for the four top-ranked input variables in 
\sixthree\ region. The plots include (a) \numjetforty, (b) \cent, (c) $H1$, 
and (d) SSLL.
The first and last bins in all figures contain the underflow and
overflow, respectively. The bottom panel displays the ratio of 
data to the total prediction. An arrow indicates that the point is off-scale. The hashed area represents the uncertainty on the background.
The dashed line shows \tth\ signal
distribution normalised to background yield. The \tth\ signal yield (solid) 
is normalised to the fitted $\mu$.}
\label{fig:postinput_lj_2} 
\end{center}
\end{figure*}

\begin{figure*}[ht!]
\begin{center}
\subfigure[]{\includegraphics[width=0.24\textwidth]{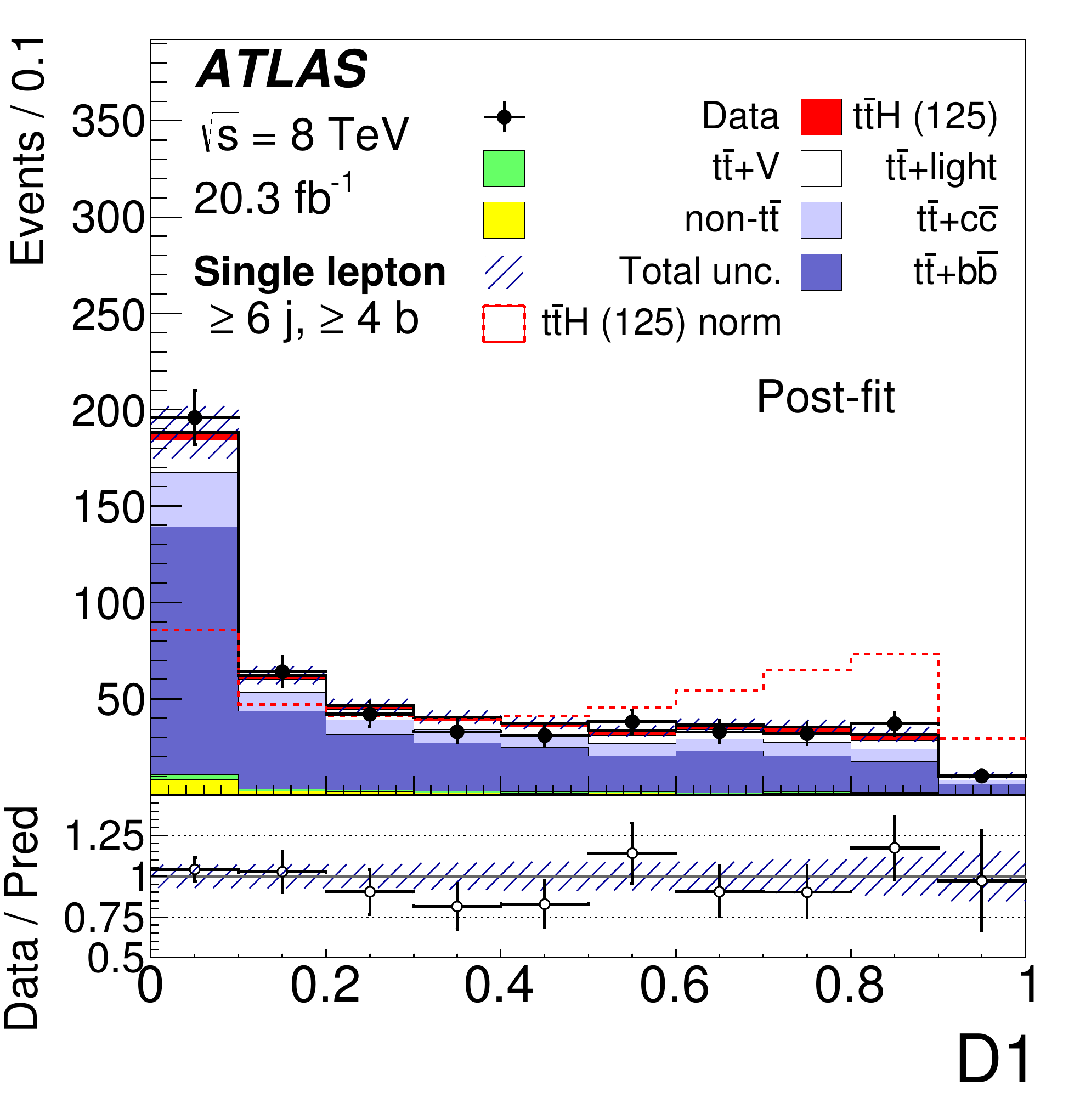}}\label{fig:postinput_lj_3a}
\subfigure[]{\includegraphics[width=0.24\textwidth]{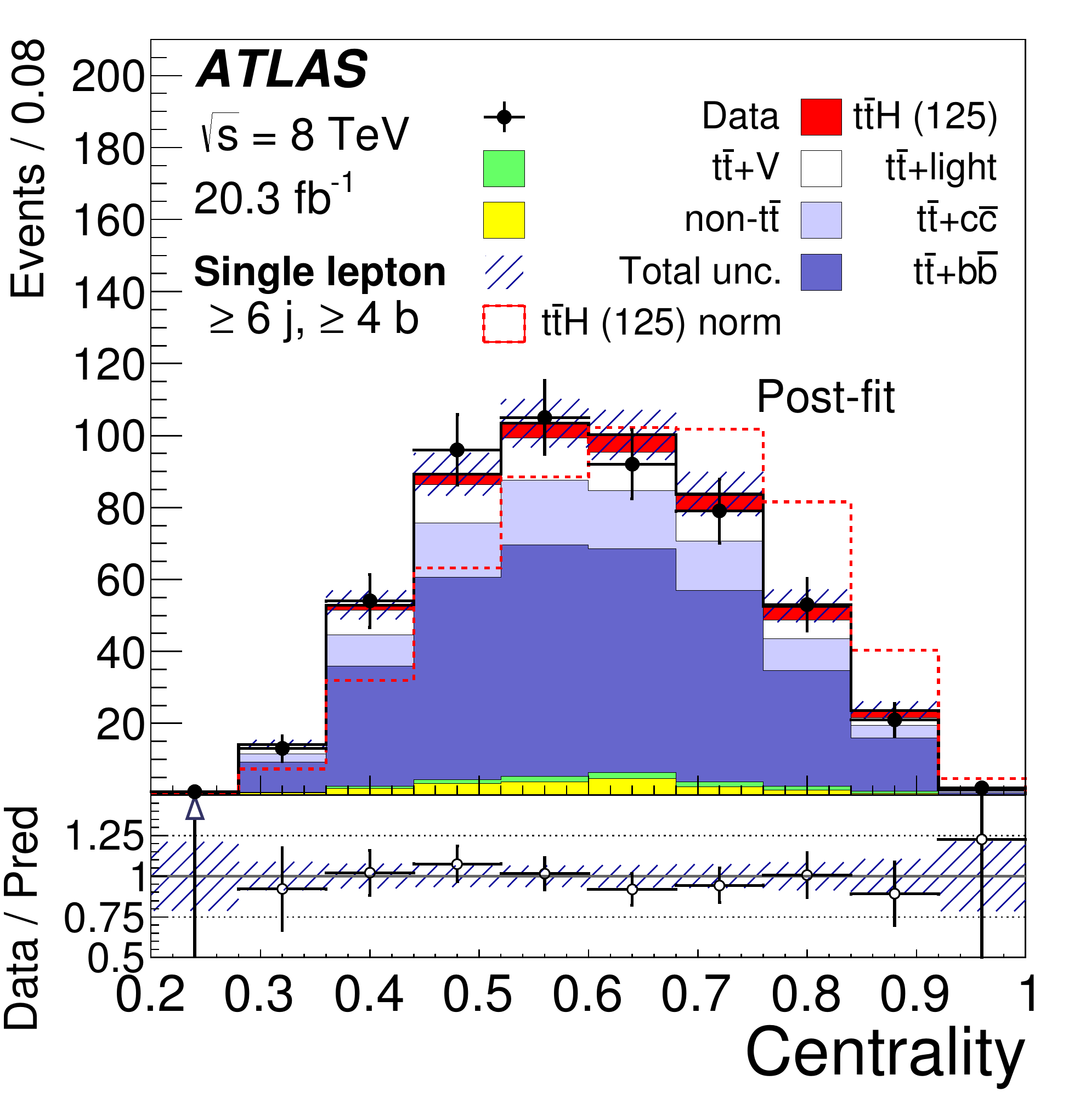}}\label{fig:postinput_lj_3b} 
\subfigure[]{\includegraphics[width=0.24\textwidth]{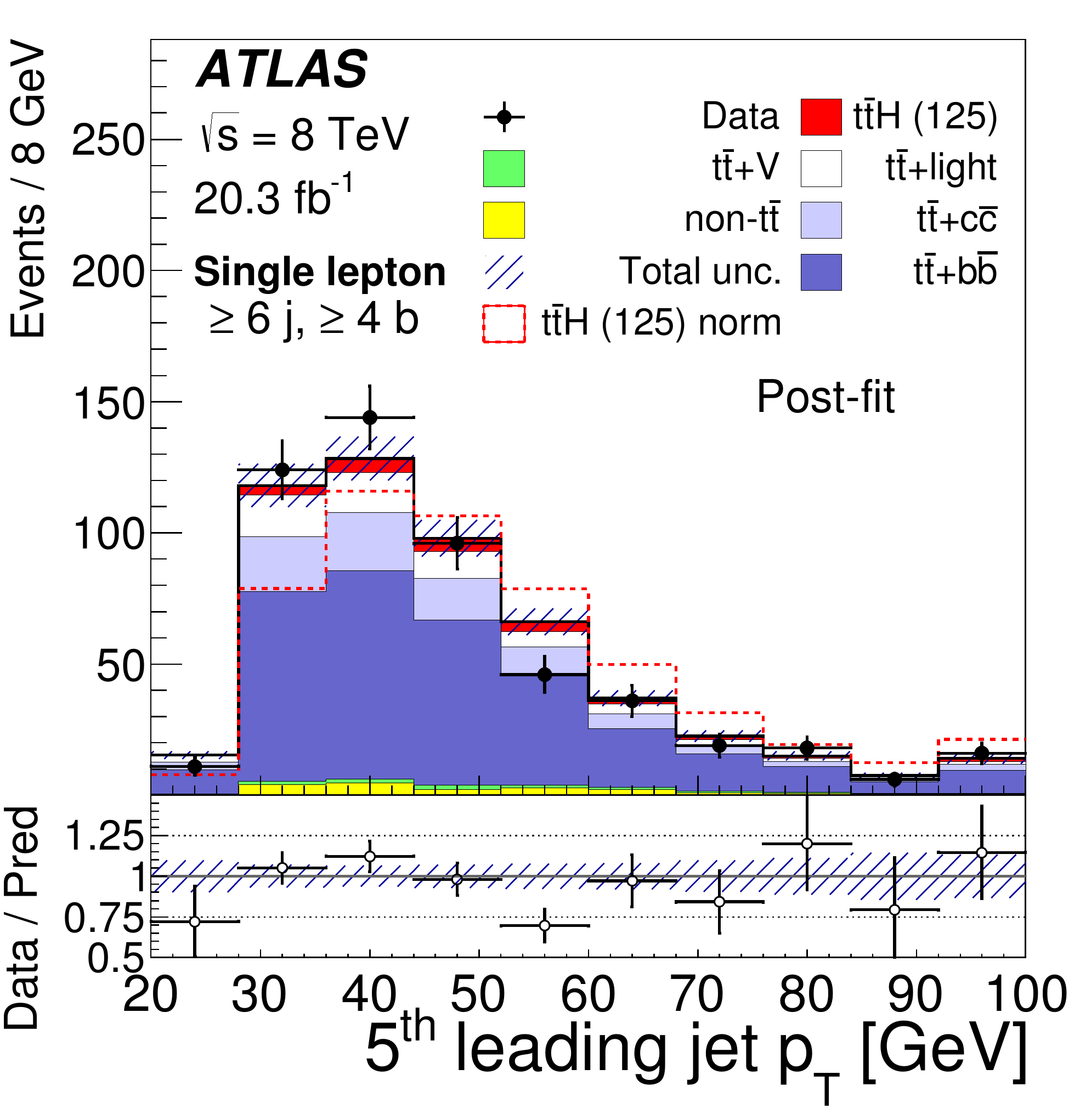}}\label{fig:postinput_lj_3c}
\subfigure[]{\includegraphics[width=0.24\textwidth]{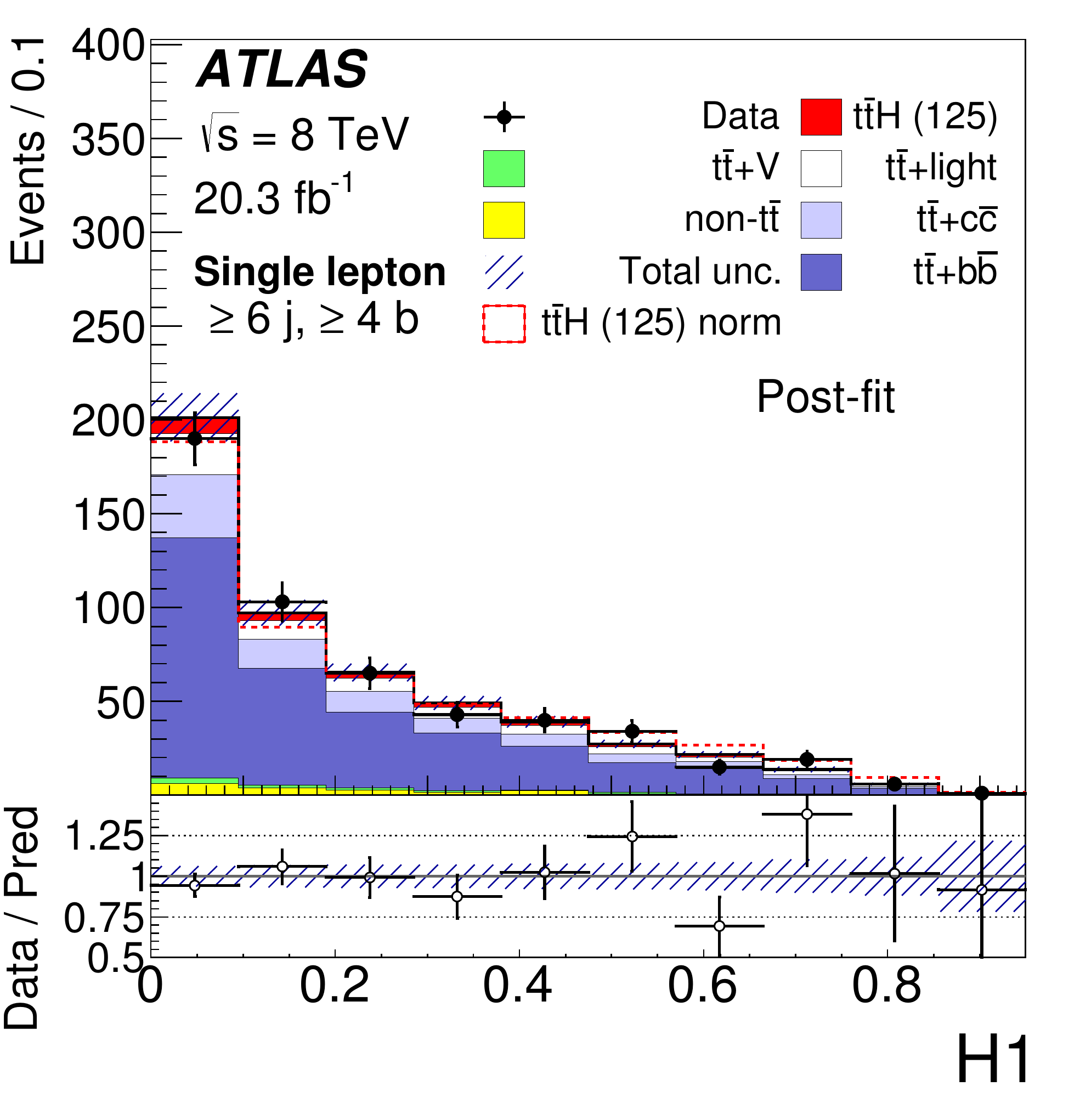}}\label{fig:postinput_lj_3d} 
\caption{Single-lepton channel: post-fit comparison of data and prediction for the four top-ranked input variables in 
\sixfour\ region. The plots include (a) $D1$, (b) \cent, (c) \ptjetfive, and (d) $H1$.
The first and last bins in all figures contain the underflow and
overflow, respectively. The bottom panel displays the ratio of 
data to the total prediction. An arrow indicates that the point is off-scale. The hashed area represents the uncertainty on the background.
The dashed line shows \tth\ signal
distribution normalised to background yield. The \tth\ signal yield (solid) 
is normalised to the fitted $\mu$.}
\label{fig:postinput_lj_3} 
\end{center}
\end{figure*}

\begin{figure*}[ht!]
\begin{center}
\subfigure[]{\includegraphics[width=0.24\textwidth]{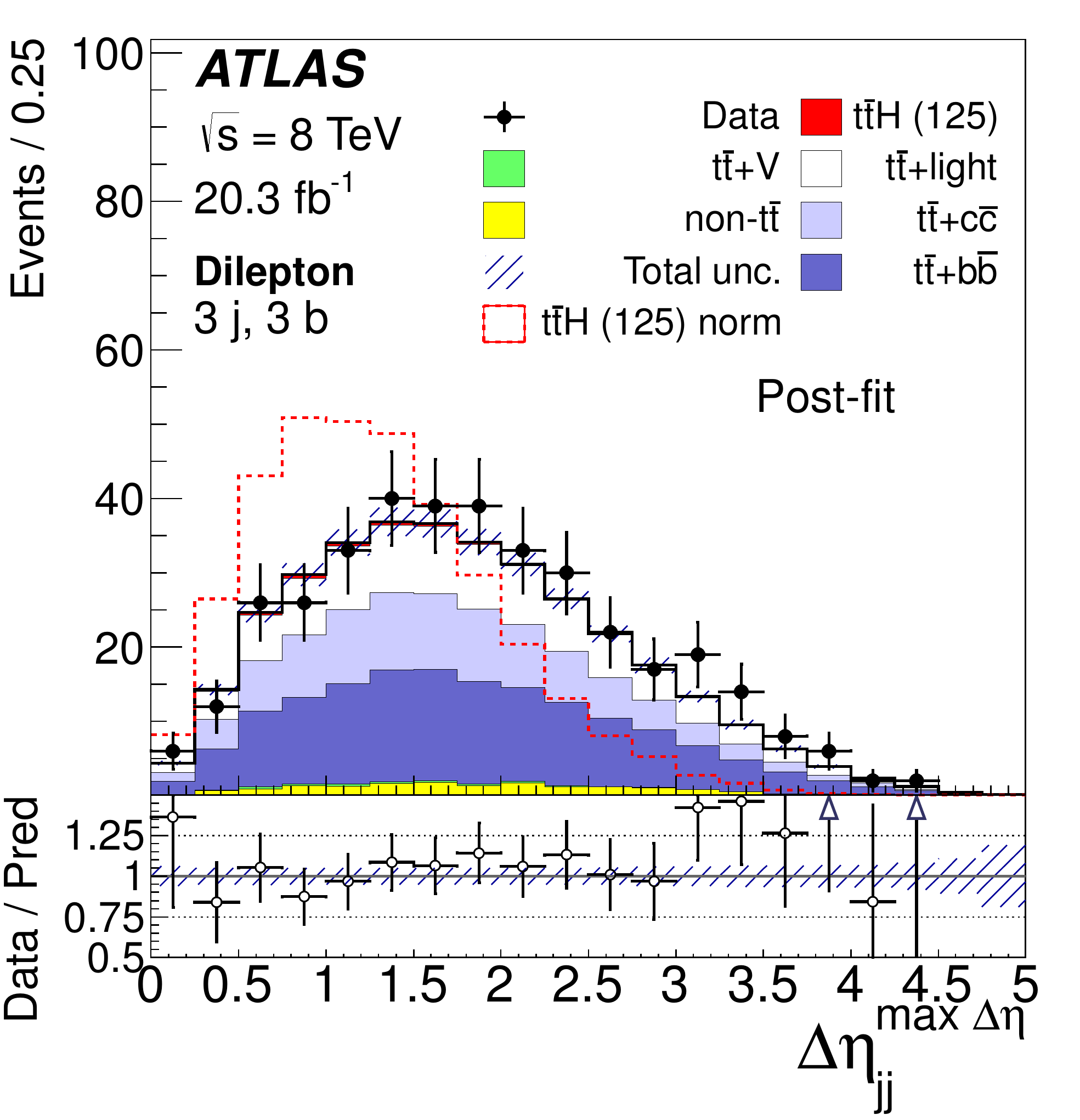}}\label{fig:postinput_dil_1a}
\subfigure[]{\includegraphics[width=0.24\textwidth]{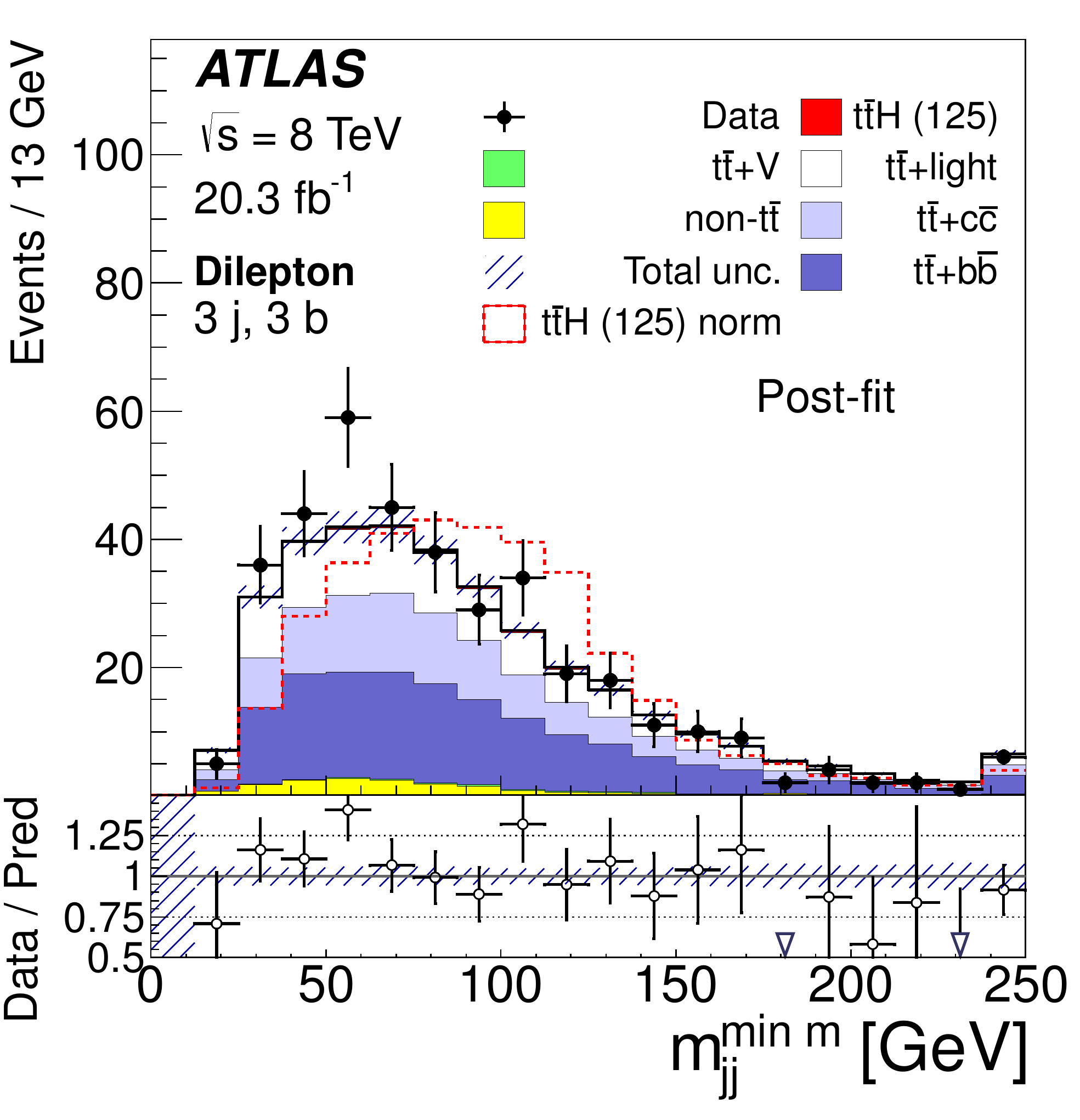}}\label{fig:postinput_dil_1b} 
\subfigure[]{\includegraphics[width=0.24\textwidth]{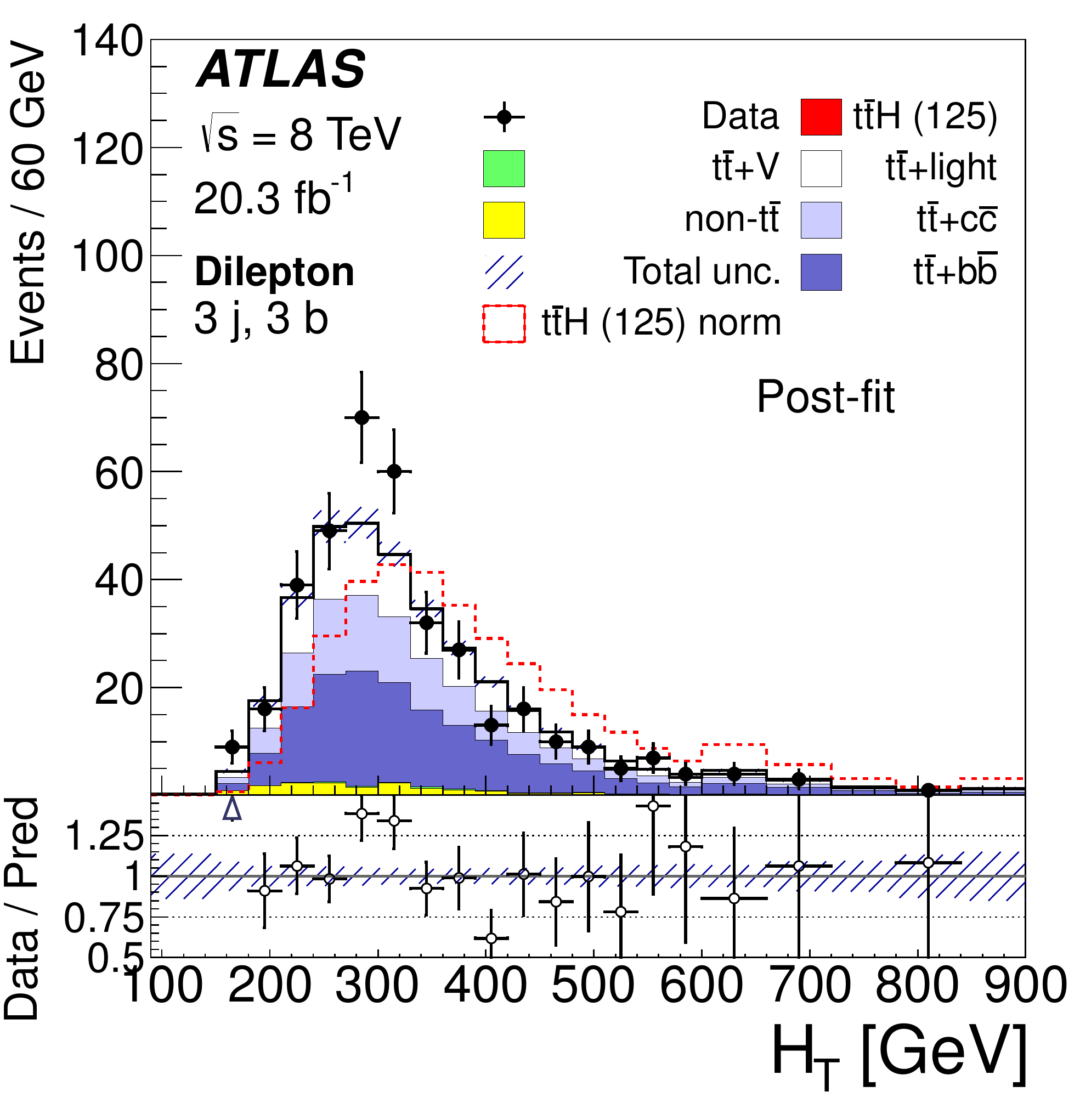}}\label{fig:postinput_dil_1c}
\subfigure[]{\includegraphics[width=0.24\textwidth]{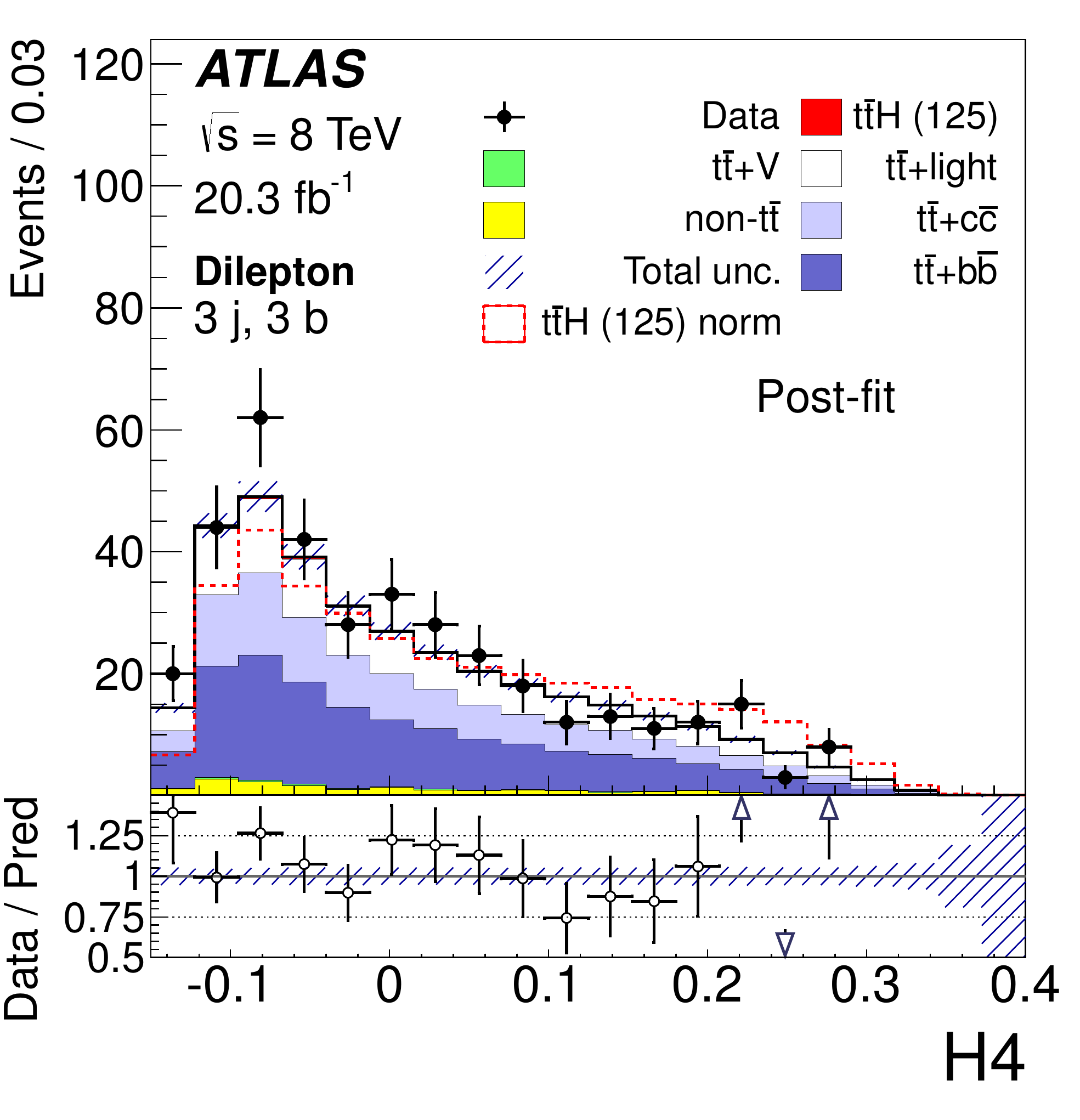}}\label{fig:postinput_dil_1d}
\caption{Dilepton channel: post-fit comparison of data and prediction for the four top-ranked input variables in the 
\threethree\ region. The plots include (a) \maxdeta, (b) \mindijetmass, (c) \htlep, and (d) $H4$.
The first and last bins in all figures contain the underflow and
overflow, respectively. The bottom panel displays the ratio of 
data to the total prediction. An arrow indicates that the point is off-scale. The hashed area represents the uncertainty on the background.
The dashed line shows \tth\ signal
distribution normalised to background yield. The \tth\ signal yield (solid) 
is normalised to the fitted $\mu$.}
\label{fig:postinput_dil_1} 
\end{center}
\end{figure*}

\begin{figure*}[ht!]
\begin{center}
\subfigure[]{\includegraphics[width=0.24\textwidth]{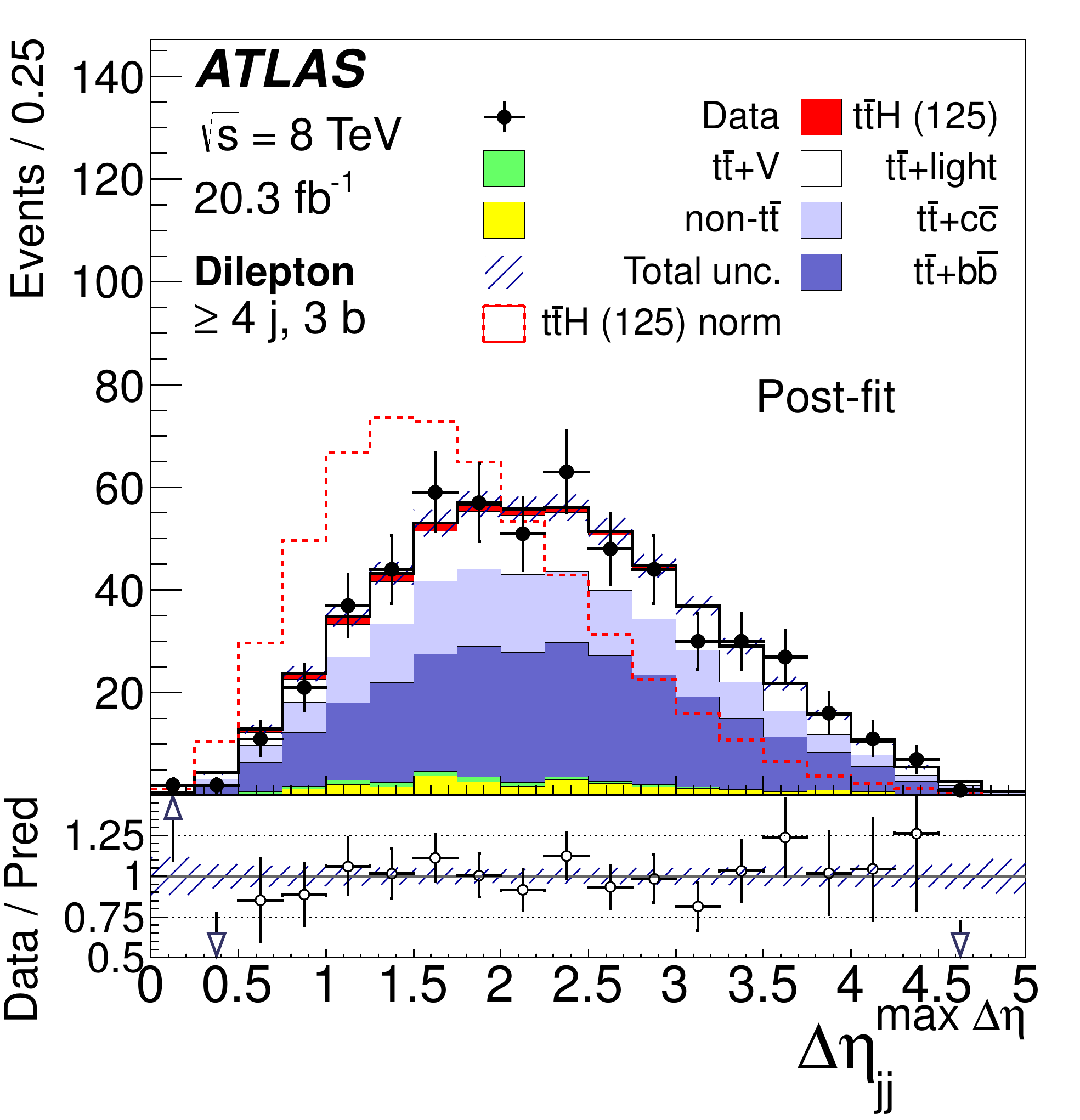}}\label{fig:postinput_dil_2a}
\subfigure[]{\includegraphics[width=0.24\textwidth]{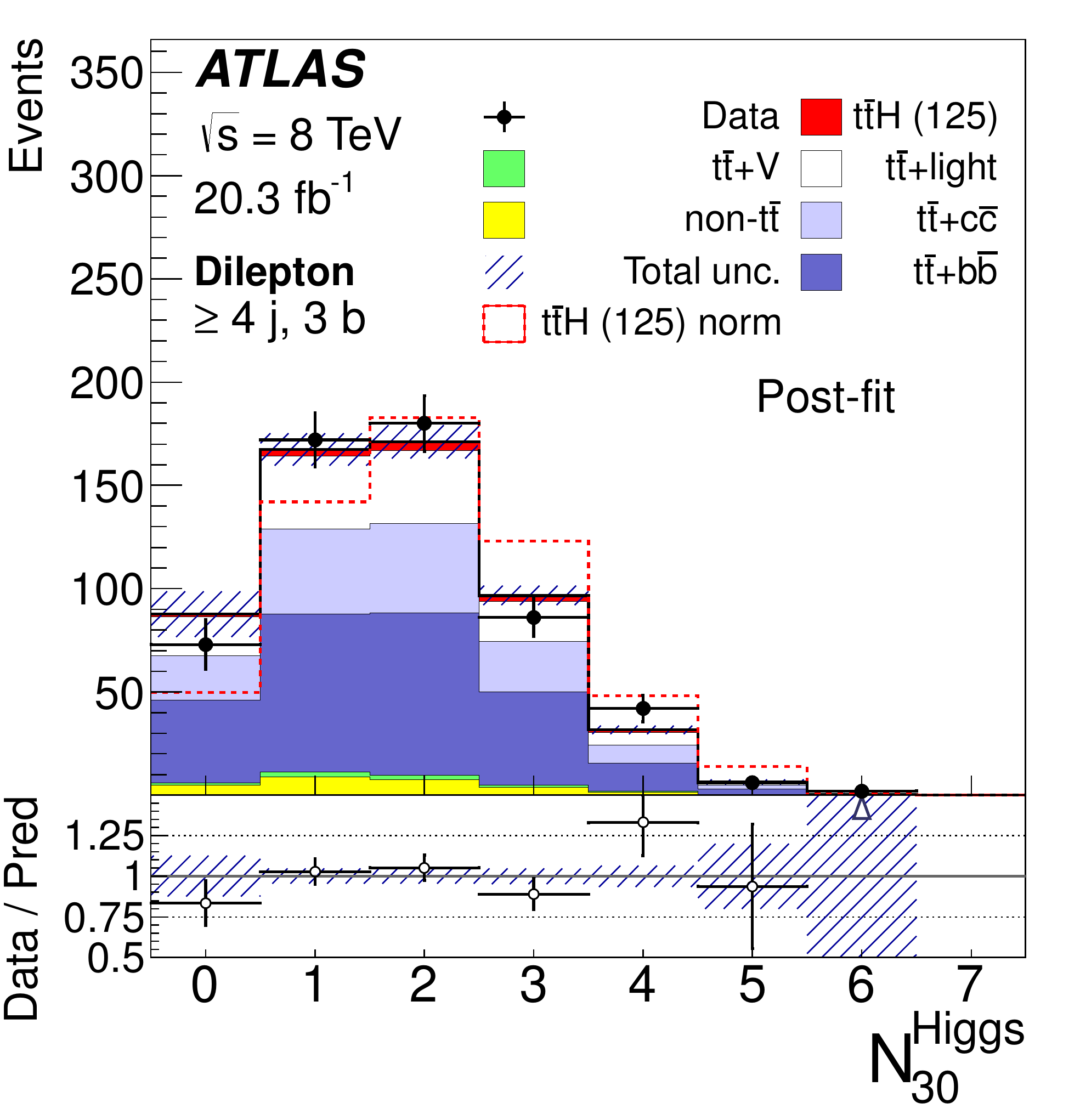}}\label{fig:postinput_dil_2b} 
\subfigure[]{\includegraphics[width=0.24\textwidth]{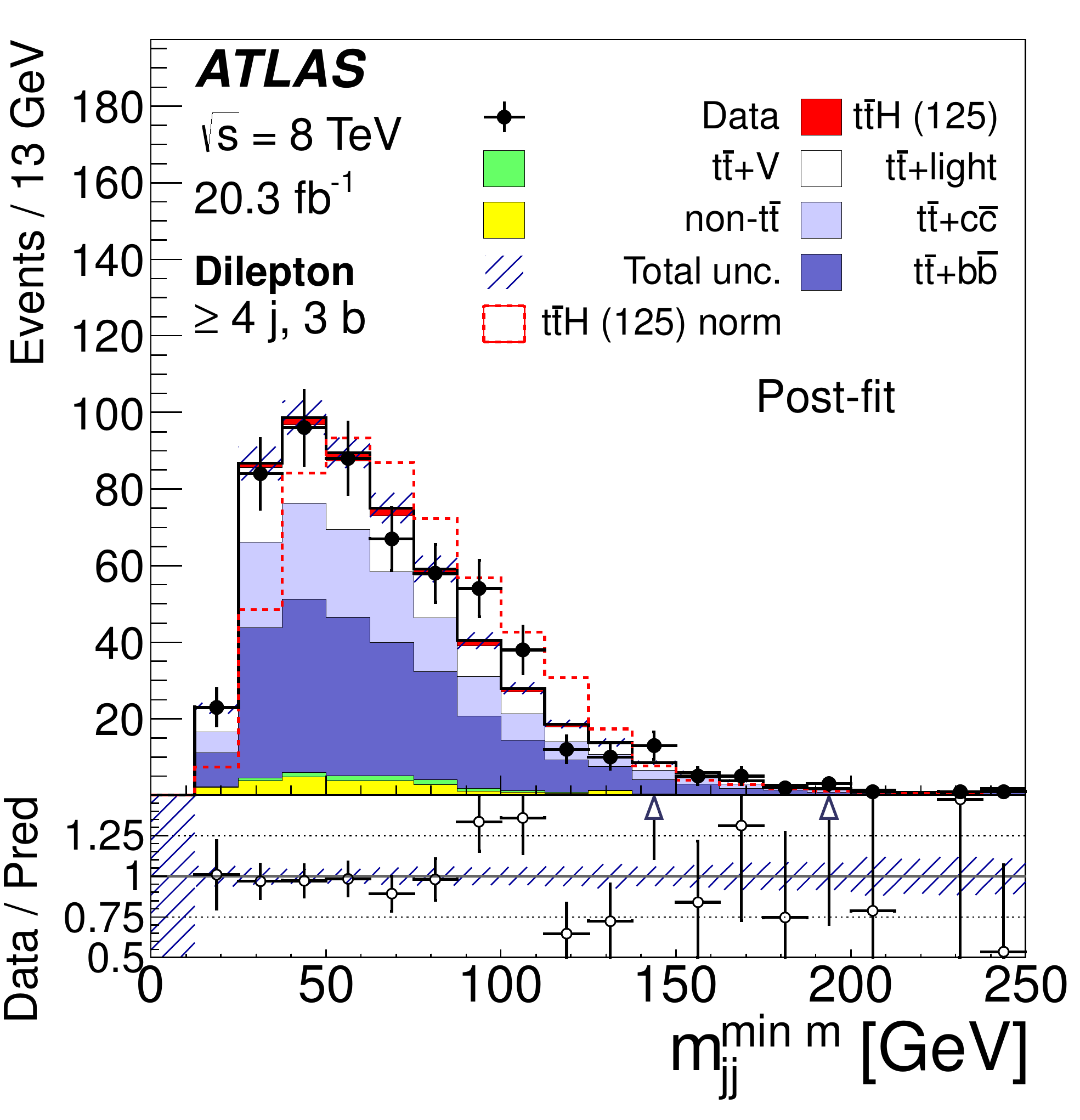}}\label{fig:postinput_dil_2c} 
\subfigure[]{\includegraphics[width=0.24\textwidth]{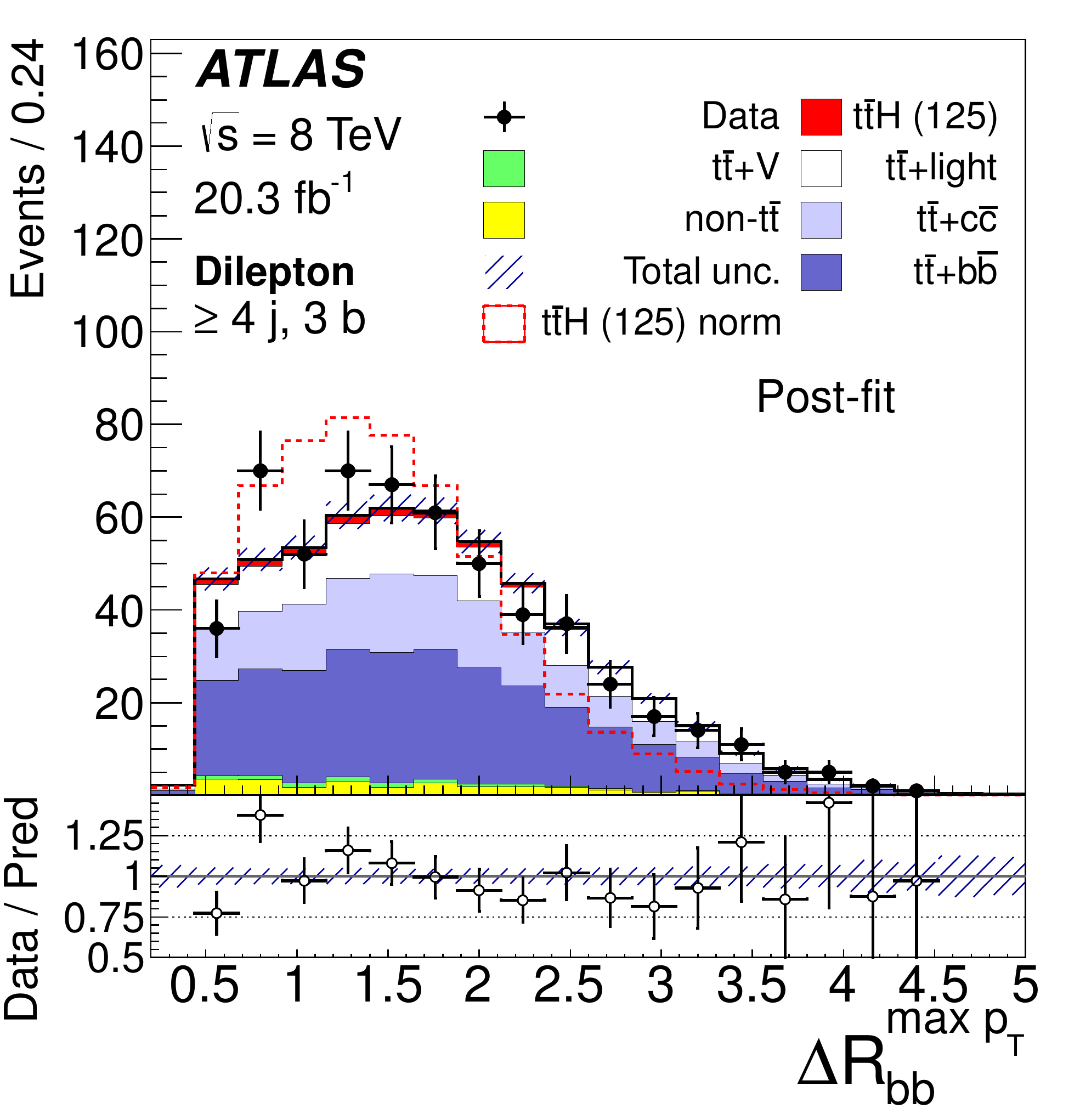}}\label{fig:postinput_dil_2d}
\caption{Dilepton channel: post-fit comparison of data and prediction for the four top-ranked input variables in the 
\fourthreedi\ region. The plots include (a) \maxdeta, (b) \nhiggsthirty, (c) \mindijetmass, and (d) \drbbmaxpt.
The first and last bins in all figures contain the underflow and
overflow, respectively. The bottom panel displays the ratio of 
data to the total prediction. An arrow indicates that the point is off-scale. The hashed area represents the uncertainty on the background.
The dashed line shows \tth\ signal
distribution normalised to background yield. The \tth\ signal yield (solid) 
is normalised to the fitted $\mu$.}
\label{fig:postinput_dil_2} 
\end{center}
\end{figure*}

\begin{figure*}[ht!]
\begin{center}
\subfigure[]{\includegraphics[width=0.24\textwidth]{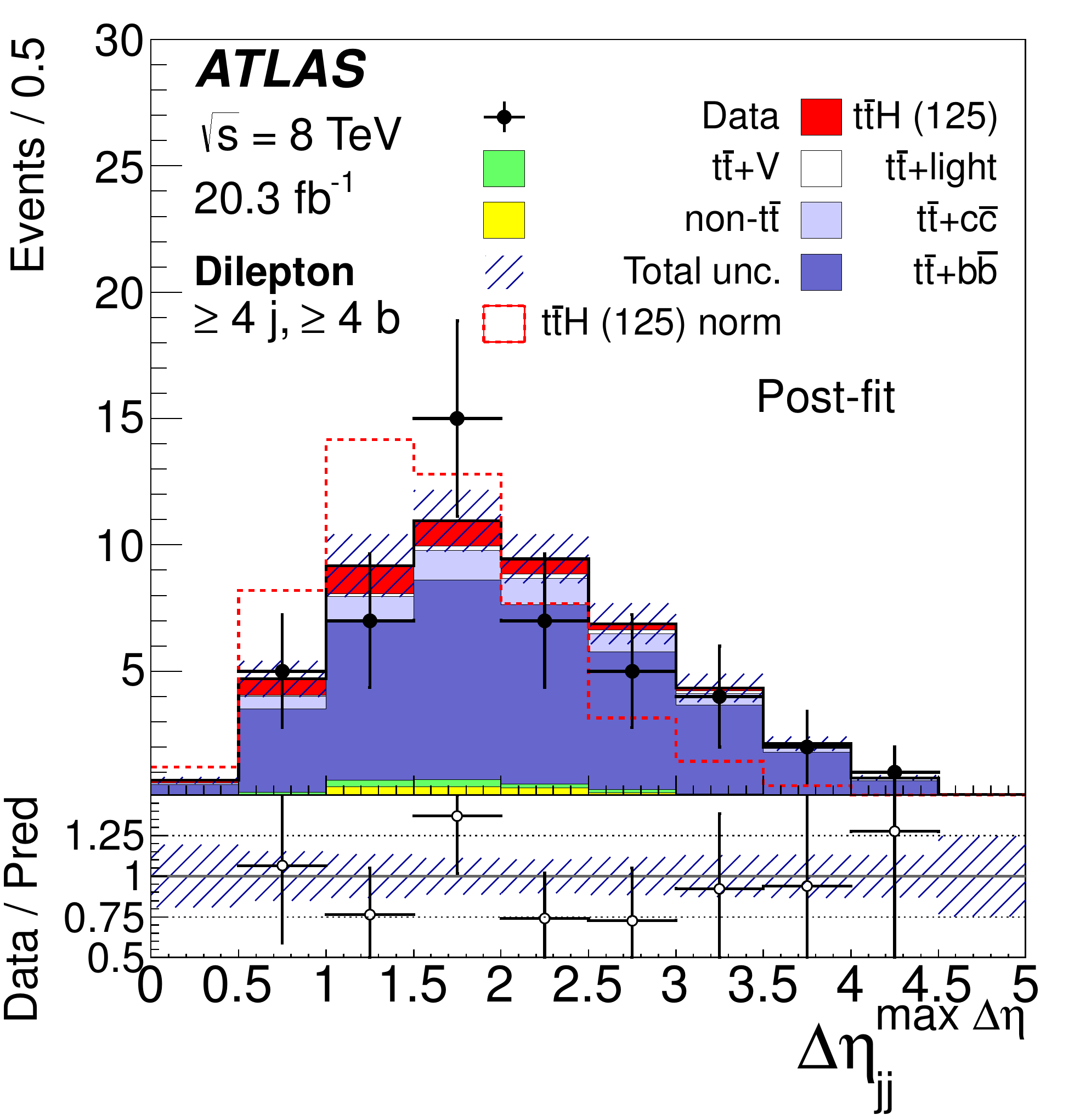}}\label{fig:postinput_dil_3a}
\subfigure[]{\includegraphics[width=0.24\textwidth]{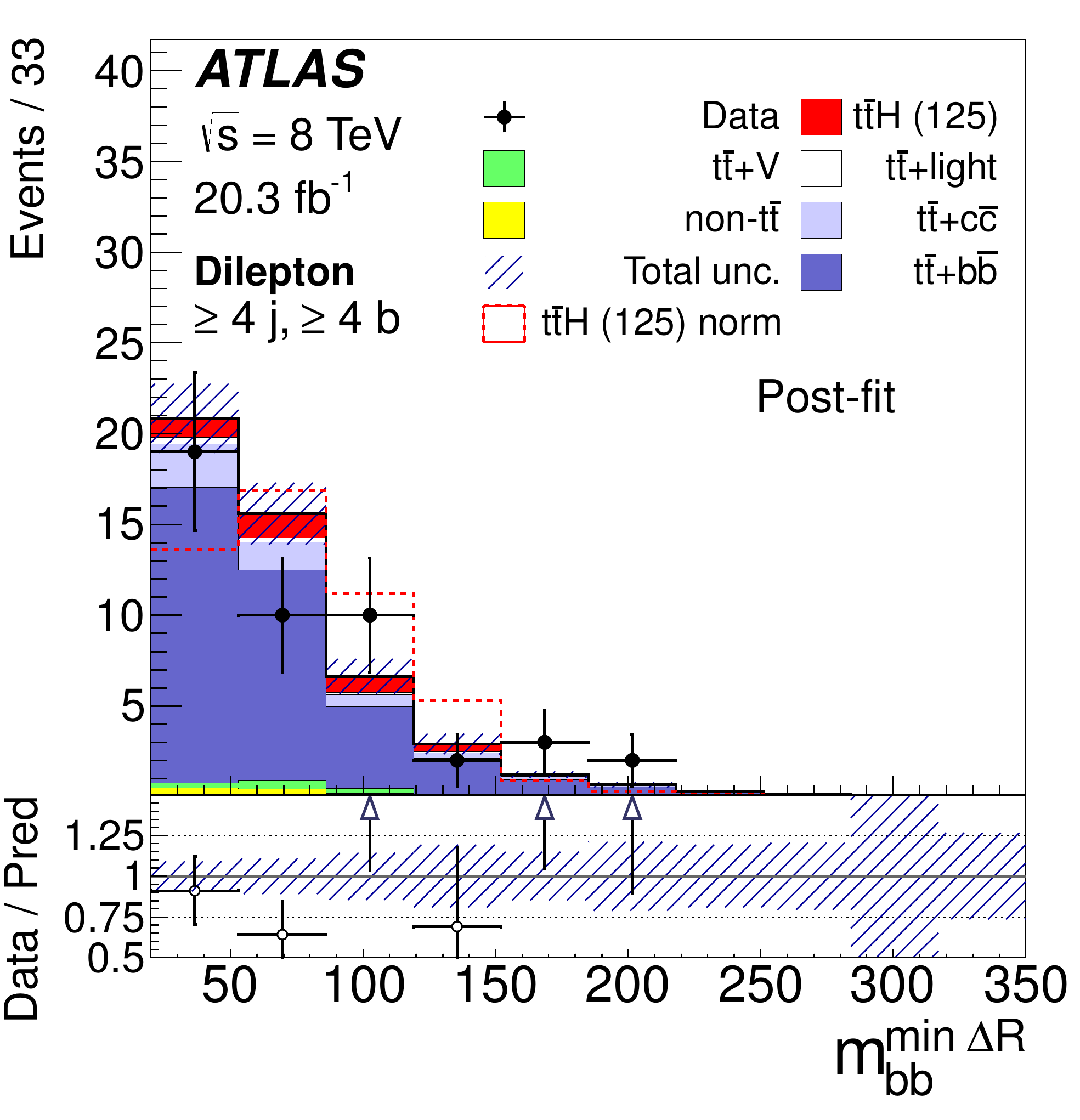}}\label{fig:postinput_dil_3b} 
\subfigure[]{\includegraphics[width=0.24\textwidth]{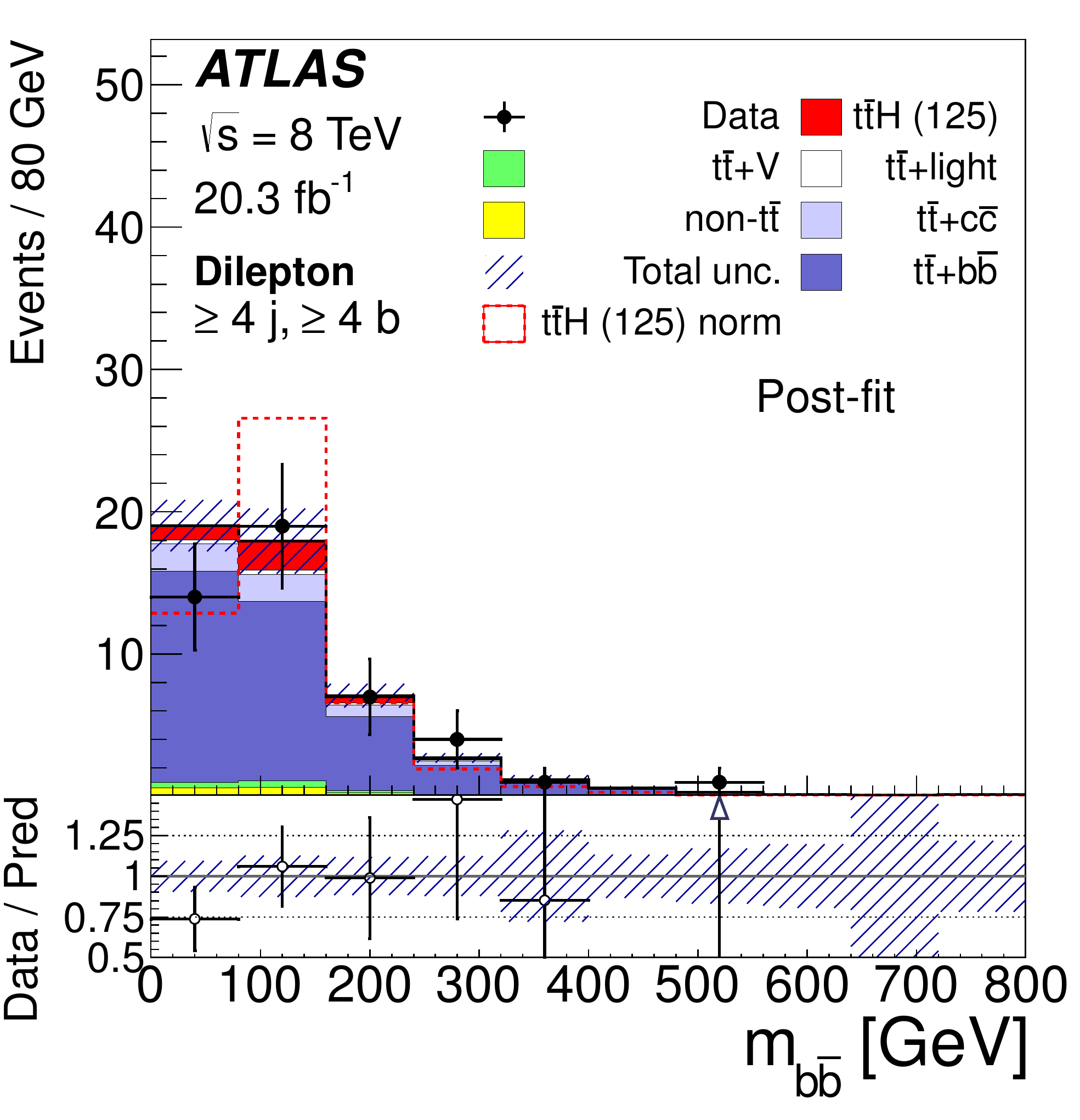}}\label{fig:postinput_dil_3c}
\subfigure[]{\includegraphics[width=0.24\textwidth]{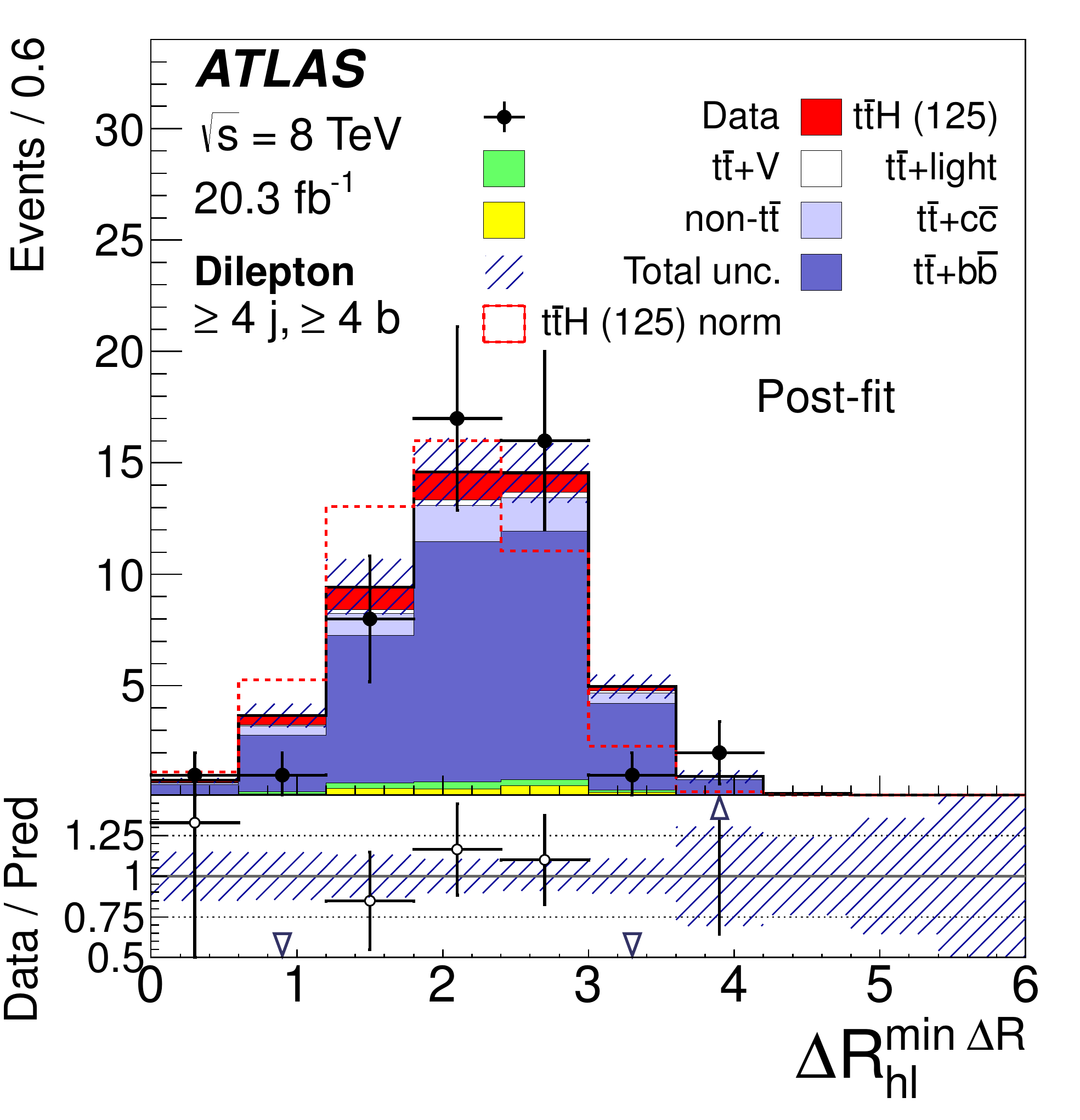}}\label{fig:postinput_dil_3d}
\caption{Dilepton channel: post-fit comparison of data and prediction for the four top-ranked input variables in the 
\fourfourdi\ region. The plots include (a) \maxdeta, (b) \mbbmindr, (c) \mbb, and (d) \mindrhl.
The first and last bins in all figures contain the underflow and 
overflow, respectively. The bottom panel displays the ratio of 
data to the total prediction. An arrow indicates that the point is off-scale. The hashed area represents the uncertainty on the background.
The dashed line shows \tth\ signal 
distribution normalised to background yield. The \tth\ signal yield (solid) 
is normalised to the fitted $\mu$.}
\label{fig:postinput_dil_3} 
\end{center}
\end{figure*}

\FloatBarrier
\onecolumn
% ATLAS Collaboration author list
% Data extracted on 27-Mar-2015 for paper reference HIGG-2013-27
\begin{flushleft}
{\Large The ATLAS Collaboration}

\bigskip

G.~Aad$^{\rm 85}$,
B.~Abbott$^{\rm 113}$,
J.~Abdallah$^{\rm 152}$,
O.~Abdinov$^{\rm 11}$,
R.~Aben$^{\rm 107}$,
M.~Abolins$^{\rm 90}$,
O.S.~AbouZeid$^{\rm 159}$,
H.~Abramowicz$^{\rm 154}$,
H.~Abreu$^{\rm 153}$,
R.~Abreu$^{\rm 30}$
% ...
\bigskip
\\
% ...
$^{11}$ Institute of Physics, Azerbaijan Academy of Sciences, Baku, Azerbaijan\\
% ...
$^{30}$ CERN, Geneva, Switzerland\\
% ...
$^{85}$ CPPM, Aix-Marseille Universit{\'e} and CNRS/IN2P3, Marseille, France\\
% ...
$^{90}$ Department of Physics and Astronomy, Michigan State University, East Lansing MI, United States of America\\
% ...
$^{107}$ Nikhef National Institute for Subatomic Physics and University of Amsterdam, Amsterdam, Netherlands\\
% ...
$^{113}$ Homer L. Dodge Department of Physics and Astronomy, University of Oklahoma, Norman OK, United States of America\\
% ...
$^{152}$ Institute of Physics, Academia Sinica, Taipei, Taiwan\\
$^{153}$ Department of Physics, Technion: Israel Institute of Technology, Haifa, Israel\\
$^{154}$ Raymond and Beverly Sackler School of Physics and Astronomy, Tel Aviv University, Tel Aviv, Israel\\
% ...
$^{159}$ Department of Physics, University of Toronto, Toronto ON, Canada\\
% ...
$^{*}$ Deceased
\end{flushleft}
% Created with ./xml2latex.py

\end{document}